\documentclass[12pt,a4paper]{report}       
\usepackage{multirow,booktabs,colortbl,tabularx}
\usepackage{makeidx}
\usepackage{latexsym}             
\usepackage{mathrsfs}                       
\usepackage{url}                                                                    
 \usepackage{enumitem}                                                                                   
 \usepackage[titletoc]{appendix}                                                                  
\usepackage{tikz}                                                  
\usetikzlibrary{shapes,arrows,positioning}  
\usetikzlibrary{calc, shapes, arrows, positioning}             
  
\tikzstyle{decision} = [diamond, draw, fill=blue!20, text width=3cm, text badly centered, node distance=3cm, inner sep=0pt]
\tikzstyle{block} = [rectangle, draw, fill=blue!20,text width=18em, text centered, rounded corners, minimum height=3em]
\tikzstyle{line} = [draw, -latex]                  
\usepackage{blkarray}
\usepackage{amsmath}                          
                              
\usepackage{amsmath,amssymb,graphicx,subfigure}

\usepackage{cite}  
\usepackage{relsize}  
              
\usepackage{setspace}    
\usepackage{amsmath, amssymb, amsthm} 
\usepackage{graphicx,color}           
\usepackage[left=3.5cm, right=2.5cm, top=2.5cm, bottom=2.5cm, includefoot, headheight=13.6pt]{geometry}
\usepackage[T1]{fontenc}                  
\usepackage[hang, small, bf, margin=20pt, tableposition=top]{caption}
\usepackage{mathpazo}
\usepackage{fancyhdr}
\usepackage{sectsty}
\usepackage{rotating}

 
\usepackage{sectsty}
\chapterfont{\large\sc\centering}
\chaptertitlefont{\centering}
\usepackage[breaklinks=true,colorlinks]{hyperref}
\usepackage[figure,table]{hypcap} 

\hypersetup{  
   bookmarksnumbered,
   pdfstartview={FitH},
   citecolor={blue},    
   linkcolor={red},     
   urlcolor={cyan},     
   pdfpagemode={UseOutlines}
}

\makeatletter    
\newcommand\org@hypertarget{}
\let\org@hypertarget\hypertarget
\renewcommand\hypertarget[2]{%
\Hy@raisedlink{\org@hypertarget{#1}{}}#2%
} \makeatother
\numberwithin{equation}{section}       

\usepackage{placeins} 

\usepackage{float} 

\begin{document}

\pagebreak
\title{ 
\LARGE{\textbf{Techniques of Model Reductions in\\
Biochemical Cell Signaling Pathways
}}\\[1cm]
\Large{A thesis submitted to the \\
Faculty of Science and Health at Koya University as a\\ partial fulfillment for the degree of Masters of Science\\ (M.Sc) in Mathematics 
} \\[.5cm]
}

\author{ \Large{By}\\ \Large{\textbf{Hemn Mohammed Rasool}}\\ \\
Having received my B.Sc. In Mathematics\\
Obtained in 2013\\
From Faculty of Science and Health/ Koya University \\
\\
Supervised by: \textbf{Dr. Sarbaz H. A. Khoshnaw} \\[.5cm]}
\maketitle 



\newpage
\begingroup
\pdfbookmark[0]{Contents}{contents}
\maketitle
\hypersetup{linkcolor=black}
    \tableofcontents
\endgroup

\newpage \vspace*{1cm}   
\pdfbookmark[0]{Acknowledgements}{acknowledgements} 
\begin{center}
\Large{\textbf{ACKNOLEDGMENT}}
\end{center}
I would like to express my special thanks of gratitude to my supervisor \textbf{( Dr. Sarbaz H. A. Khoshnaw )} who gave me the golden opportunity to do this wonderful work, which also helped me in doing a lot of research and I came to know about so many new things. I am really thankful to them.
Secondly, I wish to thank my loving and supportive wife, she helped me in all stages of my work and provided unending inspiration.
Finally, I want to say thank for my friends who helped me in my studying journey.

\newpage

\pdfbookmark[0]{Titlepage}{title} 
\maketitle
\newpage
\pdfbookmark[0]{Abstract}{Abstruse}  
\begin{center}
\large{Hemn M. Rasool}
\end{center}
    
\textbf{Abstract} 
\quad There are many mathematical models of biochemical cell signaling pathways that contain a large number of elements (species and reactions). This is sometimes a big issue for identifying critical model elements and describing the model dynamics. Thus, techniques of model reduction can be used as a mathematical tool in order to minimize the number of variables and parameters. 
 
In this thesis, we review some well known methods of model reduction for cell signaling pathways. We have also developed some approaches that provide us a great step forward in model reduction. The   
techniques are quasi-steady state approximation (QSSA), quasi-equilibrium approximation (QEA), lumping of species and entropy production analysis. They are applied on  protein translation pathways with microRNA mechanisms, chemical reaction networks, extracellular-signal-regulated kinase (ERK) pathways, NF-$\kappa$B signal transduction pathways, elongation factors EF--Tu and EF--Ts signaling pathways and Dihydrofolate reductase (DHFR) pathways.

The main aim in this thesis is to reduce the complex cell signaling pathway models. This provides one a better understanding of the dynamics of such models, and gives an accurate approximate solution. Results show that there are a good agreement between the original models and the simplified models. 

\newpage
\begingroup
\pdfbookmark[0]{List of Tables}{tables}
\maketitle
\hypersetup{linkcolor=black}
 \listoftables   
\endgroup

\newpage
\begingroup
\pdfbookmark[0]{List of Figures}{figures}
\maketitle  
\hypersetup{linkcolor=black}
    \listoffigures
\endgroup

  
\newpage   
\pagenumbering{arabic}           

\include{chapter[1]}
\chapter{Introduction} 

\section{Introduction}  
\label{general introduction}

Mathematical modeling is an important tool to describe the model dynamics for various cell signaling pathways. They can be expressed as a system of differential equations with some constant rates. While the majority of such systems are non-linear with high dimensional elements. They need some simplifications and model reductions in order to find analytical approximate solutions and describing the model behaviors. That is why in this thesis, we discuss and review the techniques of model reduction in cell signaling pathways and develop some approaches in model reduction with explanations of some new results. In addition, we use the function of deviation as a model comparison to measure the difference between the concentration species for the original and reduced models at each stage of model reduction. 

We start by giving a classic method of model reduction which is called quasi--steady state approximation(QSSA). We apply QSSA in protein translation pathways with microRNA mechanisms to classify model equations into slow and fast subsystems. Therefore, the slow manifolds and the analytical approximate solutions are calculated in different parameter values by using SBedit ToolBox in MATLAB. 

The second approach in this study is quasi--equilibrium approximation (QEA). We apply the method for simple and complex chemical reaction network models. We have also fast and slow subsystems of such models with slow manifolds and approximate solutions. More interestingly, we suggest an algorithm that contain some steps for identifying slow and fast reactions, and we write a MATLAB code to calculate the slow and fast reactions for the complex chemical reaction network. This has a great role in model reductions particularly for high dimensional systems.   

Another approach here is called lumping of compartments. This is a powerful tool to reduced the number of variables. We apply the suggested method on the linear and non--linear examples of chemical reactions, and we also apply on the complex network extra cellular-signal-regulated kinase (ERK) pathways. The number of variables is  reduced from 11 species to 8 species. Analytical approximate solutions of each species are computed for the full and reduced model using SBedit toolBox in MATLA. Interestingly, we propose a new technique to reduced the number of parameters depending on the lumping of parameters. Then, we use the suggested technique to reduce number of parameters of NF-$\kappa$B signal transduction pathways. The model includes 29 and 37 variables and parameters respectively. By using the developed technique, the number of parameters is reduced from 37 to 8. Results show that there are a good agreement between the approximate solutions of each variables in full and reduced model. 

Finally, there are another technique of model reduction that is based on eliminating the non--important reactions of complex biochemical reversible reactions. An algorithm based on the relative contribution of the entropy production of each reaction to the total entropy production was proposed in (Khoshnaw, 2015). Sometimes when we eliminate the non--important reactions some variables will be disappeared, and this is an issue for some models. Thus, we have developed this technique by linking entropy production with lumping of species. The idea is that whenever any species is disappeared we lump with one of their neighbors. In addition, we apply this developed technique on the kinetic model of elongation factors EF--Tu and EF--Ts signaling pathways to identify the non-important reactions and lumping isolated species. Results show that our developed method is more accurate compared to the previous algorithm in (Khoshnaw, 2015). Finally, we also apply the technique on the Dihydrofolate reductase (DHFR) pathways, and the number of variables and parameters are reduced from 13 and 26 to 11 and 16, respectively. The function of deviation value in all stages  is computed and the approximate solutions are given for the original and reduced model of each variables by using MATLAB. MATLAB codes of calculating approximate solutions are included in Appendix.


\section{Chemical Kinetics}   
\label{A22}      
In this section, there is a converting process of a biological system to mathematical modeling. The theory of chemical kinetics based on mass action law is used to express biological processes for mathematical modelling. The model mainly includes the following terms:  
\begin{description}[noitemsep,nolistsep]          
\item[$\bullet$] A vector of components (species)   
 $\mathcal{A}=(\mathcal{A}_{1} , \mathcal{A}_{2}, ... , \mathcal{A}_{m})$,   
 for each component $\mathcal{A}_{j}$, $j=1,2,...,m$ a non negative variable $c_{j}$ (concentration of $\mathcal{A}_{j}$, i.e.  $c_{j}=[\mathcal{A}_{j}]$) is defined; the vector of concentrations is $\mathcal{C}$. In other words, $c_{j}=\mathcal{N}_{j}/V$, where $V$ is volume, $\mathcal{N}_{j}$ is a non--negative real extensive variable (the number of molecules of that species).    
\item[$\bullet$] A vector of reactions
 $\mathcal{V}=(v_{1}, v_{2}, ... , v_{n}).$    
\item[$\bullet$] A vector of kinetic constants   
$\mathcal{K}=(k_{1}^{\mp}, k_{2}^{\mp}, ... , k_{n}^{\mp})$.  The kinetic constants depend on reaction conditions (e.g. temperature, ph, solvent, etc.)          
\end{description}
\noindent For the following $n$ reversible reactions which are represented by its stoichiometric equations
\begin{equation} 
\begin{array}{llll}  
 \mathlarger{\mathlarger{\sum\limits}}_{j=1}^m \alpha_{ij} \mathcal{A}_{j} \underset{k_{i}^{-}}{ \overset{k_{i}^{+}}{\rightleftharpoons}}\mathlarger{\mathlarger{\sum\limits}}_{j=1}^m \beta_{ij} \mathcal{A}_{j}, \quad i=1,2,...,n .
\end{array}
\label{elementry}
 \end{equation}  
The non-negative integers $\alpha_{ij}$ and $\beta_{ij}$ are called stoichiometric coefficients. To define the rate of reactions, we use the standard mass action law. The reaction rates are given below:
\begin{equation}
\begin{array}{llll}  
 v_{i} = k_{i}^{+} \mathlarger{\mathlarger{\prod\limits}}_{j=1}^m c_{j}^{\alpha_{ij}}(t)-k_{i}^{-} \mathlarger{\mathlarger{\prod\limits}}_{j=1}^m c_{j}^{\beta_{ij}}(t), \quad i=1,2,...,n ,
\end{array}
\label{rates}
 \end{equation}
where $k_{i}^{+} > 0$ and $k_{i}^{-} \geq 0$ are the reaction rate coefficients (Gorban et al., 2010, Hannemann-Tamas et al., 2013, Radulescu et al., 2008). \\         
The stoichiometric matrix is $\mathcal{G}=(\gamma_{ij}),$ where $\gamma_{ij}=\beta_{ij}-\alpha_{ij},$ for $ i=1,2,...,n$ and   $j=1,2,..,m$. The stoichiometric vector $\gamma_{i}$ is the \textit{i}th row of $\mathcal{G}$ with coordinates $\gamma_{ij}=\beta_{ij}-\alpha_{ij}$ (Yablonskii et al., 1991).
\noindent The differential equations can be used to describe the dynamics of chemical reactions. Such equations are given:
\begin{equation}
\begin{array}{llll}  
\dfrac{d\mathcal{C}}{dt}=\mathcal{W}(\mathcal{C}(t), \mathcal{K})=\mathcal{G} \enskip \mathcal{V}(\mathcal{C}(t), \mathcal{K}), \\
\mathcal{C}(0)=\mathcal{C}_{0}, \quad  t \in I\subset \mathbb{R}^{+}\cup \lbrace 0\rbrace, 
\end{array}
\label{eq1} 
 \end{equation}  
\noindent where $\mathcal{G}$ is a stoichiometric matrix of $m$ by $n$, $\mathcal{C}(0)$ is a vector of initial concentrations. The kinetic equations (\ref{eq1}) can be also expressed as follows:
\begin{equation}
\begin{array}{llll}  
\dfrac{d\mathcal{C}}{dt}=\mathlarger{\mathlarger{\sum\limits}}_{i}\gamma_{i}v_{i} .
\end{array}
\label{eq1a} 
 \end{equation} 
  
Let us give an example for stoichiometric vectors. For the SIR epidemic  disease model, this model is the most well known mathematical model for the spread of an infectious disease. The model was published in 1927 for the first time by Kermac and Mckendrick. The model may diagrammed as below. 
 
\begin{equation}  
\begin{array}{llll}
S + I{ \overset{\beta}{\rightarrow}} 2I,\\
I{ \overset{\gamma}{\rightarrow}} R,
\end{array}\label{SIR model diagram}
\end{equation} 
where $S, I$ and $R$ are susceptible, infective and removed people respectively. The parameters $ \beta$ and $\gamma$ are kinetic constants. Then, the stoichiometric vectors are given:

\begin{equation*} 
\gamma_{1}
={\left(\begin{array}{cccc}
-1\\ 1\\ 0
\end{array}\right)}, \gamma_{2}
={\left(\begin{array}{cccc}
0\\ -1\\ 1
\end{array}\right)}
\end{equation*}

and the reaction rates are $v_{1}=\beta SI$ and $v_{2}=\gamma I$. Using mass action law, we can define the system of differential equations

\begin{equation*} 
\dfrac{d}{dt}{\left(\begin{array}{cccc}
S\\I\\R
\end{array}\right)}
=\mathlarger{\mathlarger{\sum\limits}}_{i=1}^{3} \gamma_{i} v_{i}
=\gamma_{1} v_{1} + \gamma_{2} v_{2} 
\end{equation*} 

\begin{equation*}
\hspace{3cm}={\left(\begin{array}{cccc}
-1\\ 1\\ 0
\end{array}\right)} \beta SI + {\left(\begin{array}{cccc}
0\\ -1\\ 1
\end{array}\right)} \gamma I  
\end{equation*}

\begin{equation*}
\hspace{1cm}={\left(\begin{array}{cccc}
-\beta S I\\ \\ \beta S I - \gamma I \\ \gamma I
\end{array}\right)}.  
\end{equation*}

Therefore,

\begin{equation}
\begin{array}{llll}  
\dfrac{dS}{dt}=-\beta S I,\\
\dfrac{dI}{dt}=\beta S I - \gamma I,\\
\dfrac{dR}{dt}=\gamma I.
\end{array}
\label{ODE SIR}
\end{equation}
The model initial populations are $S(0)>0$, $I(0)>0$ and $R(0)\geq 0.$

Stoichiometric conservation laws are identified for system of differential equations that provide us an essential tool for reducing  chemical kinetic equations. The equation of stoichiometric conservation law is given:
\begin{equation}
\begin{array}{llll}  
\mathlarger{\mathlarger{\sum\limits}}_{j}b_{j}c_{j}=B,
\end{array}
\label{eq1ab} 
 \end{equation}
 where $B$ is a constant. We can obtain the equation (\ref{eq1ab}) from:\\ 
 \quad \quad \quad \quad $\dfrac{d}{dt}\mathlarger{\mathlarger{\sum\limits}}_{j}b_{j} c_{j}=\mathlarger{\mathlarger{\sum\limits}}_{j}b_{j}\bigg(\mathlarger{\mathlarger{\sum\limits}}_{i}\gamma_{ij}v_{i} \bigg)=\mathlarger{\mathlarger{\sum\limits}}_{i}v_{i}\bigg(\mathlarger{\mathlarger{\sum\limits}}_{j}b_{j}\gamma_{ij} \bigg)=0$, \\ if $\mathlarger{\mathlarger{\sum\limits}}_{j}b_{j}\gamma_{ij}=0$ for all reactions $i$, where $b_{j}$ for $j=1,2,...,m$ are coefficients.\\
We can notice that for each stoichiometric conservation law $B$ the coefficients $b_{i}$ for $j=1,2,...,m$ satisfy the system of linear equations $\mathlarger{\mathlarger{\sum\limits}}_{j}b_{j}\gamma_{ij}=0$ (for all $i$). In addition, the basis of the system is very useful to identify all stoichiometric conservation laws.    

\noindent For instance, if we have a particular component $\mathcal{A^{*}}$ in a model then the differential equation for the component is given:

\quad \quad \quad $\dfrac{dc_{\mathcal{A^{*}}}}{dt}=\mathlarger{\mathlarger{\sum}} \beta_{\ldots} v_{\mathcal{A^{*}},produced}-\mathlarger{\mathlarger{\sum}} \alpha_{\ldots}v_{\mathcal{A^{*}}, consumed}. $\\
\noindent This means ( rate of change of $c_{\mathcal{A^{*}}}$) = (concentration of $\mathcal{A^{*}}$ formed in all reactions)-(concentration of $\mathcal{A^{*}}$ consumed in all reactions), where $v_{\mathcal{A^{*}}}$ is the rate of formation/consumption of species $\mathcal{A^{*}}$ in a particular reaction (Khoshnaw, 2015, Singh et al., 2006).

A function of species concentrations $\mathcal{W}(\mathcal{C}(t))$ for a model is satisfied a Lipschitz condition in the concentration $\mathcal{C}(t)$ on a set $D\subset \mathbb{R}^{m}$ if a constant $L>0$ exists with $\big\vert\mathcal{W}(\mathcal{C}_{1}(t)) - \mathcal{W}(\mathcal{C}_{2}(t))\big\vert\leq L \big\vert \mathcal{C}_{1}(t) - \mathcal{C}_{2}(t)\big\vert$, for all $\mathcal{C}_{1}(t), \mathcal{C}_{2}(t)\in D$. The constant $L$ is called a Lipschitz constant for $\mathcal{W}$.


\section{Methods of Model Reduction}
There are some methods of model reduction that used for reducing biochemical reaction network elements. The techniques of model reduction here are important tools in systems
biology in order to provide our understanding of dynamics of such models and minimize the number of elements. We review some well known methods of model reduction. They are simply described in the following sections.
\subsection{Quasi-Steady State Approximation (QSSA) and Tikhonov's Theorems}
\label{A25}
In this section, we review a basic concepts of quasi-steady state approximation (QSSA) and Tikhonov's Theorems. Over the last century, the idea of the quasi--steady state was reviewed many times. In 1913 the first idea of the classical quasi--steady state approximation was suggested by Bodeustein (Bodenstein, 1913). And the extra detail of the method was then given by Briggs and Haldane in 1925 for the simplest enzyme reaction $E+S \underset{}{ \overset{}{\rightleftharpoons}} ES \overset{}{\longrightarrow}E+P$ (Briggs and Haldane, 1925). They suggested that the total concentration of enzyme ($[E]+[ES]$) is "negligibly small" in comparison with the concentration of substrate $[S]$. This condition let one to produce the well-known "Michaelis--Menten" formula (L. Michaelis, 1913). Later that, the method was more developed and improved as an significant tool to analyse the dynamics of chemical reaction mechanisms and kinetics (Christiansen, 1953, Helfferich, 1989, N. N. Semenov, 1939). The basic idea of the QSSA is generally based on the "relative smallness" of concentrations of some of the "active regents" (radicals, substrate--enzyme complexes or active components on the catalyst surface) (Aris, 1965, Briggs and Haldane, 1925, Segel and Slemrod, 1989). To define the basic equations of the method, we consider that a set of variables $\mathcal{C}(t)$ for a kinetic model can be divided into two groups: the first one is called slow variables (basics) $\mathcal{C}^{s}(t)$, and the other one is called fast variables (fast intermediate) $\mathcal{C}^{f}(t)$ (Kutumova et al., 2013). In this technique, the concentrations of slow species are supposed to be larger than the concentrations of fast species. The reaction rates of both variables are the same order, or they may occur in the same reactions. By introducing a new variable $\tilde{\mathcal{C}}^{f}(t)=\dfrac{1}{\epsilon} \mathcal{C}^{f} (t)$, where $\epsilon$ is a small parameter ($\epsilon\ll 1$), the kinetic equation (\ref{eq1}) can be split into two subsystems:
\begin{subequations} 
\begin{align}  
 \dfrac{d\mathcal{C}^{s}}{dt} =\mathcal{W}^{s} \big(\mathcal{C}^{s}(t),  \tilde{\mathcal{C}}^{f}(t), \mathcal{K}\big), \label{qssa2a} \\
 \dfrac{d\tilde{\mathcal{C}}^{f}}{dt}=\dfrac{1}{\epsilon} \mathcal{W}^{f} \big(\mathcal{C}^{s}(t), \tilde{\mathcal{C}}^{f}(t), \mathcal{K}\big). \label{qssa2b}  
\end{align} \label{qssa2} 
 \end{subequations}  
\noindent The first equation (\ref{qssa2a}) is called the slow subsystem and the other one (\ref{qssa2b}) is called the fast subsystem. The fast subsystem can be analysed and the standard singular perturbation techniques based on the Tikhonov theorem (Tikhonov, 1952, Vasil'eva, 1963) can be applied. If we have a stable dynamic of fast variables under given values of slow concentrations then the slow manifold exists. The attractive slow manifold is calculated from the algebraic equations\\ $\mathcal{W}^{f}\big(\mathcal{C}^{s}(t), \tilde{\mathcal{C}}^{f}(t), \mathcal{K}\big)=0$ when $\epsilon \longrightarrow 0$. Thus, the new system includes a smaller number of variables (species concentration) and parameters (chemical reaction constants). More explanations and applied examples of the method can be seen by the reader in (Battelli and Lazzari, 1986, Briggs and Haldane, 1925, Ciliberto et al., 2007, Conzelmann et al., 2004, Goeke et al., 2012, Gorban and Shahzad, 2011, Hanson and Schnell, 2008, Kijima and Kijima, 1983, Klonowski, 1983, Li and Li, 2013, Li et al., 2008, Pedersen et al., 2008a, Pedersen et al., 2008b, Petrov et al., 2007, Schneider and Wilhelm, 2000, Schnell, 2014, Schnell and Maini, 2000, Tzafriri and Edelman, 2004).\\

Classical singular perturbation techniques are based on a separation of variables
into fast and slow on the chosen time scale. One of the most important problems in the asymptotic methods of non--linear chemical kinetics is the problem of separation of variables. This guides us to a system of differential equations with a small parameter $\epsilon$ of the form
\begin{equation}
\begin{array}{llll}  
\dfrac{dx}{dt}=f(x, y,\epsilon),\\
\epsilon\dfrac{dy}{dt}=g(x, y, \epsilon),
\end{array}
\label{Tikh}
\end{equation} 
where $x$ is a slow variable and $y$ is a fast variable on the
$t$ time scale. One of the well--known results of asymptotic analysis for differential equations is Tikhonov's theorem. The result provide conditions on the function $g$ under which $y$ can be eliminated on the slow time--scale. It provides how well the reduced model obtained by eliminating  $y$ approximates the dynamics of the full system (Tikhonov, 1952). Tikhonov was a Soviet and Russian mathematician. He worked in a number of various fields in mathematics, and made an important contribution to topology, functional analysis, mathematical physics, and certain classes of ill--posed problems. Furthermore , he established the theory of asymptotic analysis for differential equations with small parameter in the leading derivative. Tikhonov introduced two theorems to give more information and explain more details about the systems of first--order ordinary differential equations containing small parameters in the derivatives. The first theorem contains a small parameter in some derivatives although the second theorem includes a number of parameters in some derivatives. Both theorems are defined as follows:  

\noindent \textbf{Theorem 1} (Tikhonov's first theorem (Klonowski, 1983, Tikhonov, 1952))\\
\noindent Consider a system of first--order ordinary differential equations with one small parameter $\epsilon$,
\begin{subequations}  
\begin{align}
\dfrac{dx}{dt}=f(x,z,t),\\
 \dfrac{dz}{dt}=\dfrac{1}{\epsilon} g(x,z,t),
\end{align}\label{eq2.4b}
\end{subequations}
with initial conditions 
 \begin{equation}  
\begin{array}{llll}
x(t_{0})=x_{0},\quad  z(t_{0})=z_{0}, 
\end{array}\label{eq2.5}
\end{equation}
where $x,f\in \mathbb{R}^{n}$ and $z,g \in \mathbb{R}^{s}$, $n,s\geq 1$. Putting $\epsilon\rightarrow0$ in equations (\ref{eq2.4b}), the degenerate system is obtained
 \begin{subequations}  
\begin{align}
\dfrac{dx}{dt}=f(x,z,t),\\
z=\phi(x,t),
\end{align}\label{eq2.6b}
\end{subequations}
where $z=\phi(x,t)$ is a root of the system of algebraic equations $g(x,z,t)=0$ or $g_{i}(x,z,t)=0$ for $i=1, 2, ... , s.$ The system of equations 
\begin{equation}  
\begin{array}{llll}
\dfrac{dz}{d\tau}=g(x,z,t), \quad  z(t_{0})=z_{0}
\end{array}\label{eq2.8}
\end{equation} 
is called the adjoined system (fast system) where $x$ is a parameter, and $\tau=\frac{t}{\epsilon}$. All functions we use in these theorems are continuous, and the differential equations have unique solutions. \\
\noindent The solution of equations (\ref{eq2.4b}) approaches the solution of the degenerate system (\ref{eq2.6b}) if the following conditions are satisfied:  
\begin{description}[noitemsep,nolistsep]
\item[$\bullet$] The point $ z=\phi(x,t)$ is the stable root of the adjoined system.
\item[$\bullet$] The initial values $ z_{0}$ exist in the domain of effect of the root $ z=\phi(x,t)$ with initial values $ (x_{0}, t_{0})$.
\end{description}
\noindent This theorem is also reasonable if the system of equations (\ref{eq2.4b}) depends continuously on the parameter $\epsilon$ as follows: 
\begin{subequations}  
\begin{align}
\dfrac{dx}{dt}=f(x,z,t,\epsilon),\\
\dfrac{dz}{dt}=g(x,z,t,\epsilon).
\end{align}\label{eq2.9b}
\end{subequations}
\noindent The solution of the original system (\ref{eq2.4b}) can be approximated by the solution of the degenerate system (\ref{eq2.6b}), for $ t \gg t_{d}$ where $ t_{d}=\vert \epsilon \ln\epsilon\vert$, this can be seen in (Klonowski, 1983).\\

\noindent \textbf{Theorem 2} (Tikhonov's second theorem (Klonowski, 1983, Tikhonov, 1952))\\
\noindent Consider a system of first--order ordinary differential equations with several small parameters $ \epsilon^{j}$ for $j=1, 2, ... , m$ as follows:
\begin{subequations}  
\begin{align}
\dfrac{dx}{dt}=f(x,z^{1},z^{2}, .. , z^{m},t),\\
 \dfrac{dz^{j}}{dt}=\dfrac{1}{\epsilon^{j}} g^{j}(x,z^{1},z^{2}, .. , z^{m},t), 
\end{align}\label{eq2.10b}
\end{subequations}
with initial conditions 
\begin{equation}  
\begin{array}{llll}
x(t_{0})=x_{0}, \quad z^{j}(t_{0})=z_{0}^{j},
\end{array}\label{eq2.11}
\end{equation}
where $x, f\in \mathbb{R}^{n}$ and $z^{j},g^{j} \in \mathbb{R}^{s}$, $n,s\geq 1$ for $j=1,2,...,m$. The solution of equations (\ref{eq2.10b}) with initial conditions (\ref{eq2.11}) can be given when $\epsilon^{j} \longrightarrow 0$ for $j=1,2,..,m.$ \\
\noindent Firstly, if $\epsilon^{m}\longrightarrow 0$ and putting in equations (\ref{eq2.10b}) then the singly degenerate system of first order is given 
 \begin{subequations}  
\begin{align}
\dfrac{dx}{dt}=f(x,z^{1},z^{2}, .. , z^{m},t),\\
\dfrac{dz^{j}}{dt}=\dfrac{1}{\epsilon^{j}} g^{j}(x,z^{1},z^{2}, .. , z^{m},t), \\
z^{m}=\phi ^{m}(x,z^{1},z^{2}, .. , z^{m-1},t), 
\end{align}\label{eq2.12c}
\end{subequations}
with initial conditions  
\begin{equation}  
\begin{array}{llll}
x(t_{0})=x_{0}, \quad  z^{j}(t_{0})=z_{0}^{j},\quad j=1,2,..,m-1,
\end{array}\label{eq2.12d}
\end{equation}
where $z^{m}=\phi ^{m}(x,z^{1},z^{2}, .. , z^{m-1},t)$ is a root of the system of algebraic equations $g^{m}(x,z^{1},z^{2}, .. , z^{m},t)=0$ or $g^{m}_{k}(x,z^{1},z^{2}, ... , z^{m},t)=0, k=1,2,.., s.$\\
\noindent According to the fast subsystem (\ref{qssa2b}), the adjoined system (fast system) of first order for time scale $\tau=\frac{t}{\epsilon^{m}}$ becomes
\begin{equation}   
\begin{array}{llll}
\dfrac{dz^{m}}{d\tau}=g^{m} (x,z^{1},z^{2},...,z^{m},t),\quad z^{m}(t_{0})=z^{m}_{0} .
\end{array}\label{eq2.12d}
\end{equation}
Note that in equation (\ref{eq2.12d}) $x,z^{1},z^{2},..., z^{m-1}$ are parameters, and $\tau=\frac{t}{\epsilon^{m}}$.

In mathematics, a degenerate case is ``a limiting case in which an element of a class of objects is qualitatively different from the rest of the class and hence belongs to another, usually simpler, class''. Particularly, we started out with $m+1$ differential equations (\ref{eq2.10b}), and ended up saying that one of the differential equations degenerates to an algebraic equation (\ref{eq2.12c}) when $\epsilon^{m}\rightarrow 0$. This is called the singly degenerate system of first order. A system of $m+1$ differential equations is quite different from a system of $m$
differential equations with an algebraic equation. Consequently, by putting $\epsilon^{m-1}\rightarrow0$ in the system (\ref{eq2.12c}), we obtain a system of $m-1$ differential equations with two algebraic constraints as follows:
\begin{subequations}  
\begin{align}
\dfrac{dx}{dt}=f(x,z^{1},z^{2}, .. , z^{m},t),\\
\dfrac{dz^{j}}{dt}=\dfrac{1}{\epsilon^{j}} g^{j}(x,z^{1},z^{2}, .. , z^{m},t), \\
z^{m-1}=\phi ^{m-1}(x,z^{1},z^{2}, .. , z^{m-2},t), \\
z^{m}=\phi ^{m}(x,z^{1},z^{2}, .. , z^{m-1},t), \quad j=1,2,..,m-1. 
\end{align}\label{eq2.12k}
\end{subequations} 
This is called a doubly degenerate system of first order. Similarly, a degenerate system of $k$th order can be defined with all other concepts of the $k$th order. The solution of the original system (\ref{eq2.10b}) with initial conditions (\ref{eq2.11}) approaches the solution of the degenerate ($m$ times degenerate) system if the following conditions are satisfied: 
 \begin{description}[noitemsep,nolistsep]
\item[$\bullet$] For any $j$ $(1 \leq j \leq m)$, the roots $z^{j}=\phi^{j}$ are stable roots of the adjoined equations.
\item[$\bullet$] The initial values $z^{j}_{0}$ lie in the domain of influence of the roots $z^{j}=\phi^{j}$ for initial values $(x_{0},z_{0}^{1},z_{0}^{2},...,z_{0}^{j-1},t_{o}).$
\end{description}  
For more detail see (Khoshnaw, 2015).


\subsection{Quasi-Equilibrium Approximation} 

\noindent

The history of the quasi-equilibrium approximation method is not quite clear, and it is not easy to find who proposed the method. There is a point which allows one to identify the time that the method was introduced. It can be detected that the method was not suggested before the studies of Boltzmann and Gibbs. Then, the method became very popular after the studies of Janes. The first explanation of the approach mathematically was given in (Vasil'ev et al., 1973, Vol, 1985). After that, further description and understanding of the method as a technique of model reduction has been studied in (Lee and Othmer, 2010, Noel et al., 2012). The idea of QEA was used as a model reduction technique to minimize the dimension of such systems. According to this approach, the fast reactions simply go to their equilibrium and then remain almost unchanged all the time. The validity of the approximation on quantitative terms was discussed in (Volk et al., 1977). In (Kijima and Kijima, 1983), a general procedure was proposed to simplify a complex first order chemical reaction using the principle of fast equilibrium and the steady state approximation. Recently, the method was further explained and developed by Gorban and Karlin to define invariant manifolds for physical and chemical kinetics (Gorban and Karlin, 2003). According to their study, there are two ways to construct the QE. The first one is quasi--equilibrium with respect to reactions and the other one is quasi--equilibrium with respect to species. The general formulation of QEA is based on an assumption that a set of reactions in a model is much faster than another set of reactions. The fast reactions will reach equilibrium very quickly. The initial system is given: 
\begin{equation}
\begin{array}{llll}  
\dfrac{d\mathcal{C}}{dt}=\mathlarger{\mathlarger{\sum\limits}}_{s,slow}  \mathcal{R}^{s}(\mathcal{C},\mathcal{K},t)\gamma^{s}+\dfrac{1}{\epsilon} \mathlarger{\mathlarger{\sum\limits}}_{f,fast}  \mathcal{R}^{f}(\mathcal{C},\mathcal{K},t)\gamma^{f},
\end{array}
\label{general QEA 1}  
\end{equation}
where $\epsilon$ is a small parameter ( $ 0<\epsilon\ll 1$), $\mathcal{R}^{s}$ and $\mathcal{R}^{f}$ are the function of reaction rates, $\gamma^{s}$ and $\gamma^{f}$ are stoichiometric vectors. The fast subsystem becomes
 \begin{equation}
\begin{array}{llll}  
\dfrac{d\mathcal{C}}{dt}=\dfrac{1}{\epsilon} \mathlarger{\mathlarger{\sum\limits}}_{f,fast} \mathcal{R}^{f}(\mathcal{C},\mathcal{K},t)\gamma^{f}.
\end{array}
\label{general fast part QEA 1}
\end{equation}
This approach of model reduction here is to separate variables. To do that we have to study the spaces of linear conservation laws of the original system (\ref{general QEA 1}) and of the fast subsystem (\ref{general fast part QEA 1}). Generally speaking, the system (\ref{general QEA 1}) has some conservation laws, and they are linear functions $h^{1}(\mathcal{C}), h^{2}(\mathcal{C}),..., h^{k}(\mathcal{C})$ of the concentrations that are constant in time. Particularly, we have two main cases. The first case is that if the conservation laws of the system (\ref{general fast part QEA 1}) are preserved by the original dynamics, then there is no fast--slow separation for variables (i.e. all variables of the system are either fast or constant). In this case, the system (\ref{general fast part QEA 1}) describes the dynamics of fast variables. Another case is where the fast subsystem (\ref{general fast part QEA 1}) has some more linearly independent conservation laws $h^{k+1}(\mathcal{C}), h^{k+2}(\mathcal{C}),..., h^{k+p}(\mathcal{C})$. In this case, the conservation laws are not preserved by the full system (\ref{general QEA 1}). Then, the slow variables of the system are determined by $h^{k+1}(\mathcal{C}), h^{k+2}(\mathcal{C}),..., h^{k+p}(\mathcal{C})$. On the other hand, the fast variables of the system are those $c_{i}$ such that  $(\gamma^{f})_{i} \neq 0$, for some fast reactions $f$. It can be concluded that the fast subsystem has a stable steady sate $\mathcal{C}^{*}=(c_{1}^{*}, c_{2}^{*},..., c_{m}^{*})$ satisfying $c_{j}^{*}>0$ for $j=1,2,...,m$. 
 The quasi--equilibrium manifold is given by the following algebraic equations
 \begin{equation}
\begin{array}{llll}  
 \mathlarger{\mathlarger{\sum\limits}}_{f,fast} \mathcal{R}^{f}(\mathcal{C},\mathcal{K},t)\gamma^{f}=0,\\
 h^{i}(\mathcal{C})=b_{i}, \quad 1 \leq i \leq k+p.
\end{array}
\label{general fast part QEA 1 =0}
\end{equation}
 For small parameter $\epsilon$, the equations $\mathlarger{\mathlarger{\sum\limits}}_{f,fast} \mathcal{R}^{f}(\mathcal{C},\mathcal{K},t)\gamma^{f}=0$ serve as an approximation to a slow manifold. According to the approach, there are thermodynamic restrictions of the original system and the fast subsystem, and the quasi-equilibrium appears 
to be in partial thermodynamic equilibrium. Therefore, the global stability of fast subsystems is evident, and the classical singular perturbation theory based on the Tikhonov theorem can be applied. More details and explanations of the method are given in (Gorban et al., 2010, Khoshnaw, 2015, Noel et al., 2012) . The reader can see applied examples of the method on chemical reactions in (Huang and Yong, 2013, Kijima and Kijima, 1983).\\


\subsection{Lumping of Compartments} 
\label{lumping of species}                   
One of the powerful tool to reduce the number of variables through a linear or nonlinear transformation is  lumping method (Pepiot-Desjardins and Pitsch, 2008).
During the last century, the concept of the lumping of compartments was revised many times. The use of lumped models was reported as early as 1953, with the reduced scheme being empirically determined (Okino and Mavrovouniotis, 1998). The first idea of the lumping analysis was introduced by Wei and Kuo (1969), after that Weekman (1979) has established a comprehensive overview of lumping from both theoretical and practical viewpoints. He has pointed out some of the remaining problem areas (Coxson and Bischoff, 1987). One of the main issue in systems biology is dimensional reduction for high order models. This reduction from a high dimensional differential equation system to a lower dimensional one is often achieved by lumping in their compartments, in which the new dependent variables are some functions of the original ones. To check the accuracy of the technique, we have to compare the simplified differential equations with the original equations, this can be done by calculating total differences (Li et al., 1994). Clearly the technique is worked as an effective tool for model order reduction by combining some model states based on some properties. Generally, we have three types of lumping such as:

\begin{enumerate}
\item{Generalized lumping (The states of the reduced system are linear combinations (or even nonlinear functions) of the states of the original system)},\\

\item{Proper lumping (Each state of the original system is included only in one state of the reduced system)},\\

\item{Improper lumping (states of the original system is may contribute to more than one lump) (Brochot et al., 2005, Kou and Wei, 1969)}.
\end{enumerate}

	\begin{figure}[ht]  
	\begin{center}       
		\subfigure[Generalized lump.]{%
			\includegraphics[width=0.32\textwidth]{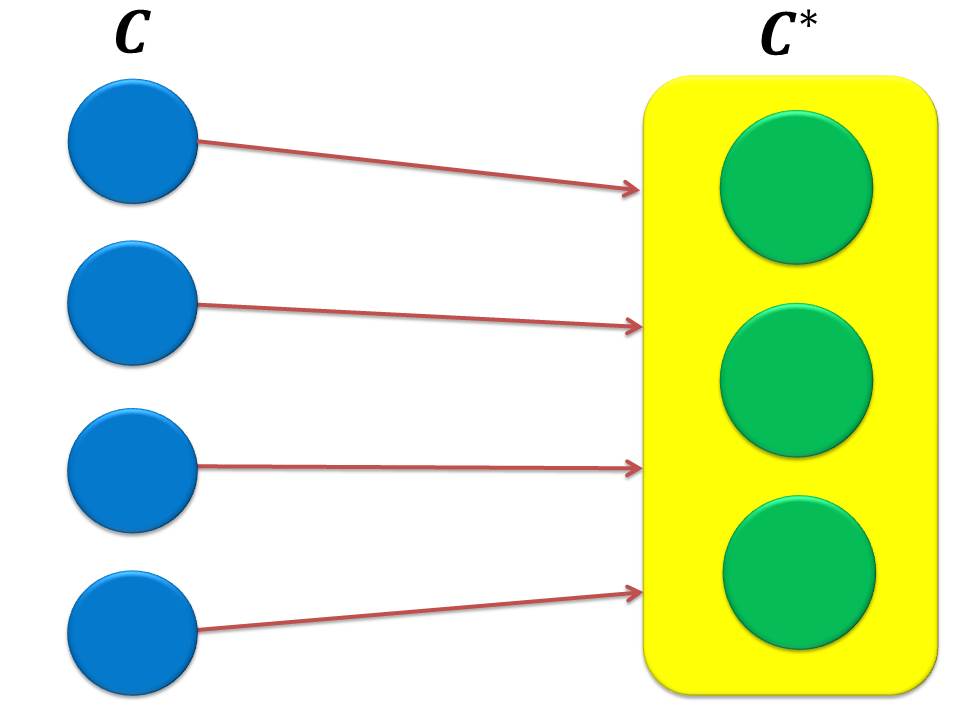}
		}		
		\subfigure[Improper lump.]{%
			\includegraphics[width=0.32\textwidth]{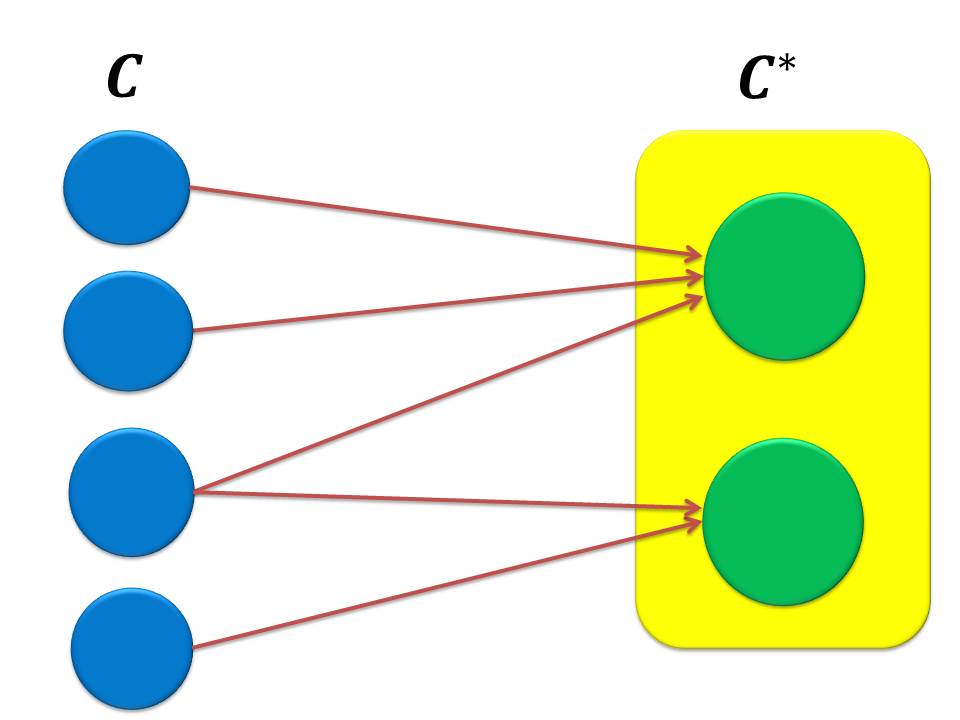}
		} 
		\subfigure[Proper lump.]{%
			\includegraphics[width=0.32\textwidth]{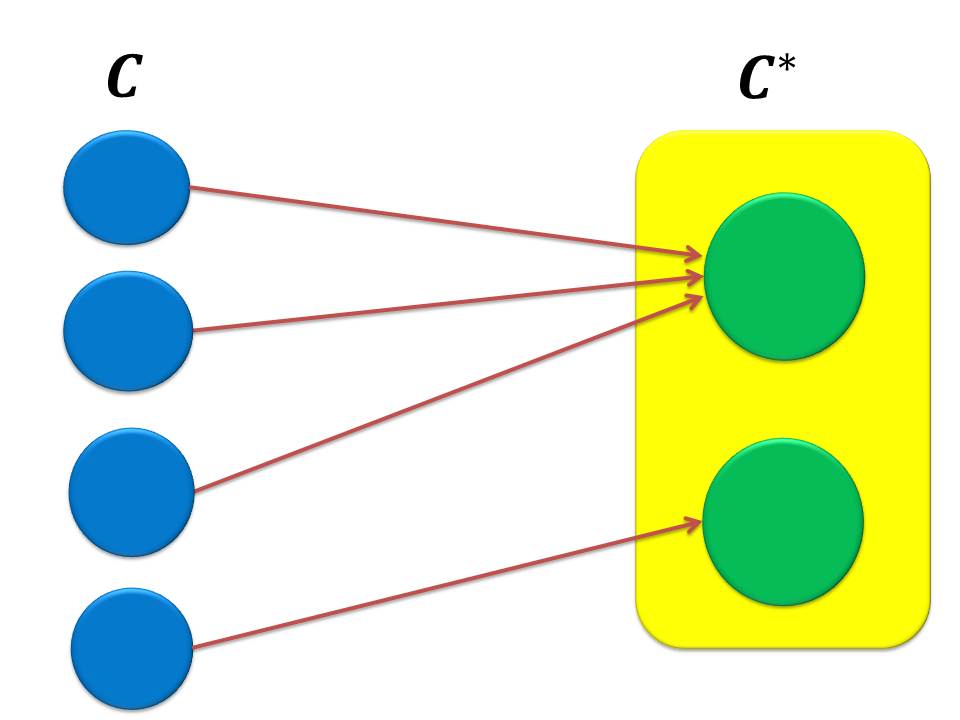}
		}
		
	\end{center}   
	\caption {Types of lumping states.}    	
\end{figure}

\noindent Proper and improper lumping are commonly used, so when the lumping transformation is linear then it is called linear lumping, otherwise it is called nonlinear lumping. The most approachable lumping schemes are linear; nonlinear transformations are also suitable but they are difficult to determine. Mathematically, linear lumping is divided into two groups, proper and improper. For proper lumping, each reactant appears in only one lump. For improper lumping, on the other hand, components may contribute to more than one lump. It is clear that the more useful and insightful is the proper lumping schemes. But generally many mathematical lumping methods are able to derive of realistic reaction networks is only improper lumps (Okino and Mavrovouniotis, 1998). In proper lumping, the union set of all the lumps must result in the whole chemical system, while the intersection set of any two lumps is the empty set (Martinez, 1990). In the lumping procedure species with similar composition and functionalities are lumped into one single representative species (Pepiot-Desjardins and Pitsch, 2008). Then, we can say that compartments in series can be lumped when they equilibrate fast, and compartments in parallel can be lumped when they have similar times scales. In other words, a group of species can be represented in the mechanism by a single variable. Consequently, combing several species as a single species   is often vital for theoretical and practical purposes. Although, some systems may have exactly lumpable, but they may not provide any practical goals (Li and Rabitz, 1990).
 
\subsection{Entropy Production Analysis} 

In this section, we introduce an important technique. This is mainly an effective relation between entropy production and lumping of species. According to amounts of entropy produced by irreversible processes, it needs the first and second laws of thermodynamics. They are studied by (Tolman and Fine, 1948). After that the total entropy production for chemical reactions was studied in (Pritchard, 1975, Pritchard et al., 1974). They assumed that the total entropy is \; " a completely monotonic function of time for several chemical reactions " . Then, more details and explanation about entropy production analysis is given in (Bykov et al., 1977, V. Dimitrov et al., 1982). Fortunately, Gorban and Karlin in 2005 developed the idea of the entropy production analysis and applied in more chemical reaction networks. In addition, the idea of entropy production used in some biological and biochemical systems in (Chang et al., 1989, Unrean and Srienc, 2011), and it was applied as a model reduction by (Kooshkbaghi et al., 2014). Recently, the technique has been used for more complex biochemical reactions (Banerjee and Bhattacharyya, 2013, Dobovisek et al., 2011, E. N. Miranda, 2010, H.Qian, 2009, Lebiedz, 2004, Lopez-Agudelo and Barragan, 2014, Martyushev and Seleznev, 2006, Prigogine, 1967, Zhang, 2014). More recently, the idea of the entropy production was more developed as a model reduction approach by by eliminating non important reactions, more details about the developed algorithm was given in (Khoshnaw, 2015).\\
For simplifying the complex kinetic equations of biochemical reactions, we improved the previous algorithm and used as amodel reduction technique here. The idea is based on relative contribution of each reaction to the total entropy production \eqref{Ent23} and lumped of isolated species. Especially, the algorithm is used to determine the non--important reactions (the least contributing reactions to the total entropy production) during the computational simulation then lumped of isolated species.


\section{Elasticity and Control Coefficients}
\noindent The power to change the state of metabolism in response to an outer signaling is called metabolic control, and it is measurable in terms of the persuasiveness of the metabolic response to external factors, without any idea about the purpose /function/mechanism of the response (Giersch, 1988). The control structure of a metabolic pathway can be quantitatively characterized by metabolic control analysis (MCA), which is a mathematical frame work for describing metabolic, signaling, and genetic pathways. MCA quantifies how variables, such as fluxes and species concentrations, depend on network parameters. In particularities able to describe how network dependent properties, called control coefficients, depend on local properties called elasticities. By means of control and elasticity coefficients, the control coefficient is the fractional change in metabolic concentration (Puigjaner et al., 1997). An important step forward to determine and analysis of the complexity of dynamic changes of species in a complex metabolic system is metabolic control analysis (Li et al., 2010, Teusink et al., 2000). There are three main types of coefficient analysis. The first one is elasticity coefficients that quantify the sensitivity of a reaction rate to the change of concentration or a parameter. The second type is flux control coefficients that measure the change of a flux along a pathway in response to a change in the rate of a reaction. The last one is concentration control coefficients that calculate the change of concentration of some metabolite species $S_{i}$ in response of a change in the rate of a reaction (Giersch, 1988).

\subsection{Elasticity}   
\noindent Elasticity coefficients are used in economics, physics, chemistry, or more generally in mathematics as a definition of point elasticity. The rate of reaction is affected by many different factors, such as pH, temperature, reactant and product concentrations and etc. The elasticity is defined by the degree to which these factors change the reaction. Elasticities in biochemistry theory called kinetic orders and describe how sensitive a reaction rate is to changes in reactant, product and effector concentrations (Kacser et al., 1995, Klipp et al., 2008, SAURO et al., 1987). The elasticity coefficient is the fractional change in the net rate for an individual substrate, with everything else is kept fixed (Puigjaner et al., 1997).
The main equation of elasticity coefficient is given

\begin{eqnarray}
E^{v_{i}}_{s_{i}} =\dfrac{\partial v}{\partial s} \dfrac{s}{v},
\label{elastisity general equation}
\end{eqnarray}

\noindent where \;$ V_{i}$ \; is reaction rate and \;$ S_{i}$\; is concentration of species. The equation of elasticity measures of the change of \; $ V_{i}$ \; in response to a change in \;$ S_{i}$ ,\; while everything else is kept fixed. If we have substrate \;{\large S},\; inhibition \;{\large I}\; and activation \;{\large A}\; in a pathway then some quantitative amounts can be considered. There is some typical values for elasticity coefficients that satisfy the following inequalities:     

$$ E^{v}_{s} =\dfrac{\partial v}{\partial s} \dfrac{s}{v} >0 \; ,\quad  E^{v}_{p} =\dfrac{\partial v}{\partial p} \dfrac{p}{v} <0.$$

\noindent That means more substrates are required to have fast rates, while more products give slower rates. In addition, if there are given inequalities
$$ E^{v}_{A} =\dfrac{\partial v}{\partial A} \dfrac{A}{v} >0 \; ,\quad  E^{v}_{I} =\dfrac{\partial v}{\partial I} \dfrac{I}{v} <0.$$
\noindent This gives us fast reaction rates required the higher activator concentration, whereas slow reaction rates are depended on the higher inhibitor concentration (Giersch, 1988).

\subsection{Control Coefficients}
\noindent A control coefficient quantifies the relative steady state change in a system variable, e.g. metabolite concentration \;{\large S}\; or pathway flux \;{\large J},\; in response to a relative change in a parameter. We have two main control coefficients, they are concentration control coefficients  and flux control coefficients (Kacser et al., 1995, Klipp et al., 2008, SAURO et al., 1987). The equation of flux control coefficients is defined below

$$ C^{J}_{v_{i}}=\dfrac{\dfrac{dJ}{dp}\dfrac{p}{J}}{\dfrac{\partial v_{i}}{\partial p}\dfrac{p}{v_{i}}}=\dfrac{d\ln(J)}{d\ln(v_{i})}=\dfrac{dJ}{dv_{i}}\dfrac{v_{i}}{J}. $$

\noindent And the equation of  concentration control coefficients is given by

$$ C^{s}_{v_{i}}=\dfrac{\dfrac{ds}{dp}\dfrac{p}{s}}{\dfrac{\partial v_{i}}{\partial p}\dfrac{p}{v_{i}}}=\dfrac{d\ln(s)}{d\ln(v_{i})}=\dfrac{dJ}{dv_{i}}\dfrac{v_{i}}{s}. $$

\noindent The flux control coefficient \; $ C^{J}_{v_{i}}$\; gives the relative small change in (a system variable) concentration with small change in pathway flux\; {\large J}\;. The word flux \;{\large J},\; also used to describe the rate of the system. Therefore, changing in the concentration can be fluctuated between increasing and decreasing (Li et al., 2010, Teusink et al., 2000). Flux coefficients usually vary from 0 to 1. The concentration control coefficient\; $C^{s}_{v_{i}}$,\; a global property of the system, and gives the relative change in metabolite concentration \;{\large S}. \;The concentration control coefficients can have large values; also can vary from negative to positive and small to large value (Li et al., 2010, Teusink et al., 2000).
There is a relationship between control coefficients and elasticity. The flux control summation theorem was discovered independently by the Kacser/Burns group and the Heinrich/Rapoport group in the early 1970s and late 1960s. The flux control summation theorem implies that metabolic fluxes are systemic properties and that their control is shared by all reactions in the system. When a single reaction changes its control of the flux this is compensated by changes in the control of the same flux by all other reactions. The two important equations are proposed as follows:

\begin{eqnarray}
\sum_{i}C^{J}_{v_{i}}=1 \quad \text{and} \quad  \sum_{i}C^{s}_{v_{i}}=0.
\label{summation theorem genaral equation}
\end{eqnarray}

\noindent  The connectivity theorems are specific relationships between elasticities and control coefficients. They are useful because they highlight the close relationship between the kinetic properties of individual reactions and the system properties of a pathway. Two basic sets of theorems exist, one for flux and another for concentrations. The concentration connectivity theorems are divided again depending on whether the system species \;${\large S_{n}} $\; is different from the local species \;${\large S_{m}}$.

 \begin{equation}
\begin{array}{llll}

\sum_{i}C^{J}_{v_{i}}\;E^{v_{i}}_{s}=0,\\
\sum_{i}C^{s_{n}}_{v_{i}}\;E^{v_{i}}_{s_{m}}=0 \quad \text{for} \quad n\neq m, \\ 
\sum_{i}C^{s_{n}}_{v_{i}}\;E^{v_{i}}_{s_{m}}=-1  \quad \text{for}\quad n= m. \\

\end{array}\label{connectivity theorem general equation}
\end{equation}


\section{Function of Deviation}
\label{function of deviation}  
The are some error formulas for finding the total differences between concentration species of the full and reduced models. Such as error integral (I)\\ $$
\mathcal{I}=\mathlarger{\mathlarger{\sum\limits}}_{i\in \mathcal{M_{I}}}  \dfrac{1}{Tn(\mathcal{M_{I}})} \mathlarger{\mathlarger{\int \limits_{0}^{T}}} \mid 1-\dfrac{x_{ir}(t)}{x_{if}(t)} \mid dt
$$ in (Rao et al., 2014), relative deviation\\
$$ \phi_{model-error} (\tilde{k},k)=\dfrac{1}{Nn_{\ell}}\mathlarger{\mathlarger{\sum\limits}}_{l=1}^N \mathlarger{\mathlarger{\sum\limits}}_{i \in \ell} \Big(\dfrac{|\tilde{x}_{i}(t_{l})-x_{i}(t_{l})|}{\tilde{w_{il}}} \Big) $$ in (Hannemann-Tamas et al., 2013), and function of deviation. We mainly use the function of deviation as an essential tool to calculate the total difference between the simplified models and the original models. The method was used for parameter estimations in (Hoops et al., 2006). This technique was developed and applied for system of chemical kinetics earlier in (Kutumova et al., 2013). By using the function of deviation, we can check the model approximation at each reduction step. Let $n$ be a number of species in the reduced model while $m$ is a number of species in the original model ($n < m$). Moreover, $c_{i}^{\mathcal{O}}(t_{ij})$ and $c_{i}^{\mathcal{R}}(t_{ij})$ are concentrations of species for the original model and the reduced model at a given time $t_{ij}$, $i=1,2,...,n$ and $j=1,2,...,p_{i}$, where $p_{i}$ is the number of points for $c_{i}(t)$. The function of deviation is given as a normalized sum of squared difference of species concentrations:
 \begin{equation}
\begin{array}{llll}
\mathcal{F}^{\mathcal{D}}(\mathcal{C}_{0},\mathcal{K},t)=\mathlarger{\mathlarger{\sum\limits}}_{i=1}^n \mathlarger{\mathlarger{\sum\limits}}_{j=1}^{p_{i}} \dfrac{\varphi_{min}}{\varphi_{i}} \Big[c_{i}^{\mathcal{O}}(t_{ij})-c_{i}^{\mathcal{R}}(t_{ij}) \Big]^{2},
\end{array}\label{SD}
\end{equation}
\noindent where  
$\varphi _{i} = \sqrt{ p_{i}^{-1}\mathlarger{\mathlarger{\sum\limits}}_{i=1}^{p_{i}} \big ( c_{i}^{\mathcal{O}}(t_{ij}) \big )^{2}}$ and $\varphi_{min}=min \left\{ {\varphi_{i}, i=1,2,..,n}\right\}$.\\ A normalizing factor $\dfrac{\varphi_{min}}{\varphi_{i}}$ is applied for model reduction to make all concentration trajectories have similar importance.   
\noindent  Calculating the value of deviation (\ref{SD}) at each reduction step is an effective tool in model reduction to check that the approximation of variables is still within allowable limits or not. According to a study published earlier in (Rao et al., 2014) the total differences between the original and reduced model only (7--8)\% and according to another recent study this value only (0.53--9.14)\% (Khoshnaw, 2015). Therefore, we use function of deviation for both types of lumping (Lumping of species and Lumping of parameters) and for entropy with lumping of species in the next chapters.
\section{Software Tools}
The best challenge of cell signaling pathways in systems biology is to understand of the dynamic behavior of variables. There are many software tools that have capabilities for modeling, visualizing, process descriptions and simulating complex cell signaling pathways. In this thesis, we use four software development tools to simulate the concentration dynamics of the species, draw the structure of biochemical reaction networks. They are defined below:

\begin{itemize}
\item {\bf{Maple}} is a computer programming language created by Maplesoft, which is a symbolic and numeric computing environment. Maple also contains other technical computing, for example data analysis, connectivity, visualization and matrix computation. We use this software tool to calculate the value of variables and to solve mathematical equations.

\item {\bf{MATLAB}} is a computer programing language created by MathWorks, which is a multi-paradigm numerical computing environment. MATLAB works in various fields such as: plotting of functions and data, matrix manipulations, implementation of algorithms, ...etc.
We use this computational tool in different purposes such as plotting variables, applying algorithms, solving system of differential equations and calculating equations and matrices.

\item {\bf{System Biology Toolbox}} is an important software tool for systems biology defined on Matlab. This offers an open and user extensible environment for cell signaling pathway models, in which to explore ideas, prototype and share new algorithms, and build applications for the analysis and simulation such models. We use the defined tool to compute numerical simulations at each reduction step, and steady state values for state variables of biochemical kinetics.

\item {\bf{PowerPoint}} is used to create a graphical representation of  all biological and biochemical process in our thesis.

\end{itemize}

\include{chapter[2]}
\chapter{Slow and Fast Subsystems for Non-linear Protein Translation Pathways}

\section{Introduction} 

\noindent Cells can be found in skin, muscles and bones. And all of those cells include billions of proteins and enzymes. Indeed, proteins are fundamental of molecular for each living creature on the Earth (Cooper, 2000). There is an important part in cells that is called microRNA (mRNA). mRNAs are a type of post-transcriptional well organized non-coding RNAs lately discovered in plants and animals. It has been shown that they regulate various biological procedures ranging from the embryotic development to the regularization of neural network model (Xu et al., 2009).  MicroRNAs (miRNAs) are 20 to 22 nucleotide RNAs that modulate the operation of eukaryotic mRNAs and have an important role in evolution, virus infection, stress responses, and cancer (Nissan and Parker, 2008). mRNAs are single-stranded RNA molecules of about 21 to 23 nucleotides in length, which modulate gene expression (Xu et al., 2009).  miRNAs function, at any rate in part, to prevent translation of mRNAs and contribute to the progress mRNA deadenylation, decapping, and $5^{'} $ to $3^{'}$ reduce of the mRNA body (Eulalio et al., 2008, Filipowicz et al., 2008, Jackson and Standart, 2007, Valencia-Sanchez et al., 2006).
There are some main functions of mRNA. The most important function of mRNA is related to gene expression regulation, and their molecules seem partly complementary to one or more mRNA molecules. They were first described in 1993 by et al (Lee et al., 1993). in the Victor Ambros lab, and still the term mRNA was only introduced in a set of three articles in 2001 (Ruvkun, 2001). As of early 2008, computational analysis by IBM proposed the existence of as many as 50 000 dissimilar mRNAs in the typical mammalian cell, each with perhaps a thousand or more possible targets (Glaser, 2008). Interestingly, microRNAs (miRNAs) are recently well thought out as key regulators of a wide variety of biological pathways, including development, differentiation and on cogenesis. Currently, remarkable advancement was made in understanding of mRNA functions, biogenesis and mechanisms of action. The RISC effector complex and mature mRNAs are incorporated, which includes as a key component an Argonaut protein. mRNAs affect gene expression by guiding the RISC complex toward particular target mRNAs. It can be seen that there is a big controversial to determine the exact mechanism of this inhibition (Zinovyev et al., 2010).
\begin{figure}[H] 
	\begin{center}             
		\subfigure{%
			\includegraphics[width=0.7\textwidth]{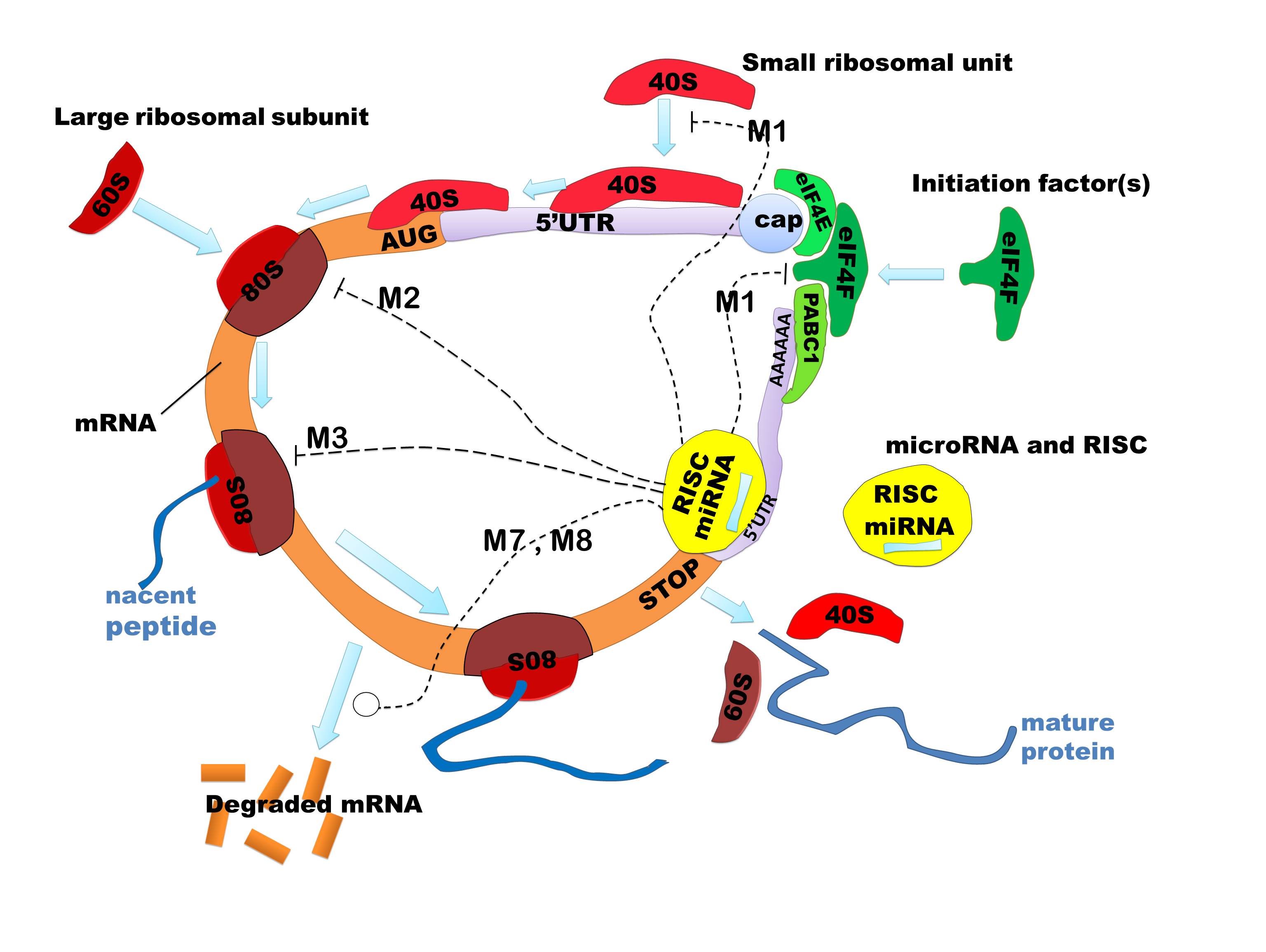}
		}             
	\end{center}   
	\caption {Protein translation process with microRNA mechanisms.}       
\end{figure}
\noindent In the last decades, many possible mechanisms of mRNA have been recognized. The most of all documented mechanisms are negative post - transcriptional regulation of mRNA by mRNA translation inhabitancy and/or mRNA rotting. Whereas, there are some possibilities show that miRNAs might also act at the decomposition stage. There are also some studies in the present literature about to determine and decide which mechanism and in which situations has a control role in living cells. It is clear that same experimental systems handling with the same pairs of miRNA and mRNA can provide contentious evidences about which is the actual mechanism of translation subdue noticed in the experiment (Zinovyev et al., 2013). 
mRNA translation is an important procedure in cell signaling pathways that can be seen in many systems of biology. In this procedure, the genetic sequences are translated from mRNA to protein by ribosome translocation, after the genetic information included in DNA is transcribed to the mRNA. There are three important components in the mRNA translation process: the mRNA (genetic template), the ribosome (assembly machinery), and the aminoacyl transfer RNAs (aa-tRNAs). 
mRNA protein translation is theoretically divided into three levels: initiation, elongation and termination. At the initiation stage, the ribosome first attaches to the mRNA then reads the mRNA codon by codon (from the 5' end of the mRNA to the 3' end). At the elongation stage, it recruits the appropriate aa-tRNA and unites the latest amino acid into the nascent muster chain, releases the discharged tRNA. At the last stage of protein translation, the completed protein from the mRNA when the ribosome reaches the end of the mRNA eventually are released (Lewin, 2007). 
There is a long history of mathematical modeling of mRNA. Then, models for mRNA have been developed in recent years with the evolution of systems and synthetic biology. The various constructs of models for mRNA translation are introduced at various levels of abstraction (Zhao and Krishnan, 2014). 
In this study, we give a detailed description for mathematical modelling of miRNA that describing the process of protein translation. We simply reviewed the previous  study of miRNA protein translation given in (Zinovyev et al., 2013). Then, we use quasi steady state approximation to separate equations into slow and fast subsystems and identifying some approximate solutions for state variables. Finally, elasticity and control coefficient are calculated for the model network in order to identify effect of reaction rates, parameters and state variables on model dynamics.

 
 \section{Model Equations of microRNA}
\noindent To explain the effect of microRNA interference with translation initiation factors, a non-linear version of the translation model was proposed in which explicitly takes into account recycling of initiation factors (eIF4F) and ribosomal subunits (40S and 60S). 
The model has seven chemical species 40S , 60S , eIF4F , F , A , R, and P, where, F=mRNA:40s, A=AUG and R=80s; And four chemical reactions, all considered to be irreversible, see Figure 2. The model variables are defined bellow: 

\FloatBarrier 
\begin{table}
\caption{State variables of microRNA pathways.}
\centering
\begin{tabular}{|c|c|p{9cm}|}
\hline 
No. & State variables & Biological meaning \\ 
\hline 
1 & 40s & Free small ribosomal subunit \\ 
\hline 
2 & eIF4F & Free initiation factor \\ 
\hline 
3 & F & State of mRNA when the small ribosomal
subunit bound to the initiation site \\ 
\hline 
4 & A & State of mRNA when the small ribosomal
subunit bound to the start codon \\ 
\hline 
5 & 60s & Free large ribosomal subunit\\ 
\hline 
6 & R & Translating ribosome \\ 
\hline
7 & Psynth & Translated protein \\ 
\hline  
\end{tabular} 
\label{table of variable mRNA}
\end{table}
\FloatBarrier 

\begin{figure}[H] 
	\begin{center}     
		\subfigure{%
			\includegraphics[width=0.8\textwidth]{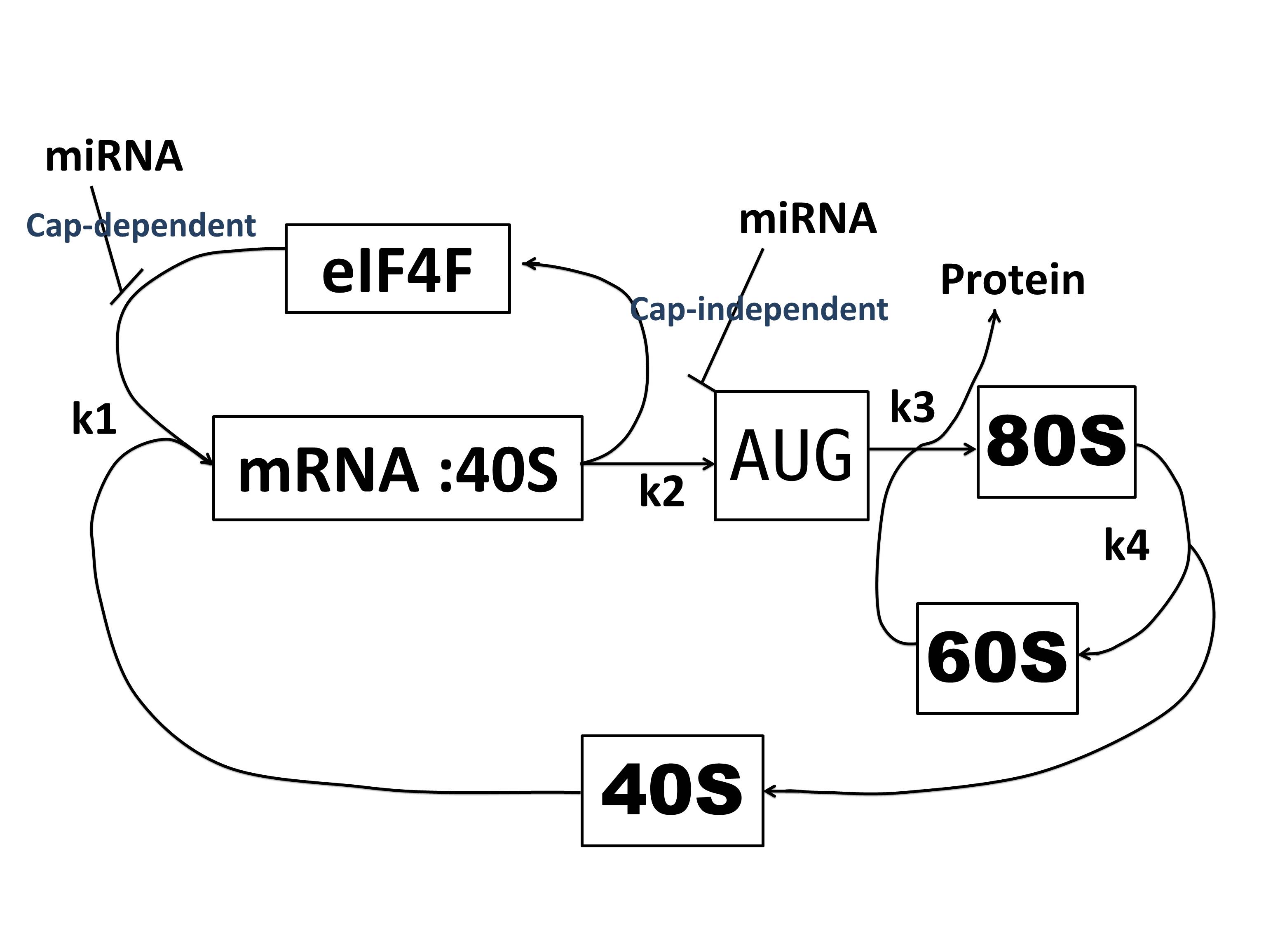}
		}             
	\end{center}   
	\caption {The model pathways for non-linear protein translation. }      
\end{figure}

And the model reaction are given below:\\
\begin{enumerate}[noitemsep,nolistsep]
\item $40S+eIF4F\rightarrow F$, assembly of the initiation complex (rate $k_{1}$ ).
\item $ F\rightarrow A$,some late and cap-independent initiation steps, such as scanning the $5^{\prime}$ UTR for the start codon A (rate $k_{2}$ ).
\item $A\rightarrow R$, assembly of ribosomes and protein translation (rate $k_{3}$).
\item $80S \rightarrow 60S + 40S$ ,recycling of ribosomal subunits (rate $k_{4}$ ).
\end{enumerate}

\noindent The model is described by the following system of nonlinear differential equations:
\begin{equation}
\begin{array}{llll}  
\dfrac{d[40S](t)}{dt}=-k_{1}[40S][eIF4F]+k_{4}[R],\\
\dfrac{d[eIF4F](t)}{dt}=-k_{1}[40S][eIF4F]+k_{2}[F],\\
\dfrac{d[F](t)}{dt}=k_{1}[40S][eIF4F]-k_{2}[F],\\
\dfrac{d[A](t)}{dt}=k_{2}[F]-k_{3}[A][60S],\\
\dfrac{d[60S](t)}{dt}=-k_{3}[A][60S]+k_{4}[R],\\
\dfrac{d[R](t)}{dt}=k_{3}[A][60S]-k_{4}[R],\\
Psynth(t)=k_{3}[A](t),
\end{array}      
\label{mod1}             
\end{equation} 
\noindent with the initial conditions 
\begin{equation}
\begin{array}{llll}   
[40S](0)=[40S]_{0}, \\

 [eIF4F](0)=[eIF4F]_{0}, \\
 
  [60S](0)=[60S]_{0},\\
  
   F(0)=A(0)=R(0)=0.
   \end{array} 
\label{ini1}
\end{equation}
\noindent The system (\ref{mod1}) has three independent stoichiometric conservation laws:
\begin{equation}
\begin{array}{llll} 
[F]+[40S]+[A]+[R]=[40S]_{0},\\
    
[F]+[eIF4F]=[eIF4F]_{0},\\

[60S]+[R]=[60S]_{0},
\end{array}
\label{cons1}
\end{equation}
\noindent where $[40S]_{0} , [60S]_{0} $\;and\;$ [eIF4F]_{0} $ are total amounts of small, big ribosomal sub units and the initiation factor respectively. The following assumptions on the model parameters and initial variable states were suggested:
\begin{equation}
\begin{array}{llll} 
k_{4} \ll k_{1}, k_{2}, k_{3} ,\\

k_{3} \gg  k_{1}, k_{2}, \\     

[eIF4F]_{0} \ll [40S]_{0},  \\

[eIF4F]_{0}<[60S]_{0} <[40S]_{0}.
\end{array}
\label{cond1}
\end{equation}
\noindent More details and descriptions about the model equations and the proposed assumptions can be found in (Zinovyev et al., 2013).  

\section{Fast and Slow Subsystems for microRNA Model}
\noindent Quasi steady state approximation is an important technique in systems biology. The method can be applied for nonlinear models in order to classify such systems into fast and slow subsystems and identify some analytical approximate solutions. More details about the QSSA method can be seen in (Khoshnaw, 2015a, Khoshnaw et al., 2016, Khoshnaw, 2015b). 
Based on conservation laws (\ref{cons1}), we can remove the following variables:
\begin{equation}
\begin{array}{llll} 
[A]=[40S]_{0}-[F]-[40S]-[R],\\ 
	    
[eIF4F]=[eIF4F]_{0}-[F],\\
	
[60S]=[60s]_{0}-[A].
\end{array}
\label{newcons1}
\end{equation}

\noindent Then, the system (\ref{mod1}) becomes 
\begin{equation}
\begin{array}{llll} 
\dfrac{d[40S](t)}{dt}=k_{1}[40S][F]-k_{1}[eIF4F]_{0}[40S]+k_{4}[R],\\
\dfrac{d[F](t)}{dt}=k_{1}[eIF4F]_{0}[40S]-k_{1}[40S][F]-k_{2}[F],\\
\dfrac{d[R](t)}{dt}=k_{3}[40S]_{0}[60S]_{0}-k_{3}[60S]_{0}[40S]-k_{3}[60S]_{0}[F]+k_{3}[40S][R]+\\
k_{3}[F][R]+k_{3}R^2 -(k_{3}[60S]_{0}+[40S]_{0}+k_{4})[R].
\end{array}
\label{newmod1} 
\end{equation}
\noindent By introducing the following new variables 

\begin{center}
$ x=\dfrac{[40S]}{[40S]_{0}} ,\quad y=\dfrac{[F]}{[eIF4F]_{0}} ,\quad z=\dfrac{[R]}{[eIF4F]_{0}}\quad and \quad \tau=k_{1}[eIF4F]_{0}\;t.$
\end{center}
\noindent The system (\ref{newmod1}) takes the form

\begin{equation}   
\begin{array}{llll}    
\dfrac{dx}{d\tau}=x(y-1)+\rho z,\\
\epsilon\dfrac{dy}{d\tau}=x(1-y)-\alpha _{1} y,\\
\epsilon\dfrac{dz}{d\tau}=\alpha _{2}(1-x)-\alpha _{3}y+\alpha _{4}xz+\epsilon \alpha _{4} yz+\epsilon z^2 -(\alpha _{3}+\alpha _{4}+\rho)z,
\end{array}\label{reducemod1}           
\end{equation}

\noindent where \quad  $\epsilon=\dfrac{[eIF4F]_{0}}{[40S]_{0}}, \quad \rho=\dfrac{k_{4}}{k_{1}[40S]_{0}},\quad \alpha _{1}=\dfrac{k_{2}}{k_{1}[40S]_{0}}, \quad \alpha _{2}=\dfrac{k_{3}[60S]_{0}}{k_{1}[eIF4F]_{0}},\\ \quad \alpha _{3}=\dfrac{k_{3}[60S]_{0}}{k_{1}[40S]_{0}} , \quad \alpha _{4}=\dfrac{k_{3}}{k_{1}}.$

\noindent According to the conditions (\ref{cond1}), \quad  $\dfrac{k_{4}}{k_{1}[40S]_{0}}\longrightarrow 0 $ \; when \; $ k_{4}\ll k_{1}$ . Then, the system (\ref{reducemod1}) is completely on the form of slow and fast subsystems with six parameters.
By applying the technique of QSSA, the system can be simplified by plugging in $\epsilon=0 $\;in the system(\ref{reducemod1}), and in the limit \; $ \epsilon\longrightarrow 0$ \;, the system takes the form

\begin{subequations}        
	\begin{align}
\dfrac{dx}{d\tau}=x(y-1), \label{fastmode1a}\\
0=x(1-y)-\alpha _{1} y, \label{fastmode1b} \\
0=\alpha _{2}(1-x)-\alpha _{3}y+\alpha _{4}xz -(\alpha _{3}+\alpha _{4})z. \label{fastmode1c}
	\end{align}\label{fastmod1}
\end{subequations}     

\noindent We can analytically solve the equations (\ref{fastmode1b}) and (\ref{fastmode1c}) for $y$ and $z$ in terms of $x$. They are given bellow\\

\begin{subequations}  
	\begin{align}
y=\dfrac{x}{\alpha _{1}+x}, \label{y interm x a}\\
z=\dfrac{\alpha _{3}x-\alpha _{2}(1-x)(\alpha _{1}+x)}{(\alpha _{4}x-\alpha _{4}-\alpha _{3})(\alpha _{1}+x)}. \label{z interm x b}
	\end{align}
\end{subequations} 

\noindent Therefore, the approximate solution for equations (\ref{reducemod1}) is sufficiently close to the manifold $\mathcal{M}_{0}$, where $\mathcal{M}_{0}$ is defined as follows
\begin{equation}  
\begin{array}{llll}
\mathcal{M}_{0}= \bigg \lbrace (x,y,z):x\in [0,1],y=\dfrac{x}{\alpha _{1}+x},
z=\dfrac{\alpha _{3}x-\alpha _{2}(1-x)(\alpha _{1}+x)}{(\alpha _{4}x-\alpha _{4}-\alpha _{3})(\alpha _{1}+x)}
\bigg \rbrace.
\end{array}\label{manifoldmod1}
\end{equation}
  
\noindent Thus, we obtain the following reduced differential equation close to the manifold $\mathcal{M}_{0}$,
\begin{equation}  
\begin{array}{llll}
\dfrac{dx}{d\tau}=\dfrac{x^2}{\alpha _{1}+x}-x.
\end{array}\label{dx/dtau}
\end{equation}

\noindent The above equation can be solved analytically. The implicit solution of the separable differential equation takes the form 
\begin{equation}  
\begin{array}{llll}
\alpha _{1} ln(x)+x=1-\alpha _{1}\tau.
\end{array}\label{xmod1}
\end{equation}

\noindent According to our proposed new variables, the equation (\ref{xmod1}) becomes 
\begin{equation}  
\begin{array}{llll}
[40S](t)=1+\dfrac{k_{2}}{k_{1}}ln([40S]_{0})-k_{2}[eIF4F]_{0}\; t-\dfrac{k_{2}}{k_{1}}ln([40S]).
\end{array}\label{[40S]mod1}
\end{equation}

\noindent The analytical solution for variables F and R  are calculated  by using equations (\ref{y interm x a}) and (\ref{z interm x b}) 

 \begin{equation}  
 \begin{array}{llll}
 [F](t)=\dfrac{[eIF4F]_{0}[40S](t)}{[40S]+\dfrac{k_{2}}{k_{1}}},
 \end{array}\label{[F]mod1}
 \end{equation}
                     
  \begin{equation}  
 \begin{array}{llll}
 [R](t)=\dfrac{[eIF4F]_{0}[60S]_{0}[40S](t)+[60S]_{0}([40S](t)-[40S]_{0})(k_{1}[40S](t)+k_{2})}{([40S](t)+\dfrac{k_{2}}{k_{1}})([40S](t)-[40S]_{0}-[60S]_{0})}.
 \end{array}\label{[R]mod1}
 \end{equation}                                                          
 
\noindent And by using equations (\ref{newcons1}), we can calculate [eIF4F], [60S] and A,          
 
    \begin{equation}  
  \begin{array}{llll}
  [eIF4F](t)=\dfrac{k_{2}[eIF4F]_{0}}{k_{1}[40S](t)+k_{2}},
  \end{array}\label{[eIF4F]mod1}
  \end{equation} 
  
    \begin{equation}  
  \begin{array}{llll}
 [60S](t)=[60S]_{0}-\\
 \dfrac{[eIF4F]_{0}[60S]_{0}[40S](t)+[60S]_{0}([40](t)-[40S]_{0})(k_{1}[40S](t)+k_{2})}{([40S](t)+\dfrac{k_{2}}{k_{1}})([40S](t)-[40S]_{0}-[60]_{0})},
  \end{array}\label{[60S]mod1}
  \end{equation} 
  
    \begin{equation}  
  \begin{array}{llll}
 [A](t)=[40S]_{0}-[40S]-\dfrac{[eIF4F]_{0}[40S](t)}{[40S]+\dfrac{k_{2}}{k_{1}}}\\
 -\dfrac{[eIF4F]_{0}[60S]_{0}[40S](t)+[60S]_{0}([40S](t)-[40S]_{0})(k_{1}[40S](t)+k_{2})}{([40S](t)+\dfrac{k_{2}}{k_{1}})([40S](t)-[40S]_{0}-[60S]_{0})}.
  \end{array}\label{[A]mod1}
  \end{equation}

\noindent  Finally, the amount of protein synthesis is given below:  

\begin{equation}
\begin{array}{cc}
Psynth(t)=k_{3}\Huge([40S]_{0}-[40S]-\dfrac{[eIF4F]_{0}[40S](t)}{[40S]+\dfrac{k_{2}}{k_{1}}}\\
-\dfrac{[eIF4F]_{0}[60S]_{0}[40S](t)+[60S]_{0}([40S](t)-[40S]_{0})(k_{1}[40S](t)+k_{2})}{([40S](t)+\dfrac{k_{2}}{k_{1}})([40S](t)-[40S]_{0}-[60S]_{0})}\Huge).
\end{array} \label{proteinmod1}
\end{equation}

\noindent It can be proved that the slow manifold $\mathcal{M}_{0}$  is normally hyperbolic and stable. We assume that the functions $ G_{1}(y,z)$ is the left side of equation (\ref{fastmod1}) and $G_{2}(y,z)$ is the left side of equation (\ref{fastmod1}). This means

\begin{eqnarray}   
G_{1}(y,z)=x(1-y)-\alpha _{1}y, \\
G_{2}(y,z)=\alpha _{2}(1-x)-\alpha _{3}y+\alpha _{4}xz-(\alpha _{3}+\alpha _{4})z.
\end{eqnarray}

\noindent Then, the Jacobian matrix of $G_{1}$ and $G_{2}$ is given:\\

\begin{equation*} 
\mathbf{J}=\dfrac{\partial(G_{1},G_{2})}{\partial(y,z)}={\left(\begin{array}{cc}
\dfrac{\partial G_{1}}{\partial y}&\dfrac{\partial G_{1}}{\partial z}\\ \\ \dfrac{\partial G_{2}}{\partial y}&\dfrac{\partial G_{2}}{\partial z}
\end{array}\right)}
\end{equation*}
\begin{equation*}
\quad \quad ={\left(\begin{array}{cc}
	-(\alpha _{1}+x) &  0 \\ 
	-\alpha _{3} & \alpha _{4}x-\alpha _{4}-\alpha _{3} 
	\end{array}\right)}.
\end{equation*}

\noindent The characteristic equation \;\; $ det(J-\lambda I)=0$ \; \; can be solved analytically to find the eigenvalues of the Jacobian matrix. We obtained the following eigenvalues:

$$\lambda _{1}=-(\alpha _{1}+x)\; \text{and}\; \lambda _{2}=\alpha _{4}x-\alpha _{4}-\alpha_{3}.$$

\noindent It is clear that the first eigenvalue is negative. It means   $ \lambda _{1}<0 $  since\\
$ x=\dfrac{[40S]}{[40S]_{0}} > 0$ ; and $\alpha_{1}=\dfrac{k_{2}}{k_{1}[40S]_{0}}>0 $ ; the other  eigenvalue is also negative $ \lambda _{2} < 0 $ \; because $ \alpha _{3}=\dfrac{k_{3}[60S]_{0}}{k_{1}[40S]_{0}} $ ; and $ \alpha _{4}=\dfrac{k_{3}}{k_{1}} $,\; and by \eqref{cond1} $ k_{3}>>k_{1} $ and $ x\in[0,1] $.\; All eigenvalues here have negative real part $ Re( \lambda _{i} ) < 0,\; \text{for} \;\; i = 1,2$. Then by definition of normally hyperbolic the slow manifold  $ \mathcal{M}_{0} $ is stable.\\
\noindent And the approximate solutions of equations \eqref{reducemod1} for different values of the small parameter $\epsilon$  can be expressed in Figure \eqref{fig:approximate solution with manifold of xy,xz}. The approximate solutions are sufficiently close to $ \mathcal{M}_{0} $ when \; $ {\large \epsilon  }$ \; becomes smaller. We have compared the species concentrations of the reduced model \eqref{dx/dtau} and the full model (dimensionless form) in Figure \eqref{fig:x and xr}


	\begin{figure}[H]
	\begin{center}     
		\subfigure[$\rho=1,$ $\alpha_{1}=1,$ $\alpha_{2}=3,$ $\alpha_{3}=2,$ $\alpha_{4}=4$]{%
			\includegraphics[width=0.48\textwidth]{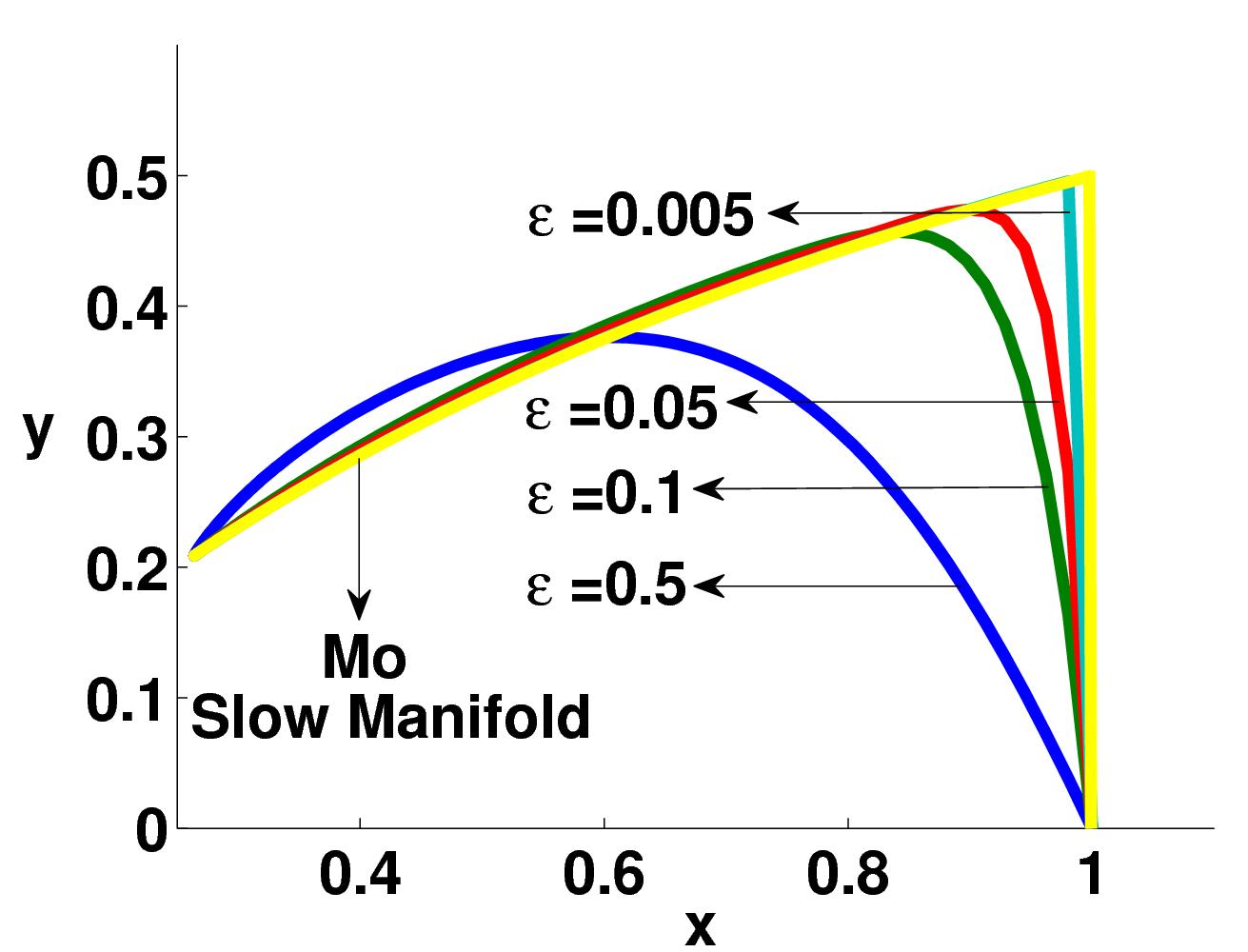}
		}		
		\subfigure[$\rho=0.001,$ $\alpha_{1}=2.2,$ $\alpha_{2}=3,$ $\alpha_{3}=2,$ $\alpha_{4}=4$]{%
			\includegraphics[width=0.48\textwidth]{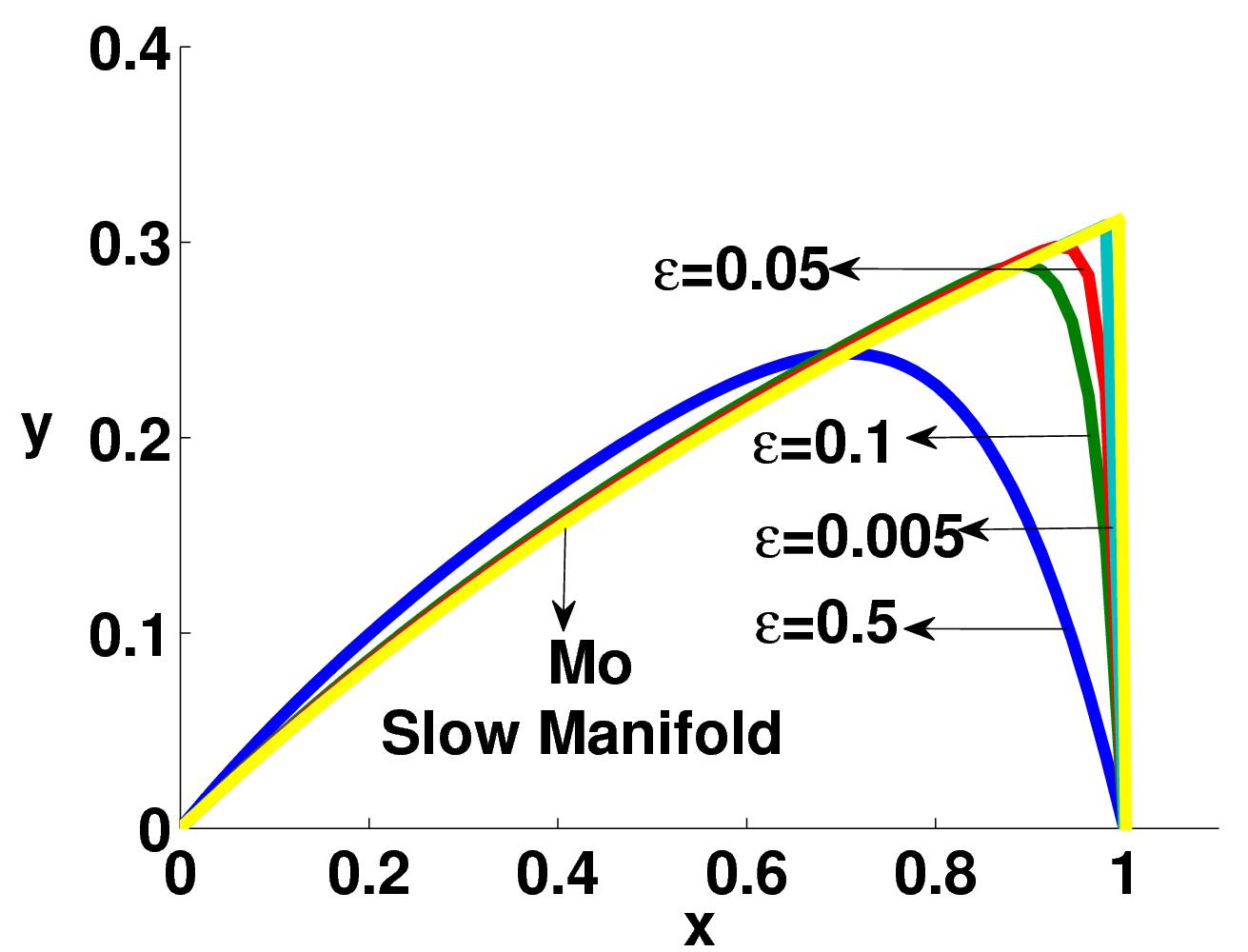}
		} \\
		\subfigure[$\rho=0.0000000001,$ $\alpha_{1}=0.0002,$\qquad \quad \quad $\alpha_{2}=25000000,$ $\alpha_{3}=0.25,$ $\alpha_{4}=500$]{%
			\includegraphics[width=0.48\textwidth]{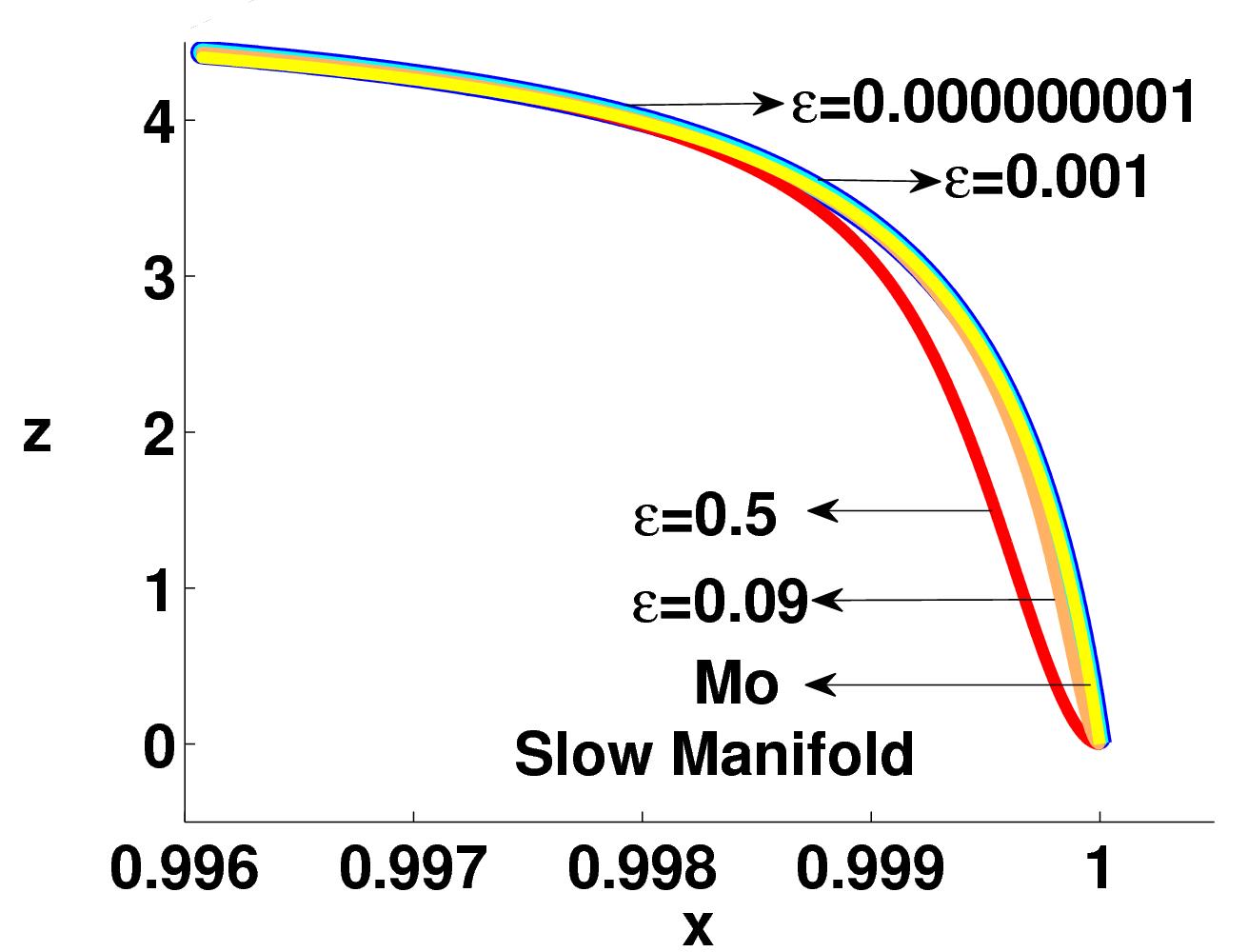}
		}
		\subfigure[$\rho=0.00000001,$ $\alpha_{1}=0.002,$ $\alpha_{2}=250,$\quad $\alpha_{3}=0.25,$ $\alpha_{4}=50$]{%
			\includegraphics[width=0.48\textwidth]{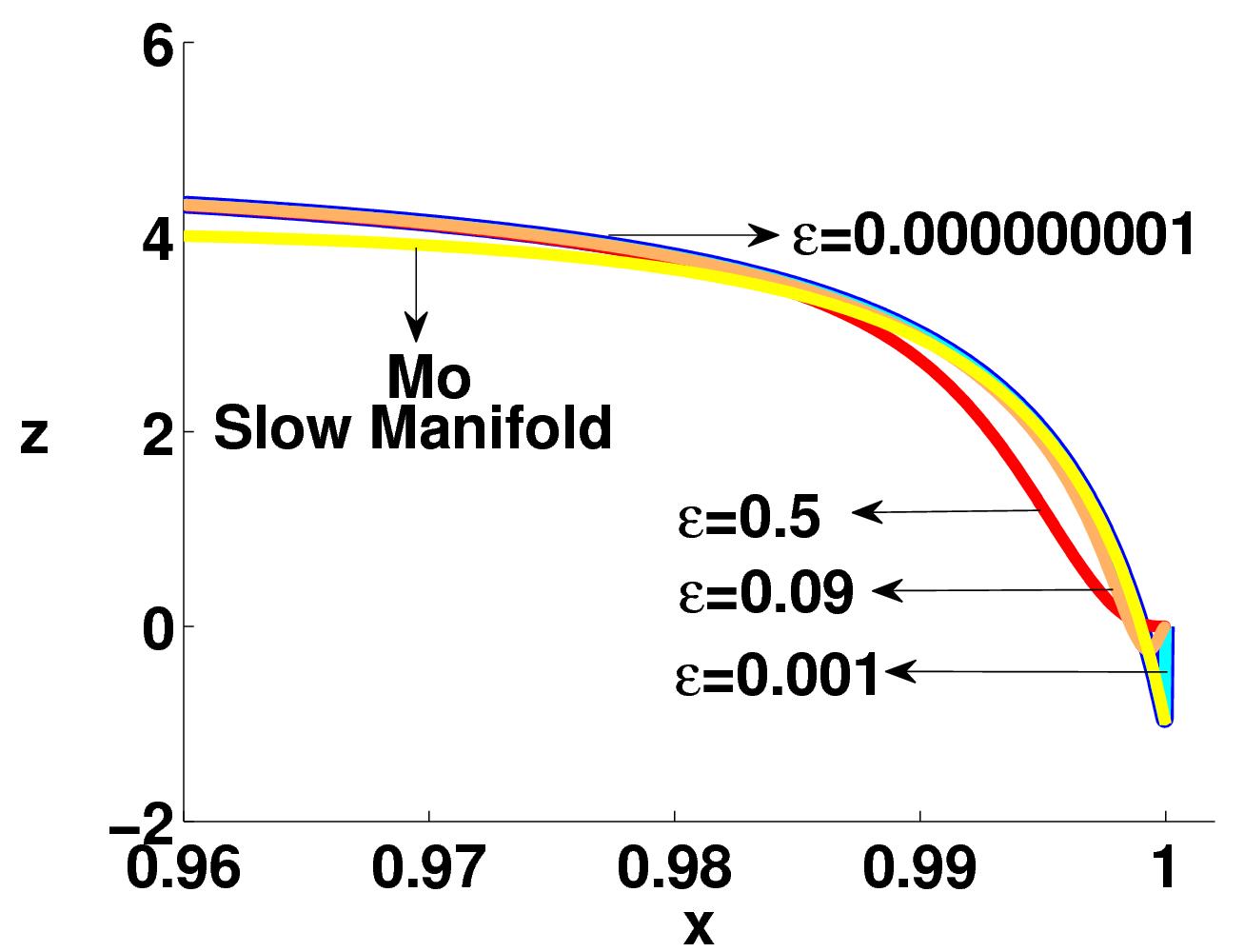}
		}
		
	\end{center}   
	\caption {Approximate solutions of equations \eqref{reducemod1} with slow manifold $\mathcal{M}_{0}$; the yellow lines are the slow manifolds and the other lines are the approximate solutions for different values of small parameter $\epsilon$, with the time interval $[0,20]$ for numerical simulations.}    	
	\label{fig:approximate solution with manifold of xy,xz}
\end{figure}


\begin{figure}[H]
	\begin{center}     
		\subfigure[$\rho=0.001,$ $\alpha_{1}=2.2,$ $\alpha_{2}=3,$ $\alpha_{3}=2,$ $\alpha_{4}=4$]{%
			\includegraphics[width=0.48\textwidth]{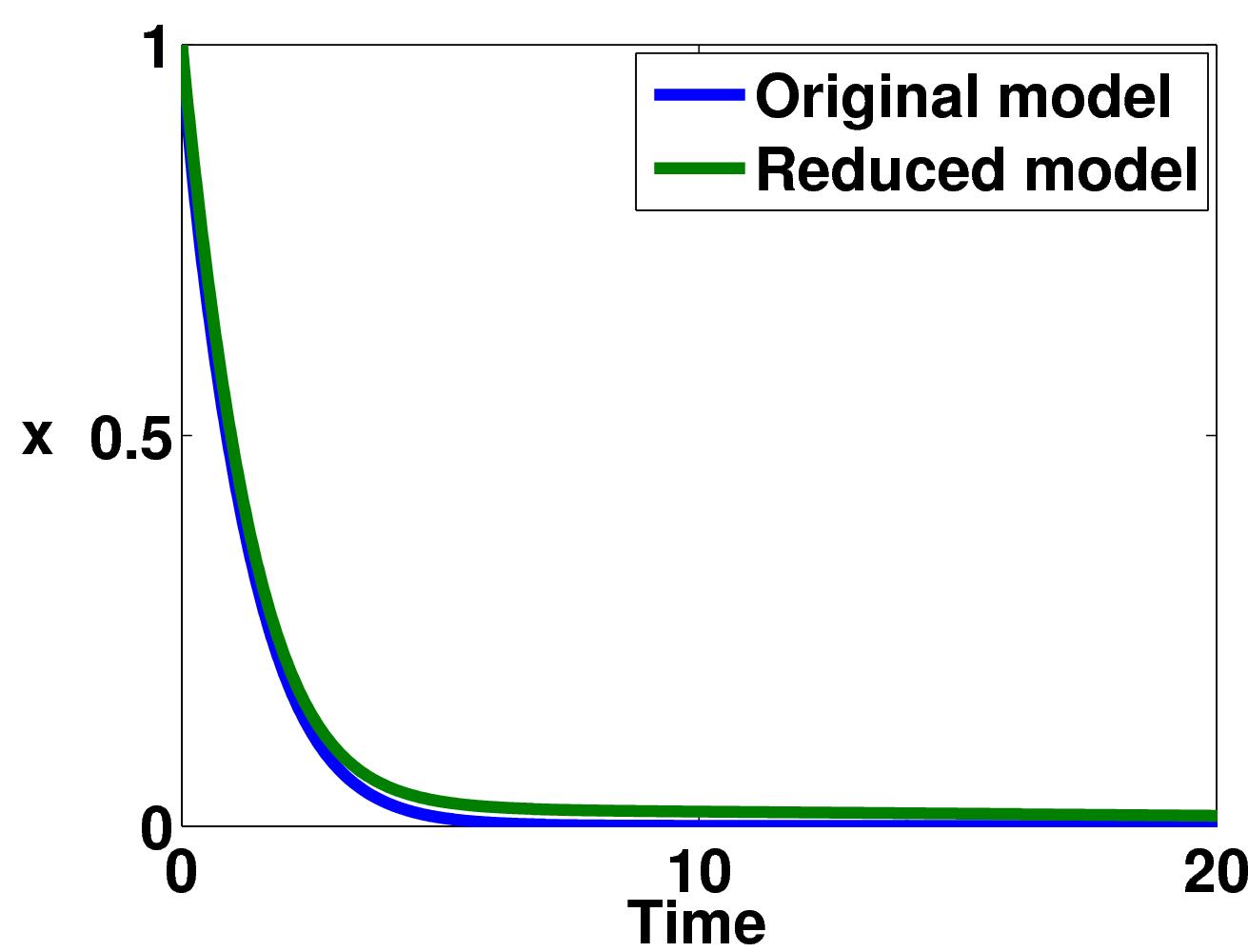}
		}
		\subfigure[$\rho=0.005,$ $\alpha_{1}=0.5,$ $\alpha_{2}=3,$ $\alpha_{3}=2,$ $\alpha_{4}=4$]{%
			\includegraphics[width=0.48\textwidth]{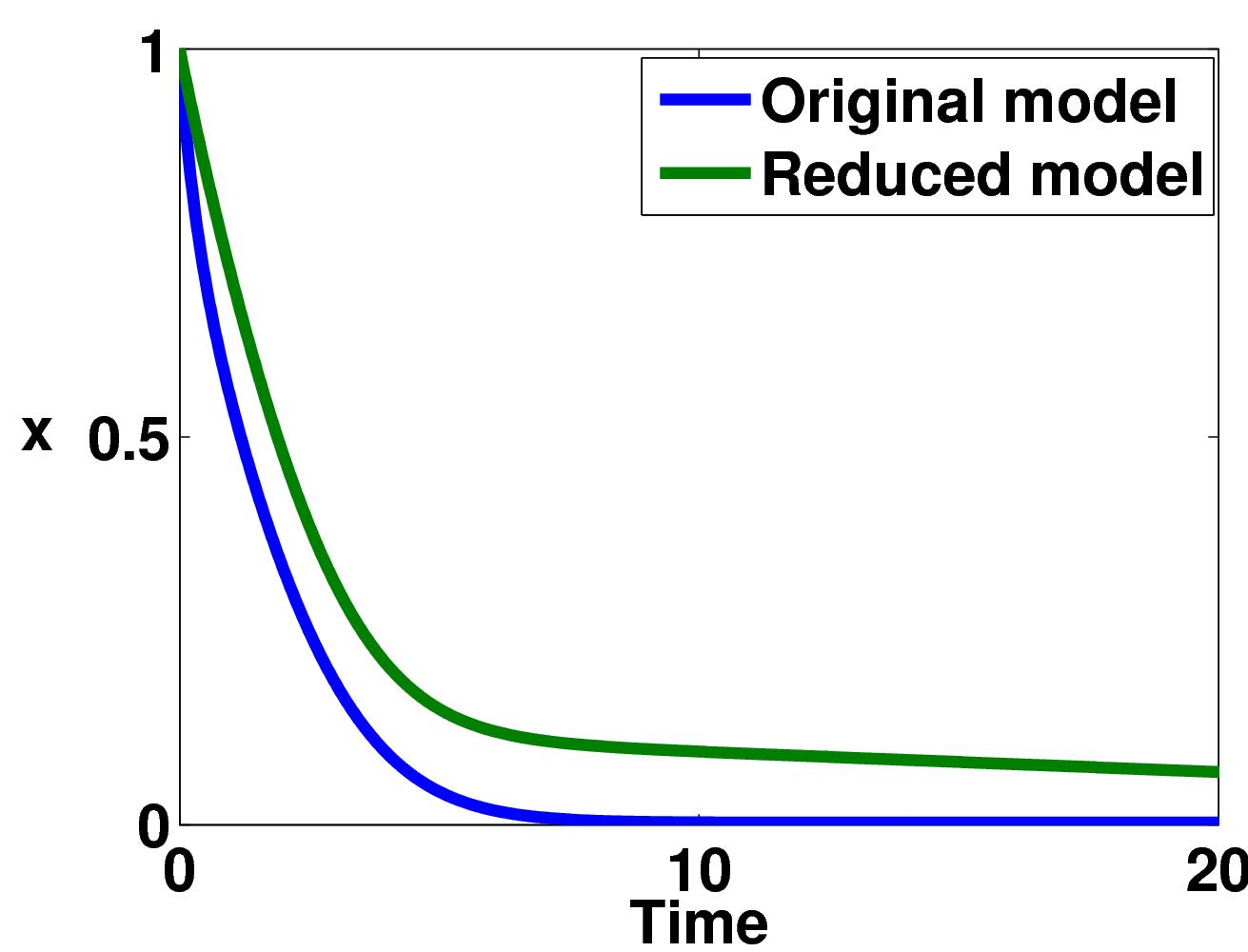}
		} 
		
	\end{center}   
	\caption {The full models (dimensionless form) (blue lines) and the reduced models (green lines) are compared in numerical simulation using the SBToolbox for Matlab;the full model \eqref{reducemod1}(dimensionless form) and reduced model \eqref{dx/dtau}.} 
	\label{fig:x and xr}   
\end{figure}

\section{Elasticity and Control Coefficients for microRNA Pathways}

Elasticity and control coefficients are an important technique to show the relationship between the fluxes and species concentrations with network parameters. Elasticity and control coefficients can be applied for linear and nonlinear models.\\

\noindent For the system \eqref{mod1}, we have the following reaction rates 

$$ v_{1}=k_{1} [40S] [eIF4F] ,\; v_{2}=k_{2}F ,\; v_{3}=k_{3}[60S]A , \; v_{4}=k_{4}R.   $$

\noindent The following elasticity equations are
calculated for the model equations \eqref{mod1} based on the equation \eqref{elastisity general equation}

\begin{eqnarray}
\begin{array}{cc}
E^{v_{1}}_{[40S]} =\dfrac{\partial v_{1}}{\partial [40S]} \dfrac{[40S]}{v_{1}}=k_{1}[eIF4F]\dfrac{[40S]}{k_{1}[40S][eIF4F]} =1, \\
E^{v_{2}}_{[40S]} =E^{v_{3}}_{[40S]} =E^{v_{4}}_{[40S]} =0,\\
E^{v_{2}}_{[eIF4F]} =E^{v_{3}}_{[eIF4F]} =E^{v_{4}}_{[eIF4F]} =0, \quad E^{v_{1}}_{[eIF4F]} =1, \\
E^{v_{1}}_{[60S]} =E^{v_{2}}_{[60S]} =E^{v_{4}}_{[60S]} =0, \quad E^{v_{3}}_{[60S]} =1,\\ 
E^{v_{1}}_{F} =E^{v_{3}}_{F} =E^{v_{4}}_{F} =0, \quad E^{v_{2}}_{F} =1, \\
E^{v_{1}}_{A} =E^{v_{2}}_{A} =E^{v_{4}}_{A} =0, \quad E^{v_{3}}_{A} =1, \\
E^{v_{1}}_{R} =E^{v_{2}}_{R} =E^{v_{3}}_{R} =0, \quad E^{v_{4}}_{R} =1. \\
\end{array} \label{elasticityresult}
\end{eqnarray}

\noindent In general, if an elasticity value is positive then an effector results in an increasing in reaction rate whereas if an elasticity value is negative then an effector results in a decrease in the reaction rate.

\noindent For the chemical reaction rate (\ref{mod1}), we assume that \;${\large [40S] } $\;,\;${\large [eIF4F]} $\;and \;${\large [60S]} $\; are fixed boundary species so that the pathway can reach a steady state. Then we have \;${\large F } $\;, \;${\large A } $\;and \;${\large R } $\; to find their control coefficients. The model has some control coefficient equations based on summation and connectivity theorem equations \eqref{summation theorem genaral equation} and \eqref{connectivity theorem general equation}, as below:

\begin{eqnarray}
\begin{array}{cccc}
C^{J}_{v_{1}}+C^{J}_{v_{2}}+C^{J}_{v_{3}}+C^{J}_{v_{4}}=1,\\
C^{F}_{v_{1}}+C^{F}_{v_{2}}+C^{F}_{v_{3}}+C^{F}_{v_{4}}=0,\\
C^{A}_{v_{1}}+C^{A}_{v_{2}}+C^{A}_{v_{3}}+C^{A}_{v_{4}}=0,\\
C^{R}_{v_{1}}+C^{R}_{v_{2}}+C^{R}_{v_{3}}+C^{R}_{v_{4}}=0,\\
C^{J}_{v_{1}}E^{v_{1}}_{F}+C^{J}_{v_{2}}E^{v_{2}}_{F}+C^{J}_{v_{3}}E^{v_{3}}_{F}+C^{J}_{v_{4}}E^{v_{4}}_{F}=0,\\
C^{J}_{v_{1}}E^{v_{1}}_{A}+C^{J}_{v_{2}}E^{v_{2}}_{A}+C^{J}_{v_{3}}E^{v_{3}}_{A}+C^{J}_{v_{4}}E^{v_{4}}_{A}=0,\\
C^{J}_{v_{1}}E^{v_{1}}_{R}+C^{J}_{v_{2}}E^{v_{2}}_{R}+C^{J}_{v_{3}}E^{v_{3}}_{R}+C^{J}_{v_{4}}E^{v_{4}}_{R}=0,\\
C^{F}_{v_{1}}E^{v_{1}}_{A}+C^{F}_{v_{2}}E^{v_{2}}_{A}+C^{F}_{v_{3}}E^{v_{3}}_{A}+C^{F}_{v_{4}}E^{v_{4}}_{A}=0 \quad \text{for}\quad F\neq A,\\
C^{F}_{v_{1}}E^{v_{1}}_{R}+C^{F}_{v_{2}}E^{v_{2}}_{R}+C^{F}_{v_{3}}E^{v_{3}}_{R}+C^{F}_{v_{4}}E^{v_{4}}_{R}=0 \quad \text{for}\quad F\neq R,\\
C^{A}_{v_{1}}E^{v_{1}}_{F}+C^{A}_{v_{2}}E^{v_{2}}_{F}+C^{A}_{v_{3}}E^{v_{3}}_{F}+C^{A}_{v_{4}}E^{v_{4}}_{F}=0 \quad \text{for} \quad A\neq F,\\
C^{A}_{v_{1}}E^{v_{1}}_{R}+C^{A}_{v_{2}}E^{v_{2}}_{R}+C^{A}_{v_{3}}E^{v_{3}}_{R}+C^{A}_{v_{4}}E^{v_{4}}_{R}=0 \quad \text{for} \quad A\neq R,\\
C^{R}_{v_{1}}E^{v_{1}}_{F}+C^{R}_{v_{2}}E^{v_{2}}_{F}+C^{R}_{v_{3}}E^{v_{3}}_{F}+C^{R}_{v_{4}}E^{v_{4}}_{F}=0  \quad \text{for} \quad R\neq F,\\
C^{R}_{v_{1}}E^{v_{1}}_{A}+C^{R}_{v_{2}}E^{v_{2}}_{A}+C^{R}_{v_{3}}E^{v_{3}}_{A}+C^{R}_{v_{4}}E^{v_{4}}_{A}=0  \quad \text{for}\quad R\neq A,\\
C^{F}_{v_{1}}E^{v_{1}}_{F}+C^{F}_{v_{2}}E^{v_{2}}_{F}+C^{F}_{v_{3}}E^{v_{3}}_{F}+C^{F}_{v_{4}}E^{v_{4}}_{F}=-1 \quad \text{for}\quad n=m,\\
C^{A}_{v_{1}}E^{v_{1}}_{A}+C^{A}_{v_{2}}E^{v_{2}}_{A}+C^{A}_{v_{3}}E^{v_{3}}_{A}+C^{A}_{v_{4}}E^{v_{4}}_{A}=-1 \quad \text{for}\quad n=m,\\
C^{R}_{v_{1}}E^{v_{1}}_{R}+C^{R}_{v_{2}}E^{v_{2}}_{R}+C^{R}_{v_{3}}E^{v_{3}}_{R}+C^{R}_{v_{4}}E^{v_{4}}_{R}=-1 \quad \text{for}\quad n=m.
\end{array}  \label{controlresult} 
\end{eqnarray}

\noindent By substituting the elasticity values equations \eqref{elasticityresult}  in to the equations \eqref{controlresult}, the following results are given   

\begin{eqnarray*}
\begin{array}{cc}
C^{J}_{v_{1}}=1, \quad C^{J}_{v_{2}}=C^{J}_{v_{3}}=C^{J}_{v_{4}}=0,\\
C^{F}_{v_{1}}=1, \; C^{F}_{v_{2}}=-1, \quad C^{F}_{v_{3}}=C^{F}_{v_{4}}=0,\\
C^{A}_{v_{1}}=1, \; C^{A}_{v_{3}}=-1, \quad C^{A}_{v_{2}}=C^{A}_{v_{4}}=0,\\
C^{R}_{v_{1}}=1, \; C^{R}_{v_{4}}=-1, \quad C^{R}_{v_{2}}=C^{R}_{v_{3}}=0.
\end{array}
\end{eqnarray*} 

\noindent According to the flux control coefficients $\; C^{J}_{v_{2}}=C^{J}_{v_{3}}=C^{J}_{v_{4}}=0 \;$, this means that second, third and the last step of reactions have not any effect on model fluxes. On the other hand, the control coefficient  $\; C^{J}_{v_{1}}=1 \;$ this give us the first reaction rate has a strong affect on the model fluxes. In other words, the model steady state fluxes are controlled by $v_{1}$.

\noindent Furthermore, concentration control coefficients quantify how variables, such as species concentrations, depend on reaction rates. In this case study, it can be more precisely concluded that there are no any relative change in F, A and R regarding to reaction rates $\; v_{2},v_{3} \;$ and $\;v_{4}.\;$  While, there is a significant change in F, A and R with respect to $\;v_{1}.\;$        

\section{Results and Discussions}
The non-linear model of miRNA protein translation has been studied that includes seven species and four parameters. Mass action law and classical chemical kinetics under constant rates are used for modelling the system. We introduce some new variables in order to reduce the number of model species and parameters. We propose QSSA to the model to analyze the fast variables and calculate slow manifolds. As a result, the analytical approximate solutions are sufficiently close to the manifolds when the given parameter $\Huge{ \epsilon }$ becomes smaller. The analytical approximate solutions here give some useful understanding about the model particularly it provides us some understanding about global dynamics. It can be also noticed that there is a good agreement between the simplified and original model dynamics.     
Results in this study show some interesting points. The first point is that how variables, such as fluxes and species concentrations, depend on network parameters. Another point is that how sensitive a reaction rate is to changes in reactant, product and effector concentrations. The proposed techniques here will be applied to a wide range of complex mRNA mechanisms. 

\include{chapter[3]}
\chapter{Chemical Reaction Networks} 

\section{Introduction of Enzyme Mechanisms}
\begin{figure}[H]  
	\begin{center}             
		\subfigure{%
			\includegraphics[width=0.9\textwidth]{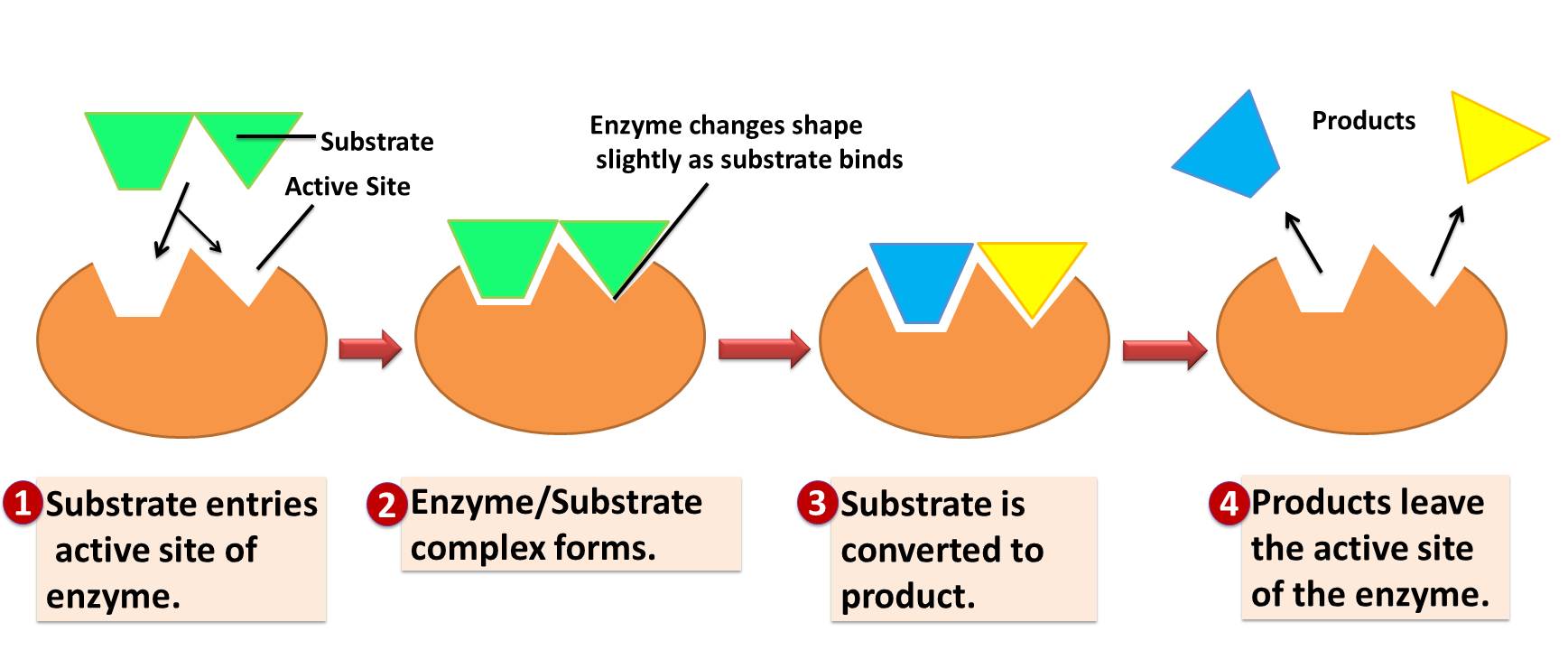}
 		}             
	\end{center}    
	\caption {According to the induced-fit model, both enzyme and substrate undergo dynamic conformational changes upon binding. The enzyme contorts the substrate into its transition state, thereby increasing the rate of the reaction.}    
 \label{enzyme work figure}
\end{figure}

The main catalyst in the enzyme reaction is an enzyme which is a protein molecule. It is clear enzymes are produced by living cells (plant, animal, and microorganism) and are absolutely required as catalysts in biochemical reactions. The specific enzyme is required in a cell for almost all reaction. A major function of enzymes in a living system is to catalyze the breaking and making of chemical bonds. Therefore, like any other catalysts, they have an important role in increasing the rate of the reaction without themselves go through permanent chemical changes. A particular protein construction helps the process of catalyst in the enzyme reactions. There is a surface in the enzyme is known as the active site which is reactions are catalyzed on that place. This site it is very important in enzymatic reactions. The main reason is that at the active site there are some chemical and physical reactions worked as catalyst for some known enzymes (Lee, 2001). \\
To understanding how enzyme works see Figure \eqref{enzyme work figure} (College, 2013).

\begin{figure}[H]  
	\begin{center}             
		\subfigure{%
			\includegraphics[width=1\textwidth]{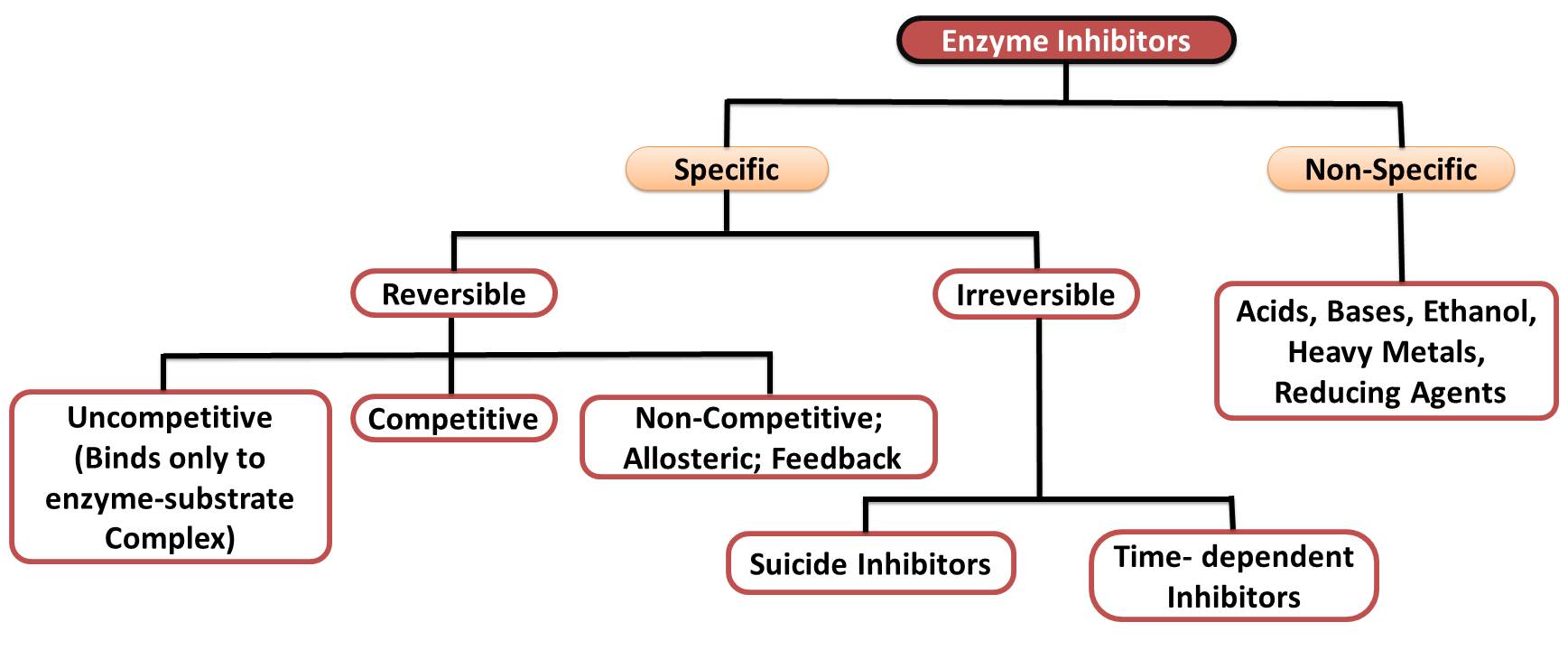}
 		}             
	\end{center}    
	\caption {A general classification of enzyme inhibitors.}
 \label{enzyme diagram figure}
\end{figure}

Even though enzymes are absolutely important for life, abnormally high enzyme activity can cause disease conditions. Hence, too active enzymes are appealing targets for the development of inhibitor molecules to diminish disease conditions (Chandra Mohan, 2013). 
For breaking the enzyme activity, there is a chemical compound which is called Enzyme inhibition.
An important branch of studying enzymatic reactions is enzyme inhibition. This is occurred in any organic or inorganic chemical compound that shared their molecules with enzyme active site. It is clear that the majority of inhibitors have same role in interactions with enzymes (Sharma, 2012).\\
Almost all enzyme inhibitors have less molecular compounds. They can be combined with enzymes in order to have enzyme-inhibitor complex. They have a great role in decreasing the reaction rates of enzymes or inhibiting the enzyme activity.     
To block the entry of substrates to the active site, the inhibitors binding with the active site of the enzyme. As an alternative, a few inhibitors can bind to a site other than the active site and cause to arise form change that prevents the entry of substrate to the active site. Based on the kind of interplay with the enzyme, inhibitor binding can be categorized as either reversible or irreversible in the Figure \eqref{enzyme diagram figure} (Chandra Mohan, 2013). 

\begin{figure}[H]  
	\begin{center}             
		\subfigure{%
			\includegraphics[width=0.9\textwidth]{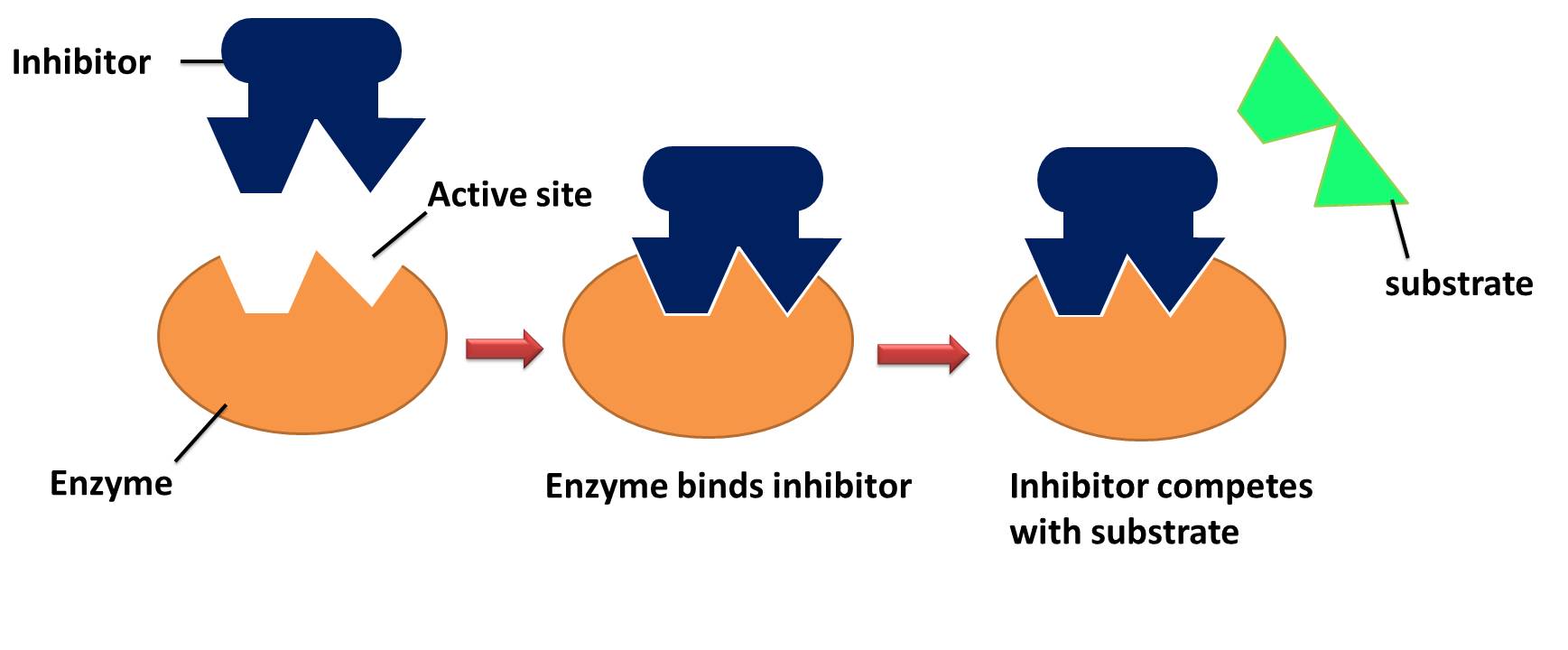}
 		}             
	\end{center}    
	\caption {Demonstrates how competitive inhibition of enzymes work.}
\label{inhibitor figure}
\end{figure}

Inhibitors play important roles in biological and clinical research.
It has also an affecting role in the reaction network such as:
Inhibitors serve as many control mechanisms in biological systems, they can regulate metabolic activities, they either block or reduce the rate of biochemical reactions, Reversible inhibitors can be used duration enzyme purification, and covalent inhibitors are commonly used to identify active site amino acids$...$etc.\\
To understanding how inhibitor with enzyme works see Figure  \eqref{inhibitor figure}.

Enzymatic activity depends on a number of factors. The most important factors that affect enzyme activity are: enzyme concentration, the amount of specific enzyme substrate, temperature, pH of the reaction medium, and the presence of activators and inhibitors (Chandra Mohan, 2013). 

There is also another enzyme that is called enzyme activators. They are molecules that bind to enzymes and increase their activity, it is working as opposite to inhibitor.
The oldest approach to realize enzyme mechanisms, and the one that remains most important, is to calculate the rate of a reaction and how it changes in response to changes in experimental parameters, a branch of knowledge known as enzyme kinetics (Szarka and Kiado, 2014).


\section{Reversible Enzyme Reactions}
Enzymatic reactions with two complex components are given bellow
\begin{equation}  
\begin{array}{llll}
E+S \underset{k^{b}_{1}}{ \overset{k^{f}_{1}}{\rightleftharpoons}} ES 
\underset{k^{b}_{2}}{ \overset{k^{f}_{2}}{\rightleftharpoons}} EP 
\underset{k^{b}_{3}}{ \overset{k^{f}_{3}}{\rightleftharpoons}} E+P
\end{array}\label{chemical reaction 1}
\end{equation} 
where $E, S, ES ,EP$ and $P$ are enzyme, substrate, enzyme--substrate complex,enzyme--product complex and product, respectively. The parameters $ k^{f}_{1}, k^{b}_{1},k^{f}_{2} ,k^{b}_{2},k^{f}_{3} $ and $k^{b}_{3}$ are kinetic constants. We define the concentrations of the species  $e=[E], s=[S], p=[P], c_{1}=[ES],c_{2}=[EP]$  (Wong et al., 2015). The reactions (\ref{chemical reaction 1}) can be expressed as a system of ODEs:

\begin{equation}  
\begin{array}{llll}
\dfrac{ds}{dt}=-k^{f}_{1} es+k^{b}_{1} c_{1},\\
\dfrac{de}{dt}=-k^{f}_{1} es+k^{b}_{1} c_{1}+k^{f}_{3} c_{2}-k^{b}_{3} pe, \\
\dfrac{dc_{1}}{dt}=k^{f}_{1} es-k^{b}_{1} c_{1}-k^{f}_{2} c_{1}+k^{b}_{2} c_{2},\\
\dfrac{dc_{2}}{dt}=k^{f}_{2} c_{1}-k^{b}_{2} c_{2}-k^{f}_{3} c_{2}+k^{b}_{3} pe,\\
\dfrac{dp}{dt}=k^{f}_{3} c_{2}-k^{b}_{3} pe,
\end{array}\label{system QEA 1}  
\end{equation} 

\noindent with the initial conditions 
\begin{equation}  
\begin{array}{llll}
e(0)=e_{0}, s(0)=s_{0}, c_{1}(0)=c_{2}(0)=p(0)=0.
\end{array}\label{initial QEA 1}
\end{equation} 
\noindent The system \eqref{system QEA 1} has two independent stoichiometric conservation laws:
\begin{equation}  
\begin{array}{llll}
e+c_{1}+c_{2}=e_{0}, \quad
s+p+c_{1}+c_{2}=s_{0}.
\end{array}\label{consrvatin QEA 1}
\end{equation} 
\noindent By applying QEA method for chemical reactions \eqref{chemical reaction 1}, if possible suppose that the first reaction ($E+S \underset{k^{b}_{1}}{ \overset{k^{f}_{1}}{\rightleftharpoons}} ES  $) becomes quasi--equilibrium when the equilibrium is fast: let  $k^{f}_{1}=\dfrac{k^{+}}{\epsilon}$ and $k^{b}_{1}=\dfrac{k^{-}}{\epsilon}$ where $ \epsilon=\dfrac{e_{0}}{s_{0}}$ then $k^{+}=\dfrac{k^{f}_{1} e_{0}}{s_{0}}$ and $k^{-}=\dfrac{k^{b}_{1} e_{0}}{s_{0}}$ \;(i.e. $k^{f}_{1}$ and $k^{b}_{1}$ are called large parameters compared to $k^{f}_{2},k^{b}_{2},k^{f}_{3}$ and $ k^{b}_{3}$). Then, the system \eqref{system QEA 1} has the form of equation (\ref{general QEA 1})
\begin{equation}  
\begin{array}{llll}
\dfrac{ds}{dt}=\dfrac{1}{\epsilon}g^{f}(s,e,c_{1},t),\\
\dfrac{de}{dt}=\dfrac{1}{\epsilon}g^{f}(s,e,c_{1},t)+g^{s}_{1}(e,c_{2},p,t),\\
\dfrac{dc_{1}}{dt}=\dfrac{-1}{\epsilon}g^{f}(s,e,c_{1},t)+g^{s}_{2}(c_{1},c_{2},t),\\
\dfrac{dc_{2}}{dt}=-g^{s}_{2}(c_{1},c_{2},t)-g^{s}_{1}(e,c_{2},p,t),\\
\dfrac{dp}{dt}=g^{s}_{1}(e,c_{2},p,t),
\end{array}\label{reduce QEA 1}
\end{equation} 
where $g^{f}(s,e,c_{1},t)=-k^{+}es+k^{-}c_{1}$ ,\; $g^{s}_{1}(e,c_{2},p,t)=k^{f}_{3} c_{2}-k^{b}_{3}pe$ \;and \\$ g^{s}_{2}(c_{1},c_{2},t)=-k^{f}_{2} c_{1}+k^{b}_{2} c_{2} $.\\ 
For $\epsilon \longrightarrow 0,$ the quasi--equilibrium approximation can be applied. Three species $S, E$ and $ES$ are participate in the fast reaction, while other components of chemical reactions are not involved in analysis of the QEA manifold. Based on the equation \eqref{general fast part QEA 1}, then the equation \eqref{reduce QEA 1} takes the form:
\begin{equation}  
\begin{array}{llll}
\dfrac{ds}{dt}=\dfrac{1}{\epsilon}(-k^{+}es+k^{-}c_{1}),\\
\dfrac{de}{dt}=\dfrac{1}{\epsilon}(-k^{+}es+k^{-}c_{1}),\\
\dfrac{dc_{1}}{dt}=\dfrac{-1}{\epsilon}(-k^{+}es+k^{-}c_{1}).
\end{array}\label{reduce fast part QEA 1}
\end{equation} 
\noindent Therefore, we obtain two slow variables which are the stoichiometric conservation laws of the fast reaction. The variables are $b_{1}(s,c_{1})=s+c_{1}$ and $b_{2}(e,c_{1})=e+c_{1}$. The first slow variable here is the sum of the free substrate and the enzyme--substrate complex, while the second one is the total amount of enzyme. More precisely, slow variables should be invariant with respect to fast motion in order to apply the Tikhnove theorem.   \\
The slow manifold for the reaction kinetics can be calculated from the algebraic equation $g^{f}(s,e,c_{1},t)=0$. This is given by  
\begin{equation}  
\begin{array}{llll}         
\mathcal{M}_{0}^{*}=\bigg \lbrace (s,e,c_{1})\in \mathbb{R}^{3}: s=\dfrac{k^{-}c_{1}}{k^{+}e} \bigg\rbrace. 
\end{array}\label{manifold QEA 1}
\end{equation}
By fixing the slow variables ($b_{1}$ and $b_{2}$), and finding the variables $s,e$ and $c_{1}$, the system (\ref{general fast part QEA 1 =0}) takes the form  
\begin{equation}  
\begin{array}{llll}
k^{+}es-k_{-}c_{1}=0,\\
s+c_{1}=b_{1},\\ 
e+c_{1}=b_{2}.
\end{array}\label{algebraic QEA 1}
\end{equation}
A quadratic equation for $c_{1}$ is obtained by using $s=b_{1}-c_{1}$ and $e=b_{2}-c_{1}$
\begin{equation}  
\begin{array}{llll}
k^{+}c_{1}^{2}-(k^{+}b_{1}+k^{+}b_{2}+k^{-})c_{1}+k^{+}b_{1}b_{2}=0.
\end{array}\label{eq c1 of QEA 1}
\end{equation}
Then equation (\ref{eq c1 of QEA 1}) can be solved analytically for $c_{1}$\\
$$c_{1}(b_{1},b_{2})=\dfrac{1}{2}\Bigg[\Big (b_{1}+b_{2}+\dfrac{k^{-}}{k^{+}}\Big)\pm \sqrt{\Big(b_{1}+b_{2}+\dfrac{k^{-}}{k^{+}} \Big)^{2}-4b_{1}b_{2}}\Bigg].$$\\
We select ``--'' for providing positive concentrations of $s,c_{1}$ and $e$, and obtaining a proper asymptotic behavior of the fast reaction: if $b_{1}\rightarrow 0$ and $b_{2}\rightarrow 0$ then $c_{1}\rightarrow 0$. Furthermore, other variables ($s$ and $e$) are obtained:\\
\quad \quad \quad $s(b_{1},b_{2})=b_{1}-\dfrac{1}{2}\Bigg[\Big (b_{1}+b_{2}+\dfrac{k^{-}}{k^{+}}\Big)- \sqrt{\Big(b_{1}+b_{2}+\dfrac{k^{-}}{k^{+}} \Big)^{2}-4b_{1}b_{2}} \Bigg]$,\\
\quad \quad \quad $e(b_{1},b_{2})=b_{2}-\dfrac{1}{2}\Bigg[\Big (b_{1}+b_{2}+\dfrac{k^{-}}{k^{+}}\Big)- \sqrt{\Big(b_{1}+b_{2}+\dfrac{k^{-}}{k^{+}} \Big)^{2}-4b_{1}b_{2}}\Bigg]$.\\

\noindent To simplify the value of our variables, we have to bring a non--linear parts in equation \eqref{eq c1 of QEA 1} to zero by using an assumption that the concentration of substrate is present in large excess compared to the total concentration of enzyme.

 \begin{equation}  
\begin{array}{llll}  
[S]\gg [ES] \quad i.e. \quad b_{1}\gg c.
\end{array}\label{condition s>>e QEA 1}  
\end{equation} 
By using condition (\ref{condition s>>e QEA 1}), equation (\ref{eq c1 of QEA 1}) is simplified as follows:
\begin{equation}  
\begin{array}{llll}
\Big(1+\dfrac{b_{2}}{b_{1}}+\dfrac{k^{-}}{k^{+}b_{1}}\Big)c_{1}=b_{2}+\mathcal{O}\Big(\dfrac{c_{1}}{b_{1}}\Big) .
\end{array}\label{eq c1 with cond. QEA1 }  
\end{equation}
The approximation of $c_{1}$ becomes
\begin{equation}  
\begin{array}{llll}
c_{1} \approx \dfrac{b_{1}b_{2}}{b_{1}+b_{2}+k^{*}} ,
\end{array}\label{value c1 appoximate QEA 1}  
\end{equation}
where $k^{*}=k^{-}/k^{+}$. The new form of the variables $s$ and $e$ become\\
\quad \quad \quad $s(b_{1},b_{2})=\dfrac{b_{1}(b_{1}+k^{*})}{b_{1}+b_{2}+k^{*}}$ and $e(b_{1},b_{2})=\dfrac{b_{2}(b_{2}+k^{*})}{b_{1}+b_{2}+k^{*}}$.

\section{Non-Competitive Inhibition Enzymatic Reactions}

The chemical reaction networks for non-competitive inhibition enzymatic reactions are defined bellow

\begin{equation}
\begin{array}{llll}
\quad \quad E\; \; +\; \; S \quad  \underset{k_{-1}}{ \overset{k_{1}}{\rightleftharpoons}} \quad ES \quad
\overset{k_{2}}{\longrightarrow} \quad E+P\\
\quad\quad + \quad\quad\quad\quad\quad\quad\quad + \\
\quad\quad\; I \quad\quad\quad\quad\quad\quad\quad\; I\\
\;k_{-3}\upharpoonleft\downharpoonright k_{3} \quad\quad\quad\;\;\; k_{-4}\upharpoonleft\downharpoonright k_{4}\\
\quad\quad EI\; \; +\; S \quad  \underset{k_{-5}}{ \overset{k_{5}}{\rightleftharpoons}} \quad ESI\;\;
\overset{k_{6}}{\longrightarrow } \quad EI+P
\end{array}\label{chemical reaction 2.1}
\end{equation} 
\noindent All reactions of the above network are simply given bellow
\begin{equation*}
\begin{array}{llll}
E+S \underset{k_{-1}}{ \overset{k_{1}}{\rightleftharpoons}} ES ,\\
ES \overset{k_{2}}\longrightarrow E+P,\\
E+I \underset{k_{-3}}{ \overset{k_{3}}{\rightleftharpoons}} EI ,\\
ES+I \underset{k_{-4}}{ \overset{k_{4}}{\rightleftharpoons}} ESI ,\\
EI+S \underset{k_{-5}}{ \overset{k_{5}}{\rightleftharpoons}} ESI ,\\
ESI \overset{k_{6}}\longrightarrow EI+P,
\end{array}
\end{equation*}

\noindent where $E, S, I, ES ,EI, ESI$ and $P$ are enzyme, substrate,inhibitor,  enzyme--substrate complex,enzyme--inhibitor complex,enzyme--substrate--inhibitor complex and product, respectively. The parameters $ k_{1}, k_{-1},k_{2} ,k_{3},k_{-3}, k_{4}, k_{-4}, k_{5}, k_{-5} $ and $k_{6}$ are kinetic constants (Chow et al., 2016). We define the concentrations of the species  $e=[E], s=[S],i=[I], p=[P], c_{1}=[ES],c_{2}=[EI]$ and $ c_{3}=[ESI]$. The reactions (\ref{chemical reaction 2.1}) can be expressed as a system of ODEs:

\begin{equation}  
\begin{array}{llll}
\dfrac{ds}{dt}=-k_{1}es+k_{-1}c_{1}-k_{5}c_{2}+k_{-5}c_{3},\\
\dfrac{de}{dt}=-k_{1}es+k_{-1}c_{1}-k_{3}ei+k_{-3}c_{2}+k_{2}c_{1}, \\
\dfrac{di}{dt}=-k_{3}ei+k_{-3}c_{2}-k_{4}c_{1}i+k_{-4}c_{3}, \\
\dfrac{dc_{1}}{dt}=k_{1}es-k_{-1}c_{1}-k_{4}c_{1}i+k_{-4}c_{3}-k_{2}c_{1},\\
\dfrac{dc_{2}}{dt}=k_{3}ei-k_{-3}c_{2}-k_{5}c_{2}s+k_{-5}c_{3}+k_{6}c_{3},\\
\dfrac{dc_{3}}{dt}=k_{5}c_{2}s-k_{-5}c_{3}+k_{4}c_{1}i-k_{-4}c_{3}-k_{6}c_{3},\\
\dfrac{dp}{dt}=k_{2}c_{1}+k_{6}c_{3},
\end{array}\label{system QEA 2.1}  
\end{equation} 
\noindent with the initial conditions 
\begin{equation}  
\begin{array}{llll}
e(0)=e_{0}, s(0)=s_{0}, i(0)=i_{0}, c_{1}(0)=c_{2}(0)=c_{3}(0)=p(0)=0.
\end{array}\label{initila cond QEA 2.1}
\end{equation} 
\noindent The system (\ref{system QEA 2.1}) has three independent stoichiometric conservation laws:
\begin{equation}  
\begin{array}{llll}
e+c_{1}+c_{2}+c_{3}=e_{0},\\
s+p+c_{1}+c_{3}=s_{0},\\
i+c_{2}+c_{3}=i_{0}.
\end{array}\label{conservation of QEA 2.1}
\end{equation} 

\noindent In order to simplify the model equations \eqref{chemical reaction 2.1}, we assume three different cases for fast reactions as they are given bellow:\\
\subsection{Case One}
Wen assume that the model network has only one fast reversible reaction. By applying QEA method for chemical reactions \eqref{chemical reaction 2.1}, if possible suppose that the first reaction ($E+S \underset{k_{-1}}{ \overset{k_{1}}{\rightleftharpoons}} ES $) becomes quasi--equilibrium when the equilibrium is fast: let  $k_{1}=\dfrac{k^{+}_{1}}{\epsilon}$ and $k_{-1}=\dfrac{k^{-}_{1}}{\epsilon}$ where $ \epsilon=\dfrac{e_{0}}{s_{0}}$ then $k^{+}_{1}=\dfrac{k_{1} e_{0}}{s_{0}}$ and $k^{-}_{1}=\dfrac{k_{-1} e_{0}}{s_{0}}$ \;(i.e. $k_{1}$ and $k_{-1}$ are called large parameters compared to $ k_{2}, k_{3} , k_{-3}, k_{4} , k_{-4}, k_{5} , k_{-5} $ and $ k_{6}$). Then, the system \eqref{system QEA 2.1} has the form of equation (\ref{general QEA 1})

\begin{equation}  
\begin{array}{llll}
\dfrac{ds}{dt}=\dfrac{1}{\epsilon}g^{f}(s,e,c_{1},t)+g^{s}_{1}(s,c_{2},c_{3},t),\\
\dfrac{de}{dt}=\dfrac{1}{\epsilon}g^{f}(s,e,c_{1},t)+g^{s}_{2}(e,i,c_{2},t)+g^{s}_{3}(c_{1},t),\\
\dfrac{di}{dt}=g^{s}_{2}(e,i,c_{2},t)+g^{s}_{4}(i,c_{1},c_{3},t),\\
\dfrac{dc_{1}}{dt}=\dfrac{-1}{\epsilon}g^{f}(s,e,c_{1},t)+g^{s}_{4}(i,c_{1},c_{3},t)-g^{s}_{3}(c_{1},t),\\
\dfrac{dc_{2}}{dt}=-g^{s}_{2}(e,i,c_{2},t)+g^{s}_{1}(s,c_{2},c_{3},t)+g^{s}_{5}(c_{3},t),\\
\dfrac{dc_{3}}{dt}=-g^{s}_{1}(s,c_{2},c_{3},t)-g^{s}_{4}(i,c_{1},c_{3},t)-g^{s}_{5}(c_{3},t),\\
\dfrac{dp}{dt}=g^{s}_{3}(c_{1},t)+g^{s}_{5}(c_{3},t),
\end{array}\label{reduce QEA 2.1}
\end{equation} 
where $g^{f}(s,e,c_{1},t)=-k^{+}_{1}es+k^{-}_{1}c_{1}$ ,\; $g^{s}_{1}(s,c_{2},c_{3},t)=-k_{5}c_{2}s+k_{-5}c_{3} ,\\ g^{s}_{2}(e,i,c_{2},t)=-k_{3}ei+k_{-3}c_{2},\; g^{s}_{3}(c_{1},t)=k_{2}c_{1} ,\; g^{s}_{4}(i,c_{1},c_{3},t)=-k_{4}c_{1}i+k_{-4}c_{3}\; $ and $\; g^{s}_{5}(c_{3},t)=k_{6}c_{3} $.\\ 
For $\epsilon \longrightarrow 0,$ the quasi--equilibrium approximation can be applied. Three species $S, E$ and $ES$ are participate in the fast reaction, while other components of chemical reactions are not involved in analysis of the QEA manifold. Based on the equation \eqref{general fast part QEA 1}, then the equation \eqref{reduce QEA 2.1} takes the form:
\begin{equation}  
\begin{array}{llll}
\dfrac{ds}{dt}=\dfrac{1}{\epsilon}(-k^{+}_{1}es+k^{-}_{1}c_{1}),\\
\dfrac{de}{dt}=\dfrac{1}{\epsilon}(-k^{+}_{1}es+k^{-}_{1}c_{1}),\\
\dfrac{dc_{1}}{dt}=\dfrac{-1}{\epsilon}(-k^{+}_{1}es+k^{-}_{1}c_{1}).
\end{array}\label{reduce fast part QEA 2.1}
\end{equation} 

\noindent Therefore, we obtain two slow variables which are the stoichiometric conservation laws of the fast reaction. The variables are $b_{1}(s,c_{1})=s+c_{1}$ and $b_{2}(e,c_{1})=e+c_{1}$. The first slow variable here is the sum of the free substrate and the enzyme--substrate complex, while the second one is the total amount of enzyme. More precisely, slow variables should be invariant with respect to fast motion in order to apply the Tikhnove theorem.   \\
The slow manifold for the reaction kinetics can be calculated from the algebraic equation $g^{f}(s,e,c_{1},t)=0$. This is given by  
\begin{equation}  
\begin{array}{llll}         
\mathcal{M}_{0}^{*}=\bigg \lbrace (s,e,c_{1})\in \mathbb{R}^{3}: s=\dfrac{k^{-}_{1}c_{1}}{k^{+}_{1}e} \bigg\rbrace. 
\end{array}\label{manifold QEA 2.1} 
\end{equation}
By fixing the slow variables ($b_{1}$ and $b_{2}$), and finding the variables $s,e$ and $c_{1}$, the system (\ref{general fast part QEA 1 =0}) takes the form  
\begin{equation}  
\begin{array}{llll}
k^{+}_{1}es-k^{-}_{1}c_{1}=0,\\
s+c_{1}=b_{1},\\ 
e+c_{1}=b_{2}.
\end{array}\label{algebraic QEA 2.1}
\end{equation}
A quadratic equation for $c_{1}$ is obtained by using $s=b_{1}-c_{1}$ and $e=b_{2}-c_{1}$
\begin{equation}  
\begin{array}{llll}
k^{+}_{1}c_{1}^{2}-(k^{+}_{1}b_{1}+k^{+}_{1}b_{2}+k^{-}_{1})c_{1}+k^{+}_{1}b_{1}b_{2}=0.
\end{array}\label{eq c1 of QEA 2.1}
\end{equation}
then equation (\ref{eq c1 of QEA 2.1}) can be solved analytically for $c_{1}$\\
$$c_{1}(b_{1},b_{2})=\dfrac{1}{2}\Bigg[\Big (b_{1}+b_{2}+\dfrac{k^{-}_{1}}{k^{+}_{1}}\Big)\pm \sqrt{\Big(b_{1}+b_{2}+\dfrac{k^{-}_{1}}{k^{+}_{1}} \Big)^{2}-4b_{1}b_{2}}\Bigg].$$\\
We select "--'' for providing positive concentrations of $s,c_{1}$ and $e$, and obtaining a proper asymptotic behavior of the fast reaction: if $b_{1}\rightarrow 0$ and $b_{2}\rightarrow 0$ then $c_{1}\rightarrow 0$. Furthermore, other variables ($s$ and $e$) are obtained:\\
\quad \quad \quad $s(b_{1},b_{2})=b_{1}-\dfrac{1}{2}\Bigg[\Big (b_{1}+b_{2}+\dfrac{k^{-}_{1}}{k^{+}_{1}}\Big)- \sqrt{\Big(b_{1}+b_{2}+\dfrac{k^{-}_{1}}{k^{+}_{1}} \Big)^{2}-4b_{1}b_{2}} \Bigg]$,\\
\quad \quad \quad $e(b_{1},b_{2})=b_{2}-\dfrac{1}{2}\Bigg[\Big (b_{1}+b_{2}+\dfrac{k^{-}_{1}}{k^{+}_{1}}\Big)- \sqrt{\Big(b_{1}+b_{2}+\dfrac{k^{-}_{1}}{k^{+}_{1}} \Big)^{2}-4b_{1}b_{2}}\Bigg]$.\\

\noindent To simplify the value of our variables, we have to bring a non--linear parts in equation \eqref{eq c1 of QEA 2.1} to zero by using an assumption that the concentration of substrate is present in large excess compared to the total concentration of enzyme.

 \begin{equation}  
\begin{array}{llll}  
[S]\gg [ES] \quad i.e. \quad b_{1}\gg c.
\end{array}\label{condition s>>e QEA 2.1}  
\end{equation} 
By using condition (\ref{condition s>>e QEA 2.1}), equation (\ref{eq c1 of QEA 2.1}) is simplified as follows:
\begin{equation}  
\begin{array}{llll}
\Big(1+\dfrac{b_{2}}{b_{1}}+\dfrac{k^{-}_{1}}{k^{+}_{1}b_{1}}\Big)c_{1}=b_{2}+\mathcal{O}\Big(\dfrac{c_{1}}{b_{1}}\Big) .
\end{array}\label{eq c1 with cond. QEA 2.1 }  
\end{equation}
The approximation of $c_{1}$ becomes
\begin{equation}  
\begin{array}{llll}
c_{1} \approx \dfrac{b_{1}b_{2}}{b_{1}+b_{2}+k^{*}} ,
\end{array}\label{value c1 appoximate QEA 1}  
\end{equation}
where $k^{*}=k^{-}_{1}/k^{+}_{1}$. The new form of the variables $s$ and $e$ become\\
\quad \quad \quad $s(b_{1},b_{2})=\dfrac{b_{1}(b_{1}+k^{*})}{b_{1}+b_{2}+k^{*}}$ and $e(b_{1},b_{2})=\dfrac{b_{2}(b_{2}+k^{*})}{b_{1}+b_{2}+k^{*}}$.\\
\subsection{Case Two}
Wen assume that the model network has only two fast reversible reactions. By applying QEA of the chemical reactions \eqref{chemical reaction 2.1}, if possible suppose that the first and the third reactions ($E+S \underset{k_{-1}}{ \overset{k_{1}}{\rightleftharpoons}} ES $\; and \; $ E+I \underset{k_{-3}}{ \overset{k_{3}}{\rightleftharpoons}} EI$) becomes quasi--equilibrium when the equilibrium is fast: let  $k_{1}=\dfrac{k^{+}_{1}}{\epsilon} ,\; k_{-1}=\dfrac{k^{-}_{1}}{\epsilon}$\; and \; $k_{3}=\dfrac{k^{+}_{3}}{\epsilon} ,\; k_{-3}=\dfrac{k^{-}_{3}}{\epsilon}$where $ \epsilon=\dfrac{e_{0}}{s_{0}}$ then $k^{+}_{1}=\dfrac{k_{1} e_{0}}{s_{0}} ,\;k^{-}_{1}=\dfrac{k_{-1} e_{0}}{s_{0}}$ \; and \;$k^{+}_{3}=\dfrac{k_{3} e_{0}}{s_{0}} ,\;k^{-}_{3}=\dfrac{k_{-3} e_{0}}{s_{0}}$ \;(i.e. $k_{1},\; k_{-1},\; k_{3}$\; and \; $k_{-3}$\; are called large parameters compared to $ k_{2}, k_{4} , k_{-4}, k_{5} , k_{-5} $ and $ k_{6}$). Then, the system \eqref{system QEA 2.1} has the form of equation (\ref{general QEA 1})

\begin{equation}  
\begin{array}{llll}
\dfrac{ds}{dt}=\dfrac{1}{\epsilon}g^{f}_{1}(s,e,c_{1},t)+g^{s}_{1}(s,c_{2},c_{3},t),\\
\dfrac{de}{dt}=\dfrac{1}{\epsilon}g^{f}_{1}(s,e,c_{1},t)+\dfrac{1}{\epsilon}g^{f}_{2}(e,i,c_{2},t)+g^{s}_{2}(c_{1},t),\\
\dfrac{di}{dt}=\dfrac{1}{\epsilon}g^{f}_{2}(e,i,c_{2},t)+g^{s}_{3}(i,c_{1},c_{3},t),\\
\dfrac{dc_{1}}{dt}=\dfrac{-1}{\epsilon}g^{f}_{1}(s,e,c_{1},t)+g^{s}_{3}(i,c_{1},c_{3},t)-g^{s}_{2}(c_{1},t),\\
\dfrac{dc_{2}}{dt}=\dfrac{-1}{\epsilon}g^{f_{2}}_{2}(e,i,c_{2},t)+g^{s}_{1}(s,c_{2},c_{3},t)+g^{s}_{4}(c_{3},t),\\
\dfrac{dc_{3}}{dt}=-g^{s}_{1}(s,c_{2},c_{3},t)-g^{s}_{3}(i,c_{1},c_{3},t)-g^{s}_{4}(c_{3},t),\\
\dfrac{dp}{dt}=g^{s}_{2}(c_{1},t)+g^{s}_{4}(c_{3},t),
\end{array}\label{reduce QEA 2.2}
\end{equation} 
where $g^{f}_{1}(s,e,c_{1},t)=-k^{+}_{1}es+k^{-}_{1}c_{1} ,\; g^{f}_{2}(e,i,c_{2},t)=-k^{+}_{3}ei+k^{-}_{-3}c_{2}$ ,\\ $g^{s}_{1}(s,c_{2},c_{3},t)=-k_{5}c_{2}s+k_{-5}c_{3} ,\;  g^{s}_{2}(c_{1},t)=k_{2}c_{1} ,\; g^{s}_{3}(i,c_{1},c_{3},t)=-k_{4}c_{1}i+k_{-4}c_{3}\; $ and $\; g^{s}_{4}(c_{3},t)=k_{6}c_{3} $.\\ 
For $\epsilon \longrightarrow 0,$ the quasi--equilibrium approximation can be applied. Five  species $S, E, I, ES$ and $EI$ are participate in the fast reaction, while other components of chemical reactions are not involved in analysis of the QEA manifold.     Based on the equation \eqref{general fast part QEA 1}, then the equation \eqref{reduce QEA 2.2} takes the form:
\begin{equation}  
\begin{array}{llll}
\dfrac{ds}{dt}=\dfrac{1}{\epsilon}(-k^{+}_{1}es+k^{-}_{1}c_{1}),\\
\dfrac{de}{dt}=\dfrac{1}{\epsilon}(-k^{+}_{1}es+k^{-}_{1}c_{1})+\dfrac{1}{\epsilon}(-k^{+}_{3}ei+k^{-}_{3}c_{2}),\\
\dfrac{di}{dt}=\dfrac{1}{\epsilon}(-k^{+}_{3}ei+k^{-}_{3}c_{2}),\\
\dfrac{dc_{1}}{dt}=\dfrac{-1}{\epsilon}(-k^{+}_{1}es+k^{-}_{1}c_{1}),\\
\dfrac{dc_{2}}{dt}=\dfrac{-1}{\epsilon}(-k^{+}_{3}ei+k^{-}_{3}c_{2}).
\end{array}\label{reduce fast part QEA 2.2}
\end{equation} 

\noindent Therefore, we obtain three slow variables which are the stoichiometric conservation laws of the fast reaction. The variables are $b_{1}(s,c_{1})=s+c_{1},\,b_{2}(e,c_{1},c_{2})=e+c_{1}+c_{2}$\; and \;$b_{3}(i,c_{2})=i+c_{2} $. The first slow variable here is the sum of the free substrate and the enzyme--substrate complex,the second one is the total amount of enzyme, while the third  one is the sum of the inhibitor and the enzyme--inhibitor. More precisely, slow variables should be invariant with respect to fast motion in order to apply the Tikhnove theorem.   \\
The slow manifold for the reaction kinetics can be calculated from the algebraic equation $g^{f}_{1}(s,e,c_{1},t)=0$\;and \; $g^{f}_{2}(e,i,c_{2})=0 $. This is given by  
\begin{equation}  
\begin{array}{llll}         
\mathcal{M}_{0}^{*}=\bigg \lbrace (s,e,c_{1})\in \mathbb{R}^{3}: s=\dfrac{k^{-}_{1}c_{1}}{k^{+}_{1}e},\; i=\dfrac{k^{-}_{3}c_{2}}{k^{+}_{3}e}\bigg\rbrace. 
\end{array}\label{manifold QEA 2.2} 
\end{equation}
\noindent By fixing the slow variables ($b_{1},b_{2}$ and $b_{3}$), and finding the variables $s,e,i,c_{1}$ and $c_{2}$, the system (\ref{general fast part QEA 1 =0}) takes the form  
\begin{equation}  
\begin{array}{llll}
k^{+}_{1}es-k^{-}_{1}c_{1}=0,\\
k^{+}_{3}ei-k^{-}_{3}c_{2}=0,\\
s+c_{1}=b_{1},\\ 
e+c_{1}+c_{2}=b_{2},\\
i+c_{2}=b_{3}.
\end{array}\label{algebraic QEA 2.2}
\end{equation}

A system equation for $c_{1}$\;and\;$c_{2}$ is obtained by using $s=b_{1}-c_{1},\;e=b_{2}-c_{1}-c_{2}$\; and \; $i=b_{3}-c_{2}$
\begin{equation}  
\begin{array}{llll}
k^{+}_{1}c_{1}^{2}+k^{+}_{1}c_{1}c_{2}-(k^{+}_{1}b_{1}+k^{+}_{1}b_{2}+k^{-}_{1})c_{1}-k^{+}_{1}b_{1}c_{2}+k^{+}_{1}b_{1}b_{2}=0,\\
k^{+}_{3}c_{2}^{2}+k^{+}_{3}c_{1}c_{2}-(k^{+}_{3}b_{2}+k^{+}_{3}b_{3}+k^{-}_{3})c_{2}-k^{+}_{3}b_{3}c_{1}+k^{+}_{3}b_{2}b_{3}=0.
\end{array}\label{eq c1,c2 of QEA 2.2}
\end{equation}
\noindent To simplify the value of our variables, we have to bring a non--linear parts in equation \eqref{eq c1,c2 of QEA 2.2} to zero by \eqref{condition s>>e QEA 2.1},
Then equation (\ref{eq c1,c2 of QEA 2.2}) is simplified as follows:
\begin{equation}  
\begin{array}{llll}
\Big(1+\dfrac{b_{2}}{b_{1}}+\dfrac{k^{-}_{1}}{k^{+}_{1}b_{1}}\Big)c_{1}+c_{2}=b_{2}+\mathcal{O}\Big(\dfrac{c_{1}c_{2}}{b_{1}}\Big),\\
\Big(\dfrac{b_{2}}{b_{1}}+\dfrac{b_{3}}{b_{1}}+\dfrac{k^{-}_{3}}{k^{+}_{3}b_{1}}\Big)c_{2}=\dfrac{b_{2}b_{3}}{b_{1}}+\mathcal{O}\Big(\dfrac{c_{1}c_{2}}{b_{1}}\Big).
\end{array}\label{eq c1,c2 with cond. QEA 2.2}  
\end{equation}
The approximation of $c_{1}$\; and \; $c_{2}$ becomes
\begin{equation}  
\begin{array}{llll}
c_{1} \approx \dfrac{b_{1}b_{2}(b_{2}+k^{*}_{3})}{(b_{1}+b_{2}+K^{*}_{1})(b_{2}+b_{3}+k^{*}_{3})},\\
c_{2} \approx\dfrac{b_{2}b_{3}}{b_{2}+b_{3}+k^{*}_{3}}.
\end{array}\label{value c1,c2 appoximate QEA 2.2}  
\end{equation}
where $k^{*}_{1}=k^{-}_{1}/k^{+}_{1}$\;and \;$k^{*}_{3}=k^{-}_{3}/k^{+}_{3}$.
Furthermore, other variables ($s,e$ \;and \;$i$) are obtained:\\
\qquad$s(b_{1},b_{2},b_{3})=b_{1}-\dfrac{b_{1}b_{2}(b_{2}+k^{*}_{3})}{(b_{1}+b_{2}+K^{*}_{1})(b_{2}+b_{3}+k^{*}_{3})} $,\\
\qquad$e(b_{1},b_{2},b_{3})=\dfrac{b_{2}(b_{2}+k^{*}_{3})(b_{2}+k^{*}_{1})}{(b_{1}+b_{2}+K^{*}_{1})(b_{2}+b_{3}+k^{*}_{3})}$,\\
\qquad $i(b_{1},b_{2},b_{3})=\dfrac{b_{3}(b_{3}+k^{*}_{3})}{b_{2}+b_{3}+k^{*}_{3}}.$\\

\subsection{Case Three}
Wen assume that the model network has three fast reversible reactions. By applying QEA of the chemical reactions \eqref{chemical reaction 2.1}, if possible suppose that the first, third and fourth reactions ($E+S \underset{k_{-1}}{ \overset{k_{1}}{\rightleftharpoons}} ES $,\; $ E+I \underset{k_{-3}}{ \overset{k_{3}}{\rightleftharpoons}} EI$\;and\;$ES+I \underset{k_{-4}}{ \overset{k_{4}}{\rightleftharpoons}} ESI$) becomes quasi--equilibrium when the equilibrium is fast: let  $k_{1}=\dfrac{k^{+}_{1}}{\epsilon} ,\; k_{-1}=\dfrac{k^{-}_{1}}{\epsilon}$,\; $k_{3}=\dfrac{k^{+}_{3}}{\epsilon} ,\; k_{-3}=\dfrac{k^{-}_{3}}{\epsilon},\;k_{4}=\dfrac{k^{+}_{4}}{\epsilon} $\;and\;$ k_{-4}=\dfrac{k^{-}_{4}}{\epsilon}$ where $ \epsilon=\dfrac{e_{0}}{s_{0}}$  then $k^{+}_{1}=\dfrac{k_{1} e_{0}}{s_{0}} ,\;k^{-}_{1}=\dfrac{k_{-1} e_{0}}{s_{0}}$ , \;$k^{+}_{3}=\dfrac{k_{3} e_{0}}{s_{0}} ,\;k^{-}_{3}=\dfrac{k_{-3} e_{0}}{s_{0}},\; k^{+}_{4}=\dfrac{k_{4} e_{0}}{s_{0}} $\;and\;$k^{-}_{4}=\dfrac{k_{-4} e_{0}}{s_{0}}$ \;(i.e. $k_{1},\; k_{-1},\; k_{3},\;k_{-3}, k_{4} , k_{-4}$\; are called large parameters compared to $ k_{2}, k_{5} , k_{-5} $ and $ k_{6}$). Then, the system \eqref{system QEA 2.1} has the form of equation (\ref{general QEA 1})

\begin{equation}  
\begin{array}{llll}
\dfrac{ds}{dt}=\dfrac{1}{\epsilon}g^{f}_{1}(s,e,c_{1},t)+g^{s}_{1}(s,c_{2},c_{3},t),\\
\dfrac{de}{dt}=\dfrac{1}{\epsilon}g^{f}_{1}(s,e,c_{1},t)+\dfrac{1}{\epsilon}g^{f}_{2}(e,i,c_{2},t)+g^{s}_{2}(c_{1},t),\\
\dfrac{di}{dt}=\dfrac{1}{\epsilon}g^{f}_{2}(e,i,c_{2},t)+\dfrac{1}{\epsilon} g^{f}_{3}(i,c_{1},c_{3},t),\\
\dfrac{dc_{1}}{dt}=\dfrac{-1}{\epsilon}g^{f}_{1}(s,e,c_{1},t)+\dfrac{1}{\epsilon} g^{f}_{3}(i,c_{1},c_{3},t)-g^{s}_{2}(c_{1},t),\\
\dfrac{dc_{2}}{dt}=\dfrac{-1}{\epsilon}g^{f_{2}}_{2}(e,i,c_{2},t)+g^{s}_{1}(s,c_{2},c_{3},t)+g^{s}_{3}(c_{3},t),\\
\dfrac{dc_{3}}{dt}=-g^{s}_{1}(s,c_{2},c_{3},t)-\dfrac{1}{\epsilon} g^{f}_{3}(i,c_{1},c_{3},t)-g^{s}_{3}(c_{3},t),\\
\dfrac{dp}{dt}=g^{s}_{2}(c_{1},t)+g^{s}_{3}(c_{3},t),
\end{array}\label{reduce QEA 2.3}
\end{equation} 
where $g^{f}_{1}(s,e,c_{1},t)=-k^{+}_{1}es+k^{-}_{1}c_{1} ,\; g^{f}_{2}(e,i,c_{2},t)=-k^{+}_{3}ei+k^{-}_{-3}c_{2} ,\\ g^{f}_{3}(i,c_{1},c_{3},t)=-k^{+}_{4}c_{1}i+k^{-}_{4}c_{3}$ ,\; $g^{s}_{1}(s,c_{2},c_{3},t)=-k_{5}c_{2}s+k_{-5}c_{3} ,\;  g^{s}_{2}(c_{1},t)=k_{2}c_{1} \; $ and $\; g^{s}_{3}(c_{3},t)=k_{6}c_{3} $.\\ 
For $\epsilon \longrightarrow 0,$ the quasi--equilibrium approximation can be applied. Six species $S, E, I,ES, EI$ and $ESI$ are participate in the fast reaction, while other components of chemical reactions are not involved in analysis of the QEA manifold. Based on the equation \eqref{general fast part QEA 1}, then the equation \eqref{reduce QEA 2.3} takes the form:

\begin{equation}  
\begin{array}{llll}
\dfrac{ds}{dt}=\dfrac{1}{\epsilon}(-k^{+}_{1}es+k^{-}_{1}c_{1}),\\
\dfrac{de}{dt}=\dfrac{1}{\epsilon}(-k^{+}_{1}es+k^{-}_{1}c_{1})+\dfrac{1}{\epsilon}(-k^{+}_{3}ei+k^{-}_{3}c_{2}),\\
\dfrac{di}{dt}=\dfrac{1}{\epsilon}(-k^{+}_{3}ei+k^{-}_{3}c_{2})+\dfrac{1}{\epsilon}(-k^{+}_{4}c_{1}i+k^{-}_{4}c_{3}),\\
\dfrac{dc_{1}}{dt}=\dfrac{-1}{\epsilon}(-k^{+}_{1}es+k^{-}_{1}c_{1}),\\
\dfrac{dc_{2}}{dt}=\dfrac{-1}{\epsilon}(-k^{+}_{3}ei+k^{-}_{3}c_{2}),\\
\dfrac{dc_{2}}{dt}=\dfrac{-1}{\epsilon}(-k^{+}_{4}c_{1}i+k^{-}_{4}c_{3}).
\end{array}\label{reduce fast part QEA 2.3}
\end{equation} 

\noindent Therefore, we obtain three slow variables which are the stoichiometric conservation laws of the fast reaction. The variables are $b_{1}(s,c_{1},c_{3})=s+c_{1}+c_{3},\\b_{2}(e,c_{1},c_{2},c_{3})=e+c_{1}+c_{2}+c_{3}$\; and \;$b_{3}(i,c_{2},c_{3})=i+c_{2}+c_{3} $. The first slow variable here is the sum of the free substrate, enzyme--substrate complex and the enzyme--substrate--inhibitor,the second one is the total amount of enzyme, while the third  one is the sum of the inhibitor ,enzyme--inhibitor and the enzyme--substrate--inhibitor. More precisely, slow variables should be invariant with respect to fast motion in order to apply the Tikhnove theorem.   \\
The slow manifold for the reaction kinetics can be calculated from the algebraic equation $g^{f}_{1}(s,e,c_{1},t)=0,\;g^{f}_{2}(e,i,c_{2})=0 $\;and\;$g^{f}_{3}(i,c_{1},c_{3},t)=0$. This is given by  
\begin{equation}  
\begin{array}{llll}         
\mathcal{M}_{0}^{*}=\bigg \lbrace (s,e,i,c_{1},c_{2},c_{3})\in \mathbb{R}^{6}: s=\dfrac{k^{-}_{1}c_{1}}{k^{+}_{1}e},\; i=\dfrac{k^{-}_{3}c_{2}}{k^{+}_{3}e},\;c_{3}=\dfrac{k^{-}_{3}k^{+}_{4}c_{1}c_{2}}{k^{+}_{3}k^{-}_{4}e}\bigg\rbrace. 
\end{array}\label{manifold QEA 2.3} 
\end{equation}
\noindent By fixing the slow variables ($b_{1},b_{2}$ and $b_{3}$), and finding the variables $s,e,i,c_{1},c_{2}$ and $c_{3}$, the system (\ref{general fast part QEA 1 =0}) takes the form  
\begin{equation}  
\begin{array}{llll}
k^{+}_{1}es-k^{-}_{1}c_{1}=0,\\
k^{+}_{3}ei-k^{-}_{3}c_{2}=0,\\
k^{+}_{4}c_{1}i+k^{-}_{4}c_{3}=0,\\
s+c_{1}+c_{3}=b_{1},\\ 
e+c_{1}+c_{2}+c_{3}=b_{2},\\
i+c_{2}+c_{3}=b_{3}.
\end{array}\label{algebraic QEA 2.3}
\end{equation}

A system of equations for $c_{1},c_{2}$\;and\;$c_{3}$ is obtained by using $s=b_{1}-c_{1}-c_{3},\\e=b_{2}-c_{1}-c_{2}-c_{3}$\; and \; $i=b_{3}-c_{2}-c_{3}$. This is given bellow:
\begin{equation}  
\begin{array}{llll}
k^{+}_{1}c^{2}_{1}+k^{+}_{1}c^{2}_{3}+k^{+}_{1}c_{1}c_{2}+2k^{+}_{1}c_{1}c_{3}+k^{+}_{1}c_{2}c_{3}-(k^{+}_{1}b_{1}+k^{+}_{1}b_{2}+k^{-}_{1})c_{1}\\
-k^{+}_{1}b_{1}c_{2}-k^{+}_{1}(b_{1}+b_{2})c_{3}+k^{+}_{1}b_{1}b_{2}=0,\\
k^{+}_{3}c^{2}_{2}+k^{+}_{3}c^{2}_{3}+k^{+}_{3}c_{1}c_{2}+k^{+}_{3}c_{1}c_{3}+2k^{+}_{3}c_{2}c_{3}-k^{+}_{3}b_{3}c_{1}-(k^{+}_{3}b_{2}+k^{+}_{3}b_{3}+k^{-}_{3})c_{2}\\
-k^{+}_{3}(b_{2}+b_{3})c_{3}+k^{+}_{3}b_{2}b_{3}=0,\\
b_{3}k^{+}_{4}c_{1}-k^{+}_{4}c_{1}c_{2}-k^{+}_{4}c_{1}c_{3}-k^{-}_{4}c_{3}=0.
\end{array}\label{eq c1,c2,c3 of QEA 2.3}
\end{equation}
\noindent To simplify the value of variables\;$c_{1},c_{2}$\;and\;$c_{3}$, we assume that all non--linear parts in equation \eqref{eq c1,c2,c3 of QEA 2.3} becomes zero by \eqref{condition s>>e QEA 2.1}.
Then the system (\ref{eq c1,c2,c3 of QEA 2.3})\;becomes a linear system for $c_{1},c_{2}$\;and\;$c_{3}$\;then it can be solve analytically for $c_{1},c_{2}$\;and\;$c_{3}$ as follows:
\begin{equation}  
\begin{array}{llll}
\Big(1+\dfrac{b_{2}}{b_{1}}+\dfrac{k^{-}_{1}}{k^{+}_{1}b_{1}}\Big)c_{1}+c_{2}+(1+\dfrac{b_{2}}{b_{1}})c_{3}=b_{2}+\mathcal{O}\Big(\dfrac{c_{1}c_{2}c_{3}}{b_{1}}\Big),\\
\dfrac{b_{3}}{b_{1}}c_{1}+\Big(\dfrac{b_{2}}{b_{1}}+\dfrac{b_{3}}{b_{1}}+\dfrac{k^{-}_{3}}{k^{+}_{3}b_{1}}\Big)c_{2}+(\dfrac{b_{2}}{b_{1}}+\dfrac{b_{3}}{b_{1}})c_{3}=\dfrac{b_{2}b_{3}}{b_{1}}+\mathcal{O}\Big(\dfrac{c_{1}c_{2}c_{3}}{b_{1}}\Big),\\
c_{1}-\dfrac{k^{-}_{4}}{k^{+}_{4}b_{1}b_{3}} c_{3}=0+\mathcal{O}\Big(\dfrac{c_{1}c_{2}c_{3}}{b_{1}}\Big).
\end{array}\label{eq c1,c2,c3 with cond. QEA 2.3}  
\end{equation}
The approximation solutions of $c_{1},\;c_{2}$\; and \; $c_{3}$ becomes
\begin{equation}  
\begin{array}{llll}
c_{1} \approx \dfrac{\alpha_{3}k^{*}_{4}}{\alpha_{1}K^{*}_{4}+\alpha_{2}b_{1}b_{3}},\\
c_{2} \approx b_{2}-(1+\dfrac{b_{2}}{b_{1}}+\dfrac{k^{*}_{1}}{b_{1}})(\dfrac{\alpha_{3}k^{*}_{4}}{\alpha_{1}K^{*}_{4}+\alpha_{2}b_{1}b_{3}})-(1+\dfrac{b_{2}}{b_{1}})(\dfrac{\alpha_{3}b_{1}b_{3}}{\alpha_{1}k^{*}_{4}+\alpha_{2}b_{1}b_{3}}),\\
c_{3} \approx \dfrac{\alpha_{3}b_{1}b_{3}}{\alpha_{1}k^{*}_{4}+\alpha_{2}b_{1}b_{3}}, 
\end{array}\label{value c1,c2 appoximate QEA 2.3}  
\end{equation}
where $k^{*}_{1}=k^{-}_{1}/k^{+}_{1},\;k^{*}_{3}=k^{-}_{3}/k^{+}_{3},\;k^{*}_{4}=k^{-}_{4}/k^{+}_{4},\\ \alpha_{1}=\dfrac{b_{3}}{b_{1}}-(\dfrac{b_{2}}{b_{1}}+\dfrac{b_{3}}{b_{1}}+\dfrac{k^{*}_{3}}{b_{1}})(1+\dfrac{b_{2}}{b_{1}}+\dfrac{k^{*}_{1}}{b_{1}}) ,\;\alpha_{2}=\dfrac{b_{2}+b_{3}}{b_{1}}-(\dfrac{b_{2}}{b_{1}}+\dfrac{b_{3}}{b_{1}}+\dfrac{k^{*}_{3}}{b_{1}})(1+\dfrac{b_{2}}{b_{1}}) ,$\;and\;$\alpha_{3}=\dfrac{b_{2}b_{3}}{b_{1}}-(\dfrac{b_{2}}{b_{1}}+\dfrac{b_{3}}{b_{1}}+\dfrac{k^{*}_{3}}{b_{1}})b_{2}$.\\
Furthermore, other variables ($s,e$ \;and \;$i$) are obtained:\\ 
\qquad$s(b_{1},b_{2},b_{3})=b_{1}-\dfrac{\alpha_{3}k^{*}_{4}}{\alpha_{1}K^{*}_{4}+\alpha_{2}b_{1}b_{3}}-\dfrac{\alpha_{3}b_{1}b_{3}}{\alpha_{1}k^{*}_{4}+\alpha_{2}b_{1}b_{3}}$,\\
\quad $e(b_{1},b_{2},b_{3})=(1+\dfrac{b_{2}}{b_{1}}+\dfrac{k^{*}_{1}}{b_{1}})(\dfrac{\alpha_{3}k^{*}_{4}}{\alpha_{1}K^{*}_{4}+\alpha_{2}b_{1}b_{3}})+(1+\dfrac{b_{2}}{b_{1}})(\dfrac{\alpha_{3}b_{1}b_{3}}{\alpha_{1}k^{*}_{4}+\alpha_{2}b_{1}b_{3}})\\
-\dfrac{\alpha_{3}k^{*}_{4}}{\alpha_{1}K^{*}_{4}+\alpha_{2}b_{1}b_{3}}-\dfrac{\alpha_{3}b_{1}b_{3}}{\alpha_{1}k^{*}_{4}+\alpha_{2}b_{1}b_{3}} $,\\
\quad $i(b_{1},b_{2},b_{3})=b_{3}-b_{2}-(1+\dfrac{b_{2}}{b_{1}}+\dfrac{k^{*}_{1}}{b_{1}})(\dfrac{\alpha_{3}k^{*}_{4}}{\alpha_{1}K^{*}_{4}+\alpha_{2}b_{1}b_{3}})\\
-(1+\dfrac{b_{2}}{b_{1}})(\dfrac{\alpha_{3}b_{1}b_{3}}{\alpha_{1}k^{*}_{4}+\alpha_{2}b_{1}b_{3}})-\dfrac{\alpha_{3}b_{1}b_{3}}{\alpha_{1}k^{*}_{4}+\alpha_{2}b_{1}b_{3}}.$\\ 


\section{An Algorithm for Identifying Slow and Fast Reactions}

Identifying slow and fast reactions become a difficult task analytically, and it may be impossible for complex biochemical reaction networks. Therefore, we need an algorithm that gives us a good step forward in identifying slow and fast reactions.\\
As a result, we propose some steps here for identifying slow and fast reactions.

\begin{enumerate}
\item[\textbf{Step One:}] Consider a chemical reaction network with $\mathit{m}$ reversible $\lbrace \mathit{r_{1},r_{2},...,r_{m}} \rbrace$ reactions and $\mathit{n} $ variables

 \begin{equation}
\begin{array}{llll}  
\mathlarger{\mathlarger{\sum\limits}}_{i=1}^{n} \alpha_{ij}A_{i} \underset{k^{-}_{j}}{\overset{k^{+}_{j}}  \rightleftharpoons }\mathlarger{\mathlarger{\sum\limits}}_{i=1}^{n} \beta_{ij}A_{i} ,\quad j=1,2,...,m.
\end{array}
\label{elementary}
 \end{equation}
 \noindent The non-negative integers $\alpha_{ij}$ and $\beta_{ij}$ are called stoichiometric coefficients.
 
\item[\textbf{Step Two:}] Use mass action law to define all forward and backward reaction rates
\begin{equation}
\begin{array}{llll}  
 v^{f}_{j}(t) = k_{j}^{+} \mathlarger{\mathlarger{\prod\limits}}_{i=1}^n [A_{i}]^{\alpha_{ij}}(t),\\
 v^{b}_{j}(t) = k_{j}^{-} \mathlarger{\mathlarger{\prod\limits}}_{i=1}^n [A_{i}]^{\alpha_{ij}}(t), \quad j=1,2,...,m ,
\end{array}
\label{rates}
 \end{equation}
 where $k_{j}^{+} > 0$ and $k_{j}^{-} \geq 0$ are the reaction rate coefficients.
 
     \item[\textbf{Step Three:}] Compute $ |v_{j}(t)|=|v^{f}_{j}(t)-v^{b}_{j}(t)|$ \quad for $t\in T\subset R $.
 
 \item[\textbf{Step Four:}] If $ |v_{j}(t)|<\epsilon$,\;for $j=1,2,...,p$.\;where \;$p\leqslant m$ \;and\; $0<\epsilon\ll 1,$\;then the fast reactions are
 \begin{equation}  
\begin{array}{llll}         
R^f=\bigg \lbrace v_{j}:|v_{j}(t)|<\epsilon,\;j=1,2,...,p.\;p\leqslant m \bigg\rbrace. 
\end{array}\label{R^{f}}
\end{equation}
and the slow reactions are
 \begin{equation}  
\begin{array}{llll}         
R^{s}=\bigg \lbrace v_{j}:|v_{j}(t)|\nless\epsilon,\;j=1,2,...,k.\;k\leqslant m \bigg\rbrace. 
\end{array}\label{R^{s}}
\end{equation}
where $k+p=m$, and $R=R^{f}\cup R^{s}$ ; $R$ is a set of all reactions, $R^{f}$ is a set of all fast reactions and $R^{s}$ is a set of all slow reactions.
\end{enumerate}

The above steps can be also expressed in the following Flowcharts:
\begin{center}
    \resizebox{0.8 \linewidth}{!}{%
      \begin{tikzpicture}[node distance = 3cm, auto]
        \node[block,text width =7cm]                  (A1){Given m reversible reactions\\
        $\mathlarger{\mathlarger{\sum\limits}}_{i=1}^{n} \alpha_{ij}A_{i} \underset{k^{-}_{j}}{\overset{k^{+}_{j}}  \rightleftharpoons }\mathlarger{\mathlarger{\sum\limits}}_{i=1}^{n} \beta_{ij}A_{i}$\\
        $for  \quad j=1,2,...,m.$\\
        $R=(r_{1},r_{2},...,r_{m})\in R^{m} $ };
        \node[block, below of=A1,node distance = 5cm,text width =9cm]   (A2){Define all forward and backward reaction rates\\
          $v^{f}_{j}(t) = k_{j}^{+} \mathlarger{\mathlarger{\prod\limits}}_{i=1}^n [A_{i}]^{\alpha_{ij}}(t)$\\
 $v^{b}_{j}(t) = k_{j}^{-} \mathlarger{\mathlarger{\prod\limits}}_{i=1}^n [A_{i}]^{\alpha_{ij}}(t)$};
 
       \node[decision, below of=A2, node distance = 6cm, text width =3cm ] (A3){$v_{j}=|v^{f}_{j}-v^{b}_{j}|$\\ If $ Max(v_{j})<\epsilon$\\ for \;$ 0<\epsilon\ll 1$\\$t\in T\subset R$};

        \node[block,below left of=A3, node distance = 10cm,text width =7cm ]   (A4){The set of slow reactionns\\
        $R^{s}=\bigg \lbrace v_{j}:|v^{f}_{j}-v^{b}_{j}|\nless\epsilon\bigg\rbrace $\\ for \;$j=1,2,...,k.\;k\leqslant m $\\
        $R^{s}\subseteq R^{k} \subseteq R^{m} $};        
          \node[block,below right of=A3, node distance = 10cm,text width =8cm ]   (A5){The set of fast reactions \\
          $R^f=\bigg \lbrace v_{j}:|v^{f}_{j}-v^{b}_{j}|<\epsilon\bigg\rbrace $\\ for \; $j=1,2,...,p.\;p\leqslant m $\\$ R^{f}\subseteq R^{p}\subseteq R^{m}$};  
   
        \path[line,line width=0.05cm] (A1) --          (A2); 
        \path[line,line width=0.05cm] (A2) --          (A3); 
        \path[line,line width=0.05cm] (A3) -|          (A4);
        \path[line,line width=0.05cm] (A3) -|          (A5);
      \end{tikzpicture}%
    }%
  \end{center}   
\quad \quad \textbf{The Flowchart for identifying slow and fast reactions.}
\section{Results and Discussions}
The quasi equilibrium approximation (QEA) is an important tool of model reduction for reversible chemical reactions. Simply, the idea of this method is that fast reactions go their equilibrium very quickly. We applied the QEA technique first on the simple enzyme reactions \eqref{chemical reaction 1}. It reduced from $5$ and $6$ to $3$ and $3$ variables and parameters respectively. Then, we calculated some analytical solutions of the model variables. Another example here is that we applied the idea of QEA on the non-competitive inhibition enzymatic reactions \eqref{chemical reaction 2.1}, this model consists of $7$ variables and $10$ parameters. The model has also three conservations. The model has been solved in three different cases. In the first case, we supposed that the first reaction becomes quasi equilibrium when the equilibrium is fast, then we introduced a new variables for scaling and making a small parameter $\epsilon$ for separate the original system into slow and fast subsystem. When $\epsilon \rightarrow 0$, the full system is reduced to $3$ variables and $2$ parameters. We calculated the slow manifolds that provide us the behavior dynamics of the of the slow variables. After that the model solutions are calculated based on the conservations laws and fixing the slow variables ($b_{1}$ and $b_{2}$). In the second case, we supposed that the first and third reactions are fast, and by the same procedure us we gave before the slow manifolds and the model solutions are calculated analytically.\\ 
Finally, we supposed that the first, third and fourth reactions are  fast reactions, we applied all steps as we mentioned before then we found the slow manifolds and the model solutions analytically. The only problem about this technique is that for complex cell signalling pathways with high dimensional elements is quite difficult to identify slow and fast reactions. Thus, we suggested an algorithm only for identifying the slow and fast reaction in a complex model first and then applying the idea of QEA.     


\include{chapter[4]}
\chapter{Lumping of Compartments}

\section{Mathematical Formulation for Lumping of Compartments}
We consider a system of differential equations for a chemical network as follows:
 \begin{equation}
\begin{array}{llll}  
\dfrac{dC}{dt}=H(C,P),
\end{array}
\label{general system 1}
\end{equation}
\noindent
where $C$ is a vector of state variables and $ C \in R^{n} $, $P$ is a vector of chemical constants (parameters), $ P \in R^{m}$ or $P=\Big( p_{1},p_{2},...,p_{m} \Big)$, and $ H=\Big( h_{1},h_{2},...,h_{n} \Big)$.
\newline 
We suppose that $H(C,P)$ is a linear function. Therefore, the function $H$ can be written as  $H(C,P)=K C,$ where $K$ is a stoichiometric matrix of reaction rates. Then the system \eqref{general system 1} becomes
 \begin{equation}
\begin{array}{llll}  
\dfrac{dC}{dt}=K C.
\end{array}
\label{general linear system 1}
\end{equation} 
By introducing a vector of new variables
$C^{*}=(c^{*}_{1},c^{*}_{2},...,c^{*}_{n_{1}}), n_{1}\leq n$,\; where each component of $C^{*}$ is defined below

$$ c_{i}^{*}=\sum_{j\in J} c_{j},$$
where $J=\lbrace 1,2,...,n \rbrace$ and $i=1,2,...,n_{1}$. This is called lumping of compartments. We define a lumping matrix M as follows:

$$M_{n_{1},n}=\bordermatrix{
                   &c_1&c_2&\ldots &c_n\cr
                c^{*}_{1}&a_{1,1} &  a_{1,2}  & \ldots & a_{1,n}\cr
                c^{*}_{2}& a_{2,1}  &  a_{2,2} & \ldots & a_{2,n}\cr
                \vdots& \vdots & \vdots & \ddots & \vdots\cr
                c^{*}_{n_{1}}& a_{n_{1},1}  &   a_{n_{1},2}       &\ldots & a_{n_{1},n}},$$

where $a_{ij}\in\lbrace 0,1 \rbrace$ for $i=1,2,...,n_{1}$\; and $j=1,2,...,n$. There is an important equation that is called lumping transformation:
\begin{equation}
\begin{array}{llll}  
 C^{*}=M C.
\end{array}
\label{TransformationA}
\end{equation}
From equation \eqref{TransformationA}, the set of original states $C$ can be calculated as follows:
\begin{equation}
\begin{array}{llll}  
C=M^{+} C^{*},
\end{array}
\label{TransformationB}
\end{equation} 
where $M^{+}$ is pseudo inverse of $M$, such that $MM^+=I$. Multiplying both sides of equation (\ref{general linear system 1}) by $M$, the system becomes
 \begin{equation}
\begin{array}{llll}  
\dfrac{dC^{*}}{dt}=K^{*}C^{*} = H^{*}(C^{*},P^{*}) ,
\end{array}
\label{reduced system 1}
\end{equation}
where $K^{*}=M K M^{+}$ and $C^{*}\in R^{n_{1}}, n_{1}\leq n$ and $P^{*}\in R^{m_{1}}, m_{1}\leq m$. The equation \eqref{reduced system 1} is called reduced model of the system \eqref{general linear system 1}.


\noindent
If $H(C,P)$ a is nonlinear function then equation \eqref{general system 1} becomes,
\begin{equation}
\begin{array}{llll}  
\dfrac{dC}{dt}=K C + G(C),
\end{array}
\label{general nonlinear system 1}
\end{equation}
\newline
where $G(C)$ is a non--linear term. By applying the above procedure, the equation \eqref{general nonlinear system 1} takes the following form
\begin{equation}
\begin{array}{llll}    
\dfrac{dC^{*}}{dt}=K^{*} C^{*} + M G(M^{+} C^{*}).
\end{array}
\label{reduced nonlinear  system +G(C) 1}
\end{equation}
Thus, the equation \eqref{reduced nonlinear  system +G(C) 1} is called reduced model.
Furthermore, the reduced model \eqref{reduced nonlinear  system +G(C) 1} can be also written as follows
\begin{equation}
\begin{array}{llll}  
\dfrac{dC^{*}}{dt}=M H(M^{+} C^{*}).
\end{array}\label{reduced nonlinear system 1}
\end{equation}


\section{Applications}
The proposed technique here plays an important role in model reductions. Particularly, this method can be used for model reductions in chemical reaction networks and cell signalling pathways. We apply this technique in linear and nonlinear chemical reaction models in order to reduce the number of state variables.

\subsection{Linear Networks}
The idea of lumping species can be simply used for linear chemical networks. We consider a linear network with three species and six parameters. 
\begin{equation}  
\begin{array}{llll}
 \quad \quad \; \; \;  p_{0}\uparrow\\
\quad C_{2} { \overset{p_{1}}\longleftarrow} C_{1} 
\underset{p_{4}}{ \overset{p_{3}}\rightleftharpoons} C_{3}\\
p_{2}\downarrow \quad \quad \quad \; p_{5}\downarrow
\end{array}\label{linear network diagram 1}
\end{equation}
Then, the stoichiometric matrix and sate variables of the network are given, respectively. 
 \begin{equation*} 
K={\left(\begin{array}{cccc}
-(p_{0}+p_{1}+p_{3})&0&p_{4}\\ \\ p_{1}&-p_{2}&0\\ \\p_{3}&0&-(p_{4}+p_{5}),
\end{array}\right)}, C={\left(\begin{array}{cccc}
C_{1}\\ \\ C_{2}\\ \\C_{3}
\end{array}\right)} .
\end{equation*}
Using mass action law, the system of ODE's for linear network \eqref{linear network diagram 1} is given below
\begin{equation} 
\begin{array}{llll}   
\dfrac{dC_{1}}{dt}=-(p_{0}+p_{1}+p_{3})C_{1}+p_{4}C_{3},\\
\dfrac{dC_{2}}{dt}=p_{1}C_{1}-p_{2}C_{2},\\
\dfrac{dC_{3}}{dt}=p_{3}C_{1}-(p_{4}+p_{5})C_{3},\\
\end{array} \label{ODE orginal simple example linear}
\end{equation}
with initial conditions $C_{1}(0)=5,\;C_{2}(0)=C_{3}(0)=0$,\;and chemical reaction constants $p_{0}=1,\;p_{1}=8,\;p_{2}=25,\;p_{3}=10,\;p_{4}=15,\;p_{5}=20.$

We use the following proper lumping for the linear network \eqref{linear network diagram 1}
\begin{figure}[H]  
	\begin{center}             
		\subfigure{%
			\includegraphics[width=0.4\textwidth]{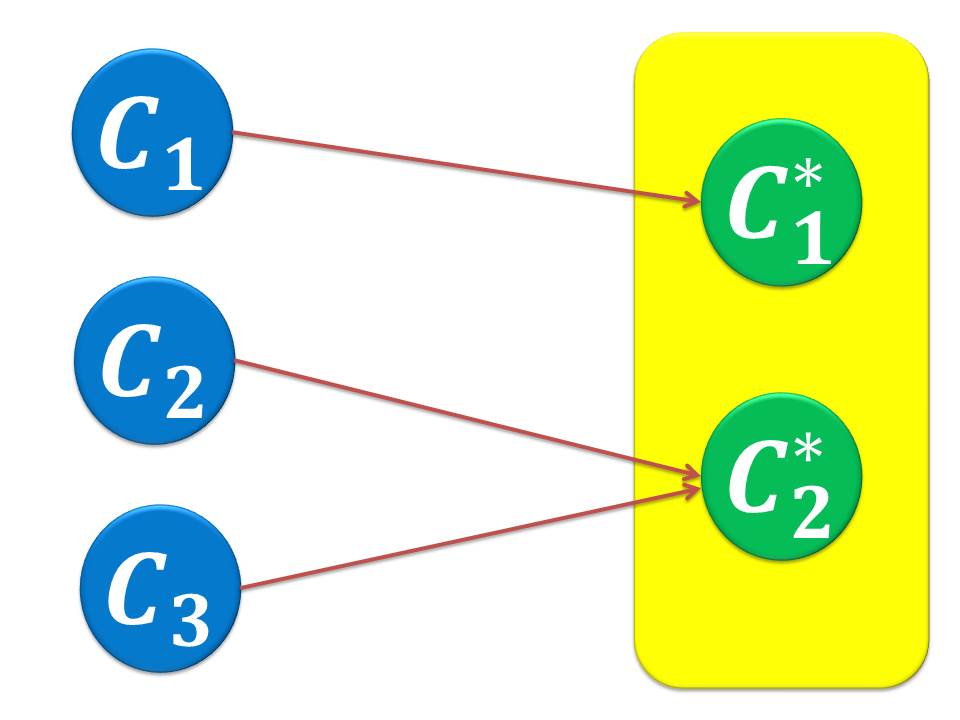}
 		}             
	\end{center}    
	\caption {Lumping species of the linear network \eqref{linear network diagram 1}. }     
\end{figure} 

The proper lumping here is simply presented as  $C^{*}_{1}=C_{1},$\;and $C^{*}_{2}=C_{2}+C_{3}$. Then, the model network takes the following form
\begin{equation}  
\begin{array}{llll}
\uparrow\\
  C^{*}_{1} {\rightleftharpoons} C^{*}_{2}\\
\quad \quad \; \; \downarrow 
\end{array}\label{reduce linear network diagram 1}
\end{equation}
The lumping matrix $M$ is given 
 \begin{equation*} 
M={\left(\begin{array}{cccc}
1&0&0\\ \\ 0&1&1
\end{array}\right)} .
\end{equation*}
The pseudo-inverse of $M$ is calculated as follows
 \begin{equation*} 
M^{+}={\left(\begin{array}{cccc}
1&0\\ \\ 0&\dfrac{1}{2}\\ \\0&\dfrac{1}{2}
\end{array}\right)}.
\end{equation*}
By using equation \eqref{reduced system 1}, the reduced model then becomes
\begin{equation} 
\dfrac{d}{dt}{\left(\begin{array}{cccc}
C^{*}_{1}\\C^{*}_{2}
\end{array}\right)}
={\left(\begin{array}{cccc}
-(p_{0}+p_{1}+p_{3})C^{*}_{1}+\dfrac{1}{2}p_{4}C^{*}_{2}\\ \\(p_{1}+p_{3})C^{*}_{1}-\dfrac{1}{2}(p_{2}+p_{4}+p_{5})C^{*}_{2}
\end{array}\right)}.
\label{ode reduced linewar example limping}
\end{equation} 

\begin{figure}[H]  
	\begin{center}         
		\subfigure{%
			\includegraphics[width=0.32\textwidth]{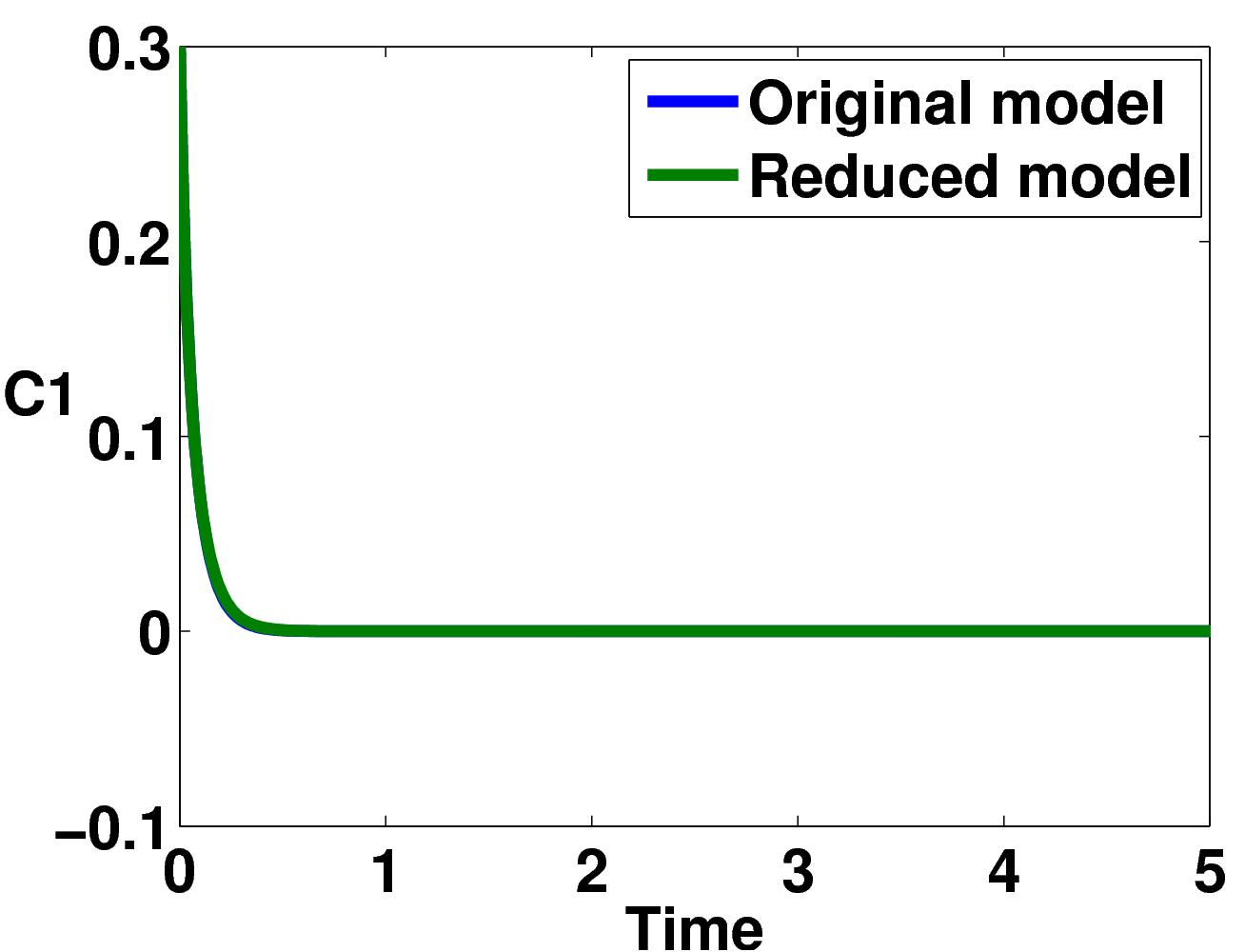}
		}  		
		\subfigure{%
			\includegraphics[width=0.32\textwidth]{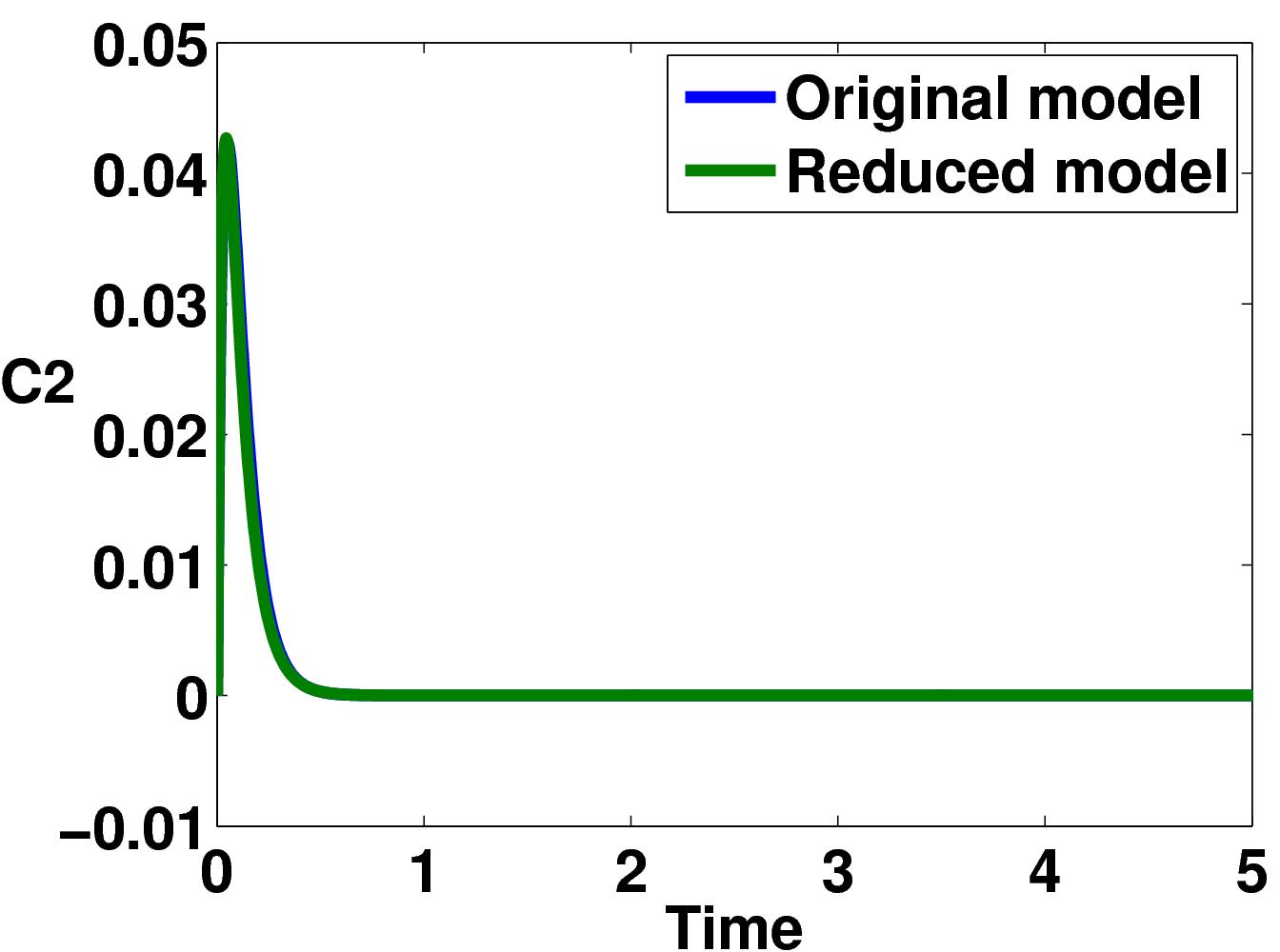}
		}   
		\subfigure{%
			\includegraphics[width=0.32\textwidth]{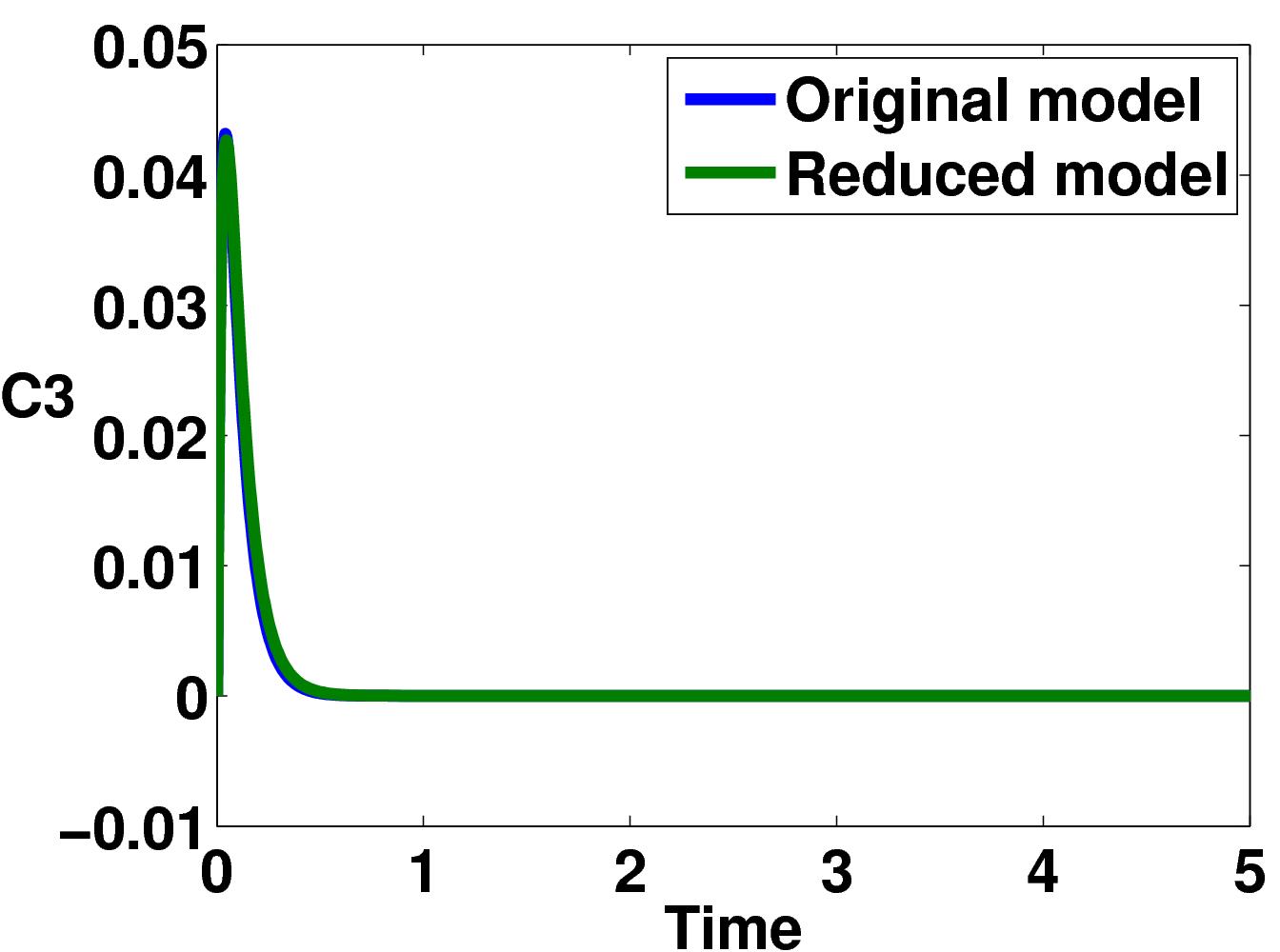}
		}
		
	\end{center}   
	\caption {Computational simulations of $C_{1},\;C_{2}\;\text{and} \; C_{3}$\; in original \eqref{ODE orginal simple example linear} and reduced \eqref{ode reduced linewar example limping} system; the blue lines are the original and the green lines are the reduced approximate solutions, with the time interval $[0,5]$ for numerical simulations.}    	
	\label{fig:solution of linear simple example}
\end{figure}

We use computational simulations for comparing the dynamics of the state variables in original and reduced models. It can be concluded that there is a good agreement between the original and the reduced model for initial conditions and parameters used in numerical simulations; see Figure \eqref{fig:solution of linear simple example}.

\subsection{Non--Linear Networks}
The proposed technique can also be used for non--linear chemical networks. We consider a non--linear network with four variables and two parameters
\begin{equation}  
\begin{array}{llll}
C_{1} { \overset{p_{1}}{\longrightarrow}} C_{2}+C_{3} 
{ \overset{p_{2}}{\longrightarrow}} C_{4} .
\end{array}\label{nonlinear network diagram 1}
\end{equation}
By using mass action law, the system of ODE's of the model \eqref{nonlinear network diagram 1} is given
\begin{equation}
\begin{array}{llll}  
\dfrac{dC_{1}}{dt}=-p_{1}C_{1},\\
\dfrac{dC_{2}}{dt}=p_{1}C_{1}-p_{2}C_{2}C_{3},\\
\dfrac{dC_{3}}{dt}=p_{1}C_{1}-p_{2}C_{2}C_{3},\\
\dfrac{dC_{4}}{dt}=p_{2}C_{2},
\end{array} \label{ODE orginal simple example nonlinear}
\end{equation}
with initial conditions $C_{1}(0)=10,\;C_{2}(0)=C_{3}(0)=C_{4}(0)=0$,\;and chemical reaction parameters $p_{1}=6,\;p_{2}=5.$

We use the following proper lumping for the linear network \eqref{nonlinear network diagram 1},
 
\begin{figure}[H]  
	\begin{center}             
		\subfigure{%
			\includegraphics[width=0.4\textwidth]{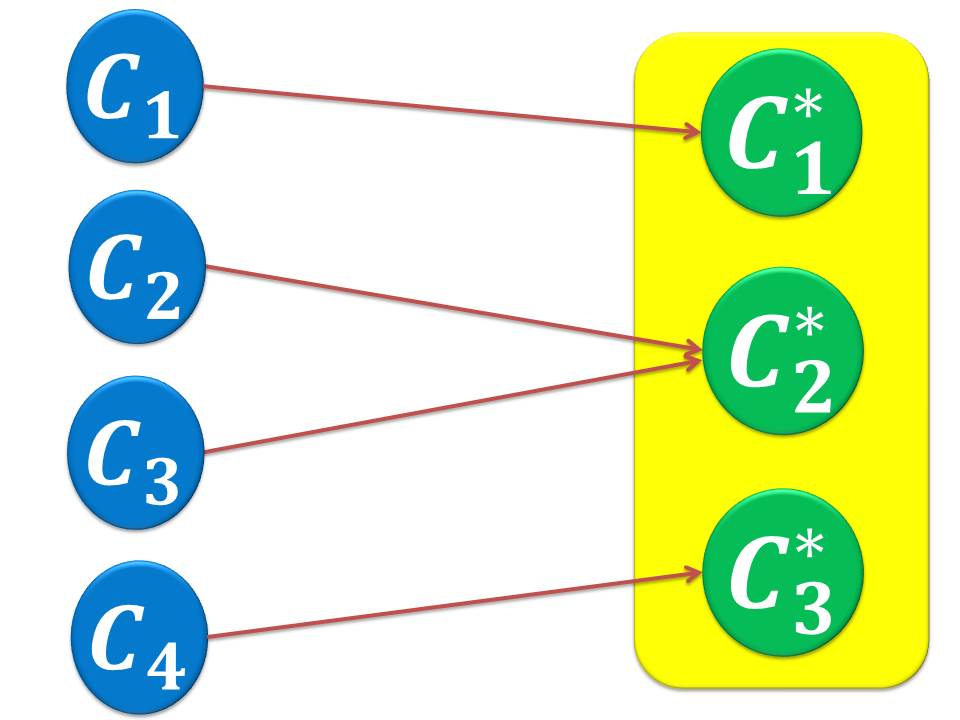}
 		}             
	\end{center}    
	\caption {Proper lumping species of the non--linear network \eqref{nonlinear network diagram 1}. }     
\end{figure} 

The proper lumping here is simply used as
 $C^{*}_{1}=C_{1},C^{*}_{2}=C_{2}+C_{3}\;$ and \;$ C^{*}_{3}=C_{4}$. Then, the model network takes the following form

\begin{equation}  
\begin{array}{llll}
C^{*}_{1} {\longrightarrow} C^{*}_{2} 
{\longrightarrow} C^{*}_{3} 
\end{array}.
\end{equation}

The lumping matrix $M$ is given
 \begin{equation*} 
M={\left(\begin{array}{cccc}
1&0&0&0\\ \\ 0&1&1&0\\ \\0&0&0&1
\end{array}\right)}.
\end{equation*}
The pseudo--inverse of $M$ is calculated as follows
 \begin{equation*} 
M^{+}={\left(\begin{array}{cccc}
1&0&0\\ \\ 0&\dfrac{1}{2}&0\\ \\0&\dfrac{1}{2}&0\\ \\0&0&1
\end{array}\right)}.
\end{equation*} 
Using the equation \eqref{TransformationB}, we obtain the following relations 
 \begin{equation*} 
{\left(\begin{array}{cccc}
C_{1}\\ \\C_{2}\\ \\C_{3}\\ \\C_{4}
\end{array}\right)}
={\left(\begin{array}{cccc}
1&0&0\\ \\ 0&\dfrac{1}{2}&0\\ \\0&\dfrac{1}{2}&0\\ \\0&0&1
\end{array}\right)}
{\left(\begin{array}{cccc}
C^{*}_{1}\\ \\C^{*}_{2}\\ \\C^{*}_{3}
\end{array}\right)}
={\left(\begin{array}{cccc}
C^{*}_{1}\\ \\ \dfrac{C^{*}_{2}}{2}\\ \\ \dfrac{C^{*}_{2}}{2}\\C^{*}_{3}
\end{array}\right)}.
\end{equation*}

By using equation \eqref{reduced nonlinear system 1}, the reduced model then becomes
\begin{equation} 
\dfrac{d}{dt}{\left(\begin{array}{cccc}
C^{*}_{1}\\ \\ C^{*}_{2}\\ \\C^{*}_{3}
\end{array}\right)}
={\left(\begin{array}{cccc}
-p_{1}C^{*}_{1}\\ \\2p_{1}C^{*}_{1}- \dfrac{p_{2}}{2} (C^{*}_{2})^{2}\\ \\ \dfrac{p_{2}}{4}(C^{*}_{2})^{2}
\end{array}\right)}.
\label{nonlinear simple example lumping reduced ode}
\end{equation}

\begin{figure}[H]  
	\begin{center}     
		\subfigure{%
			\includegraphics[width=0.48\textwidth]{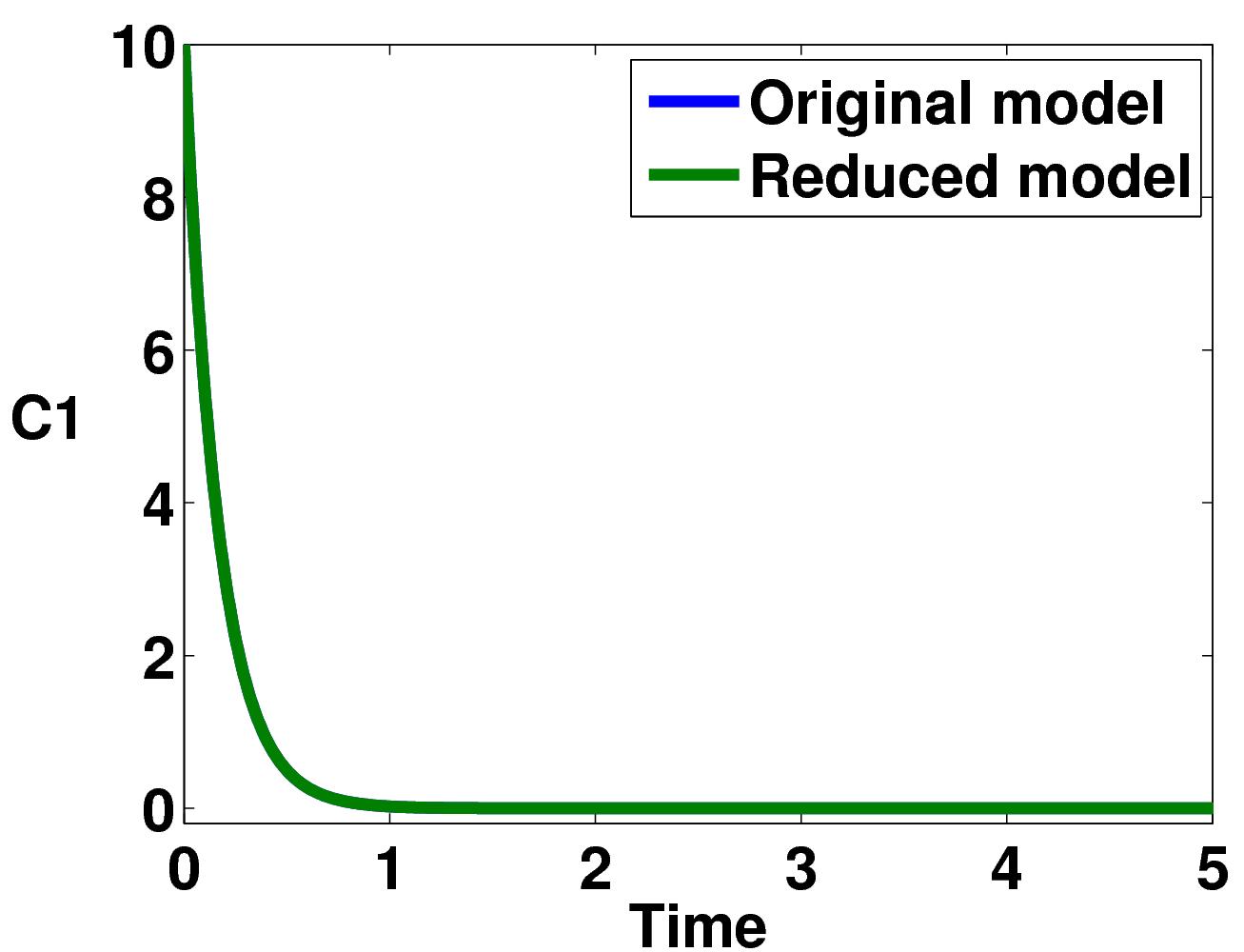}
		}		
		\subfigure{%
			\includegraphics[width=0.48\textwidth]{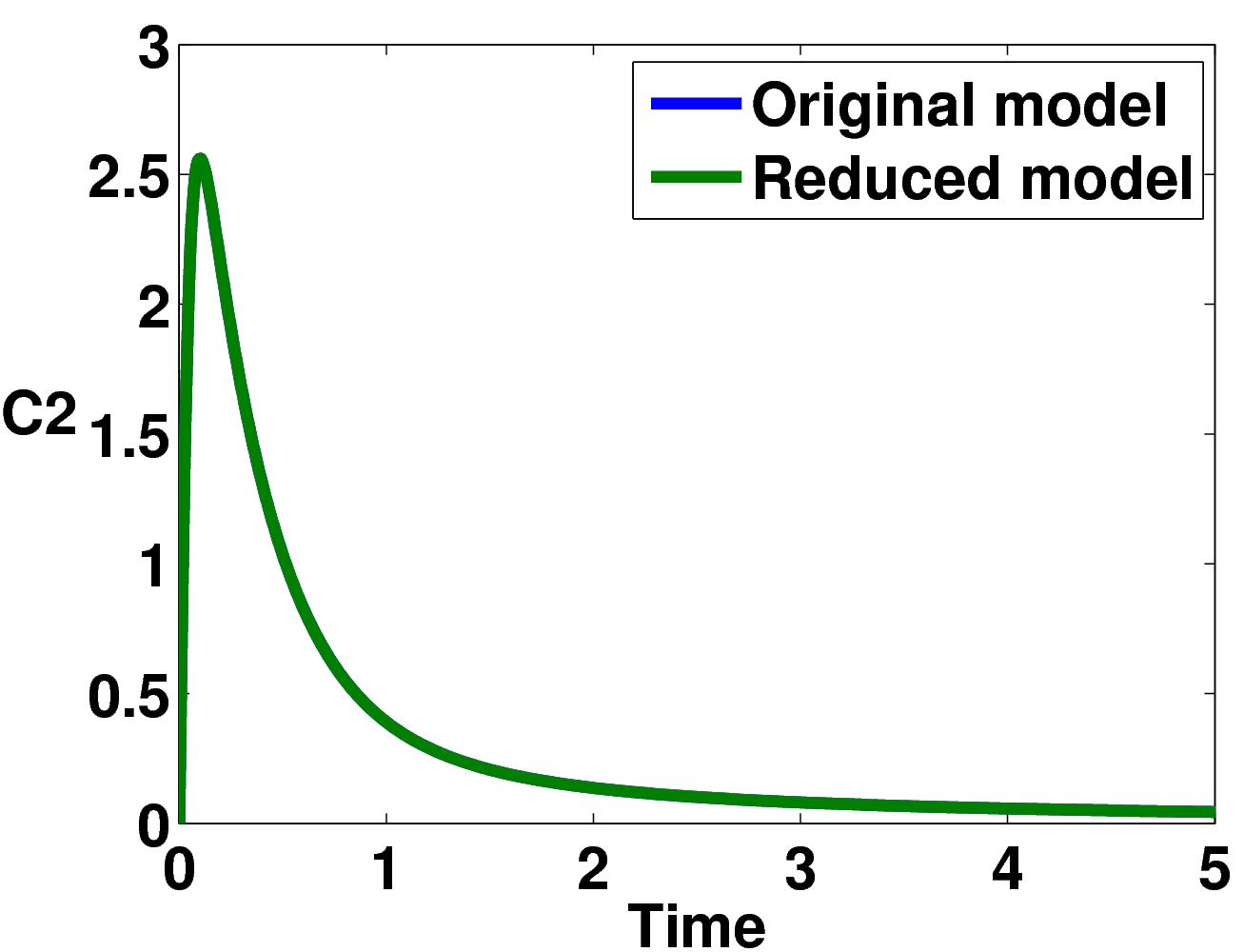}
		} \\
		\subfigure{%
			\includegraphics[width=0.48\textwidth]{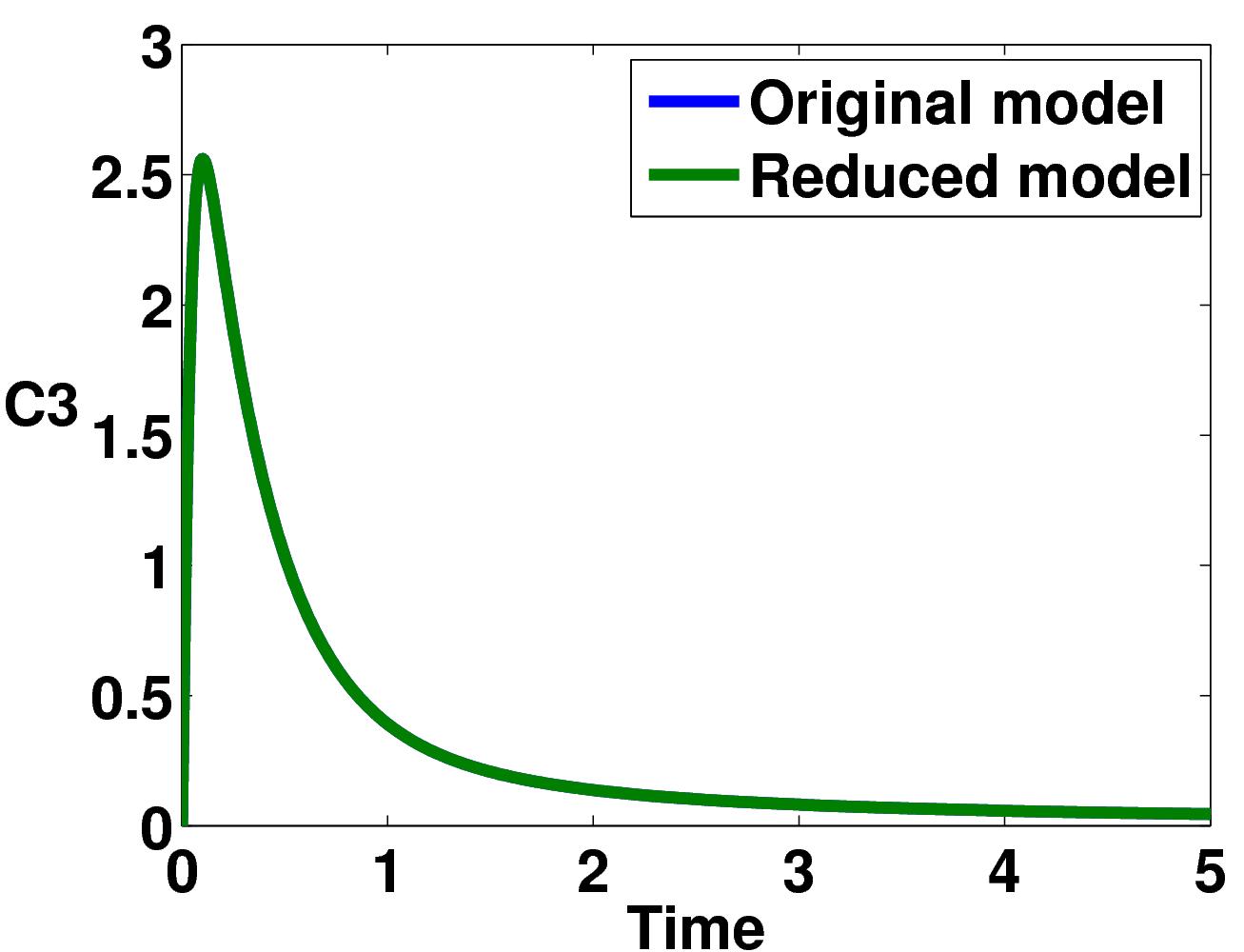}
		}
     	\subfigure{%
			\includegraphics[width=0.48\textwidth]{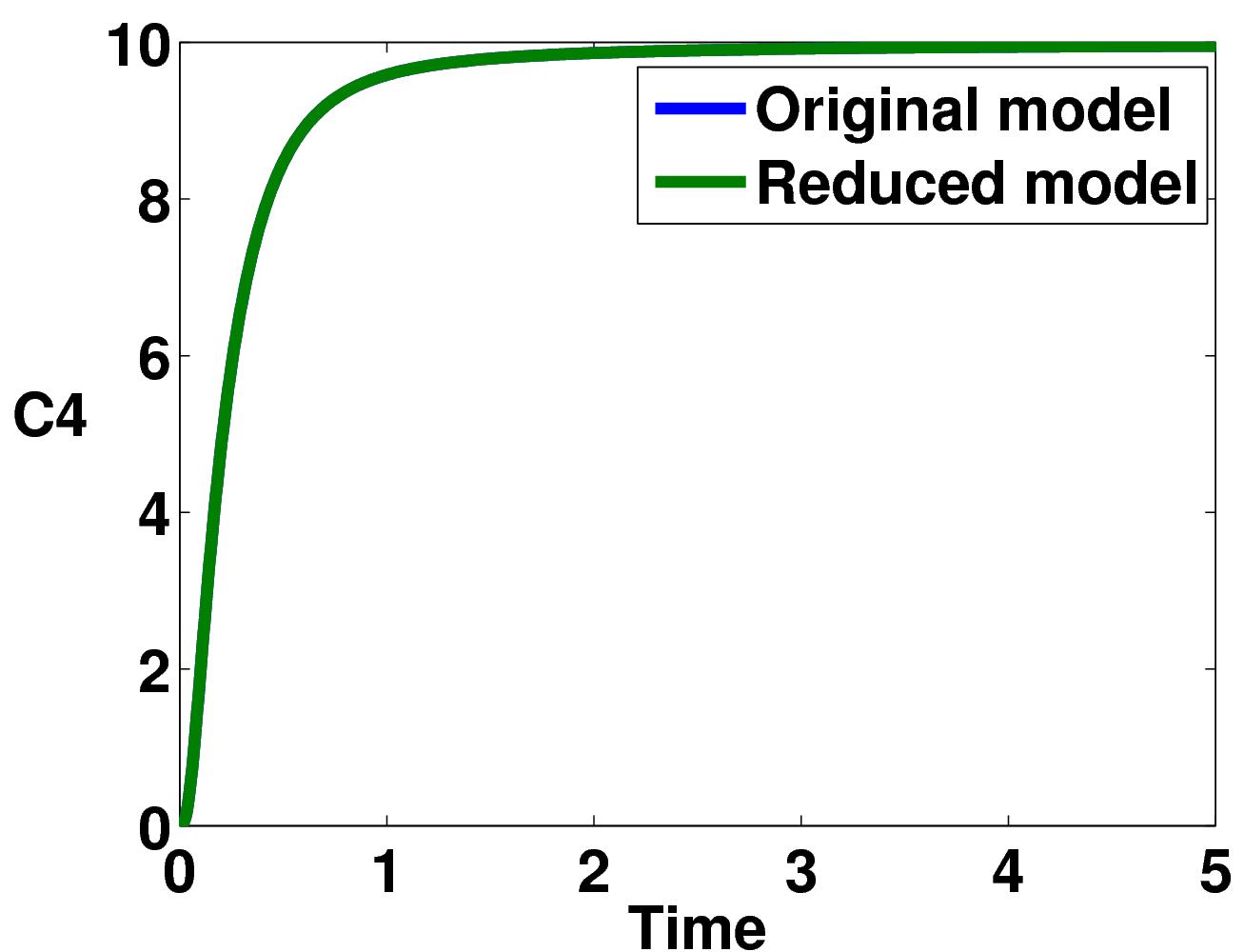}
		}
	\end{center}   
	\caption {Numerical simulations for $C_{1},\;C_{2},\;C_{3}\;\text{and} \; C_{4}$\; in original \eqref{ODE orginal simple example nonlinear} and reduced \eqref{nonlinear simple example lumping reduced ode} system; the blue lines are the original and the green lines are the reduced approximate solutions, with the time interval $[0,5]$ for computational simulations.}    	
	\label{fig:solution of nonlinear simple example}  
\end{figure}

\noindent We use computational simulations for comparing the dynamics of the state variables in original and reduced models. It can be concluded that there is a good agreement between the original and the reduced model for initial conditions and parameters used in numerical simulations, see Figure \eqref{fig:solution of nonlinear simple example}.

\newpage
\subsection{ERK Signalling Pathways }
\begin{figure}[H]  
\begin{center}     
		\subfigure{%
			\includegraphics[width=1\textwidth]{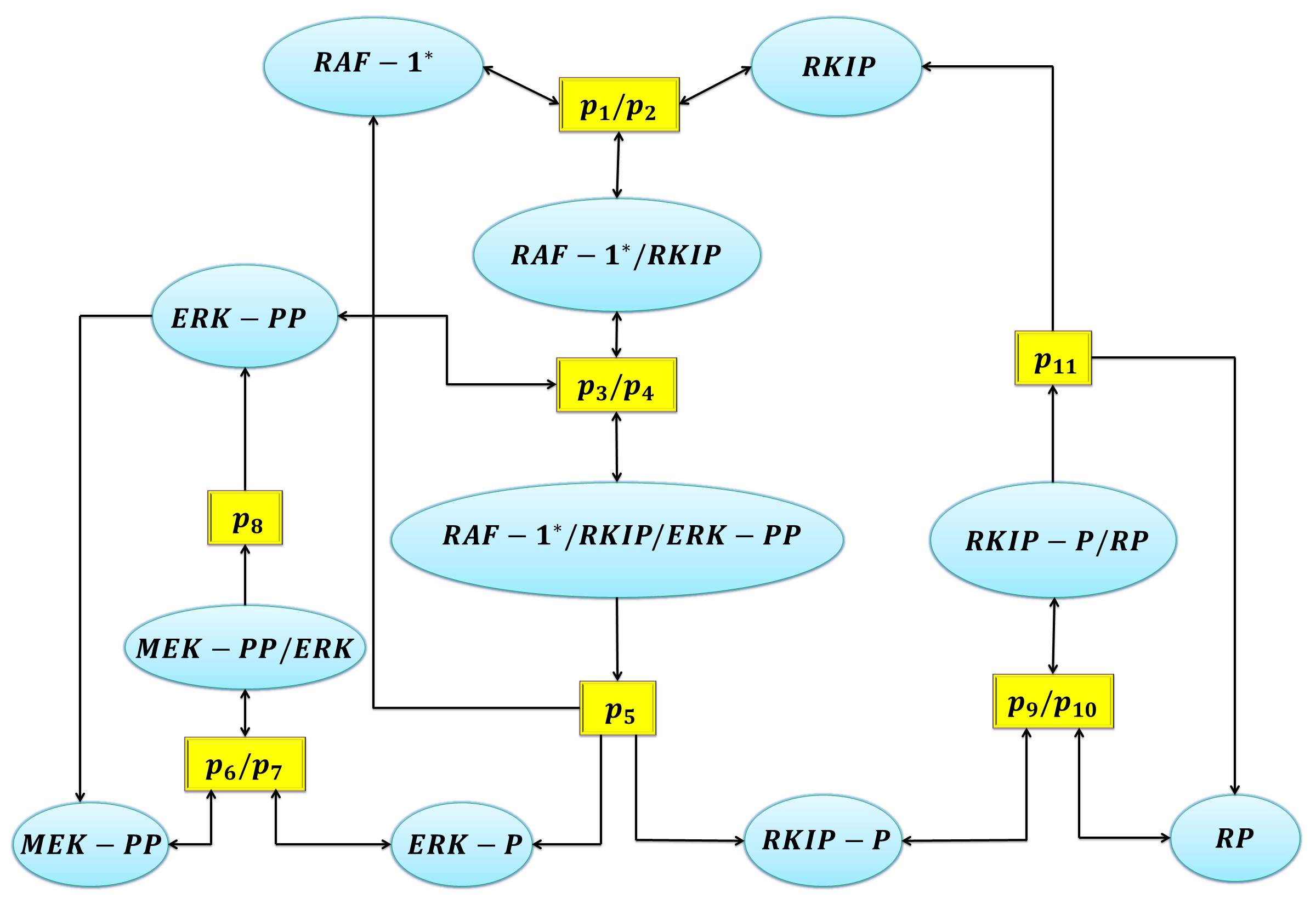}
		}		
	\end{center}   
	\caption {Graphical representation of the ERK signaling pathways.}    	
	\label{ERK model diagram} 
\end{figure}

\noindent
In cell signaling pathways, extracellular-signal-regulated kinase (ERK) pathway can be identified as an important case of the mitogen activated protein kinase (MAPK) pathway. The given pathway is sometimes known as the Ras-Raf-MEK-ERK pathway. This is often occurred as a chain of proteins in the cell. The signaling pathways have a greater role for transformation a signal between receptors and DNA in the cell. Receptors are located on the cell membrane that receive signals from outside cells (Orton et al., 2005).  The MEK-ERK pathway consists of chain of proteins that can be joined to adjacent proteins via phosphate groups. They work as an $"on"$ and $"off"$ switch. 
Such signaling pathways are also connected with some human diseases. The well-known diseases MAPK signaling in cancer (McCubrey et al., 2007). It is obvious that Ras and B-Raf   are occurred   in many cancers of MAPK signaling pathways. Such  proteins play a role in cell division and  differentiation. 
Some steps of tumor development are also affected by the ERK signaling pathway. This is happened when a protein is mutated and it is fixed in the $"on"$ or $"off"$ position. The pathway components were initially investigated in cancer cells. In cancer treatments, some drugs are used that reverse the $"on"$ or $"off"$ switch (Ramos, 2008, Shaul and Seger, 2007, Yao and Seger, 2009).
In point of view, controlling is an important process in cell differentiation and proliferation. One of the main scientific interests in cell signaling is understanding the reaction mechanism. RKIP plays on the behavior of this pathway this achieved by the experimental investigation. In fact, ERK is a complex signaling pathway and includes a set of variables and parameters.  Figure \eqref{ERK model diagram} only shows a part of the ERK pathway, it considers the subset of the ERK pathway regulated by RKIP(Raf kinase inhibitor protein). Here, each node of the scheme is labeled and the corresponding protein is denoted. For example, ERK-P and MEK-PP are proteins, but MEK-PP/ERK is a complex built-up from the fifth and seven. The suffix $-P$ and $–PP$ denote phosphorylated and double phosphorylated proteins, respectively. The concentration of each signaling component is denoted by $\lbrace c_{i}:i = 1, 2, . . . , 11\rbrace$. Moreover, reaction rate  constants are denoted by $\lbrace p_{i}:i = 1, 2, . . . , 11 \rbrace$. Here, $\lbrace c_{i}:i = 1, 2,..., 11 \rbrace$ are state variables representing concentrations of the proteins $Raf-1_, RKIP, Raf-1_/RKIP, Raf-1_/RKIP/ERK-PP, ERK-P, RKIP-P, MEK-PP, MEK-PP/ERK, ERK-PP, RP$ and $RKIP-P/RP$ respectively, and $p_{i} (i = 1, 2, . . . , 11)$ are corresponding model coefficients (reaction rate constants) (Petrov et al., 2007). The chemical reaction network of  the ERK signaling pathways is given, see Figure \eqref{ERK model diagram}.
There are also a set of data for state variables and parameters, see Tables \eqref{table of parameter ERK} and \eqref{table of variable ERK}.

\FloatBarrier 
\begin{table}
\caption{Summary of parameter values for ERK signalling pathways.}
\centering
\begin{tabular}{|p{2cm}|c|c|c|c|c|c|c|c|c|c|c|} 
\hline 
Parameters & $p_{1}$ & $p_{2}$ & $p_{3}$ & $p_{4}$ & $p_{5}$ & $p_{6}$ & $p_{7}$ & $p_{8}$ & $p_{9}$ & $p_{10}$ & $p_{11}$ \\ 
\hline 
Estiamate values & 0.191 & 0.09 & 0.433 & 0.93 & 5 & 0.031 & 0.95 & 4 & 0.9 & 10 & 7 \\ 
\hline 
\end{tabular} 
\label{table of parameter ERK}
\end{table}

\begin{table}
\caption{Stationary values of state variables for ERK signalling pathways.}
\centering
\begin{tabular}{|c|c|c|c|}
\hline 
No. & State variables & Symbols & Stationary values \\ 
\hline 
1 & $RAF-1^{*}$ & $C_{1}$ & 0.01 \\ 
\hline 
2 & $RKIP$ & $C_{2}$ & 0.1 \\ 
\hline 
3 & $RAF-1^{*}/RKIP$ & $C_{3}$ & 0.4 \\ 
\hline 
4 & $RAF-1^{*}/RKIP/ERK-pp$ & $C_{4}$ & 0.4 \\ 
\hline 
5 & $ERK-P$ & $C_{5}$ & 0.1 \\ 
\hline 
6 & $RKIP-P$ & $C_{6}$ & 0.05 \\ 
\hline 
7 & $MEK-PP$ & $C_{7}$ & 0.55 \\ 
\hline 
8 & $MEK-PP/ERK$ & $C_{8}$ & 0.5 \\ 
\hline 
9 & $ERK-PP$ & $C_{9}$ & 0.4 \\ 
\hline 
10 & $RP$ & $C_{10}$ & 0.19 \\ 
\hline 
11 & $RKIP-P/RP$ & $C_{11}$ & 0.1 \\ 
\hline 
\end{tabular} 
\label{table of variable ERK}
\end{table}
\FloatBarrier 

\noindent Then the biochemical diagram \eqref{ERK model diagram} is represented mathematically by the following system of nonlinear differential equations

\begin{equation}
     \begin{array}{llll}  
\dfrac{dC_{1}}{dt}=-p_{1}C_{1}C_{2}+p_{2}C_{3}+p_{5}C_{4},\\
\dfrac{dC_{2}}{dt}=-p_{1}C_{1}C_{2}+p_{2}C_{3}+p_{11}C_{11},\\
\dfrac{dC_{3}}{dt}=p_{1}C_{1}C_{2}-p_{2}C_{3}-p_{3}C_{3}C_{9}+p_{4}C_{4},\\
\dfrac{dC_{4}}{dt}=p_{3}C_{3}C_{9}-p_{4}C_{4}-p_{5}C_{4},\\
\dfrac{dC_{5}}{dt}=p_{5}C_{4}-p_{6}C_{5}C_{7}+p_{7}C_{8},\\
\dfrac{dC_{6}}{dt}=p_{5}C_{4}-p_{9}C_{6}C_{10}+p_{10}C_{11},\\
\dfrac{dC_{7}}{dt}=-p_{6}C_{5}C_{7}+p_{7}C_{8}+p_{8}C_{8},\\
\dfrac{dC_{8}}{dt}=p_{6}C_{5}C_{7}-p_{7}C_{8}-p_{8}C_{8},\\
\dfrac{dC_{9}}{dt}=-p_{3}C_{3}C_{9}+p_{4}C_{4}+p_{8}C_{8},\\
\dfrac{dC_{10}}{dt}=-p_{9}C_{6}C_{10}+p_{10}C_{11}+p_{11}C_{11},\\
\dfrac{dC_{11}}{dt}=p_{9}C_{6}C_{10}-p_{10}C_{11}-p_{11}C_{11}.
     \end{array} \label{ODE ERk signaling}
\end{equation}

\noindent In this chemical reaction pathways, we can apply the proposed technique of model reduction. This is for minimizing the number of state variables. Therefore, we take some cases of species lumping with total differences between the reduced and original models, see Table \eqref{table lumping of ERK}.

\FloatBarrier 
\begin{table}
\caption{Applying lumping technique for the model signalling pathways \eqref{ODE ERk signaling}, six different cases are used with their total error and remaining variables in the reduced model.}
\begin{tabular}{|c|p{5.5cm}|p{2.5cm}|p{2cm}|p{2cm}|}
\hline 
Cases & Lumping species & Total differences & Remaining variables &Lumping types\\ 
\hline 
Case\;1 & $C^{*}_{1}=C_{1}+C_{2}+C_{5}+C_{7}$,\;$C^{*}_{2}=C_{3}+C_{6}+C_{9}+C_{10}$,\;$C^{*}_{3}=C_{4}+C_{8}+C_{11}$&71.44\% &\;\;\quad 3&Proper\\ 
\hline
Case\;2 & $C^{*}_{1}=C_{1}+C_{2}$,\;$C^{*}_{2}=C_{3}+C_{6}+C_{9}$,\;$C^{*}_{3}=C_{4}+C_{8}+C_{11}$,\;$C^{*}_{4}=C_{5}+C_{7}+C_{10}$ &123.72\% &\;\;\quad4&Proper\\ 
\hline 
Case\;3 & $C^{*}_{1}=C_{1}+C_{2}$,\;$C^{*}_{2}=C_{3}+C_{9}$,\;$C^{*}_{3}=C_{4}+C_{8}+C_{11}$,\; $C^{*}_{4}=C_{5}+C_{7}$,\;$C^{*}_{5}=C_{6}+C_{9}+C_{10} $&31.38\% &\;\;\quad 5&Improper\\ 
\hline 
Case\;4 & $C^{*}_{1}=C_{1}+C{2},\;C^{*}_{2}=C_{3}$,\;$C^{*}_{3}=C_{5}+C{7}$,\;$C^{*}_{4}=C_{4}+C{8}+C_{11}$,\;$C^{*}_{5}=C_{6}+C{9},\;C^{*}_{6}=C_{10}$ & 83.19\% & \;\;\quad 6&Proper \\ 
\hline 
Case\;5 & $C^{*}_{1}=C_{1}+C_{2}$,\;\;\;$C^{*}_{2}=C_{3}+C_{9},$\; $C^{*}_{3}=C_{4}$,\;$C^{*}_{4}=C_{6}+C_{10},\;C^{*}_{5}=C_{11}$,\;$C^{*}_{6}=C_{5}+C_{7},\;C^{*}_{7}=C_{8}$ & 9.87\%  &\;\;\quad 7&Proper \\ 
\hline 
Case\;6 & $C^{*}_{1}=C_{1}+C_{2},\;C^{*}_{2}=C_{3}$,\; $C^{*}_{3}=C_{4}+C_{8},\;C^{*}_{4}=C_{5}$,\; $C^{*}_{5}=C_{6}+C_{9},\;C^{*}_{6}=C_{7}$,\;$C^{*}_{7}=C_{10},$\;$C^{*}_{8}=C_{11} $ & 4.54\% &\;\;\quad 8 &Proper\\ 
\hline 
\end{tabular}  
\label{table lumping of ERK}
\end{table}  
\FloatBarrier 

\begin{figure}[H]  
	\begin{center}     
		\subfigure{%
			\includegraphics[width=0.32\textwidth]{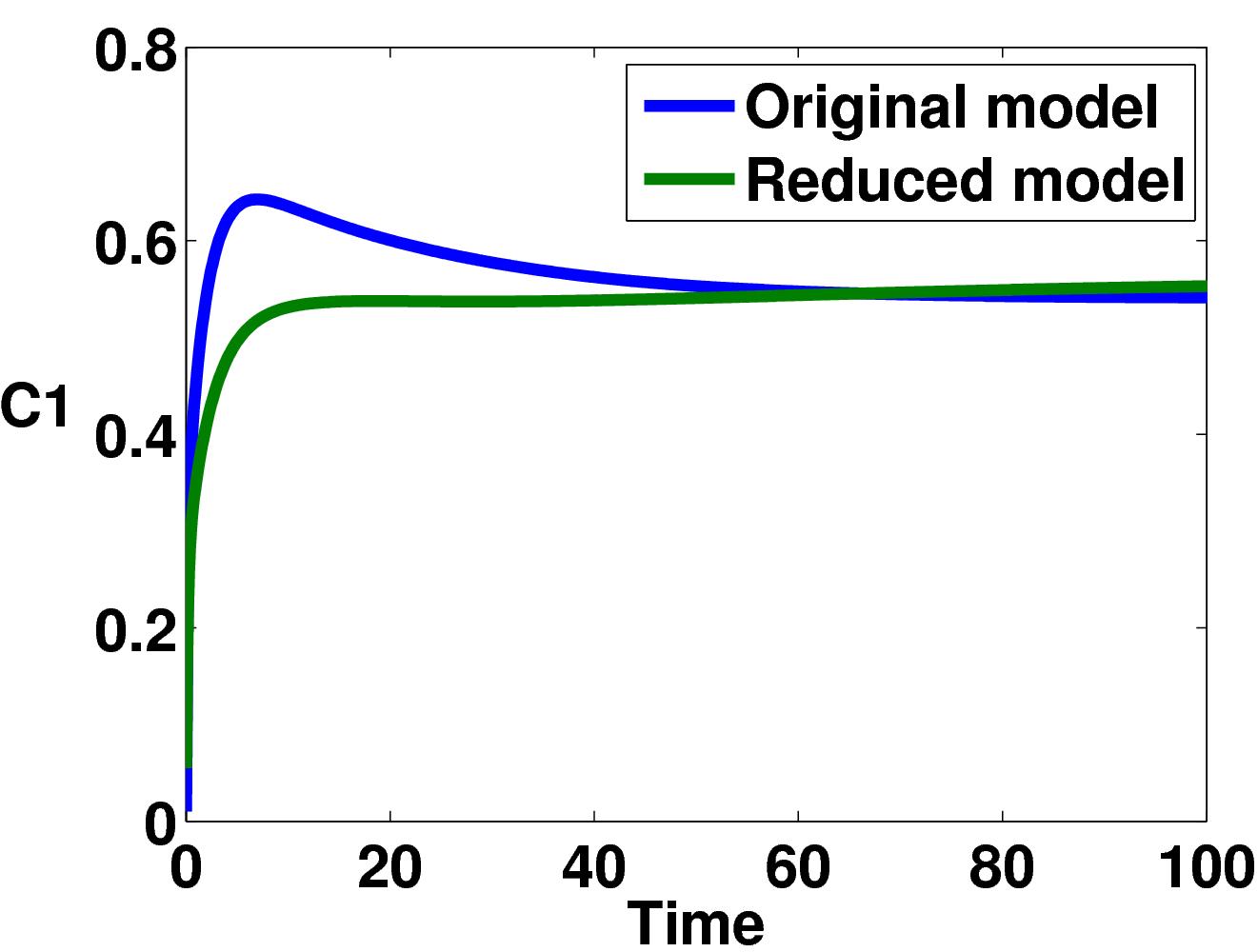}
		}		
		\subfigure{%
			\includegraphics[width=0.32\textwidth]{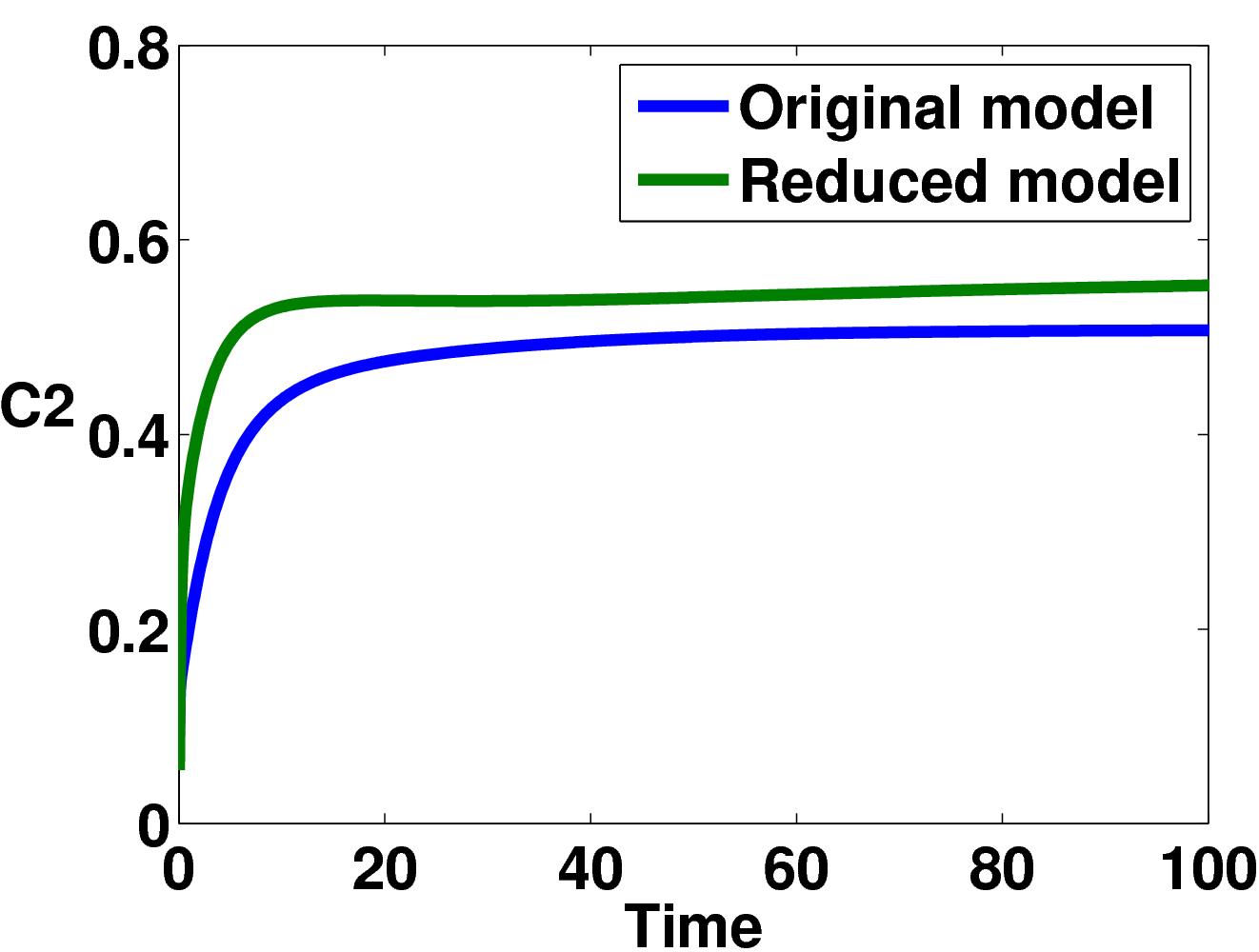}
		} \\
		\subfigure{%
			\includegraphics[width=0.32\textwidth]{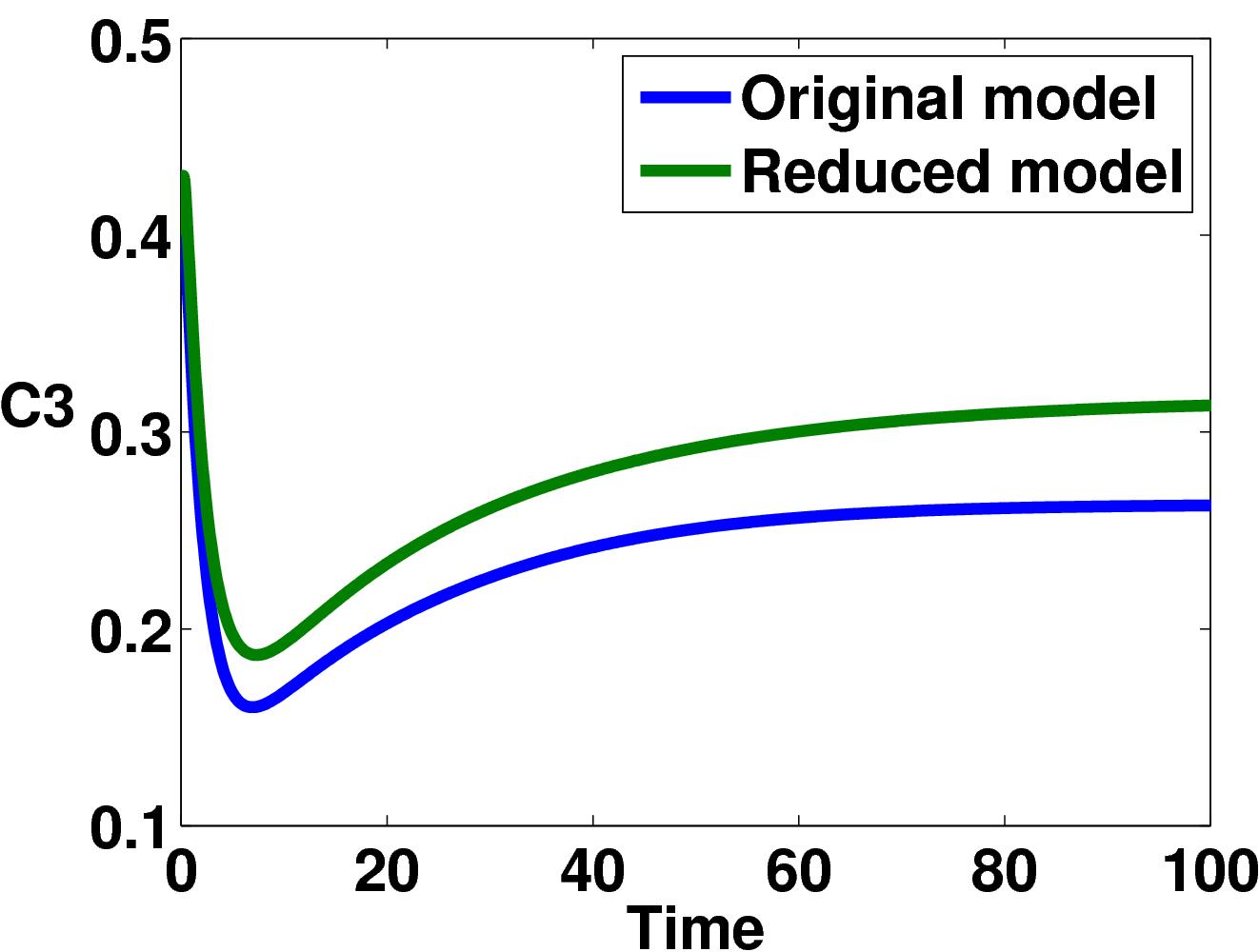}
		}
     	\subfigure{%
			\includegraphics[width=0.32\textwidth]{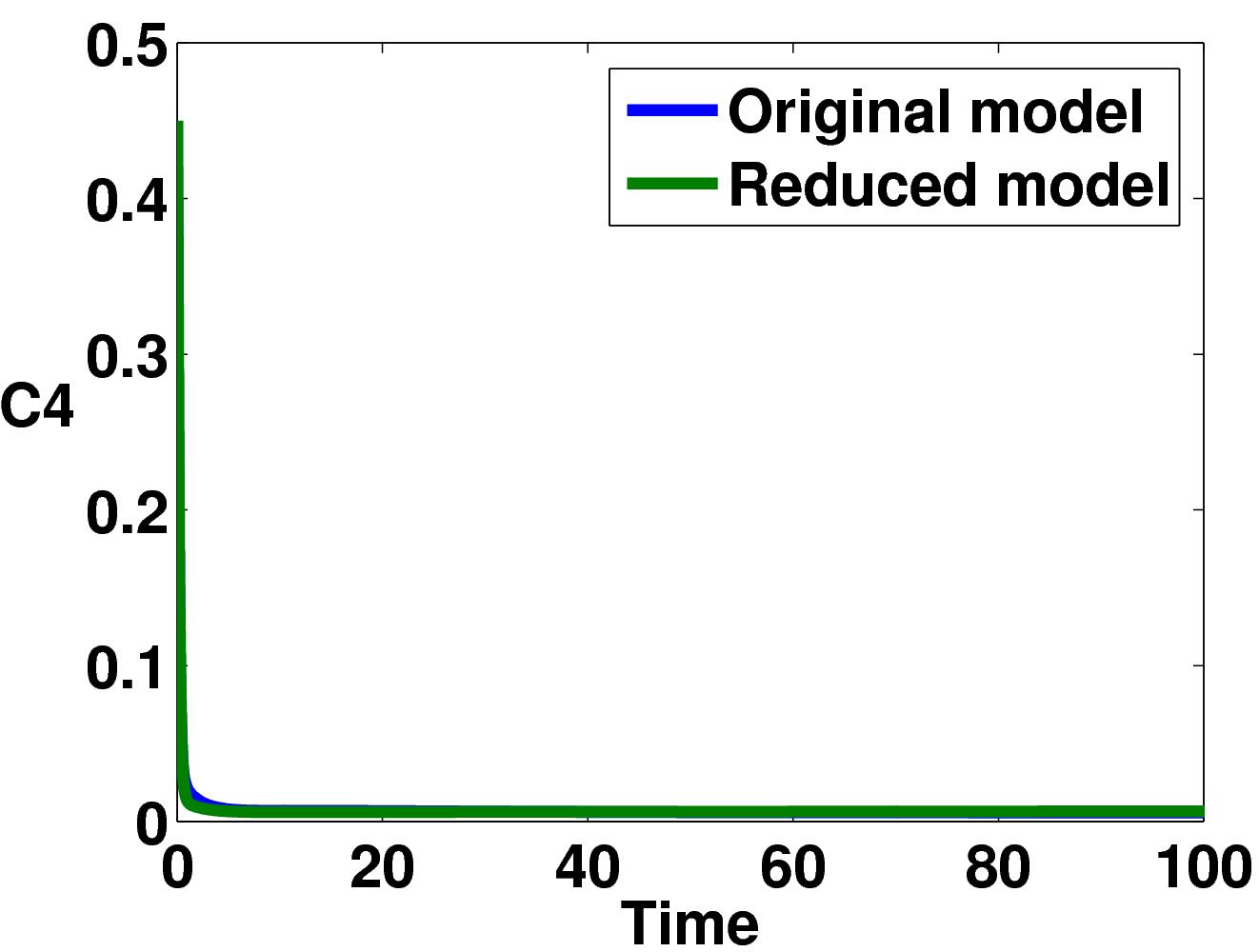}
		} 
		\subfigure{%
			\includegraphics[width=0.32\textwidth]{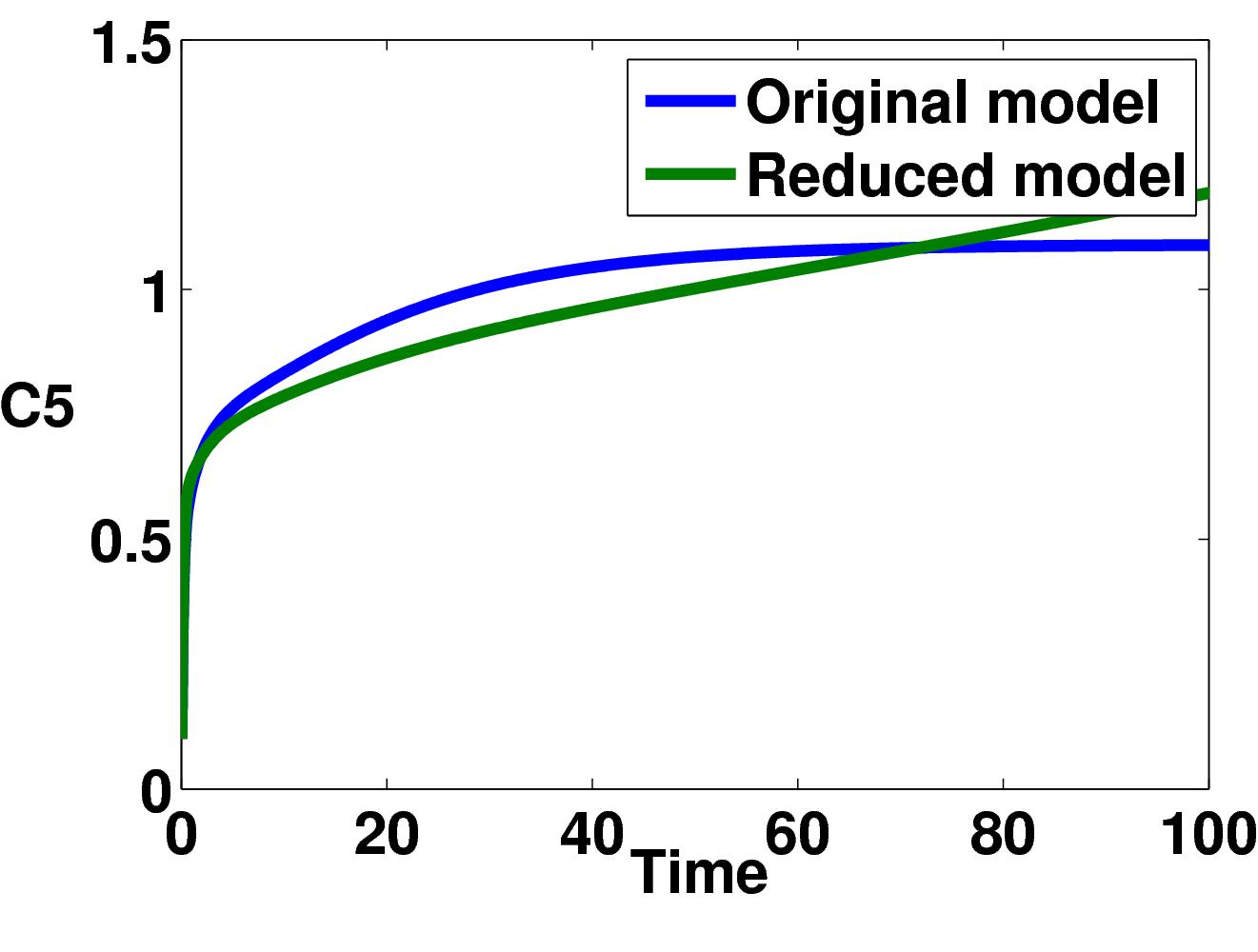}
			}
		\subfigure{%
			\includegraphics[width=0.32\textwidth]{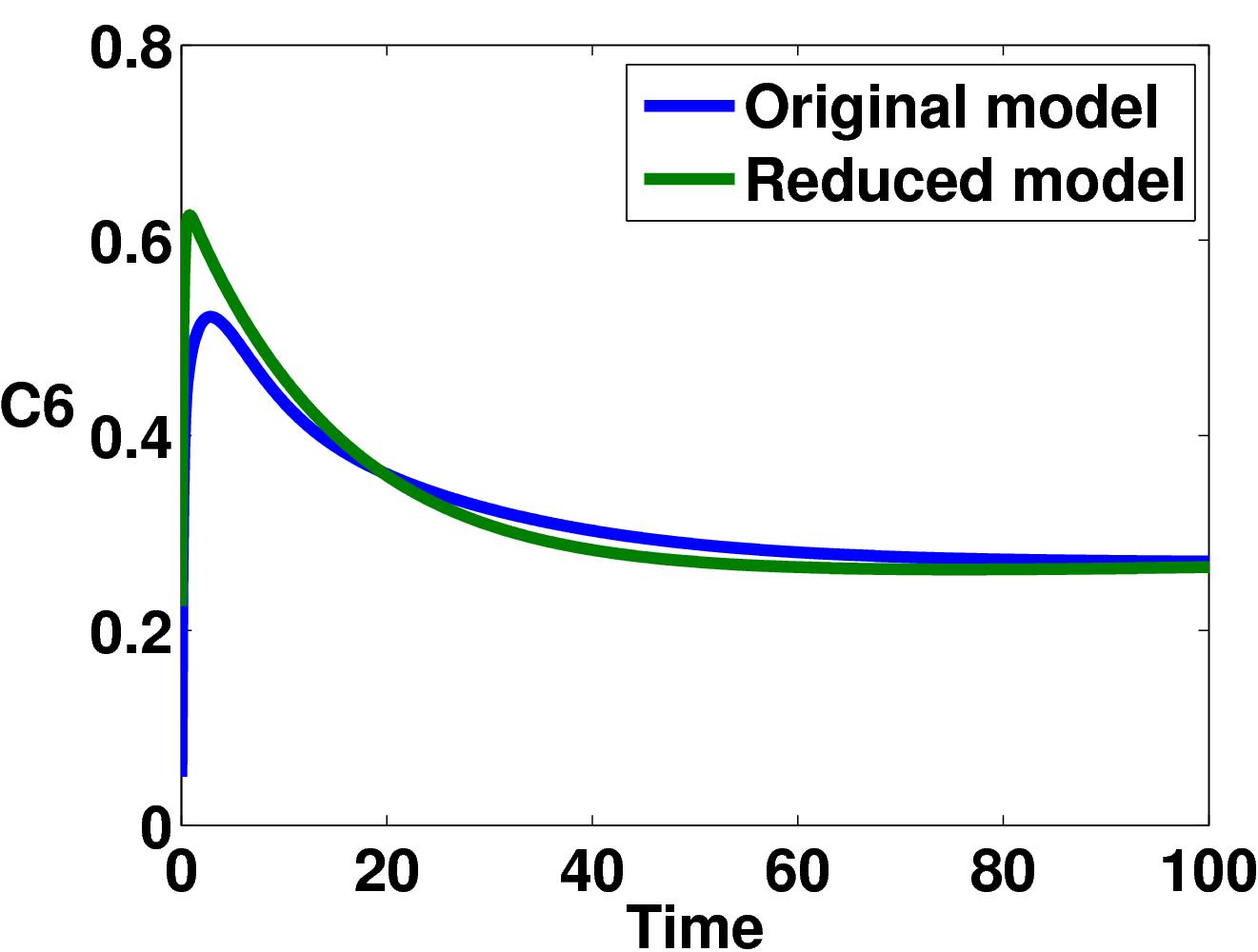}
	    }
		\subfigure{%
			\includegraphics[width=0.32\textwidth]{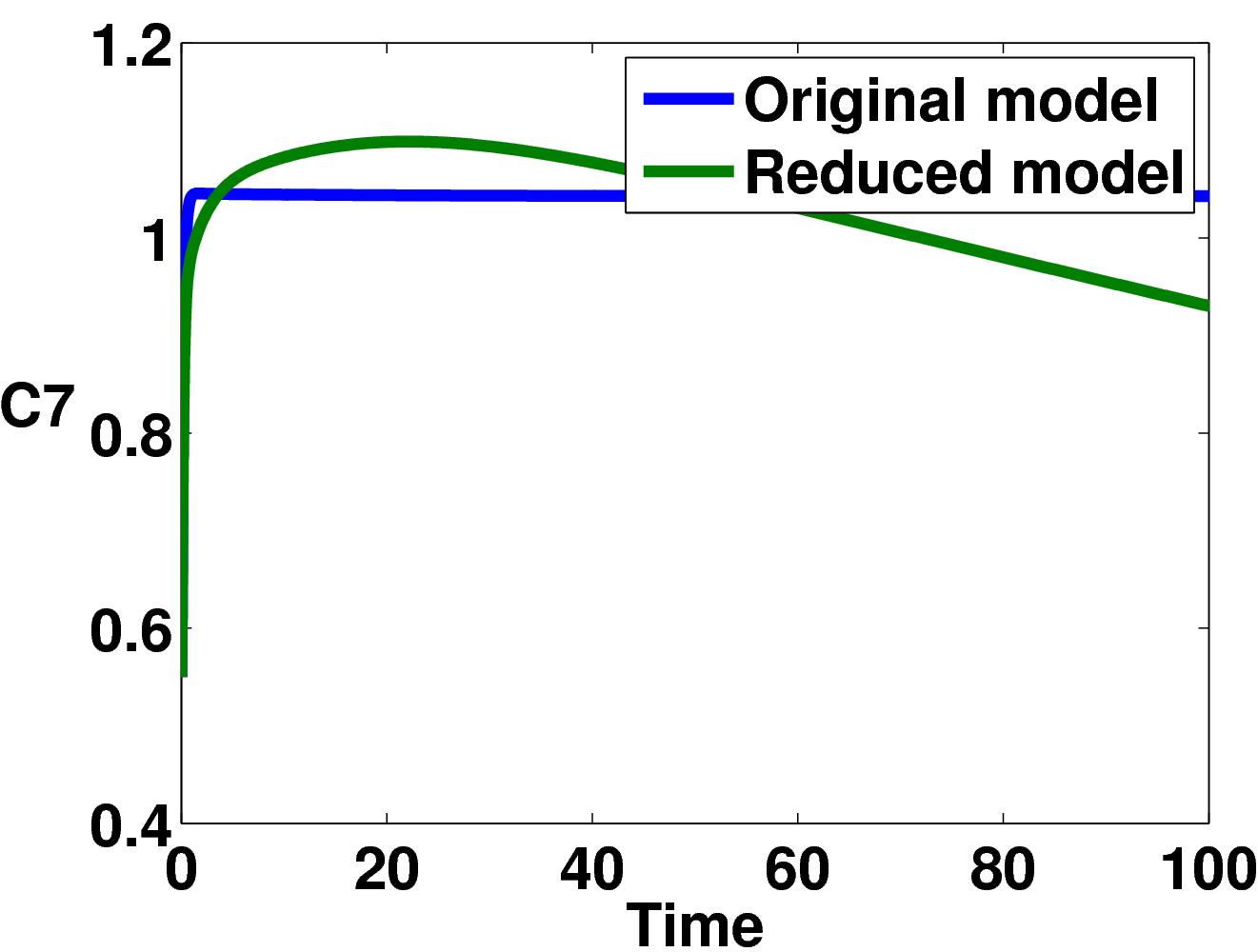}
			}
		\subfigure{%
			\includegraphics[width=0.32\textwidth]{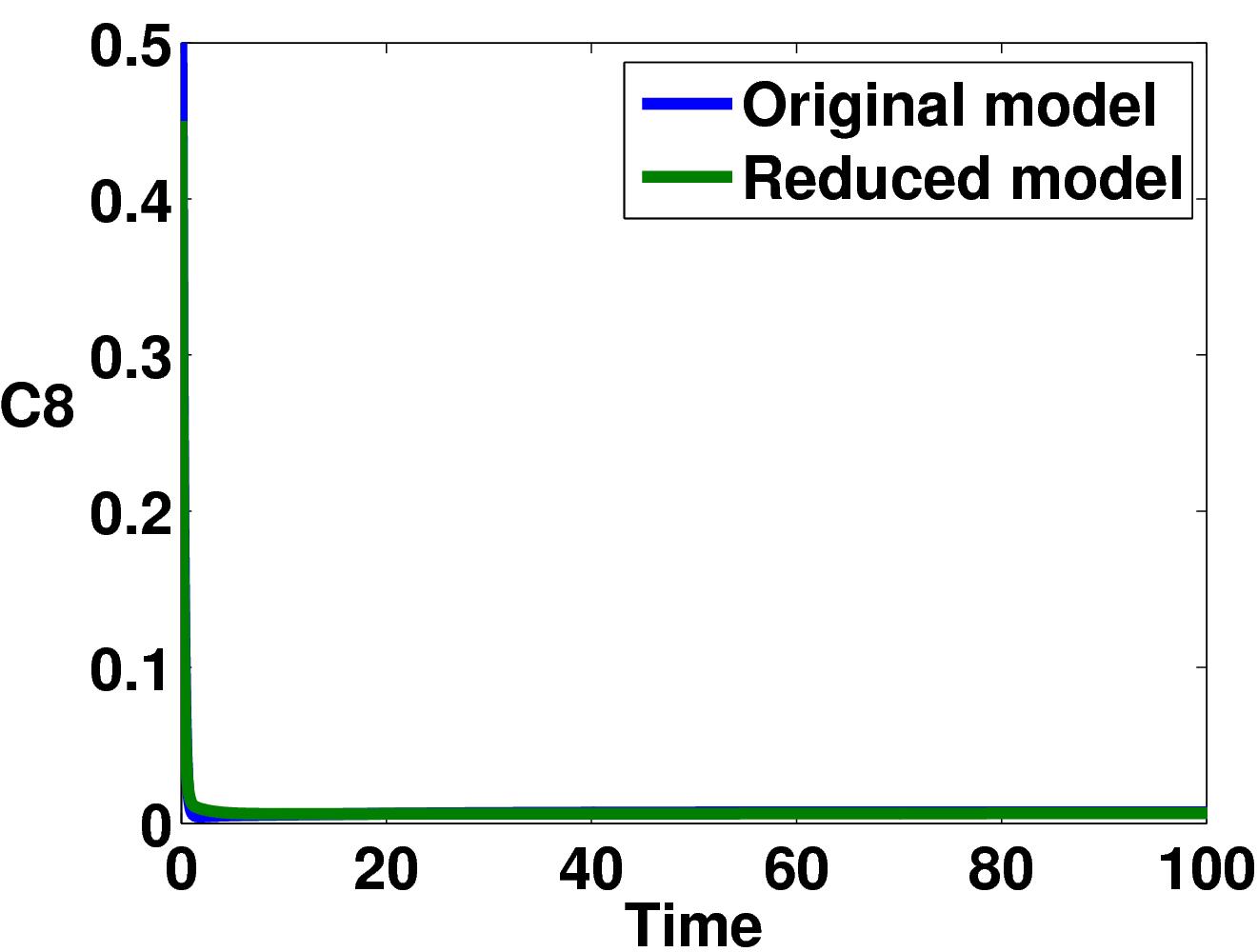}
			}
		\subfigure{%
			\includegraphics[width=0.32\textwidth]{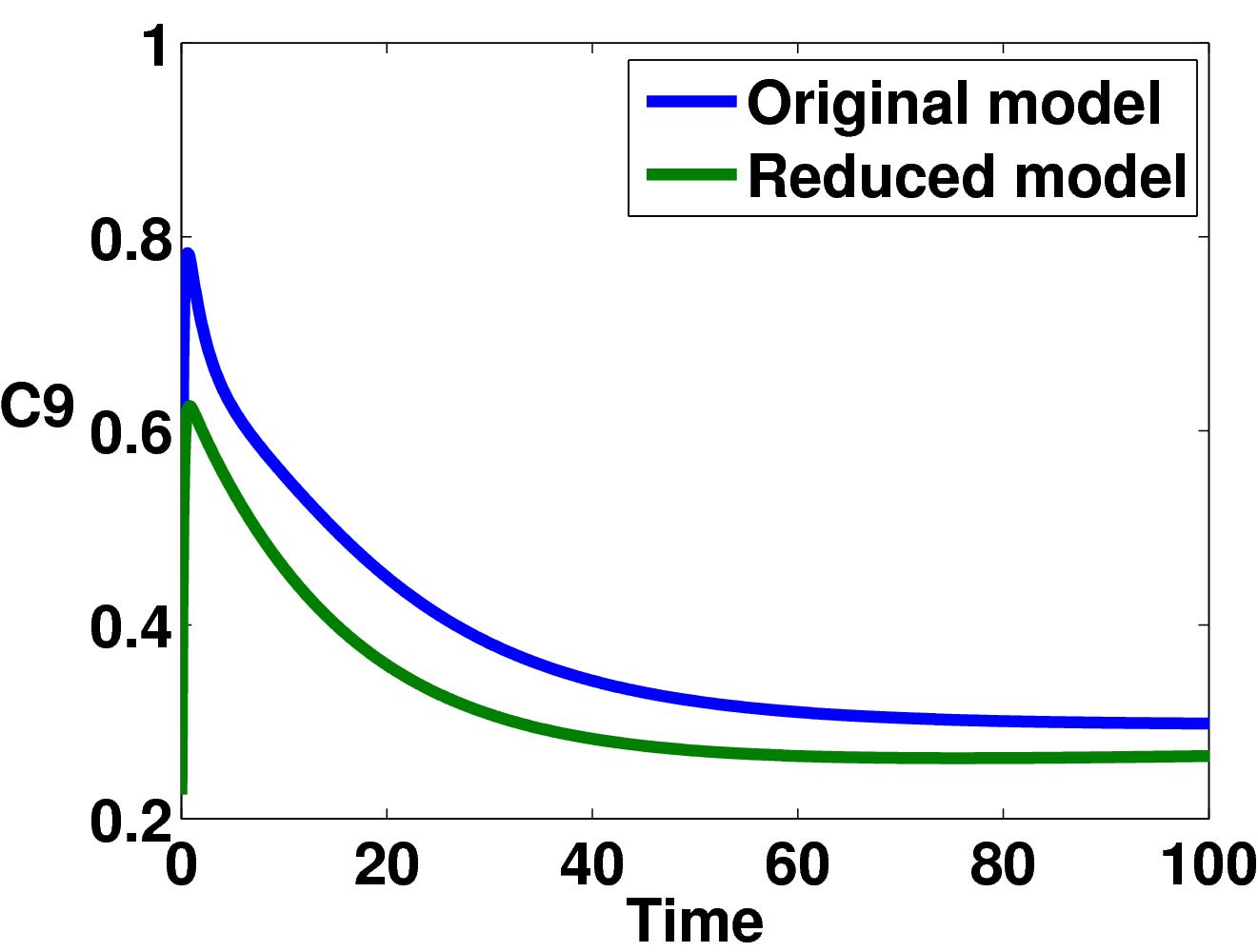}
			}
		\subfigure{%
			\includegraphics[width=0.32\textwidth]{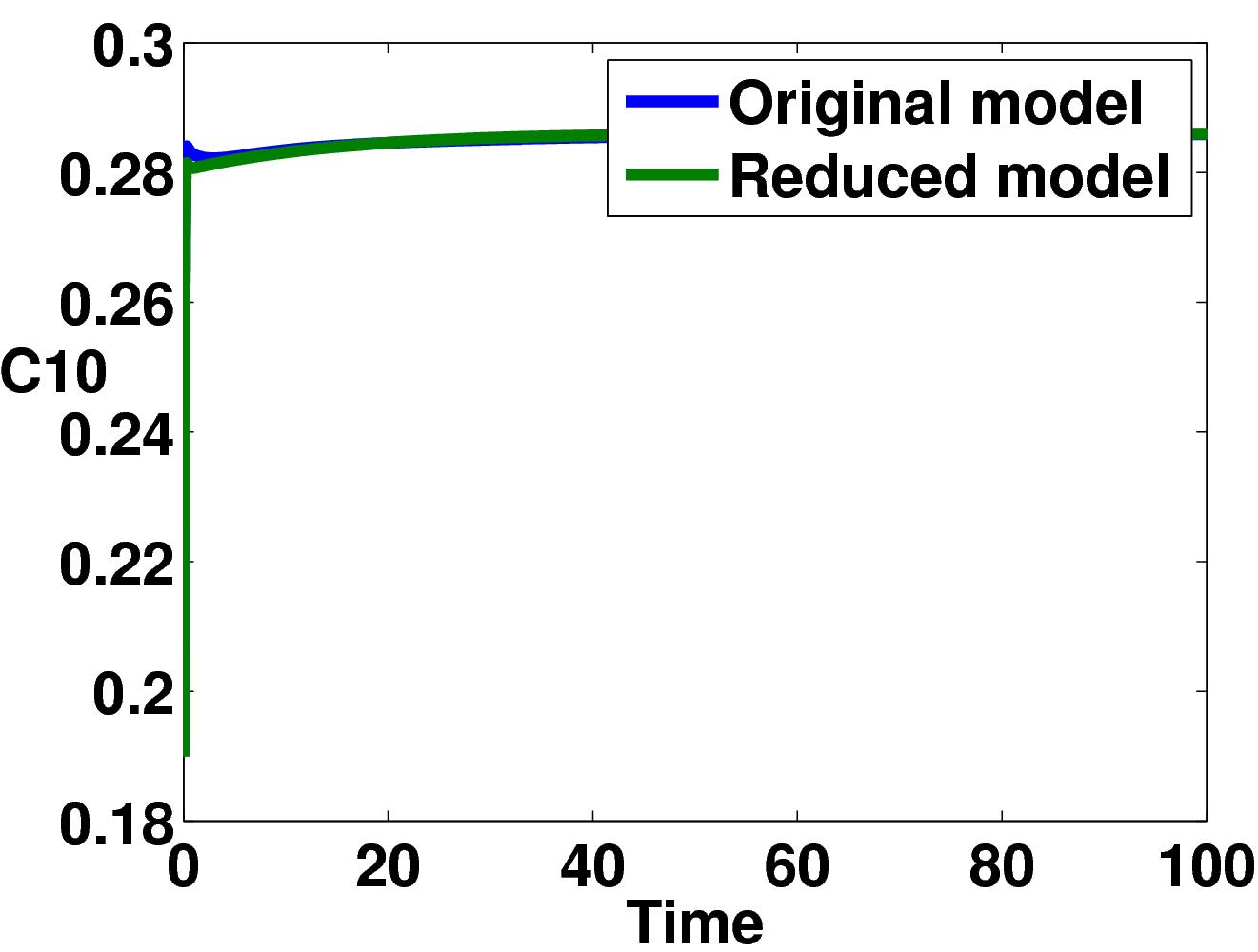}
			}
		\subfigure{%
			\includegraphics[width=0.32\textwidth]{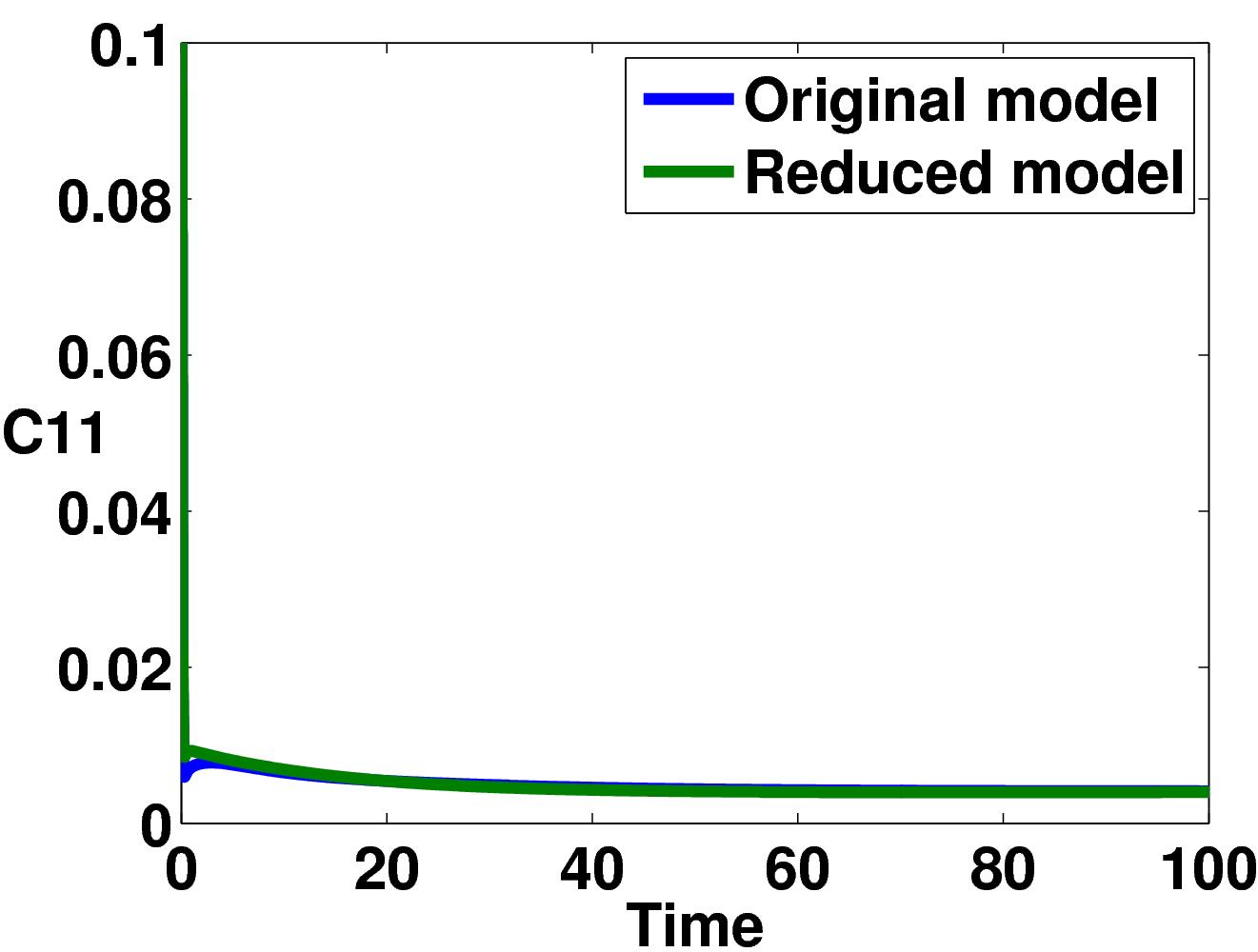}
		}
	\end{center}   
	\caption {Numerical simulations $\lbrace c_{i}:i=1, 2, .., 11\rbrace$ in original and reduced system of ERK signaling pathways; the blue lines are the original and the green lines are the reduced approximate solutions, with the time interval $[0,100]$ in computational simulations.}    	
	\label{fig:solution of ERK}
\end{figure}

\noindent We use some computational simulations in order to compared the original and reduced model. This is for state variables $\lbrace c_{i}:i=1, 2, .., 11\rbrace$ in case 6, see Figure \eqref{fig:solution of ERK}. 


\subsection{Results and Discussions}
The lumping of compartments is an effective tool for model reduction, especially for complex cell signaling pathways. Here we have applied the suggested technique on some chemical reaction mechanisms. Firstly, the suggested approach has been applied on two chemical chains. The first one is a linear and the second one is non linear chemical chain. Their variables are minimized from $3$ to $2$ and $4$ to $3$ variables respectively. Figures \eqref{fig:solution of linear simple example} and \eqref{fig:solution of nonlinear simple example} showed a good agreement between the original and reduced model. \\
After that, we have also applied the proposed method on the ERK signaling pathways, which includes $11$ variables and parameters. There are $6$ different cases of model reduction as shown in Table \eqref{table lumping of ERK}. According to the value of deviation and the number of elements, there are two effective cases, they are case $5$ and case $6$. As it is clear that in case $5$ the number of variables reduced from $11$ to $7$ and the value of deviation is only $9.87 \%$, and in case $6$ the number of variables reduced to $8$ and the value of deviation is smaller which is $4.54 \%$. Finally, the approximate solutions of the original and reduced model for case $6$ computed in computational simulations, this is illustrated in Figure \eqref{fig:solution of ERK}. It can be seen that the approximate solutions of original and reduced model are very close.

\newpage
\newpage
\section{Lumping of Parameters}
In this section by following (Brochot et al., 2005, Kou and Wei, 1969), we propose a new technique of model reduction to reduce the number of parameters (constants). Consider a system of differential equations for a chemical network as follows:
 \begin{equation}
\begin{array}{llll}  
\dfrac{dC}{dt}=H(C,P),
\end{array}
\label{lumping:general system 1p}
\end{equation}
\noindent
where $C$ is a vector of state variables, $ C \in R^{n} $,$P$ is a vector of chemical parameters(constants), $ P \in R^{m}, P=(p_{1},p_{2},...,p_{m})$, and $ H=\Big(h_{1},h_{2},...,h_{n}\Big)$.\\
\noindent We assume that all parameters are included in this interval
\begin{equation}
\begin{array}{llll}  
 p_{j} \in \Big[ \beta_{1},\beta_{n} \Big] \subseteq R^{+}, \text{for} 
 \; j=1,2,...,m .
\end{array}
\label{PartitionA}
\end{equation}

In order to choose the best way of lumping parameters,we divide the given interval \eqref{PartitionA} into sub-intervals as follows 
\begin{equation}
\begin{array}{llll}  
 \Big [ \beta_{1},\beta_{n} \Big ]=\mathlarger{\mathlarger{\bigcup}}_{i=1}^{n-1}\big [ \beta_{i},\beta_{i+1} \big ].
\end{array}
\label{PartitionB}
\end{equation} 

The proposed intervals may not equally spaces, it can be selected  with the condition that $\vert \beta_{i+1} - \beta_{i} \vert < \alpha$ ; $ \alpha \in R^{+} \cup \lbrace 0 \rbrace $.\\

Then, we introduce a vector of new parameters
$P^{*}=\big(p^{*}_{1},p^{*}_{2},...,p^{*}_{m_{1}}\big),\; m_{1}\leq m$,\; where each element of $P^{*}$ is defined below
\begin{equation}
\begin{array}{llll}  
p_{i}^{*}=\mathlarger{\mathlarger{\sum\limits}}_{j\in J} p_{j},
\end{array}
\label{PartitionC}
\end{equation}
where $J=\lbrace 1,2,...,m \rbrace$ and $i=1,2,...,m_{1}$. This is called lumping of parameters (constants). We define a lumping matrix $M$ as follows:
\begin{equation} 
M=\bordermatrix{
                   &p_1&p_2&\ldots &p_m\cr
                p^{*}_{1}&a_{11} &  a_{12}  & \ldots & a_{1m}\cr
                p^{*}_{2}& a_{21}  &  a_{22} & \ldots & a_{2m}\cr
                \vdots& \vdots & \vdots & \ddots & \vdots\cr
                p^{*}_{m_{1}}& a_{m_{1} 1}  &   a_{m_{1} 2}       &\ldots & a_{m_{1} m}}.
\label{PartitionD}
\end{equation}
where $a_{ij}\in\lbrace 0,1 \rbrace$ for $i=1,2,...,m_{1}$\; and $j=1,2,...,m$.\\
There is an important equation that is called lumping transformation of parameters 
\begin{equation}
\begin{array}{llll}  
P^{*}=M P.
\end{array}
\label{PartitionE}
\end{equation}
From equation \eqref{PartitionE}, the set of original parameters $P$ can be calculated as follows:
\begin{equation}
\begin{array}{llll}  
P=M^{+}P^{*},
\end{array}
\label{PartitionF}
\end{equation}
where $M^{+}$ is pseudo-inverse of $M$ such that $M M^{+}=I$.

Therefore, the equation \eqref{lumping:general system 1p} takes the form
\begin{equation}
\begin{array}{llll}  
\dfrac{dC}{dt}=H(C,P^*).
\end{array}
\label{lumping:reduced system 1p}
\end{equation}
The equation \eqref{lumping:reduced system 1p} is called reduced model of the system with less parameters.
\section{Applications}
We proposed a new technique here that plays an important role in model reductions. This approach can be used for model reductions in chemical reaction networks and cell signalling pathways. We apply this technique in linear and non--linear models in order to reduce the number of parameters. 

\subsection{Linear Example}
The idea of lumping parameters can be simply used for linear chemical networks. We consider a linear network with three species and five parameters.
\begin{equation}  
\begin{array}{llll}
\quad C_{2} \underset{p_{12}}{ \overset{p_{21}}\rightleftharpoons} C_{1} 
\underset{p_{31}}{ \overset{p_{13}}\rightleftharpoons} C_{3}\\
\quad \quad \; \; p_{10}\downarrow  
\end{array}\label{lumping:linear network diagram 1p}
\end{equation} 
The system of ODE's of the linear network becomes
\begin{equation} 
\begin{array}{llll}       
\dfrac{dC_{1}}{dt}=-(p_{10}+p_{12}+p_{13})C_{1}+p_{21}C_{2}+p_{31}C_{3},\\
\dfrac{dC_{2}}{dt}=p_{12}C_{1}-p_{21}C_{2},\\
\dfrac{dC_{3}}{dt}=p_{13}C_{1}-p_{31}C_{3},\\
\end{array} \label{lumping:ODE original example p}
\end{equation}
with initial conditions $C_{1}(0)=0.5,\;C_{2}(0)=0.2$\; and\;$C_{3}(0)=0.3$,\;and reaction parameters $p_{10}=1,\;p_{12}=0.1,\;p_{21}=0.2,\;p_{13}=0.1,\;p_{31}=0.2$.
\newline
In this example, we lump the given parameters as follows: \\ 
 $P^{*}_{1}=P_{10},$\;$P^{*}_{2}=P_{12}+P_{21}$,\;and$P^{*}_{3}=P_{13}+P_{31}$. The model network then takes the form
\begin{equation}  
\begin{array}{llll}
\quad C_{2} { \overset{p^*_{2}}\longleftrightarrow} C_{1} 
{ \overset{p^*_{3}}\longleftrightarrow} C_{3}\\
\quad \quad \; \;\; p^{*}_{1}\downarrow  
\end{array}\label{lumping:reduce linear network diagram 1p}
\end{equation}
In addition, the lumping matrix $M$ is
 \begin{equation*} 
M={\left(\begin{array}{ccccccc}
1&0&0&0&0\\ \\ 0&1&1&0&0  \\  \\ 0&0&0&1&1
\end{array}\right)},
\end{equation*}
with pseudo--inverse $M^{+}$ given below: 
 \begin{equation*} 
M^{+}={\left(\begin{array}{ccccc}
1&0&0\\ \\ 0&\dfrac{1}{2}&0\\ \\ 0&\dfrac{1}{2}&0\\ \\ 0&0&\dfrac{1}{2}\\ \\ 0&0&\dfrac{1}{2}
\end{array}\right)} .
\end{equation*}
The lumping transformation equation is given 
 \begin{equation*} 
{\left(\begin{array}{cccccc}
P_{10}\\ \\P_{12}\\ \\P_{21}\\ \\P_{13}\\ \\P_{31}
\end{array}\right)}
={\left(\begin{array}{cccc}
1&0&0\\ \\ 0&\dfrac{1}{2}&0\\ \\ 0&\dfrac{1}{2}&0\\ \\ 0&0&\dfrac{1}{2}\\ \\ 0&0&\dfrac{1}{2}
\end{array}\right)}
{\left(\begin{array}{cccc}
P^{*}_{1}\\ \\P^{*}_{2}\\ \\P^{*}_{3}
\end{array}\right)}
={\left(\begin{array}{cccc}
P^{*}_{1}\\ \\ \dfrac{P^{*}_{2}}{2}\\ \\ \dfrac{P^{*}_{2}}{2}\\ \\ \dfrac{P^{*}_{3}}{2}\\ \\ \dfrac{P^{*}_{3}}{2}
\end{array}\right)}.
\end{equation*}
Thus, the reduced model for the original model \eqref{lumping:ODE original example p} becomes
\begin{equation} 
\begin{array}{lll}
\dfrac{dC_{1}}{dt}=-p^{*}_{1}C_{1}-\dfrac{1}{2}p^{*}_{2}(C_{1}-C_{2})-\dfrac{1}{2}p^{*}_{3}(C_{1}-C_{3}),\\
\dfrac{dC_{2}}{dt}=\dfrac{1}{2}p^{*}_{2}(C_{1}-C_{2}),\\
\dfrac{dC_{3}}{dt}=\dfrac{1}{2}p^{*}_{3}(C_{1}-C_{3}),
\end{array}\label{lumping:ODE example reduced p}  
\end{equation}
with $p^{*}_{1}=1,\;p^{*}_{1}=0.3$\;and \;$p^{*}_{1}=0.3$. It can be concluded that the reduced system \eqref{lumping:ODE example reduced p} has 3 parameters while the original system \eqref{lumping:ODE original example p} has 5 parameters.  

\subsection{Mathematical Model for NF-$\kappa$B Signal Transduction Pathways}
 
An important self-protection mechanism in the body is called inflammation. This has a great role to prevent the spread of infectious diseases. There is an  example of transcription factor which is called nuclear factor-$\kappa$B (NF-$\kappa$B). This works an essential role in immune cells for inflammation process. There are some target genes identified for NF-$\kappa$B, for instance, TNF-$\alpha$ and IL-10 are two pro- and anti-inflammatory cytokines. It is clear that I$\kappa$B$\alpha$ (sequesters free NF-$\kappa$B)  and A20 (inactivates IKK)  are also NF-$\kappa$B responsive genes.  

Interestingly, it can be found that several mathematical models with computational simulations of inflammatory signaling pathways have been suggested, for example, the IL-6 signal transduction pathway model  and  the TNF-$\alpha$ signaling
pathway model. The suggested models describe the dynamical analysis of signaling pathways initiated by a single pro-inflammatory cytokine. Recently, a mathematical model has been developed to show interactions between IL-6 (pro-inflammatory) and IL-10 (anti-inflammatory). More recently, another computational model has been proposed to describe an interaction between the beginning synthesized pro-inflammatory (TNF-$\alpha$) and anti-inflammatory (IL-10). 

More interestingly, the interaction between the pro- and anti-inflammatory signaling is not well-understood (Maiti et al., 2014, Nathan, 2002). Therefore, the suggested model plays an important step forward for  modeling the interaction between pro- and anti-inflammatory signaling mediators that is important in inflammation and maintaining homeostasis. The mathematical model developed here is a combination of an inflammatory module and an anti-inflammatory module. The suggested model is established by representing biochemical reactions in the signal pathways, then the model equations are given as a set of non-linear ordinary differential equations. The signaling pathways model here includes  29 state variables  and 37 parameters . Each differential equation represents the rate of change of the concentration of a particular protein involved in the pathway; see Tables \eqref{table of initial condition of inflammatory} and \eqref{table of parameter of inflammatory}. Readers can see more details in (Maiti et al., 2014).
\begin{figure}[H]  
\begin{center}     
		\subfigure{%
			\includegraphics[width=1.1\textwidth]{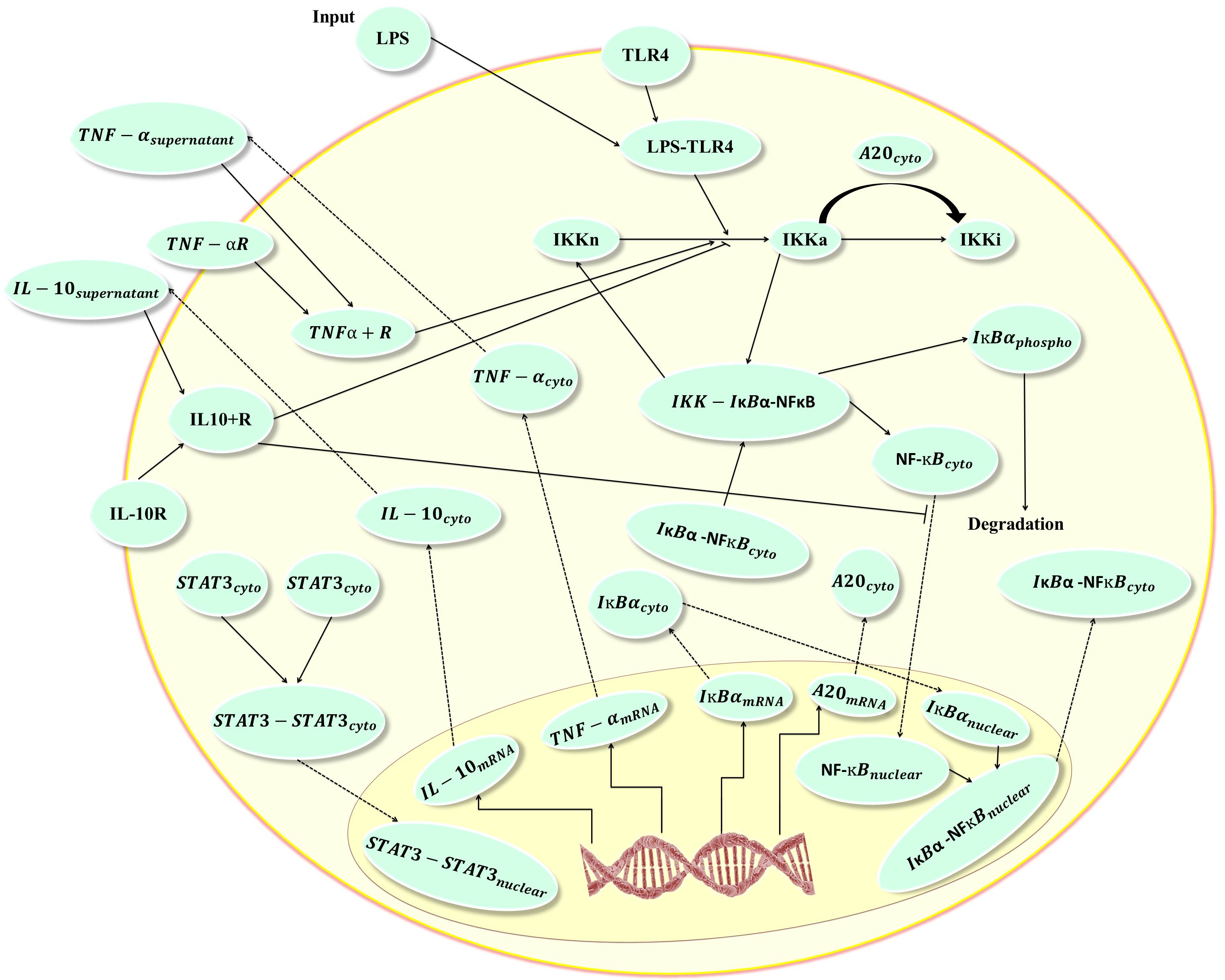}
		}			
	\end{center}   
	\caption {Implemented reaction network for the LPS-induced NF-$\kappa$B signal transduction pathways with TNF-$\alpha$ (positive) and IL-10 (negative) feedback regulation.}    	
	\label{model diagram} 
\end{figure}
\FloatBarrier 
\begin{table}
\caption{State variables and their initial values for NF-$\kappa$B signal transduction pathways.}

\begin{tabular}{|c|p{5.5cm}|p{2cm}|p{4cm}|}
\hline 
No. & State variables & Symbols & Initial values, $\mu M$ \\ 
\hline 
1 & $TLR4$ & $x_{1}$ & 0.1 \\ 
\hline 
2 & $LPS-TLR4$ & $x_{2}$ & 0 \\ 
\hline 
3 & $IL-10_{supernatant}$ & $x_{3}$ & 0.0000046 \\ 
\hline 
4 & $IL-10R$ & $x_{4}$ & 0.1 \\ 
\hline 
5 & $IL10-IL10R$ & $x_{5}$ & 0 \\ 
\hline 
6 & $TNF-\alpha_{supernatant}$& $x_{6}$ & 0 \\ 
\hline 
7 & $TNF-\alpha R$ & $x_{7}$ & 0.1 \\ 
\hline 
8 & $TNF\alpha - TNF\alpha R$ & $x_{8}$ & 0 \\ 
\hline 
9 & $IKK_{neutral}$ & $x_{9}$ & 0.2 \\ 
\hline 
10 & $IKK_{active}$ & $x_{10}$ & 0 \\ 
\hline 
11 & $IKK_{inactive}$ & $x_{11}$ & 0 \\ 
\hline 
12 & $I\kappa B\alpha -NF\kappa B_{cyto}$ & $x_{12}$ & 0.25 \\ 
\hline 
13 & $IKK-I\kappa B\alpha NF\kappa B$ & $x_{13}$ & 0 \\ 
\hline 
14 & $NF\kappa B_{cyto}$ & $x_{14}$ & 0.003 \\ 
\hline 
15 & $NF\kappa B_{nuclear}$ & $x_{15}$ & 0 \\ 
\hline 
16 & $I\kappa B\alpha_{phospho}$ & $x_{16}$ & 0 \\ 
\hline 
17 & $ A20_{mRNA}$ & $x_{17}$ & 0 \\ 
\hline 
18 & $A20_{cyto}$ & $x_{18}$ & 0.0048 \\ 
\hline 
19 & $I\kappa B\alpha_{mRNA}$ & $x_{19}$ & 0 \\ 
\hline 
20 & $I\kappa B\alpha_{cyto}$ & $x_{20}$ & 0.0025 \\ 
\hline 
21 & $I\kappa B\alpha_{nuclear}$ & $x_{21}$ & 0 \\ 
\hline 
22 & $I\kappa B\alpha -NF\kappa B_{nuclear}$ & $x_{22}$ & 0 \\ 
\hline 
23 & $IL-10_{mRNA}$ & $x_{23}$ & 0 \\ 
\hline 
24 & $IL-10_{cyto}$ & $x_{24}$ & 0 \\ 
\hline 
25 & $TNF-\alpha_{mRNA}$ & $x_{25}$ & 0 \\ 
\hline 
26 & $TNF-\alpha_{cyto}$ & $x_{26}$ & 0 \\ 
\hline 
27 & $STAT3_{cyto}$ & $x_{27}$ & 0.592 \\ 
\hline 
28 & $STAT3-STAT3_{cyto}$ & $x_{28}$ & 0 \\ 
\hline 
29 & $STAT3-STAT3_{nuclear}$ & $x_{29}$ & 0 \\ 
\hline 
\end{tabular}  \label{table of initial condition of inflammatory}
\end{table}
\FloatBarrier 

\FloatBarrier 
\begin{table}
\caption{List of parameters for NF-$\kappa$B signal transduction pathways.}

\begin{tabular}{|p{0.5cm}|c|c|c|}
\hline 
No. & Parameters & Descriptions & Values  \\ 
\hline 
1 & $kv$ & Nuclear: Cytoplasmic (Volume) & 1.17  \\ 
\hline 
2 & $kf_{1}$ & LPS binding to receptor & 0.264 \\ 
\hline 
3 & $kr_{1}$ & Dissociation of LPS+receptor complex & 0.00125  \\ 
\hline 
4 & $kf_{2}$ & IL-10 binding to receptor & 0.00025  \\ 
\hline 
5 & $kr_{2}$ & Dissociation of IL-10+receptor complex & 0.000611  \\ 
\hline 
6 & $kf_{3}$ & TNF-$\alpha$ binding to receptor & 0.0025  \\ 
\hline 
7 & $kr_{3}$ & Dissociation of TNF-$\alpha$+receptor complex & 0.00125  \\ 
\hline 
8 & $kf_{4}$ & I$\kappa$Ba and NF-$\kappa$B association & 0.0025 \\ 
\hline 
9 & $kfi$ & IKK activation & 0.00162 \\ 
\hline 
10 & $kk_{1}$ & Inactivation of IKK by A20 & 0.00025  \\ 
\hline 
11 & $kk_{3}$ & Association of IKK with $I\kappa B\alpha -NF\kappa B$ & 1.0  \\ 
\hline 
12 & $ti3$ & Catalytic breakdown of$IKK-I\kappa B\alpha -NF\kappa B$ & 0.000172 \\ 
\hline 
13 & $iln$ & NF-$\kappa$B nuclear import & 0.00152 \\ 
\hline 
14 & $a20_{trans}$ & A20 translation & 0.5 \\ 
\hline 
15 & $kdeg_{A20}$ & Degradation of A20 protein & 0.0003  \\ 
\hline 
16 & $i\kappa b\alpha_{trans}$ & I$\kappa$B$\alpha$ translation & 0.5 \\ 
\hline 
17 & $kdeg_{I\kappa B\alpha}$ & Degradation of phosphorylated I$\kappa$B$\alpha$ & 0.000128 \\ 
\hline 
18 & $il10_{trans}$ & IL-10 translation & 0.5 \\ 
\hline 
19 & $ksec_{IL10}$ & Secretion of IL-10 from cytoplasm to supernatant & 0.0000203 \\ 
\hline 
20 & $kdeg_{IL-10sup}$ & Degradation of IL-10 in supernatant & 0.000074 \\ 
\hline 
21 & $tnf\alpha_{trans}$ & TNF-$\alpha$ translation & 0.5 \\ 
\hline 
22 & $ksec_{TNF\alpha}$  & Secretion of TNF-$\alpha$ from cytoplasm to supernatant & 0.0000516 \\ 
\hline 
23 & $kdeg_{TNF\alpha sup}$  & Degradation of TNF-$\alpha$ in supernatant & 0.0000746 \\ 
\hline 
24 & $Dn$ & Degradation of intracellular cytokine & 0.0104 \\ 
\hline 
25 & $iki$ & I$\kappa$B$\alpha$ nuclear import & 0.001  \\ 
\hline 
26 & $eki$ & I$\kappa$B$\alpha$ nuclear export & 0.0005 \\ 
\hline 
27 & $eni$ & I$\kappa$B$\alpha$-NF$\kappa$B nuclear export & 0.01 \\ 
\hline 
28 & $k_{1}$ & STAT3 activation and dimerization & 0.0154 \\ 
\hline 
29 & $k_{2}$ & Dissociation of STAT3 dimer & 0.000033 \\ 
\hline 
30 & $i_{stat3}$ & STAT3 dimer nuclear import & 0.0000356 \\ 
\hline 
31 & $Sm$ & Transcription due to NF-$\kappa$B & 0.1 \\ 
\hline 
32 & $Sm_{-}il10$ & IL-10 Translation due to STAT3 & 1.5 \\ 
\hline 
33 & $p$ & Transcription parameter & 0.005 \\ 
\hline 
34 & $Dm$ & Degradation of mRNA & 0.0104 \\ 
\hline 
35 & $C$ & Maximum NF-$\kappa$B concentration in nucleus & 0.108 \\ 
\hline 
36 & $C_{STAT3}$ & Maximum STAT3 concentration in nucleus & 0.05 \\ 
\hline 
37 & $IL10_{-}IL10R_{max}$ & IL10-IL10R maximum concentration & 0.00000256 \\ 
\hline 
\end{tabular}  \label{table of parameter of inflammatory}
\end{table}
\FloatBarrier 

Then the system of differential equations of the network \eqref{model diagram} is given bellow:
\begin{equation} 
\begin{array}{lllllllllllllll}

\dfrac{dx_{1}}{dt}=-kf_{1}[LPS] \; x_{1}+kr_{1} \; x_{2}, \\

\dfrac{dx_{2}}{dt}=kf_{1}[LPS] \; x_{1}-kr_{1} \; x_{2}, \\

\dfrac{dx_{3}}{dt}=-kf_{2} \; x_{3} \; x_{4}+kr_{2} \; x_{5}+\dfrac{0.36}{200}\; ksec_{IL10} \;x_{24}-kdeg_{IL-10sup} \; x_{3},\\

\dfrac{dx_{4}}{dt}=-kf_{2} \; x_{3} \; x_{4}+kr_{2} \; x_{5}, \\

\dfrac{dx_{5}}{dt}=kf_{2} \; x_{3} \; x_{4}-kr_{2} \; x_{5}, \\

\dfrac{dx_{6}}{dt}=-kf_{3} \; x_{6} \; x_{7}+kr_{3} \; x_{8}+\dfrac{0.36}{200}\; ksec_{TNF \alpha} \; x_{26}-kdeg_{TNF\alpha sup} \; x_{6}, \\

\dfrac{dx_{7}}{dt}=-kf_{3} \; x_{6} \; x_{7}+kr_{3} \; x_{8} ,\\

\dfrac{dx_{8}}{dt}=kf_{3} \; x_{6} \; x_{7}-kr_{3} \; x_{8} , \\

\dfrac{dx_{9}}{dt}=-kfi \; \; k_{in} (x_{2}+x_{8})x_{9}+ti_{3} \; x_{13} ,\\

\dfrac{dx_{10}}{dt}=kfi \; \; k_{in} (x_{2}+x_{8})x_{9}-kk_{3} \; K_{in} \; x_{10} \; x_{12}-kk_{1} \; x_{10} \; x_{18} ,\\

\dfrac{dx_{11}}{dt}=kk_{1} \; x_{10} \; x_{18} ,\\

\dfrac{dx_{12}}{dt}=kf_{4} \; x_{14} \; x_{20}+eni \; kv \; x_{22}-kk_{3} \; k_{in} \; x_{10} \; x_{12},\\

\dfrac{dx_{13}}{dt}=kk_{3} \; k_{in} \; x_{10} \; x_{12}-ti_{3} \; x_{13},\\

\dfrac{dx_{14}}{dt}=-kf_{4} \; x_{14} \; x_{20}+ti_{3} \; x_{13}-iln \; k_{in} \; x_{14},\\

\dfrac{dx_{15}}{dt}=iln \; k_{in} \; \dfrac{x_{14}}{kv}-kf_{4} \; x_{15} \; x_{21},\\

\dfrac{dx_{16}}{dt}=ti_{3} \; x_{13}-kdeg_{I\kappa B\alpha} \; x_{20}-x_{16},\\

\dfrac{dx_{17}}{dt}=sm \; p \; \dfrac{x_{15}}{C+x_{15}}-Dm \; x_{17}, \\

\dfrac{dx_{18}}{dt}=a20_{trans} \; x_{17}-kdeg_{A20} \; x_{18} ,\\

\dfrac{dx_{19}}{dt}=sm \; p \; \dfrac{x_{15}}{C+x_{15}}-Dm \; x_{19},\\

\dfrac{dx_{20}}{dt}=-kf_{4} \; x_{14} \; x_{20}+ikb\alpha_{trans} \; x_{19}-iki \; x_{20}+eki \; kv \; x_{21} ,\\

\dfrac{dx_{21}}{dt}=-kf_{4} \; x_{15} \; x_{21}+iki \; \dfrac{x_{20}}{kv}-eki \; x_{21},\\

\dfrac{dx_{22}}{dt}=kf_{4} \; x_{15} \; x_{21}-eni \; x_{22},\\

\dfrac{dx_{23}}{dt}=0.4 \; sm \; p \; \dfrac{x_{15}}{C+x_{15}}+0.6 \; sm_{-}il10 \; p \; \dfrac{x_{29}}{C_{-}STAT3+x_{29}}-Dm \; x_{23},\\

\dfrac{dx_{24}}{dt}=il10_{trans} \; x_{23}-ksec_{IL10} \; x_{24}-Dn \; x_{24},\\

\dfrac{dx_{25}}{dt}=sm \; p \; \dfrac{x_{15}}{C+x_{15}}-Dm \; x_{25},\\

\end{array}\label{lumping of parameter:ODE inflammatory model}  
\end{equation}

\begin{equation*} 
\begin{array}{lllllllllllll}

\dfrac{dx_{26}}{dt}=tnf\alpha_{trans} \; x_{25}-ksec_{TNF\alpha}\; x_{26} -Dn \; x_{26},\\

\dfrac{dx_{27}}{dt}=-2 \; k_{1} \; x_{5} \; x_{27}^2 +2 \; K_{2} \; x_{28},\\

\dfrac{dx_{28}}{dt}=k_{1} \; x_{5} \; x_{27}^2 - K_{2} \; x_{28}-i_{stat3} \; x_{28}+eni \; kv \; x_{29}  ,\\

\dfrac{dx_{29}}{dt}=i_{stat3} \; \dfrac{x_{28}}{kv}-eni \; x_{29} ,

\end{array} \label{lumping of parameter:ODE inflammatory model}  
\end{equation*} 

where, $ k_{in}=max[(1-\dfrac{x_{5}}{[IL10-IL10R_{max}]},0)]. $

\newpage
\begin{table}
\caption{Applying lumping technique for the model NF-$\kappa$B signal transduction pathways \eqref{lumping of parameter:ODE inflammatory model}, four different cases are used with their total error and remaining parameters in the reduced model.}
\begin{tabular}{|c|p{2.3cm}|p{2.5cm}|p{2.2cm}|p{1.7cm}|p{2.5cm}|}
\hline 
Cases & Lumping parameters & Total differences & Remaining parameters & Lumping types & Selecting parameters \\ 
\hline 
Case 1 & $\Omega_{1}$ & $2.55 * 10^{-11}$ & 8 & Proper & Sub intervals \\ 
\hline 
Case 2 & $\Omega_{2}$ & $2.52 * 10^{-11}$ & 9 & Improper & Sub intervals \\ 
\hline 
Case 3 & $\Omega_{3}$ & $1.86 * 10^{-7}$ & 13 & Proper & Randomly \\ 
\hline 
Case 4 & $\Omega_{4}$ & $2.98 * 10^{-8}$ & 20 & Proper & Neighboring \\ 
\hline
\end{tabular} 
\label{table:lumping of NF-kB}
\end{table}
\FloatBarrier 
Where the lumping parameters are defined as follows:\\
$\Omega_{1}= \lbrace P_{1}^{*}=Sm_{-}il10, \; P_{2}^{*}=kv+kk_{3}, \; P_{3}^{*}=kf_{1}+a20_{trans}+i\kappa b\alpha_{trans}+il10_{trans}+tnf\alpha_{trans}+sm+C, \; P_{4}^{*}=Dn+eni+Dm+C_{STAT3}, \; P_{5}^{*}=kr_{1}+kf_{3}+kr_{3}+kf_{4}+kfi+iln+iki+p, \; P_{6}^{*}=kf_{2}+kr_{2}+kk_{1}+ti_{3}+kdeg_{A20}+kdeg_{I\kappa B\alpha}+eki, \; P_{7}^{*}=ksec_{IL10}+kdeg_{IL-10sup}+ksec_{TNF\alpha}+kdeg_{TNF\alpha sup}+k_{2}+i_{stat3}, \; P_{8}^{*}=IL10_{-}IL10R_{max} \rbrace$,
\newline
$\Omega_{2}=\lbrace P_{1}^{*}=kv, \; P_{2}^{*}=Sm_{-}il10,\; P_{3}^{*}=kv+kk_{3},\; P_{4}^{*}=kf_{1}+a20_{trans}+i\kappa b\alpha_{trans}+il10_{trans}+tnf\alpha_{trans}+Sm+C,\; P_{5}^{*}=Dn+eni+Dm+C_{STAT3},\; P_{6}^{*}=kr_{1}+kf_{3}+kr_{3}+kf_{4}+kfi+iln+iki+p,\; P_{7}^{*}=kf_{2}+kr_{2}+kk_{1}+ti_{3}+kdeg_{A20}+kdeg_{I\kappa B\alpha}+eki,\; P_{8}^{*}=ksec_{IL10}+kdeg_{IL-10sup}+ksec_{TNF\alpha}+kdeg_{TNF\alpha sup}+k_{2}+i_{stat3},\; P_{9}^{*}=IL10_{-}IL10R_{max} \rbrace$,
\newline
$\Omega_{3}=\lbrace P_{1}^{*}=kf_{1}+kr_{1},\; P_{2}^{*}=kf_{2}+kr_{2}+ksec_{IL10}+kdeg_{IL_{-}10sup},\; P_{3}^{*}=kf_{3}+kr_{3}+ksec_{TNF\alpha}+kdeg_{TNF\alpha sup},\; P_{4}^{*}=il10_{trans}+Dn,\; P_{5}^{*}=kfi+kk_{1}+kk_{3}+ti_{3}+iln,\; P_{6}^{*}=kdeg_{I\kappa B\alpha},\; P_{7}^{*}=Sm+p+Dm,\; P_{8}^{*}=a20_{trans}+kdeg_{A20},\; P_{9}^{*}=iki+eki,\; P_{10}^{*}=i\kappa b\alpha_{tran}+kf_{4}+Sm_{-}il10+tnf\alpha_{trans},\; P_{11}^{*}=k_{1}+k_{2},\; P_{12}^{*}=kv+i_{stat3}+eni,\; P_{13}^{*}=C+C_{STAT3}+IL10_{-}IL10R_{max} \rbrace$,
\newline
$\Omega_{4}=\lbrace P_{1}^{*}=kf_{1}+kr_{1},\; P_{2}^{*}=kf_{2}+kr_{2},\; P_{3}^{*}=kf_{3}+kr_{3},\; P_{4}^{*}=ksec_{IL10}+kdeg_{IL_{-}10sup},\; P_{5}^{*}=kfi+kk_{1}+kk_{3}+iln,\; P_{6}^{*}=kdeg_{I\kappa B\alpha}+ti_{3},\; P_{7}^{*}=Sm,\; P_{8}^{*}=a20_{trans}+kdeg_{A20},\; P_{9}^{*}=iki+eki,\; P_{10}^{*}=i\kappa b\alpha_{tran},\; P_{11}^{*}=kf_{4},\; P_{12}^{*}=Sm_{-}il10+p+Dm,\; P_{13}^{*}=il10_{trans}+Dn,\; P_{14}^{*}=tnf\alpha_{trans}+ksec_{TNF\alpha}+kdeg_{TNF\alpha sup},\; P_{15}^{*}=k_{1}+k_{2},\; P_{16}^{*}=i_{stat3}+eni,\; P_{17}^{*}=kv,\; P_{18}^{*}=C,\; P_{19}^{*}=C_{STAT3},\; P_{20}^{*}=IL10_{-}IL10R_{max} \rbrace$.

\begin{figure}[H]  
	\begin{center}     
		\subfigure{%
			\includegraphics[width=0.2\textwidth]{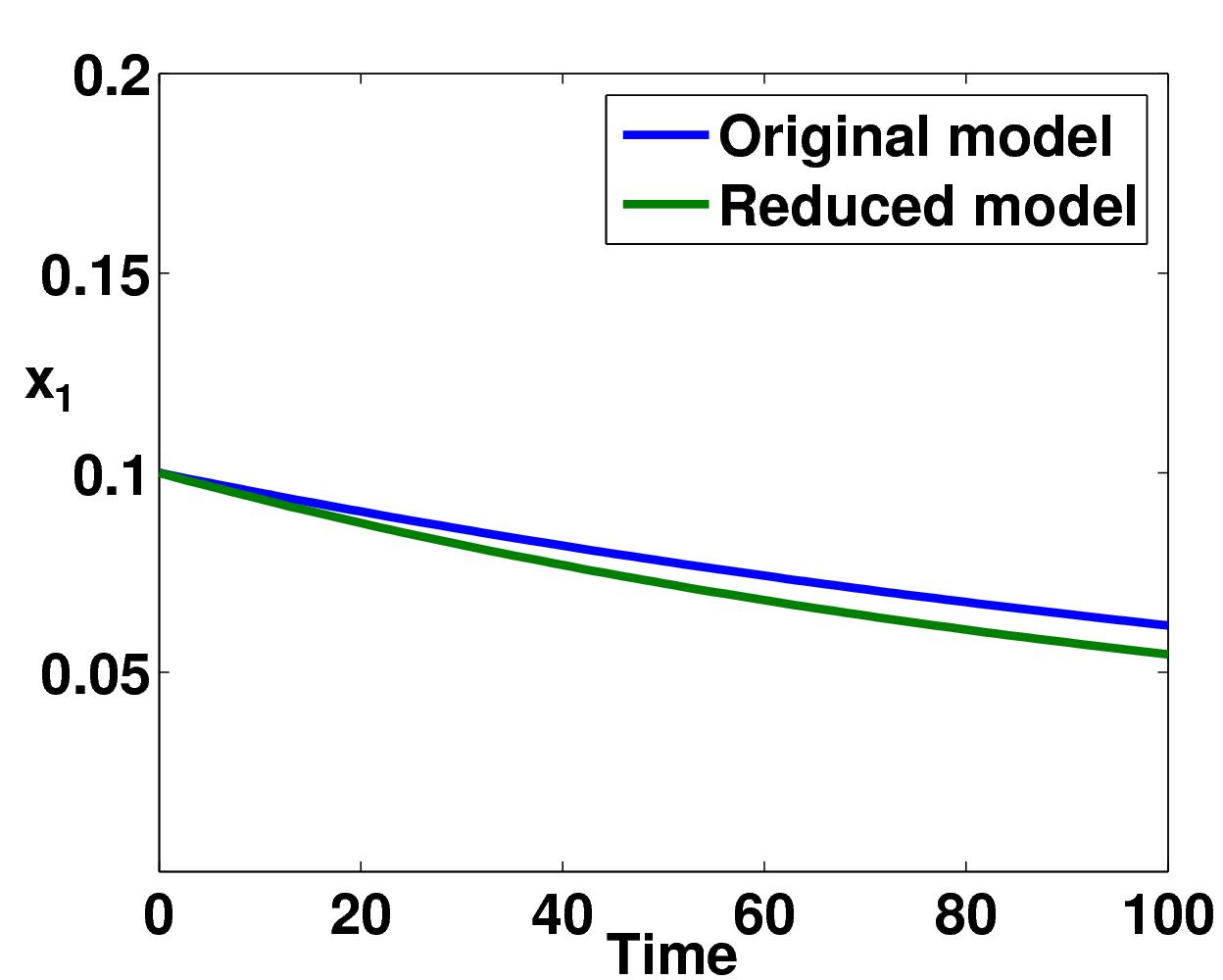}
		}		\\
		\subfigure{%
			\includegraphics[width=0.2\textwidth]{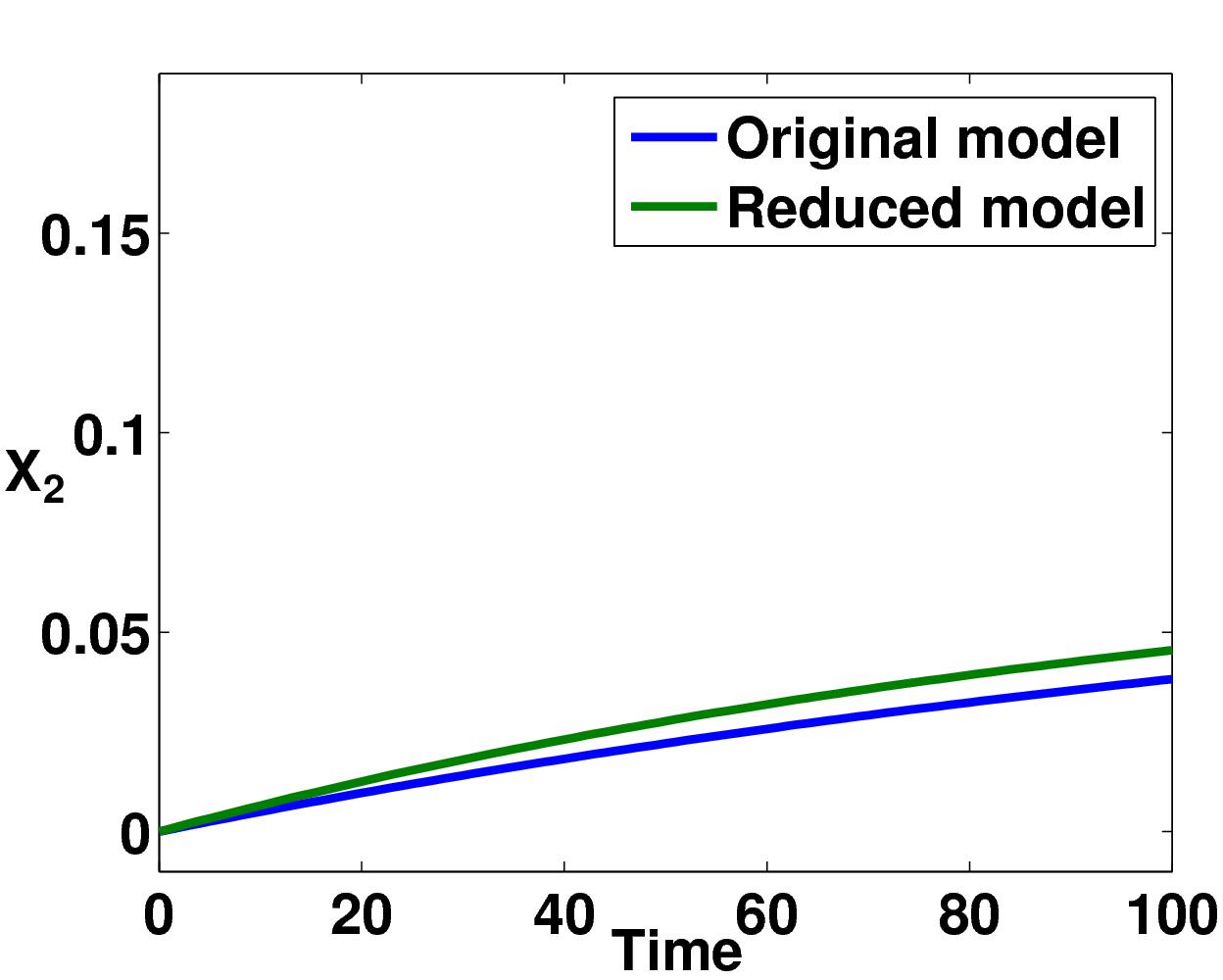}
		} 
		\subfigure{%
			\includegraphics[width=0.2\textwidth]{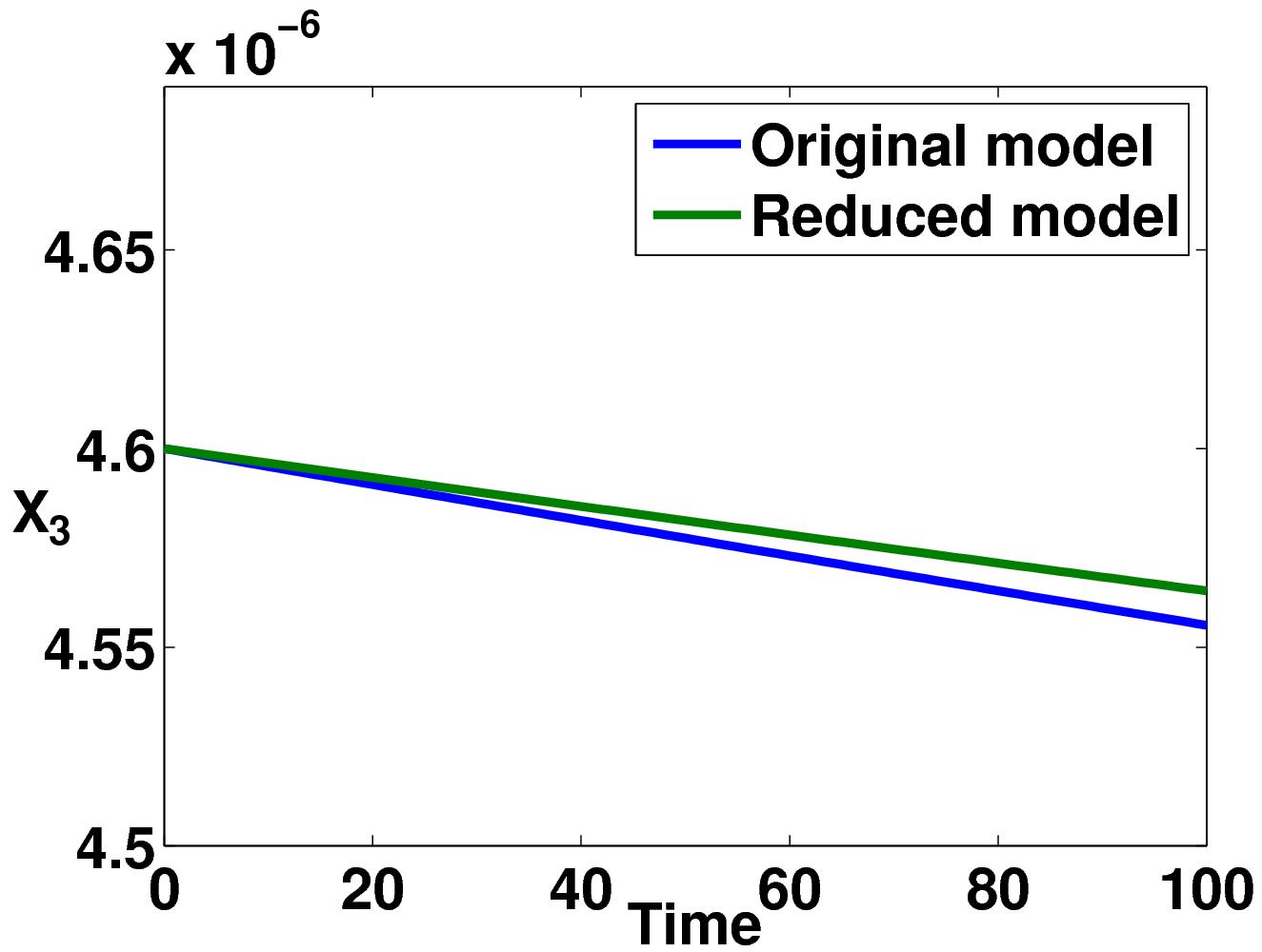}
		}
     	\subfigure{%
			\includegraphics[width=0.2\textwidth]{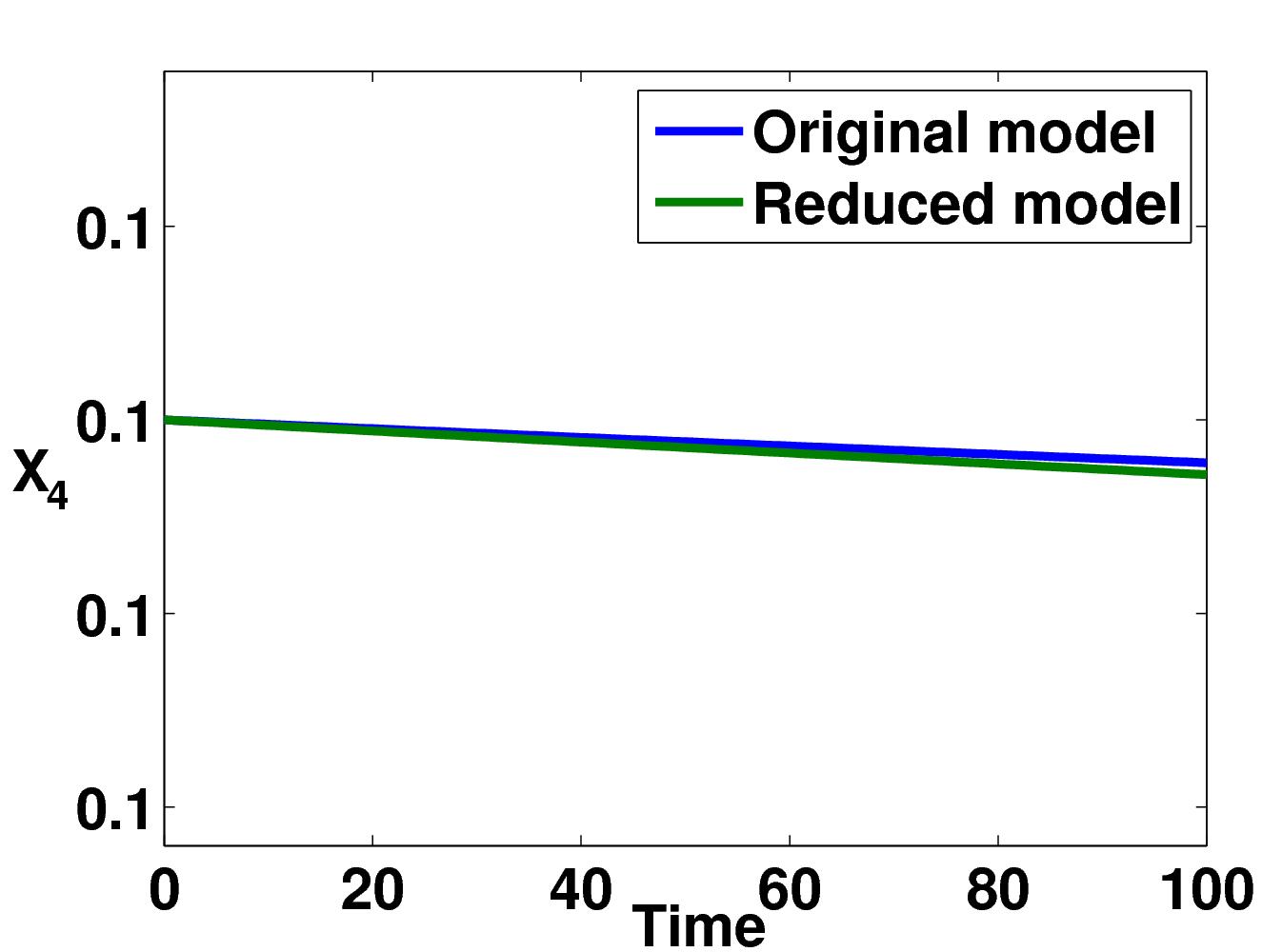}
		}
		     	\subfigure{%
			\includegraphics[width=0.2\textwidth]{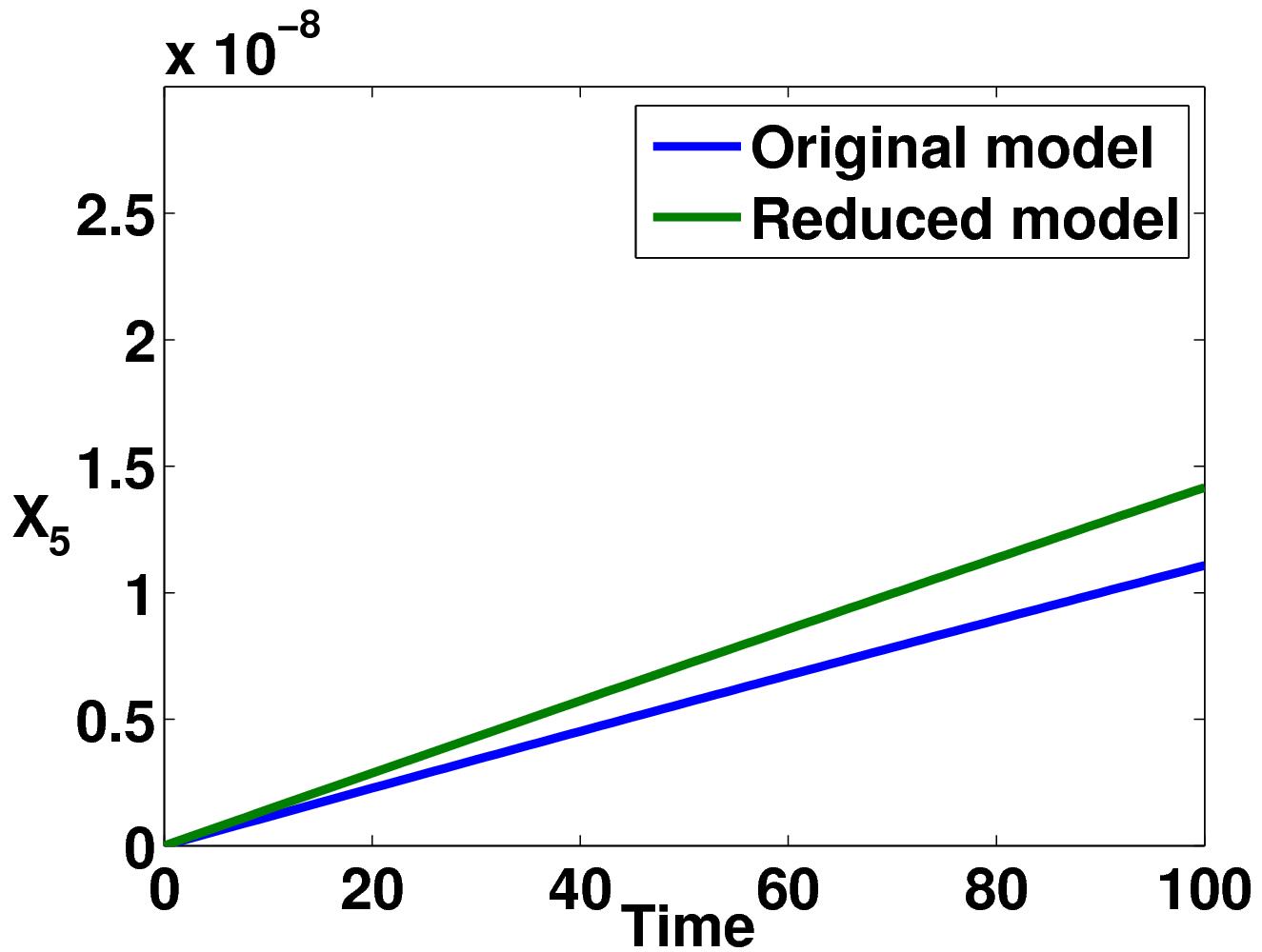}
		} 
		     	\subfigure{%
			\includegraphics[width=0.2\textwidth]{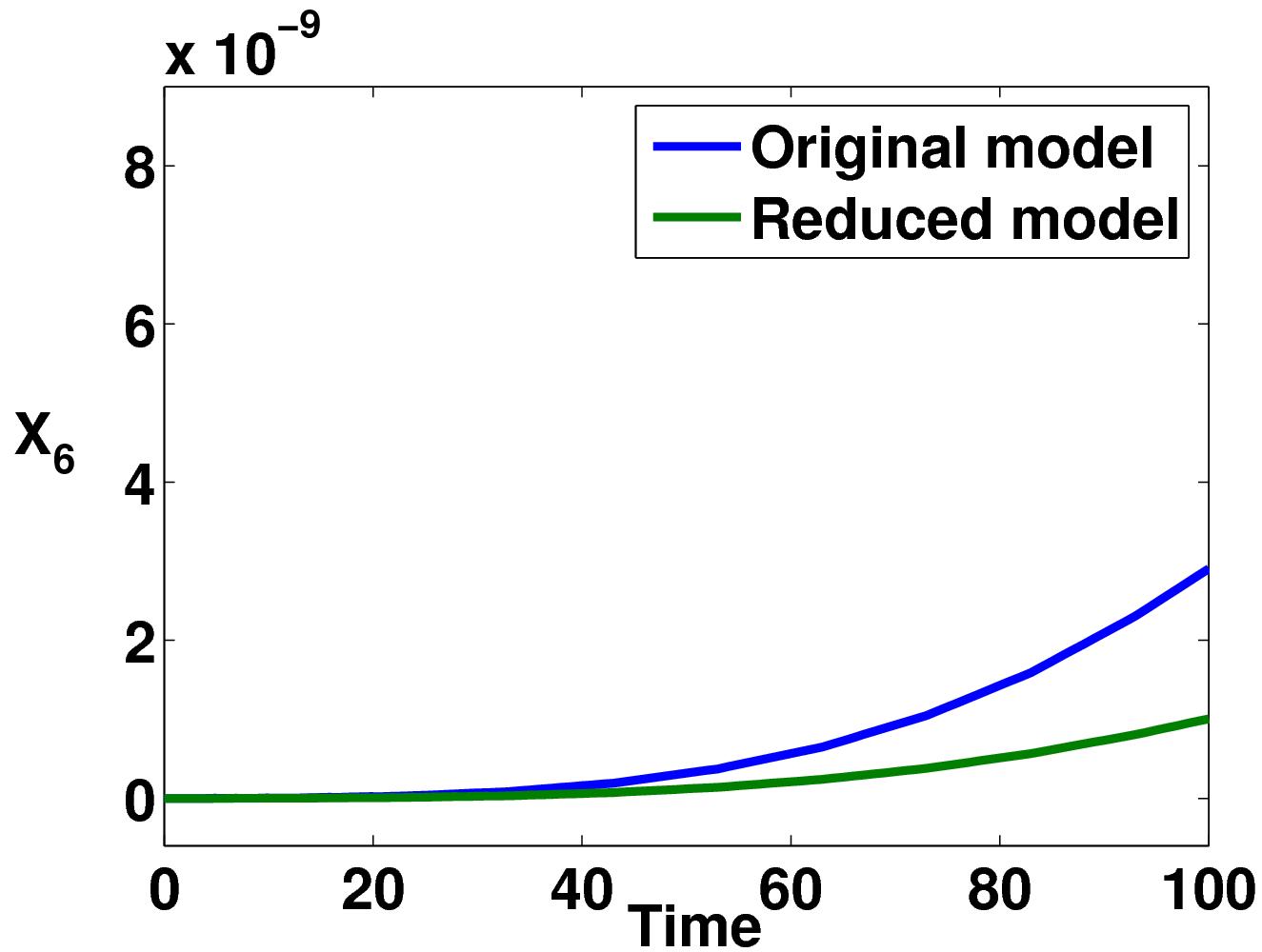}
		} 
		     	\subfigure{%
			\includegraphics[width=0.2\textwidth]{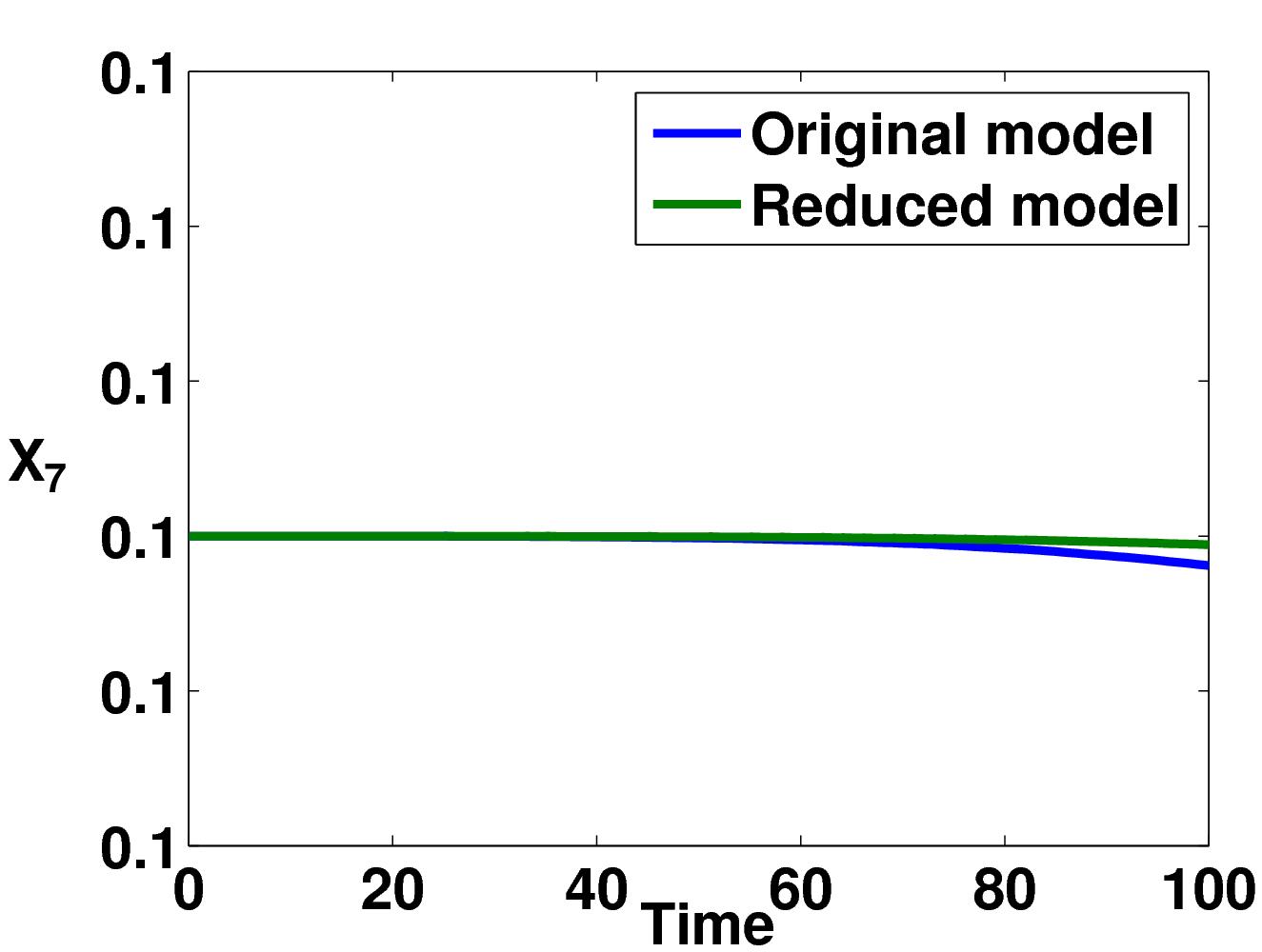}
		} 
		     	\subfigure{%
			\includegraphics[width=0.2\textwidth]{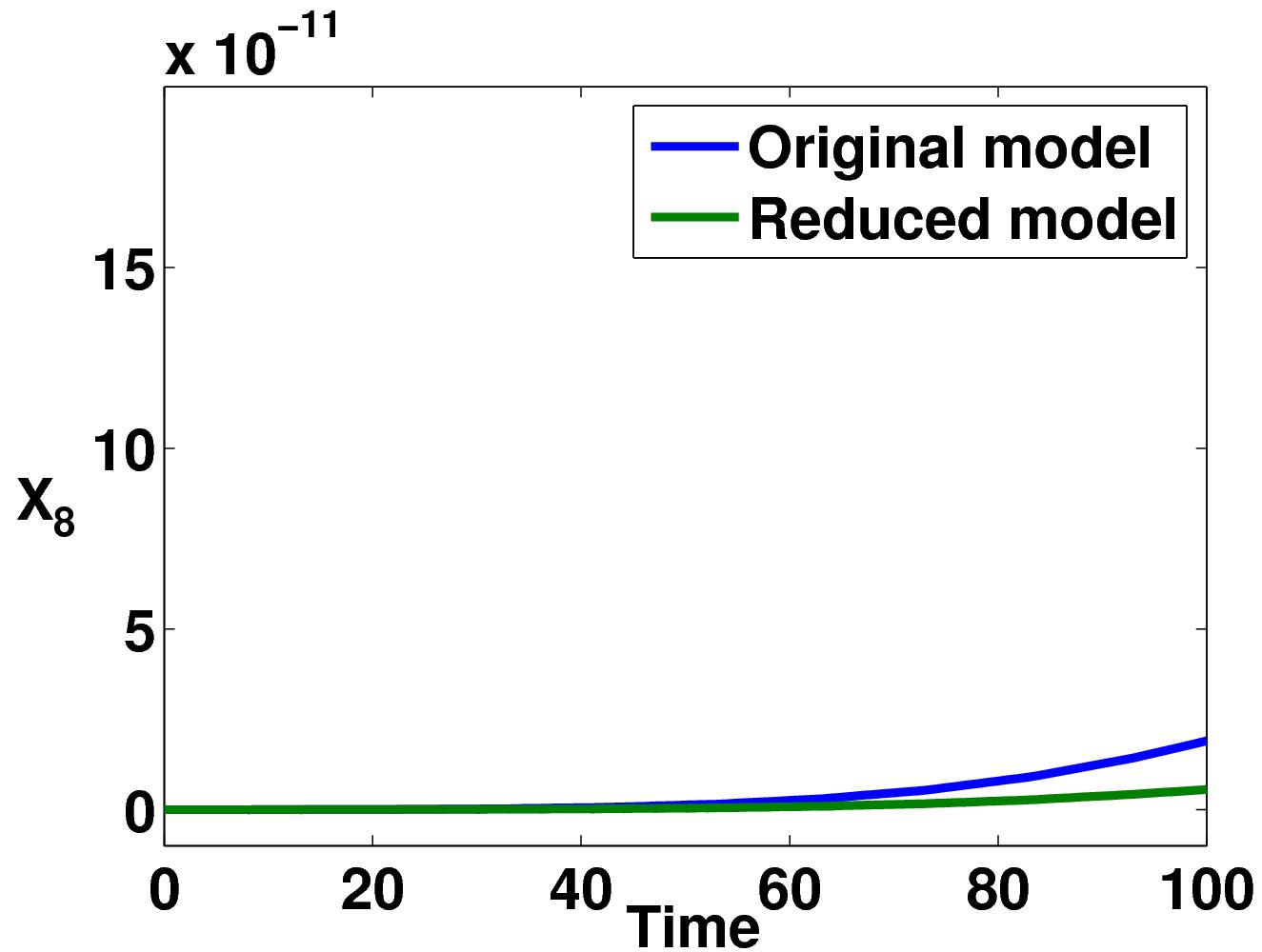}
		} 
		     	\subfigure{%
			\includegraphics[width=0.2\textwidth]{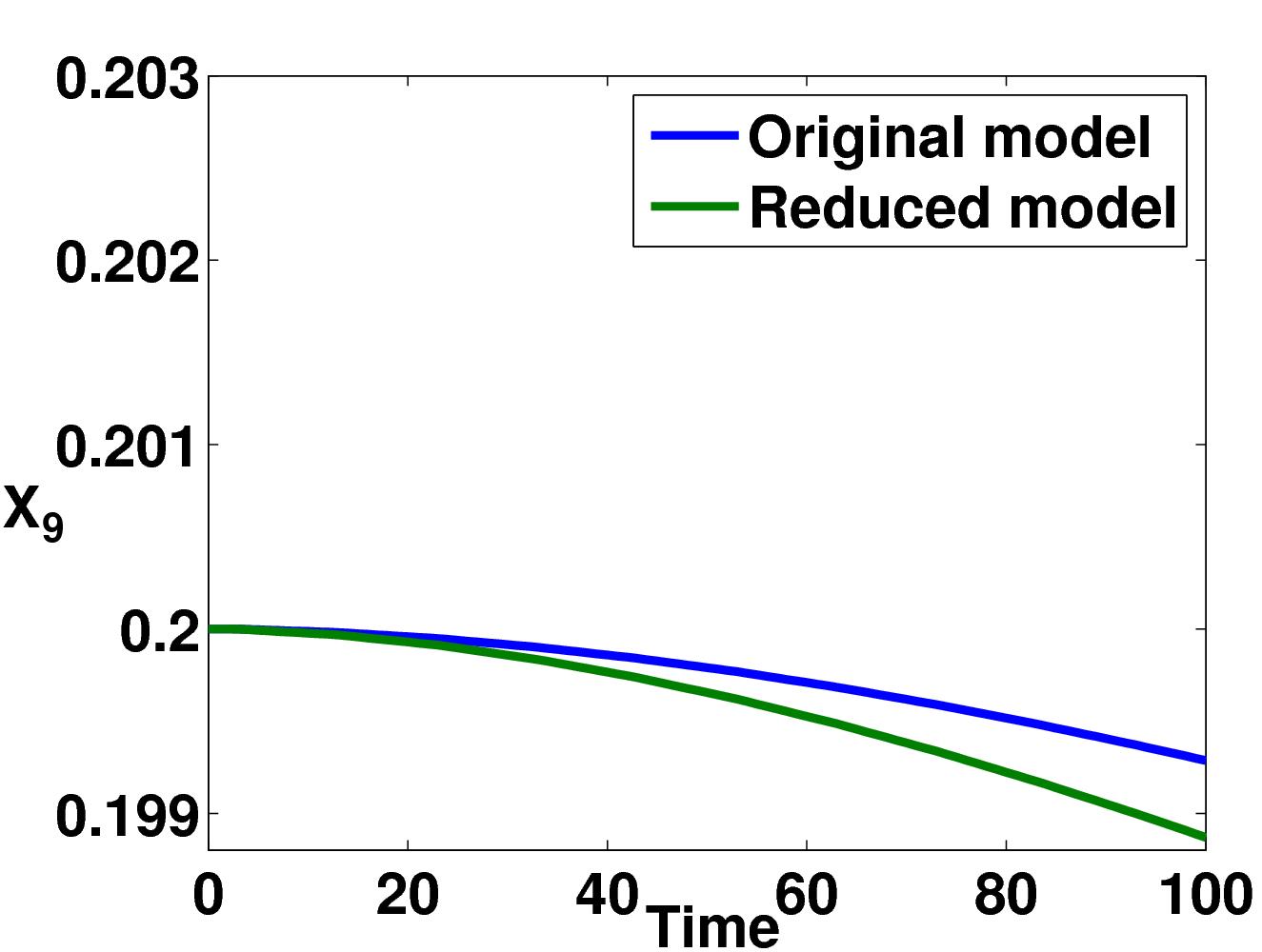}
		}  
		     	\subfigure{%
			\includegraphics[width=0.2\textwidth]{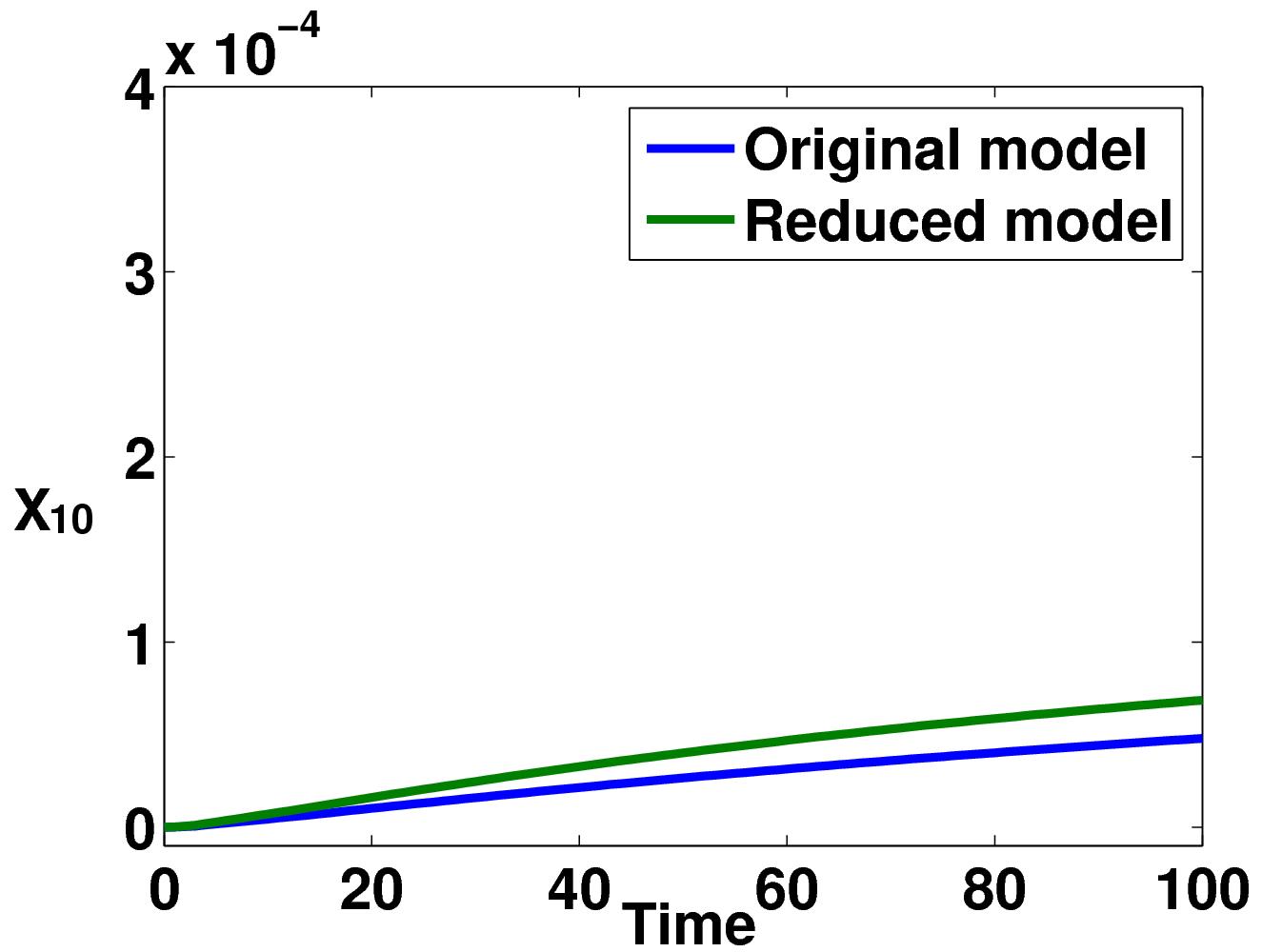}
		} 
		     	\subfigure{%
			\includegraphics[width=0.2\textwidth]{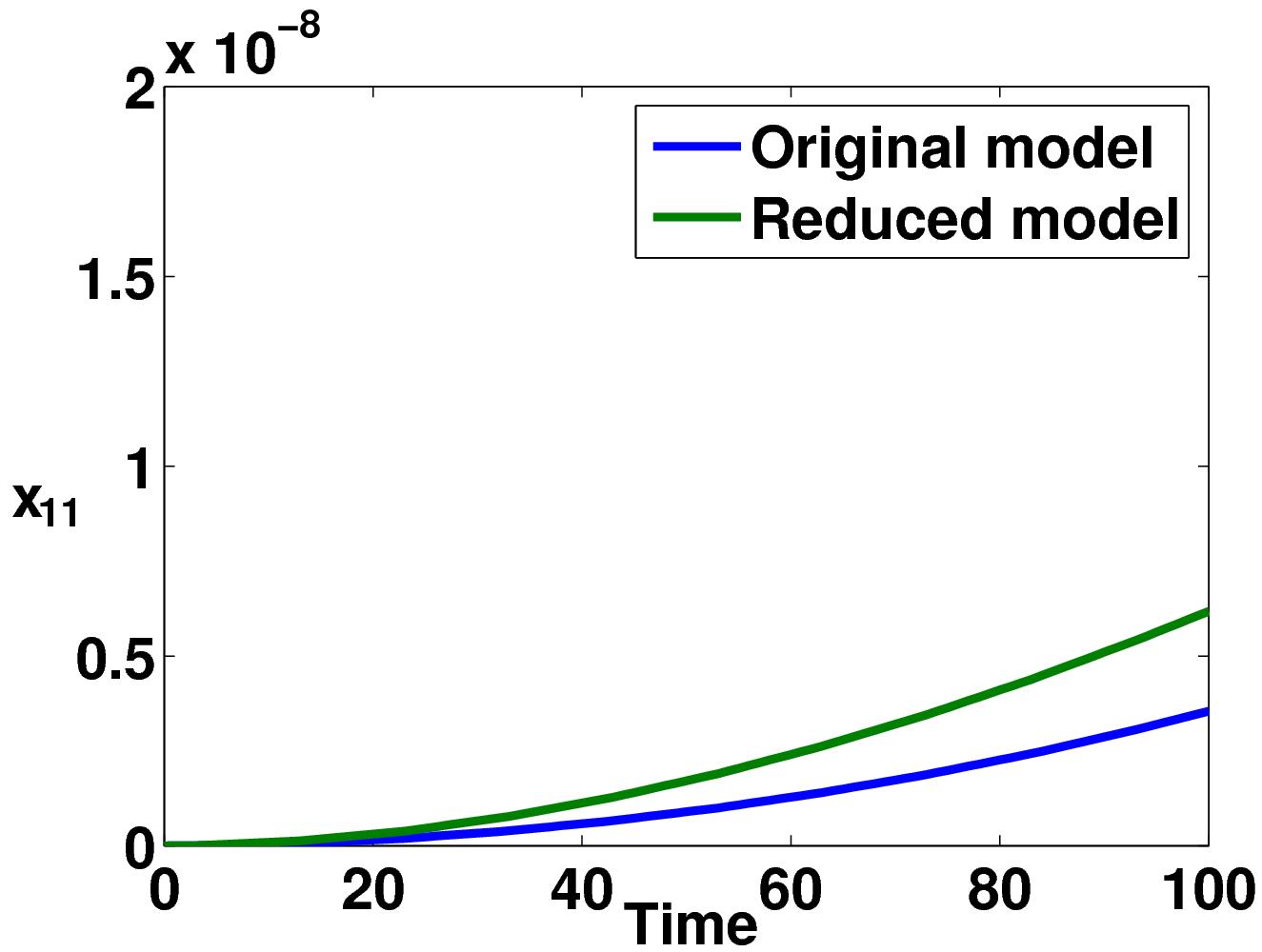}
		} 
				     	\subfigure{%
			\includegraphics[width=0.2\textwidth]{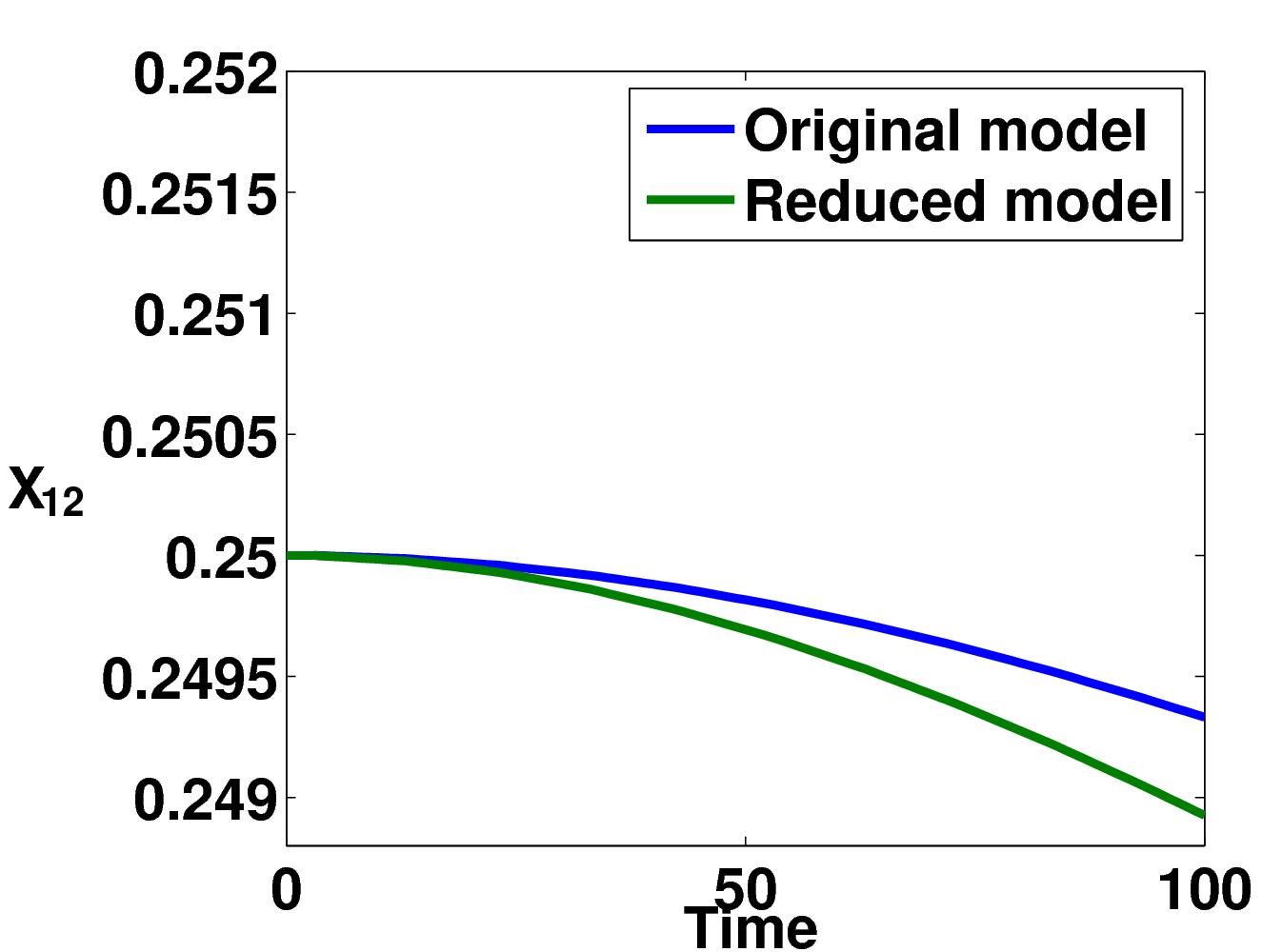}
		} 
		     	\subfigure{%
			\includegraphics[width=0.2\textwidth]{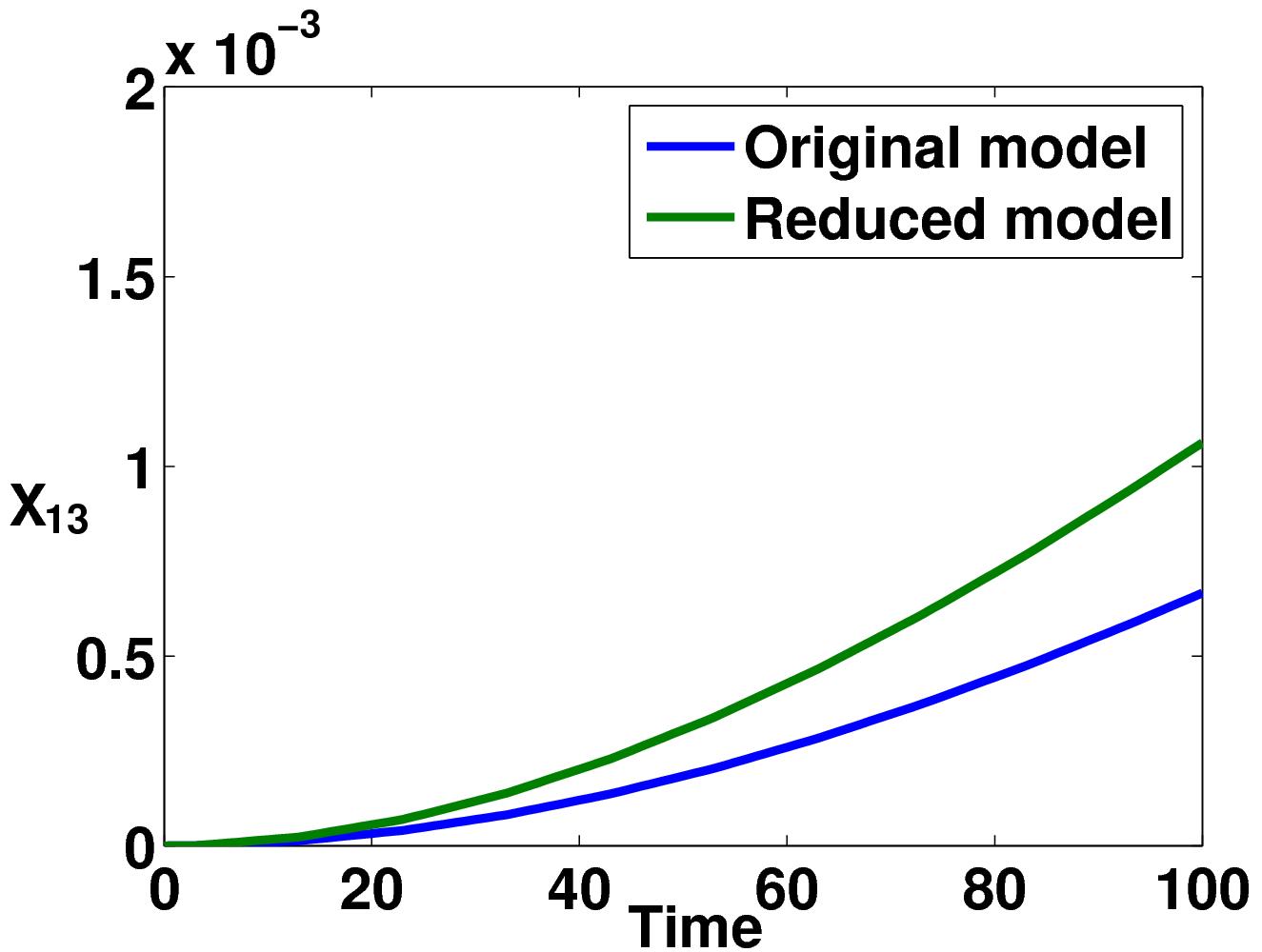}
		} 
		     	\subfigure{%
			\includegraphics[width=0.2\textwidth]{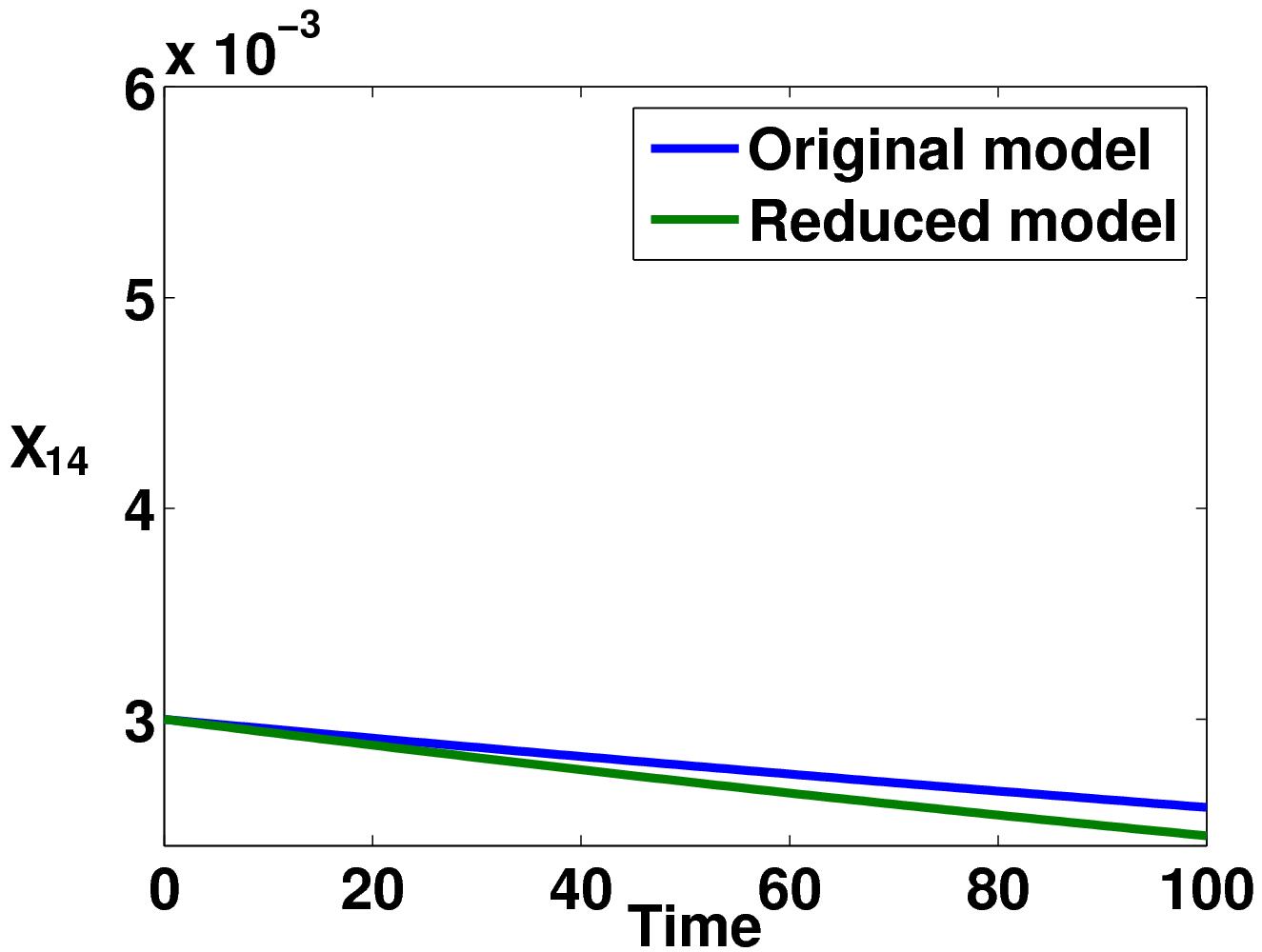}
		} 
		     	\subfigure{%
			\includegraphics[width=0.2\textwidth]{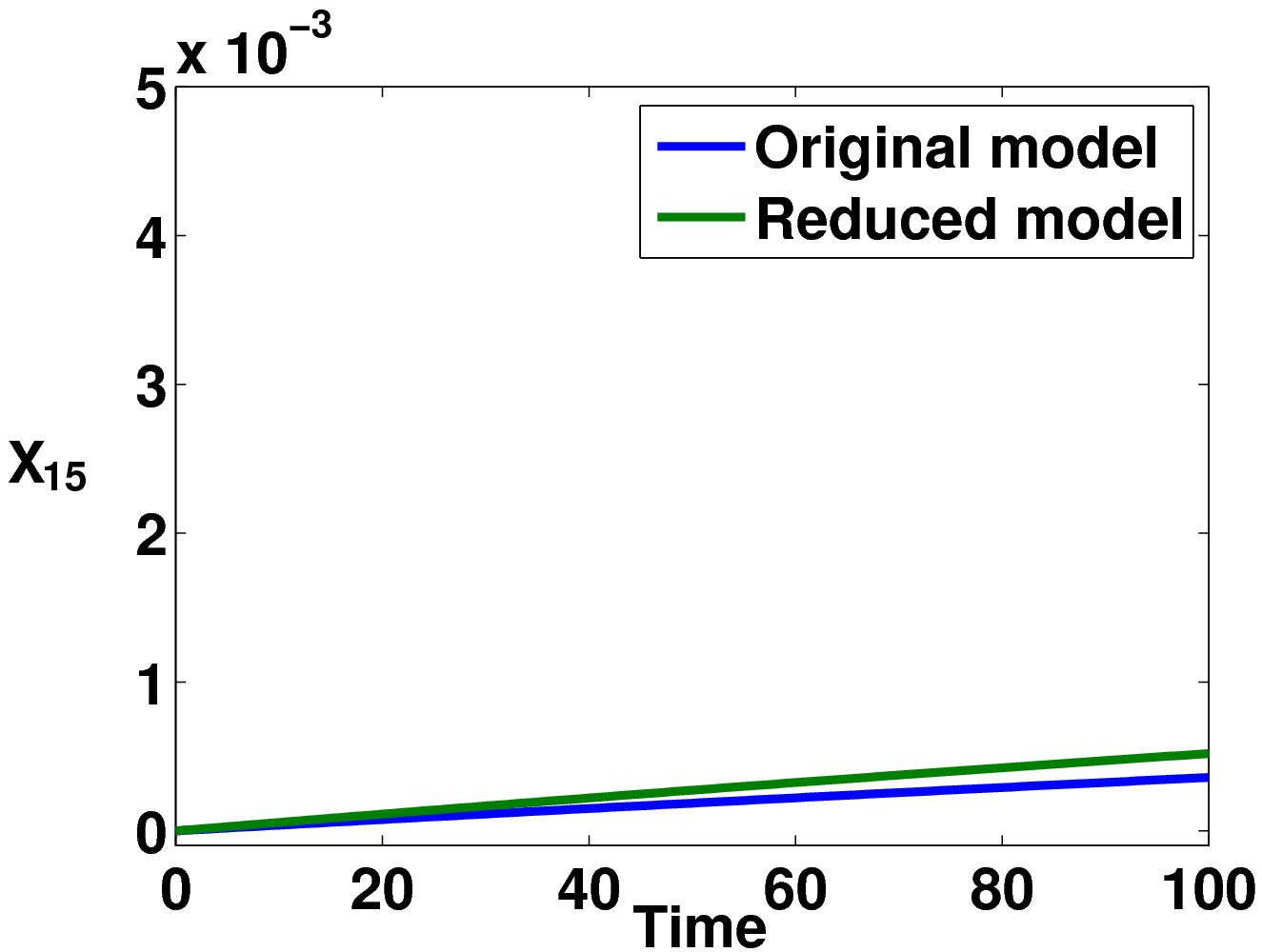}
		}  
		     	\subfigure{%
			\includegraphics[width=0.2\textwidth]{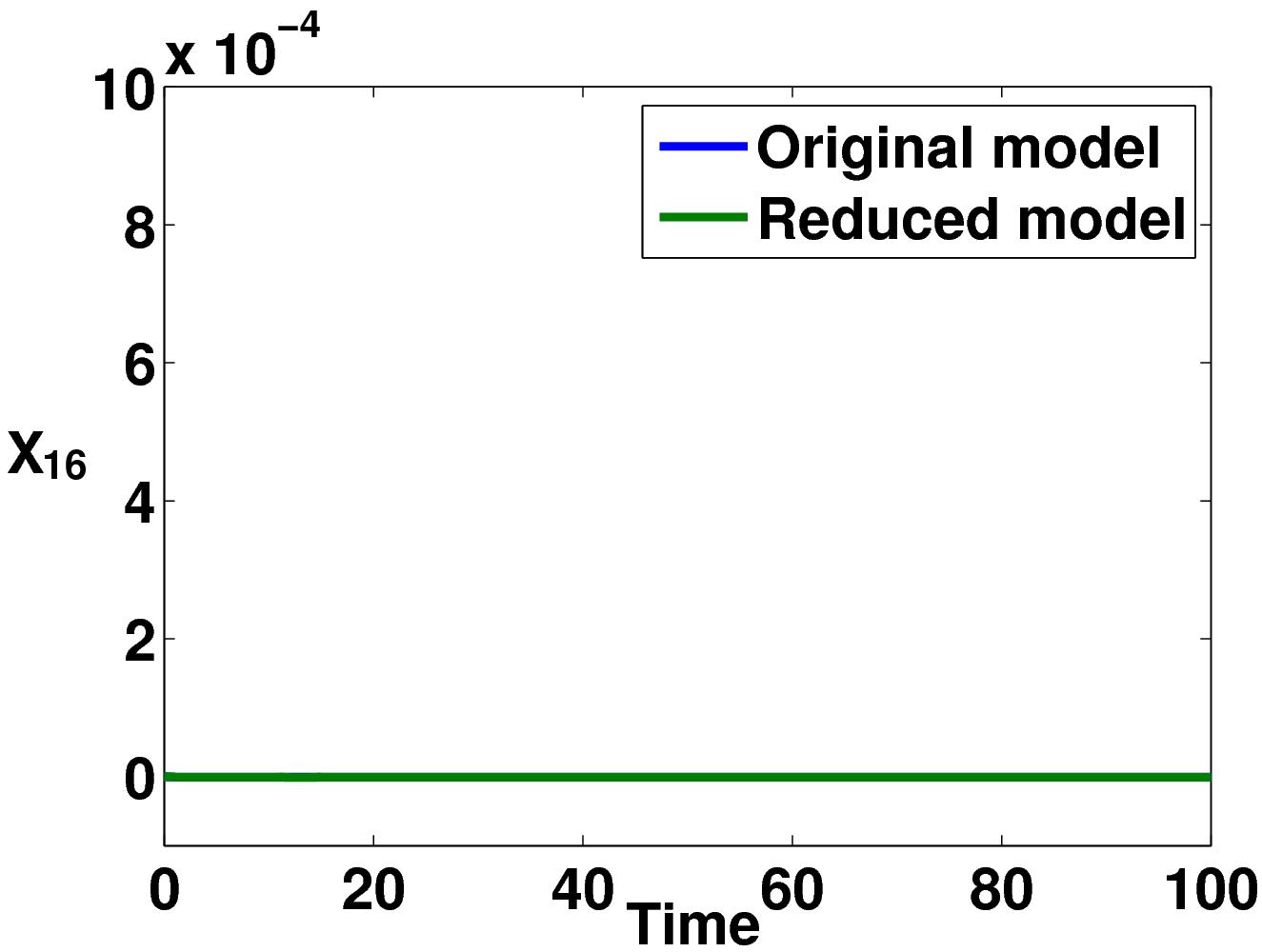}
		} 
		     	\subfigure{%
			\includegraphics[width=0.2\textwidth]{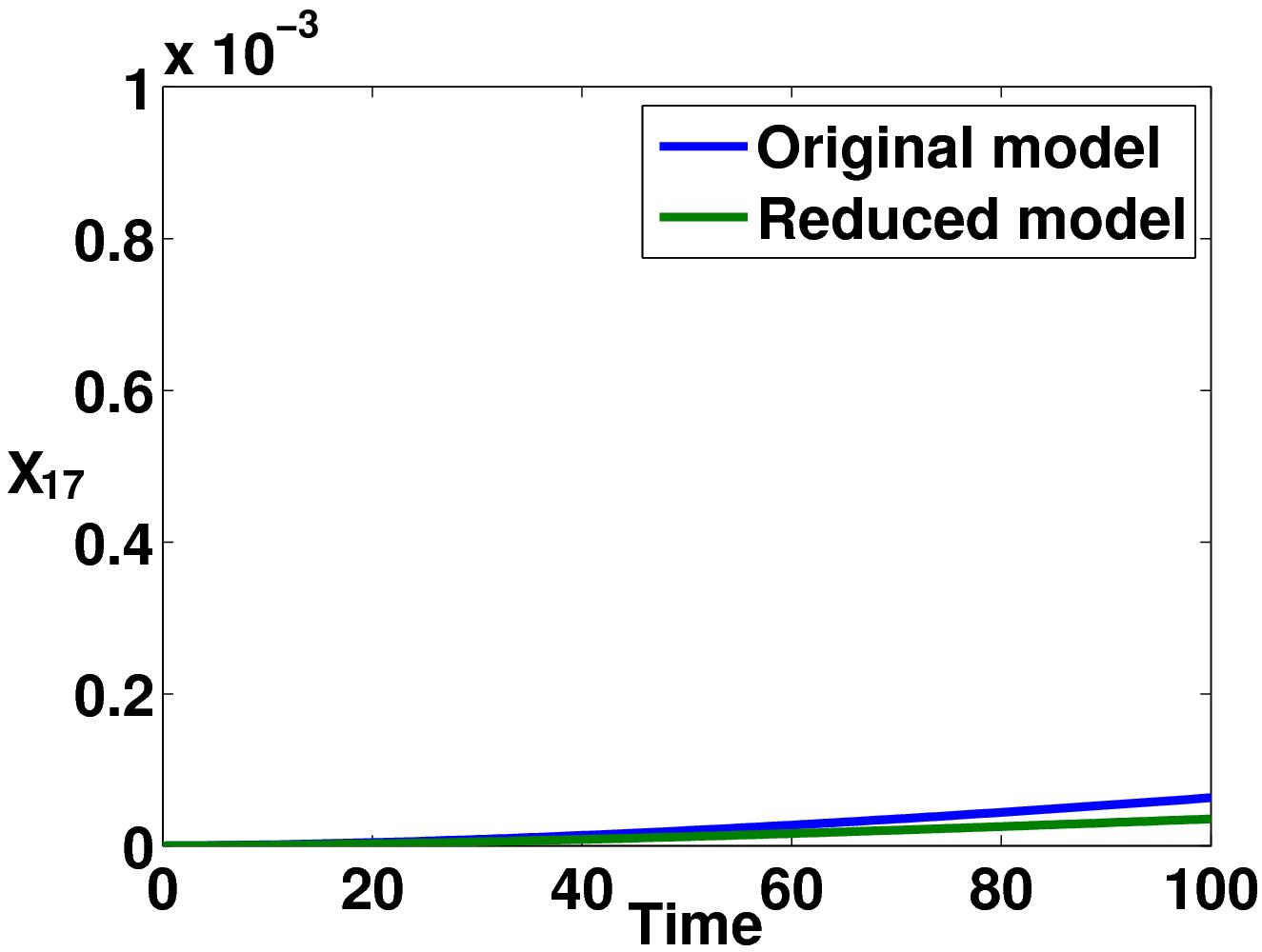}
		}
				     	\subfigure{%
			\includegraphics[width=0.2\textwidth]{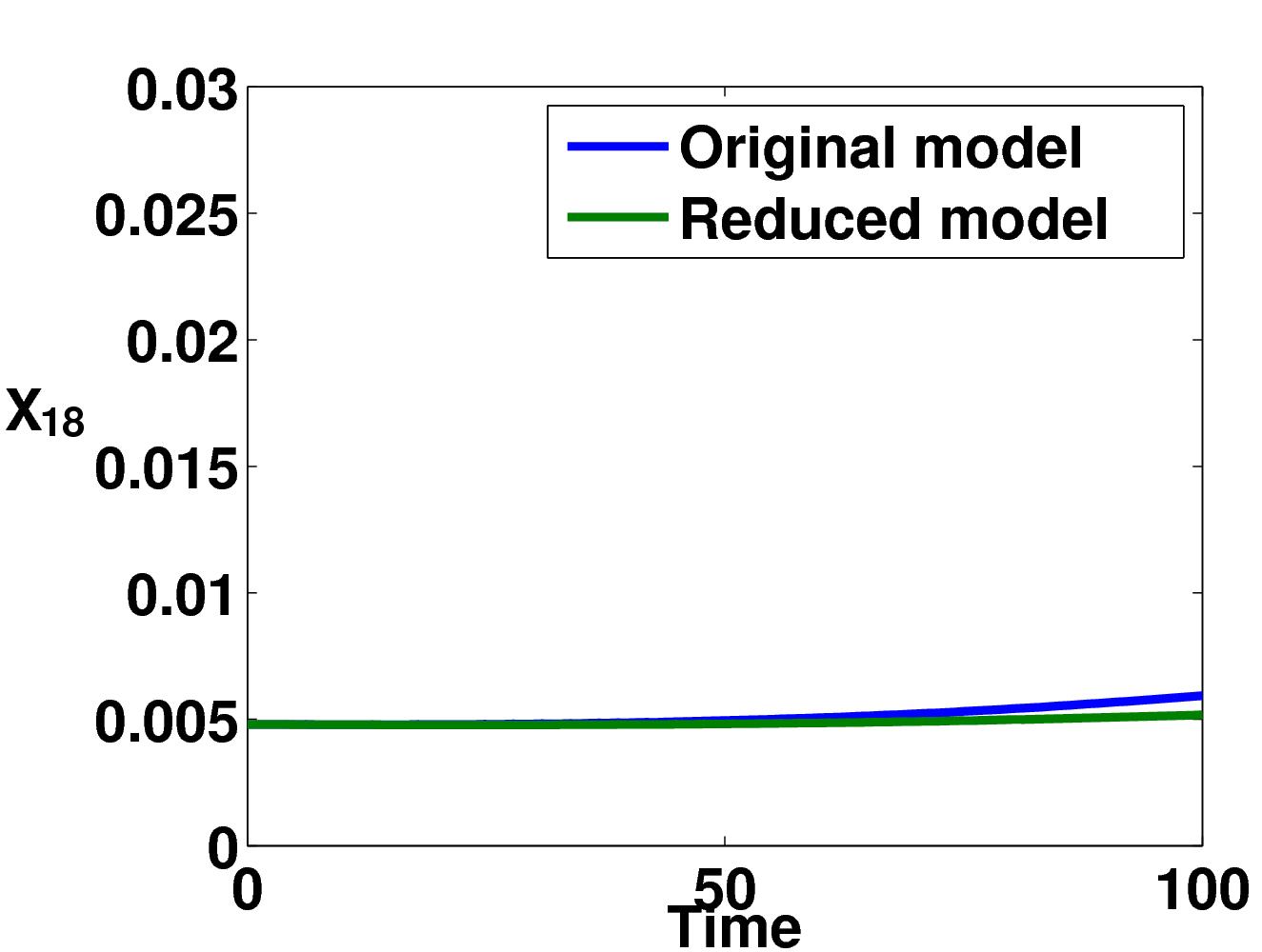}
		} 
		     	\subfigure{%
			\includegraphics[width=0.2\textwidth]{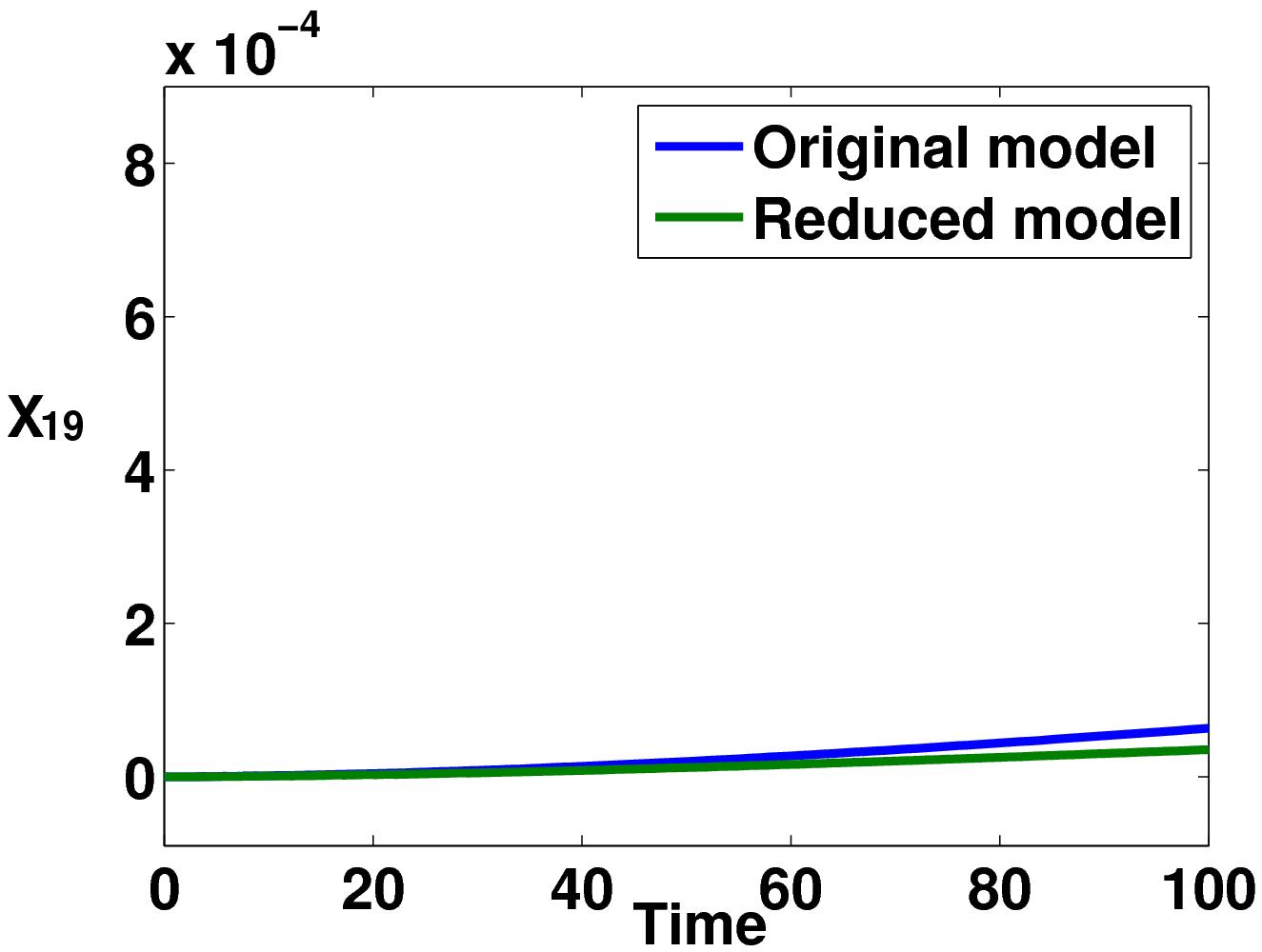}
		} 
		     	\subfigure{%
			\includegraphics[width=0.2\textwidth]{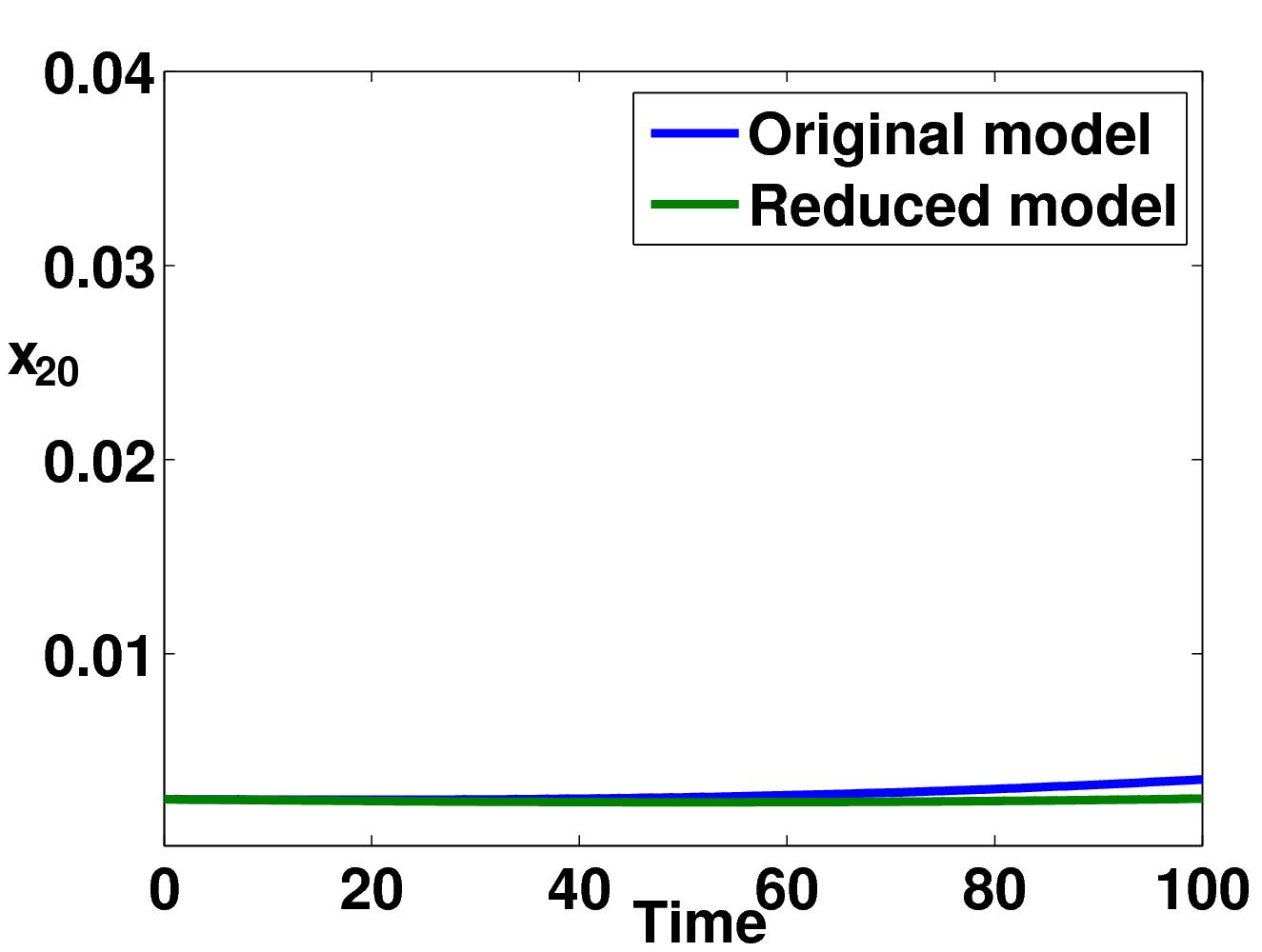}
		} 
		     	\subfigure{%
			\includegraphics[width=0.2\textwidth]{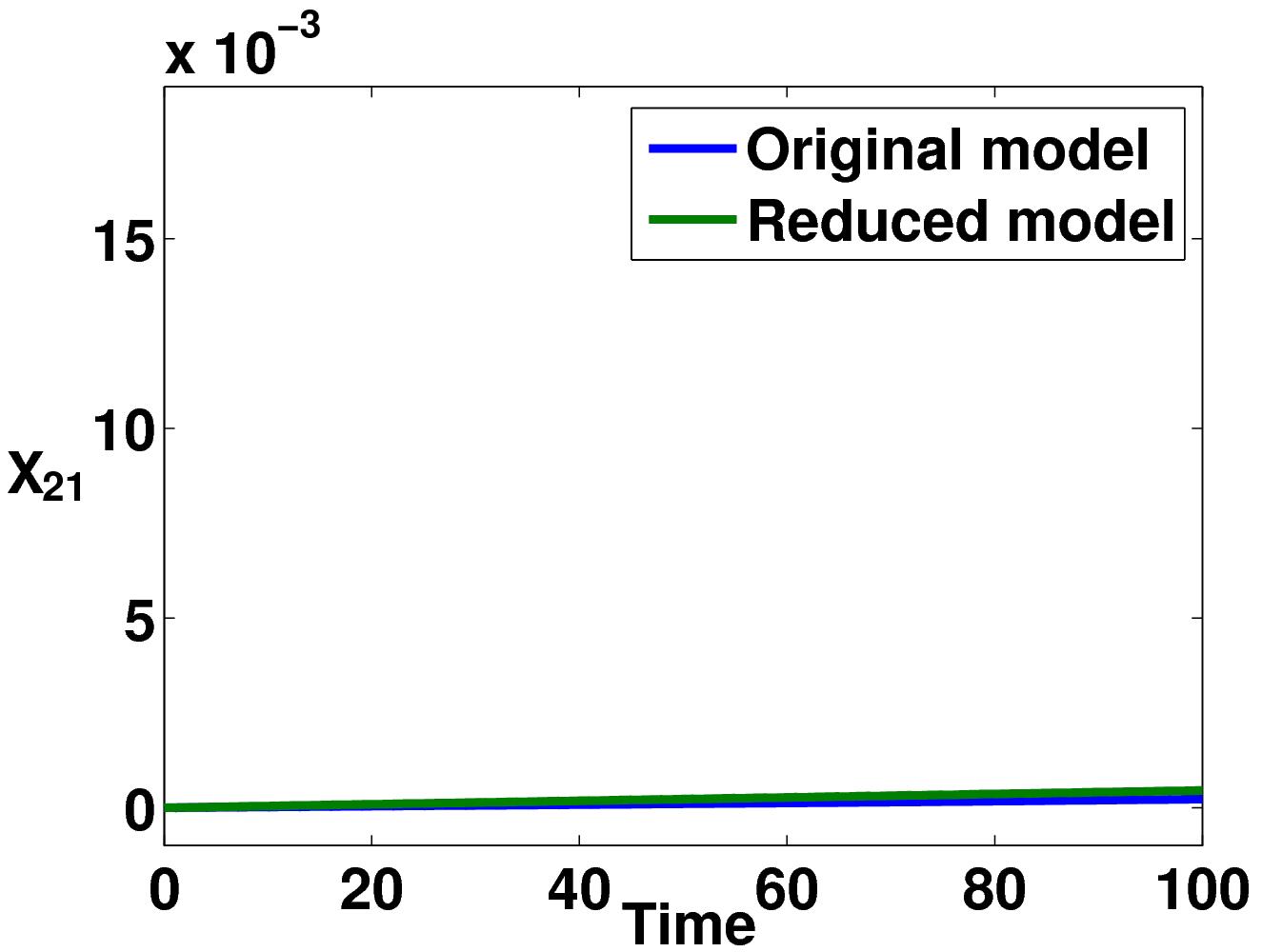}
		}  
		     	\subfigure{%
			\includegraphics[width=0.2\textwidth]{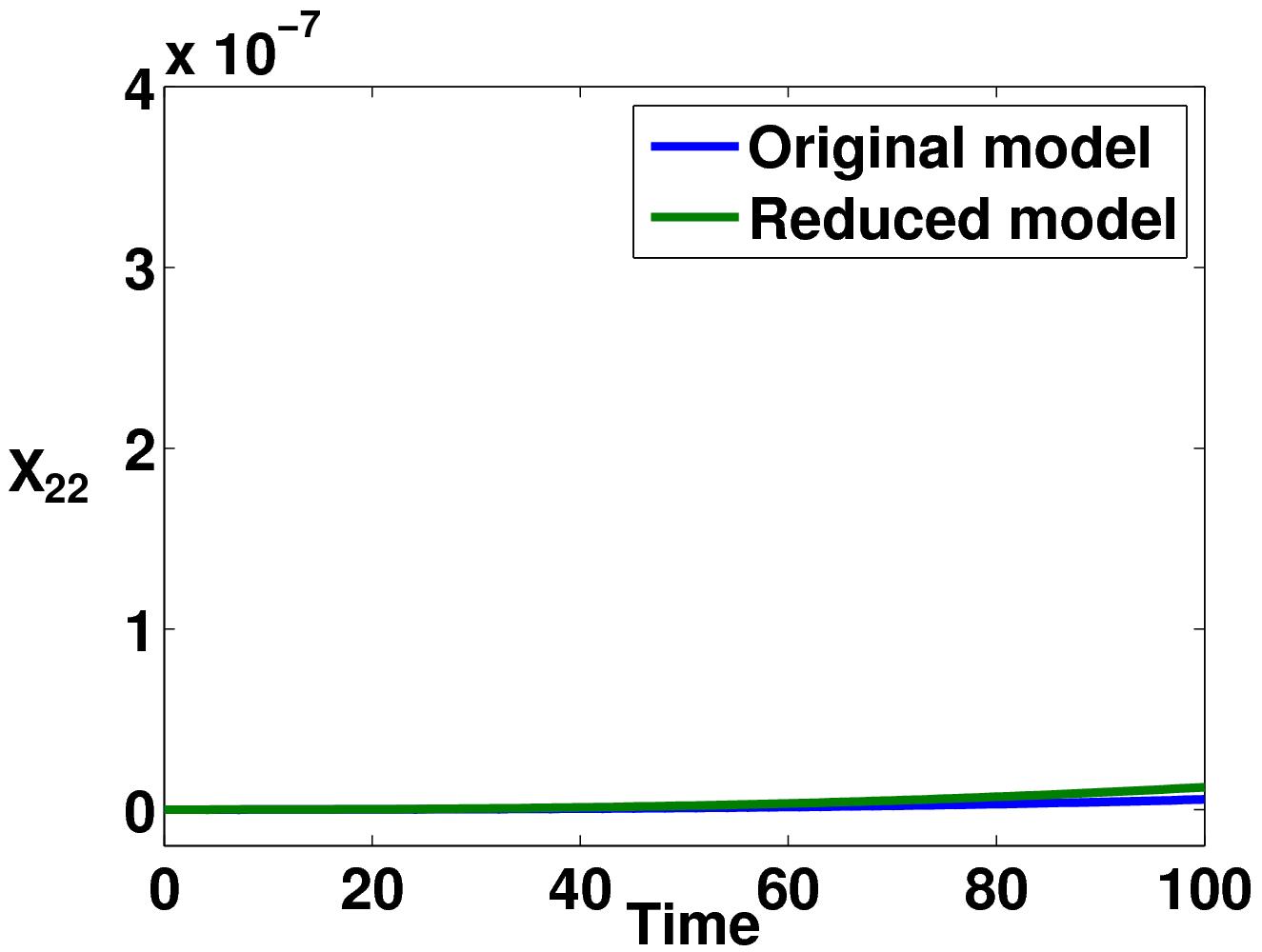}
		} 
		     	\subfigure{%
			\includegraphics[width=0.2\textwidth]{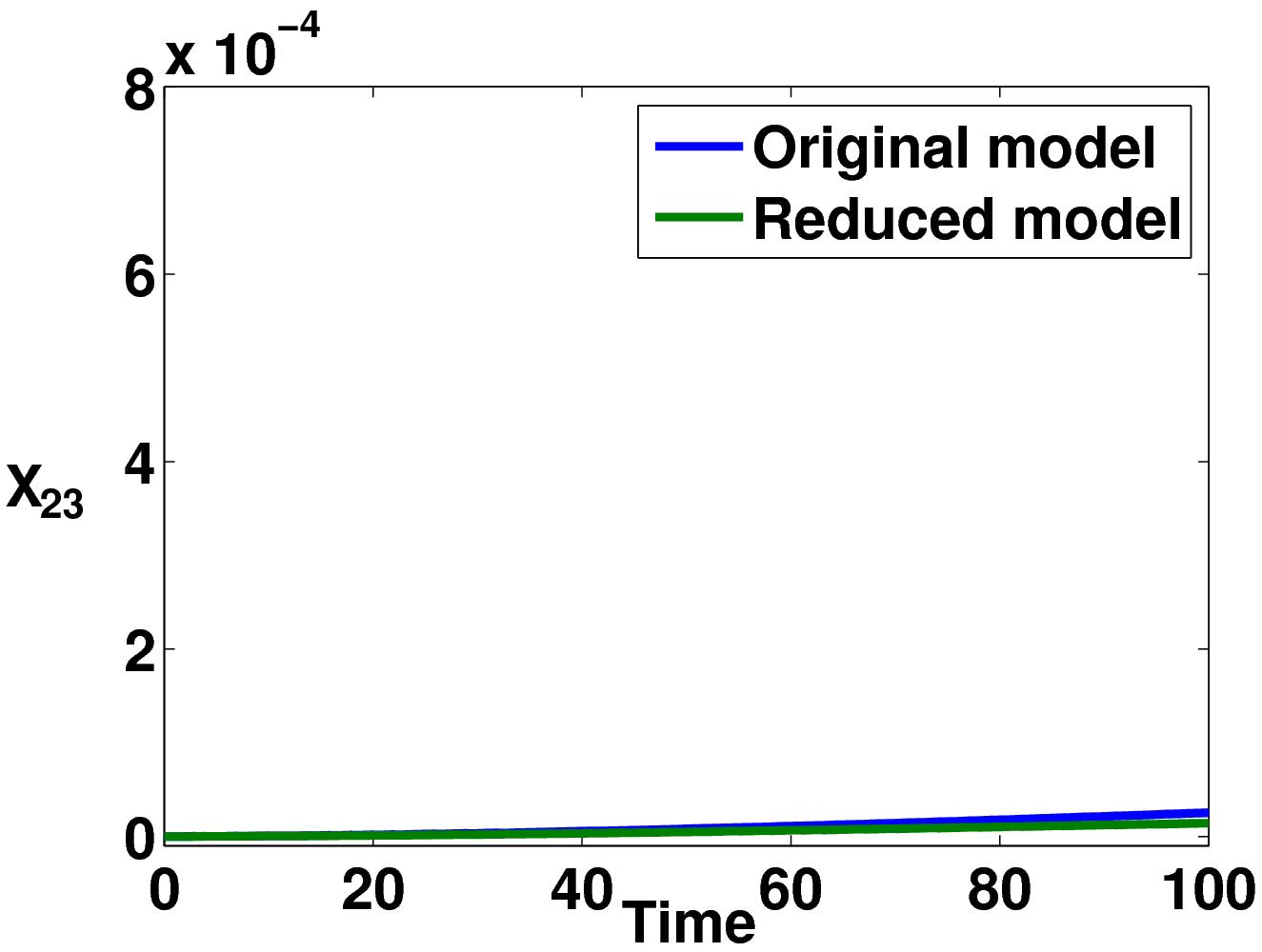}
		}	
				     	\subfigure{%
			\includegraphics[width=0.2\textwidth]{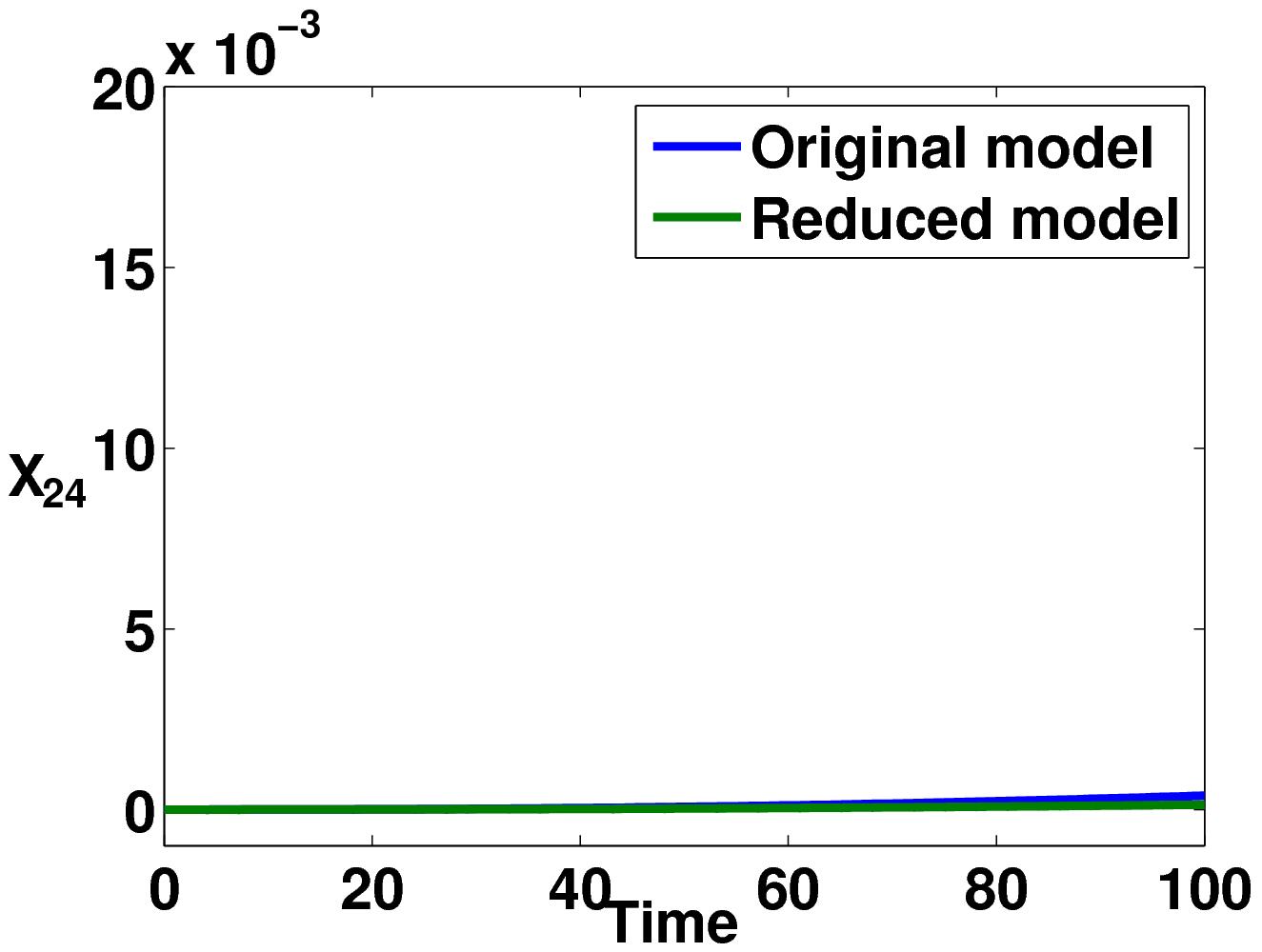}
		}
				     	\subfigure{%
			\includegraphics[width=0.2\textwidth]{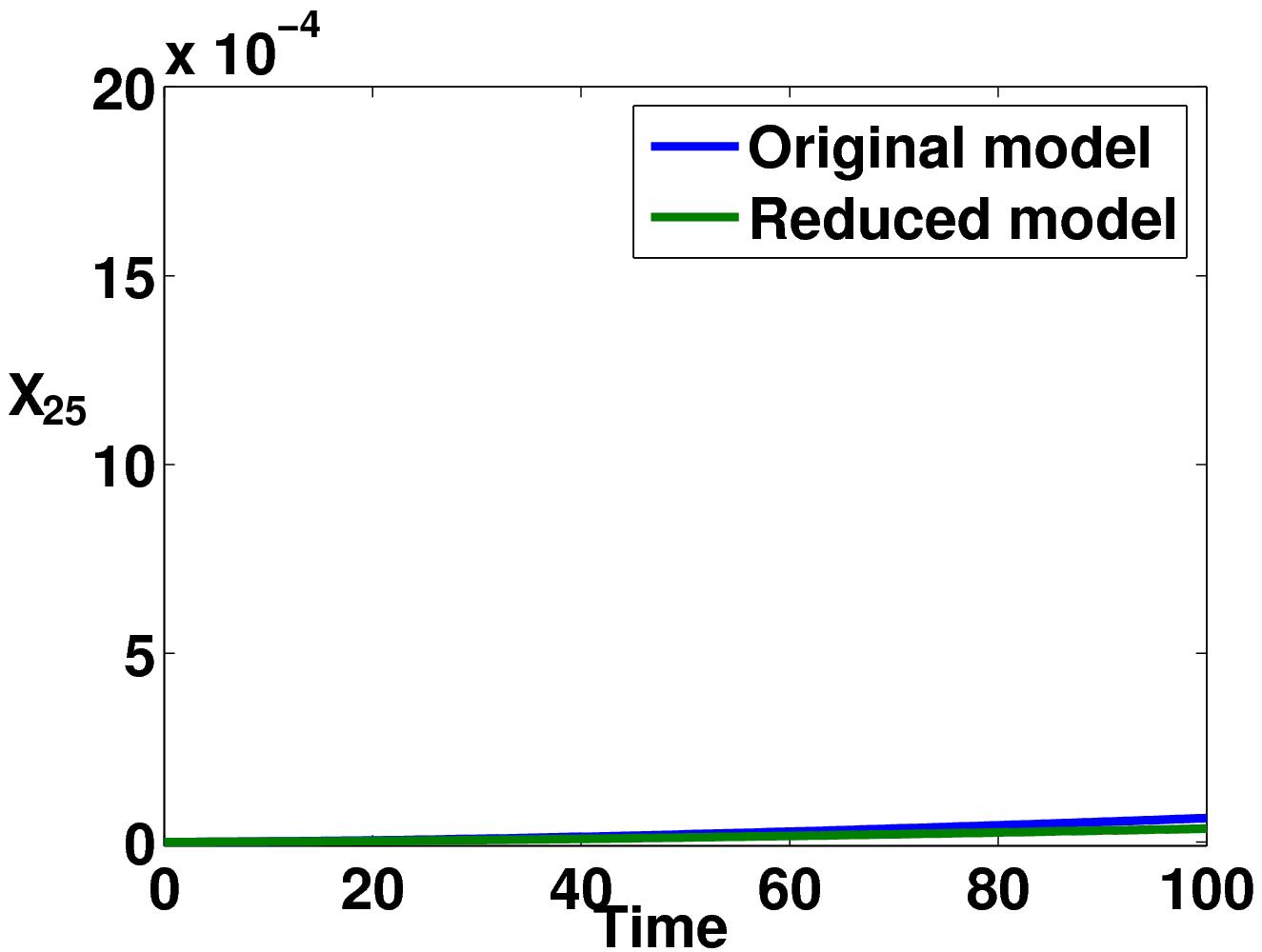}
		}
				     	\subfigure{%
			\includegraphics[width=0.2\textwidth]{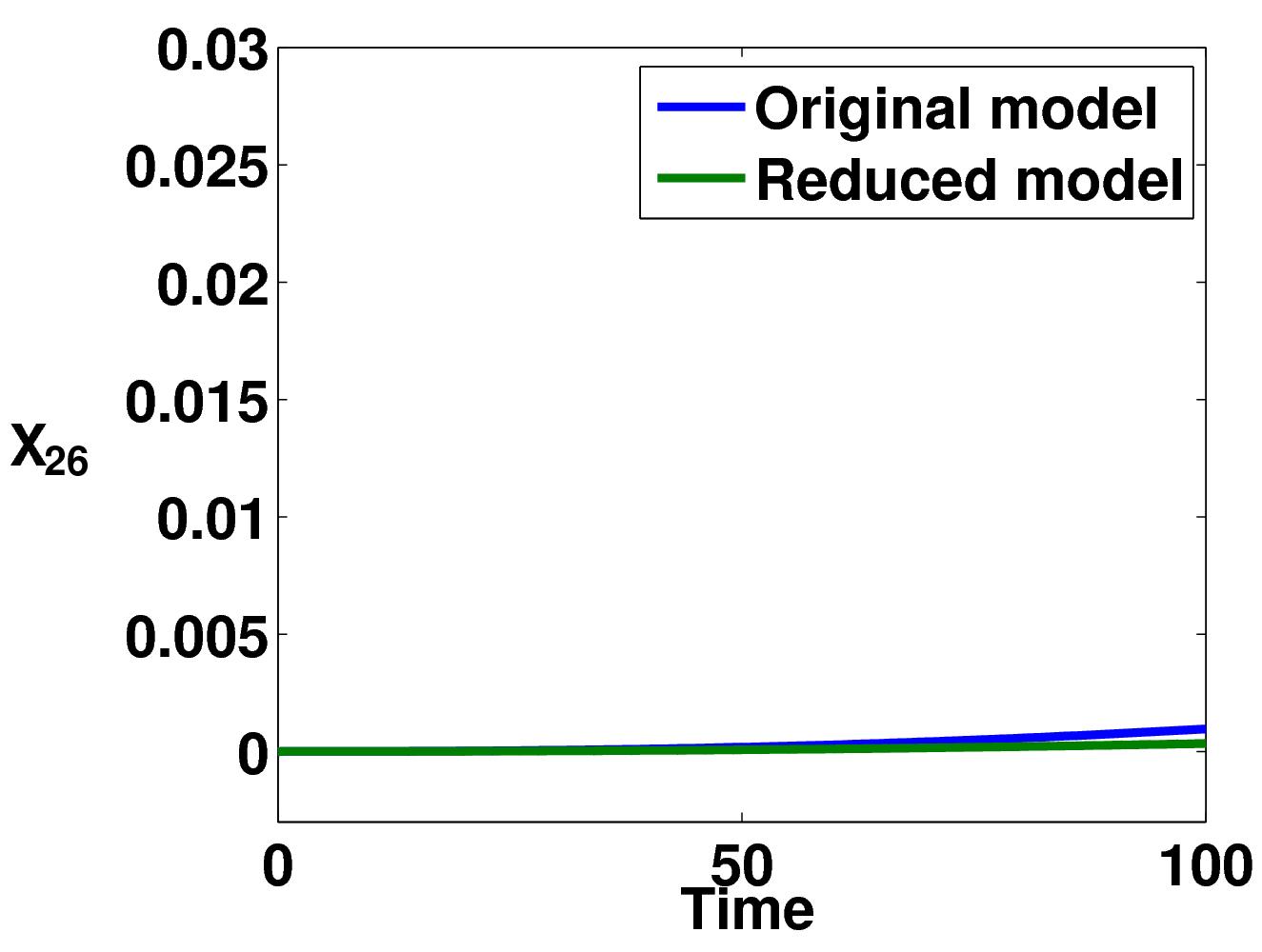}
		}
				     	\subfigure{%
			\includegraphics[width=0.2\textwidth]{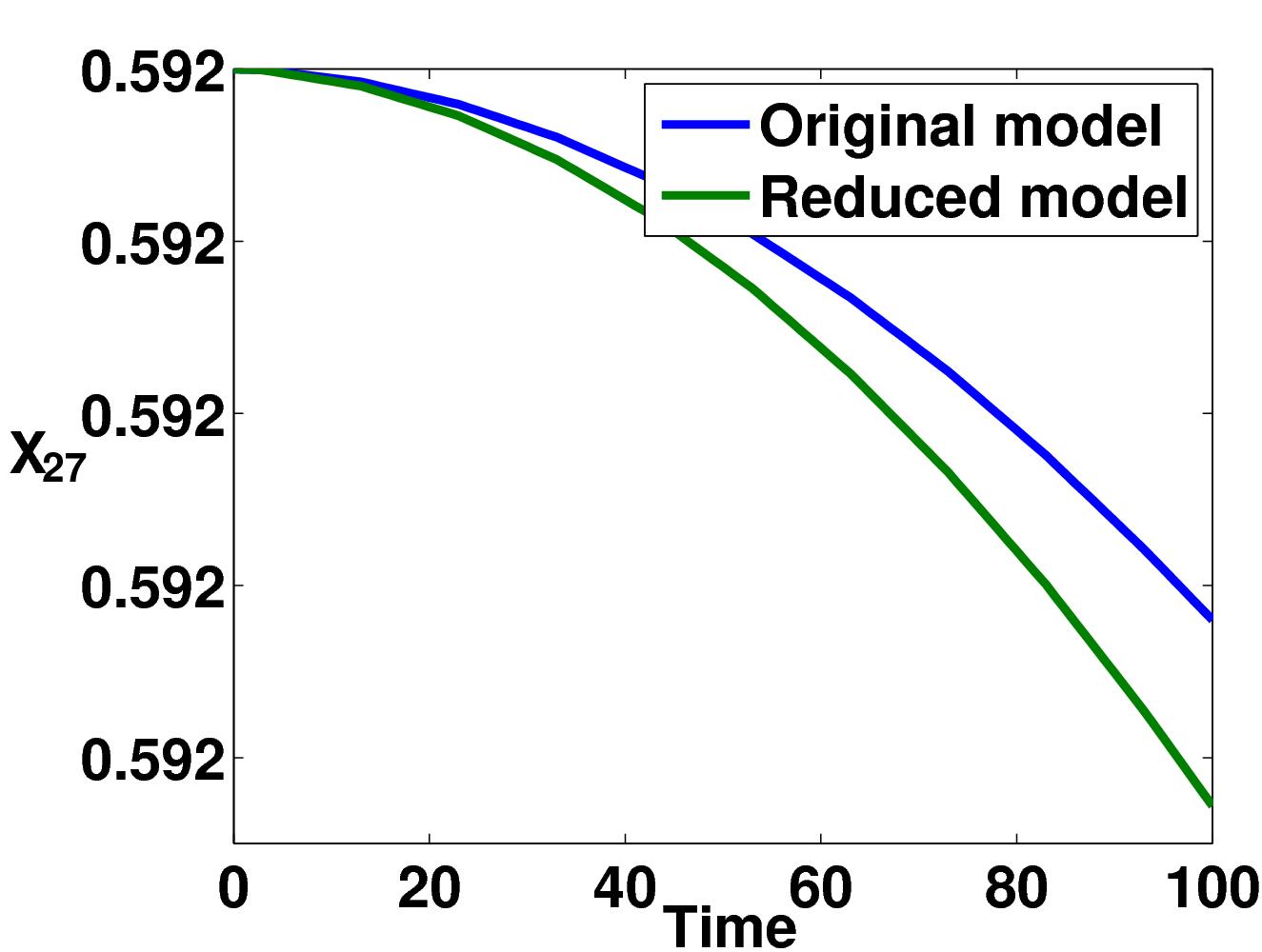}
		}
				     	\subfigure{%
			\includegraphics[width=0.2\textwidth]{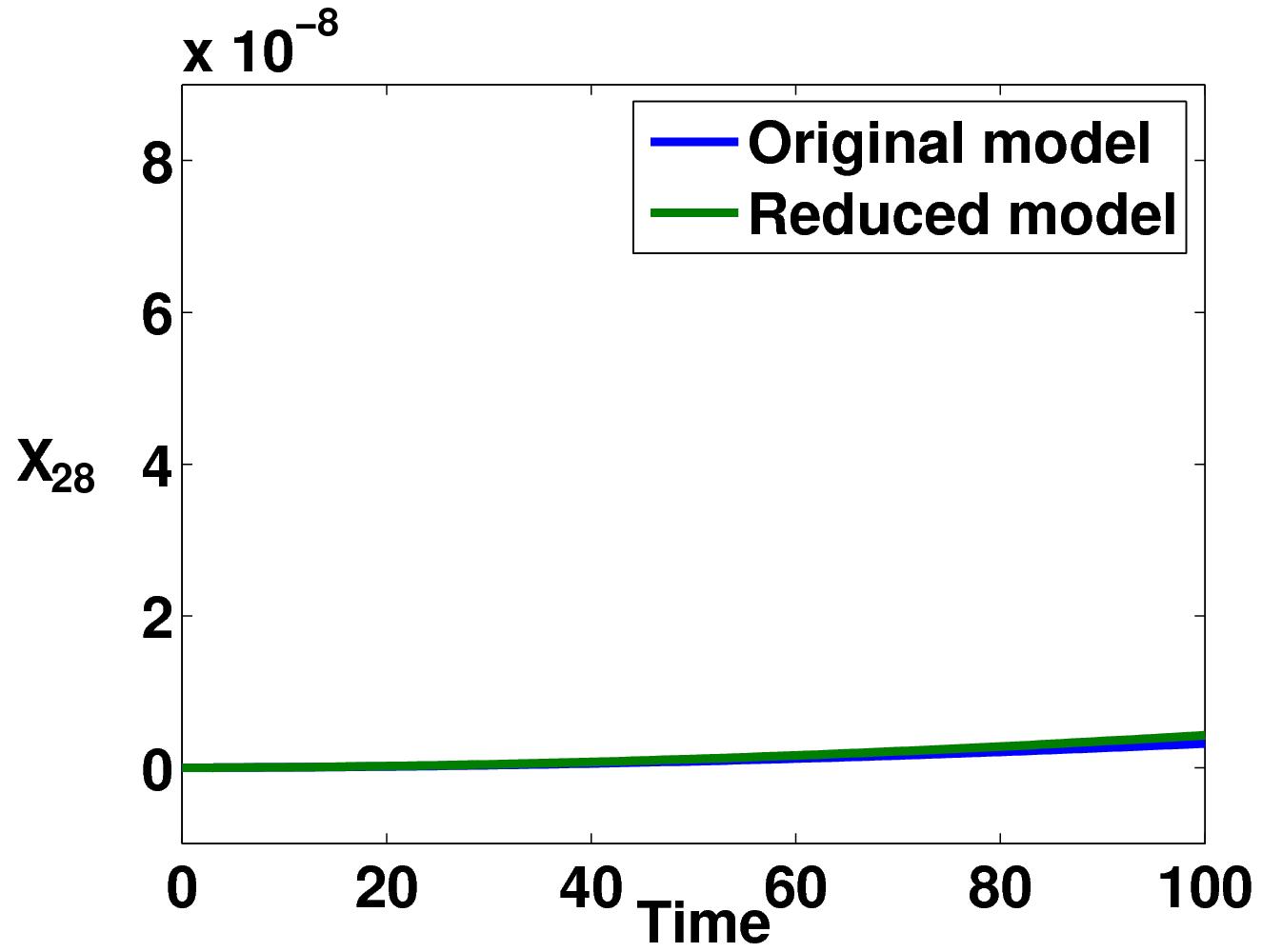}
		}
				     	\subfigure{%
			\includegraphics[width=0.2\textwidth]{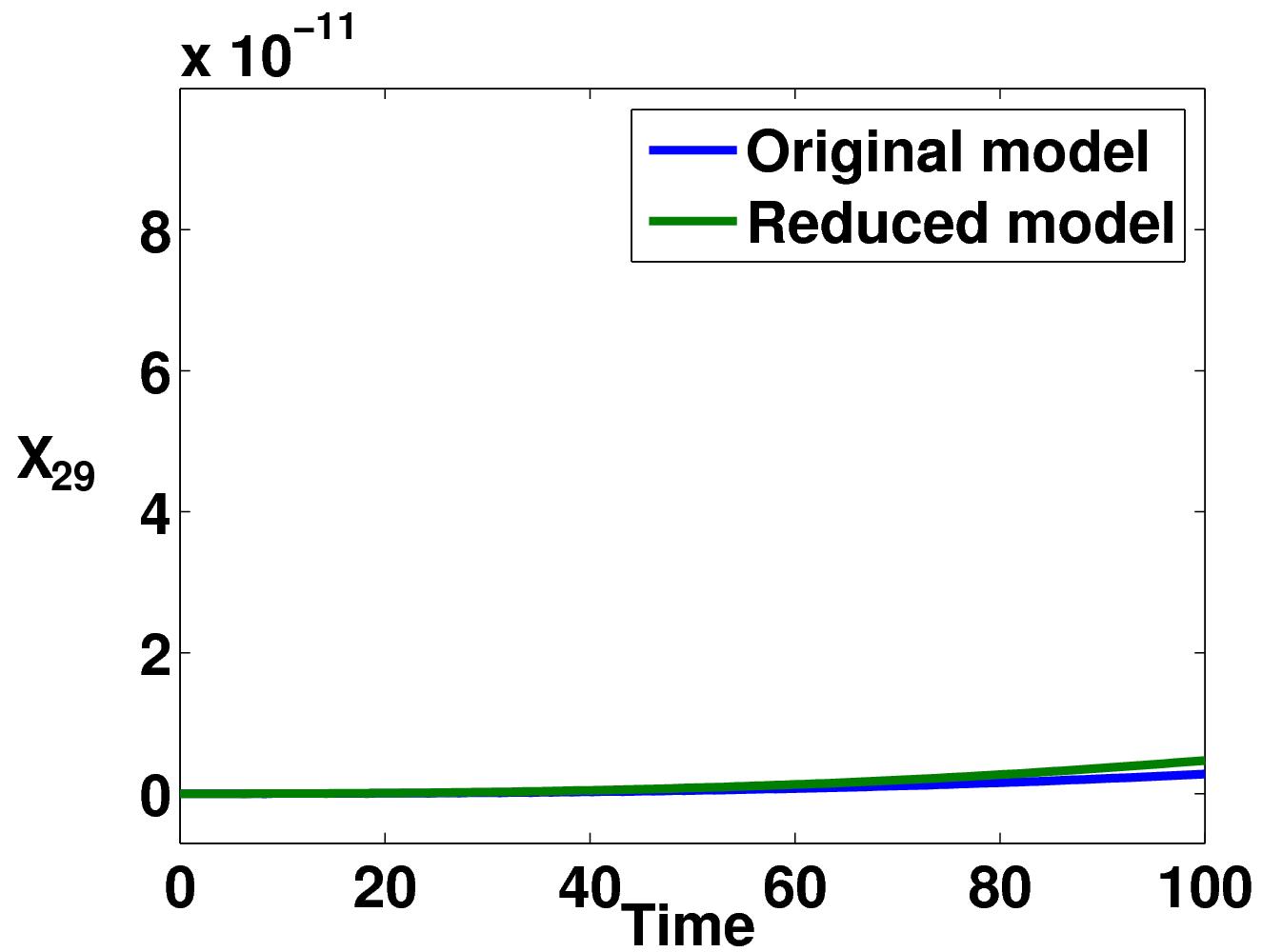}
		}	
	\end{center}   
	\caption {Numerical simulations of state variables in original and reduced system of NF-$\kappa$B signal transduction pathways; the blue lines are the original and the green lines are the reduced approximate solutions, with the time interval $[0,100]$ in computational simulations.}    	
	\label{fig:solution of inflammatory}
\end{figure}

\noindent We use some computational simulations in order to compared the original and reduced model. This is for state variables $ C_{i},\; i=1,2,...,29. $ in case 1; see Figure \eqref{fig:solution of inflammatory}.


\subsection{Results and Discussions}

According to the basic idea of lumping of species, we have proposed a technique based on lumping of parameters.
The proposed technique plays an important role in minimizing complex biochemical reaction networks. We have applied the approach for liner and non-linear examples of chemical reaction networks. The first example is a simple model of linear chemical reactions. The model is reduced from $5$ to $3$ parameters based on the suggested technique. The second model in this work is a complex cell signalling pathways. This is called NF-$\kappa$B signal transduction pathways. We use three different ways for selecting lumping parameters such as sub-interval selection, randomly selection and neighboring selection. Interestingly, the number of parameters of NF-$\kappa$B signal pathways is minimized from $37$ to ($20, 13, 9$, and $8$) in four different cases; see Table \eqref{table:lumping of NF-kB}. Results in case one show the fewer number of parameters with minimum error. Computational simulations are calculated using Matlab for initial parameters in Table \eqref{table of parameter of inflammatory} and initial state variables in Table \eqref{table of initial condition of inflammatory}. The blue lines represent the numerical solutions for the original systems and the green lines show the approximate solutions for the reduced system; see Figure \eqref{fig:solution of inflammatory}. It can be clearly seen that there are a good agreement between the original and reduced model solutions. The simplified model of NF-$\kappa$B signal transduction pathways helps to study the full model and describing the model dynamics. Some effective results are obtained based on the proposed technique. Firstly, the suggested approach here importantly plays in minimizing the number of parameters and in calculating analytical approximate solutions. Secondly, the reliability and accuracy of model reduction technique are usually computed by looking at the model reduction error. Another result is that selecting a set of parameters for lumping is easier than the classical techniques of lumping elements. Furthermore, the proposed method can further be developed and applied to a wide range of complex NF-$\kappa$B signal transduction pathway mechanisms and high dimensional cell signalling models in systems biology.

\include{chapter[5]}
\chapter{A Model Reduction Approach Based on Entropy Production and Lumping of Species} 

\section{Mathematical Formulations of Entropy Production}
The general equation of entropy production can be given bellow:
\newline

\begin{equation}     
\begin{array}{llll} 
d\mathcal{S}=d^{ex}\mathcal{S}+d^{in}\mathcal{S},
\end{array}\label{Ent1}
\end{equation} 
\noindent where $d^{ex}S$ and $d^{in}S$ are the change of the system entropy with regards to the interactions surrounding and inside the system, respectively.\\
\noindent Suppose that there is a model of chemical reactions consisting of $m$ species and $n$ reversible chemical reactions, and let the number of moles (particles) of the species presented by a vector $\mathcal{N}=\big(\mathcal{N}_{1},\mathcal{N}_{2},...,\mathcal{N}_{m}\big)$. The chemical reactions are given:      
\begin{equation}  
\begin{array}{llll}
 \mathlarger{\mathlarger{\sum\limits}}_{j=1}^m \alpha_{ij} \mathcal{N}_{j} \underset{k_{i}^{b}}{ \overset{k_{i}^{f}}{\rightleftharpoons}} \mathlarger{\mathlarger{\sum\limits}}_{j=1}^m \beta_{ij} \mathcal{N}_{j}, \quad i=1,2,...,n,
\end{array}\label{Ent2}
\end{equation} 
\noindent where $k_{i}^{f}$ and $k_{i}^{b}$ are the forward and backward reaction constants, respectively; $\alpha_{ij}$ and $\beta_{ij}$ are the stoichiometric coefficients of the \textit{j}th species in the \textit{i}th reaction for the reactants and products. The reaction rate of the \textit{i}th reaction is defined by the mass action law
\begin{equation}  
\begin{array}{llll}
v_{i}=v_{i}^{f}-v_{i}^{b}=k_{i}^{f} \mathlarger{\mathlarger{\prod\limits}}_{j=1}^m c_{j}^{\alpha_{ij}}(t) -k_{i}^{b}  \mathlarger{\mathlarger{\prod\limits}}_{j=1}^m c_{j}^{\beta_{ij}}(t),
\end{array}\label{Ent3}
\end{equation} 
\noindent where $c_{j}$ is the concentration of the \textit{j}th species ( i.e. $c_{j}=\mathcal{N}_{j}/V$, V is volume). 
\noindent A kinetic equation of the \textit{j}th species in a closed system can be written as a system of ordinary differential equations:        
\begin{equation}  
\begin{array}{llll}
\dfrac{dc_{j}}{dt}= \mathlarger{\mathlarger{\sum\limits}}_{i=1}^{n_{1}}\gamma_{ij}v_{i}, \quad j=1,2,...,m, \quad 1\leq n_{1} \leq n.
\end{array}\label{Ent7}
\end{equation} 
\noindent Equation (\ref{Ent7}) can be rewritten in vector form using the reactor volume $V$
\begin{equation}  
\begin{array}{llll}
\dfrac{d\mathcal{N}}{dt}=V\dfrac{d\mathcal{C}}{dt}=f(\mathcal{N}),
\end{array}\label{Ent8}
\end{equation}  
\noindent where $\mathcal{C}$ is a vector of concentration of species.\\
In chemistry, a quantity that measures the amount of substance that is being changed in an equilibrium reaction is called the extent of reaction. Most of the time is denoted by the Greek letter $\xi$. There is also a relation between the change of the \textit{j}th species in the \textit{i}th reaction and the change in the extent of the reaction $\xi_{i}$ as follows:
\begin{equation}  
\begin{array}{llll}
d\mathcal{N}_{j}=\mathlarger{\mathlarger{\sum\limits}}_{i=1}^{n}\gamma_{ij}d\xi_{i},
\end{array}\label{Ent9}
\end{equation}   
\noindent where 
 \begin{equation}  
\begin{array}{llll}
\dfrac{d\xi_{i}}{dt}=V(v_{i}^{f}-v_{i}^{b}).
\end{array}\label{Ent10}
\end{equation}
\noindent The equation of de Donder affinity of the \textit{i}th reaction is given by
 \begin{equation}  
\begin{array}{llll}
\psi_{i}=- \mathlarger{\mathlarger{\sum\limits}}_{j=1}^m \gamma_{ij}\mu_{j}.
\end{array}\label{Ent13}
\end{equation}
\newline The same quantity (\ref{Ent13}) can be changed from the principle of detailed balance, and it becomes
 \begin{equation}  
\begin{array}{llll} 
\psi_{i}=R_{c}T \ln\bigg(\dfrac{v_{i}^{f}}{v_{i}^{^{b}}} \bigg).
\end{array}\label{Ent14}
\end{equation} 
\noindent The formula of the changes in the number of moles of the \textit{j}th species in a system is defined by
 \begin{equation}  
\begin{array}{llll}
d\mathcal{N}_{j}=d^{in}\mathcal{N}_{j}+d^{ex}\mathcal{N}_{j},
\end{array}\label{Ent15}
\end{equation}
\noindent where $d^{in}\mathcal{N}_{j}$ and $d^{ex}\mathcal{N}_{j}$ are the change in the number of moles with irreversible reactions and the system exterior, respectively.\\
For open systems, the change in system entropy with the system exterior and chemical reactions are given:
 \begin{equation}  
\begin{array}{llll}
d^{ex}\mathcal{S}=\dfrac{1}{T}\Big(dU+P dV-\mathlarger{\mathlarger{\sum\limits}}_{j=1}^m \mu_{j} d^{ex}\mathcal{N}_{j} \Big),
\end{array}\label{Ent16}
\end{equation}
 \begin{equation}  
\begin{array}{llll}
d^{in}\mathcal{S}=- \dfrac{1}{T}\mathlarger{\mathlarger{\sum\limits}}_{j=1}^m \mu_{j} d^{in}\mathcal{N}_{j}.
\end{array}\label{Ent17}
\end{equation}
For closed systems, the system entropy with its surroundings becomes zero, i.e. $d^{ex}\mathcal{S}=0$ and therefore $d\mathcal{S}=d^{in}\mathcal{S}$, and then the entropy production for a chemical reaction is of the form 
 \begin{equation}  
\begin{array}{llll}
\dfrac{d^{in}\mathcal{S}}{dt}=-\dfrac{1}{T} \mathlarger{\mathlarger{\sum\limits}}_{j=1}^m \mu_{j} \dfrac{d\mathcal{N}_{j}}{dt},
\end{array}\label{Ent18}
\end{equation}
where
 \begin{equation}  
\begin{array}{llll}
\dfrac{d\mathcal{N}_{j}}{dt}= \mathlarger{\mathlarger{\sum\limits}}_{i=1}^n \gamma_{ij}\dfrac{d\xi_{i}}{dt}.
\end{array}\label{Ent19}
\end{equation}
Therefore, the equation for the entropy production is given as:
\begin{equation}  
\begin{array}{llll}
\dfrac{d^{in}\mathcal{S}}{dt}=-\dfrac{1}{T} \mathlarger{\mathlarger{\sum\limits}}_{j=1}^m \mu_{j}  \mathlarger{\mathlarger{\sum\limits}}_{i=1}^n \gamma_{ij}\dfrac{d\xi_{i}}{dt}.
\end{array}\label{Ent20}
\end{equation}
By using equation (\ref{Ent13}), the entropy production becomes
\begin{equation}  
\begin{array}{llll}
\dfrac{d^{in}\mathcal{S}}{dt}=\dfrac{1}{T} \mathlarger{\mathlarger{\sum\limits}}_{i=1}^n \psi_{i} \dfrac{d\xi_{i}}{dt}.
\end{array}\label{Ent21}
\end{equation} 
By using equations (\ref{Ent10}) and (\ref{Ent14}), the total entropy production per unit volume is given as a semi-definite function
\begin{equation}  
\begin{array}{llll}
\dfrac{d^{in}\mathcal{S}}{dt}=R_{c} \mathlarger{\mathlarger{\sum\limits}}_{i=1}^n E^{p}_{i}(t) \geq 0 \quad \forall t\in I\subset \mathbb{R},
\end{array}\label{Ent22}   
\end{equation} 
where
\begin{equation}  
\begin{array}{llll}
E^{p}_{i}(t)=(v_{i}^{f}-v_{i}^{b})\ln\bigg(\dfrac{v_{i}^{f}}{v_{i}^{b}}\bigg), \quad  i=1,2,...,n.
\end{array}\label{Ent22a}   
\end{equation}
In a result, there is a strong relationship between the entropy production of each reaction and the total entropy production of the system at time $t$,
\begin{equation}  
\begin{array}{llll}
\mathcal{R}_{i}^{c}(t)=\dfrac{E^{p}_{i}(t)}{ \mathlarger{\mathlarger{\sum\limits}}_{i=1}^n E^{p}_{i}(t)} , \quad \forall t \in I\subset \mathbb{R}, \quad i=1,2,...,n.
\end{array}\label{Ent23}
\end{equation} 
\section{Relative Contribution and Lumping of Species Algorithm}
\label{A52}
In this section, we will discussing an official way to model reduction for biochemical reaction networks, which is based on the entropy production analysis and depended on the idea of the relative contribution of entropy production of each reaction to the total entropy production with lumping of isolated species in a model. So we suggested some steps of reduction for recognizing the non-important reactions .Where we remove non-important reactions there may be some isolated species then we try to lump isolated species with one of their neighbors. It can be a applicable way to deal with many issues including identifying critical model elements and simplifying complex biochemical reversible reactions to smaller size. The term $v_{i}^{f}/v_{i}^{b}$ for $i=1,2,...,n$ in the logarithm of (\ref{Ent22a}) must be strictly positive. The proposed algorithm is stopped and can not be used in two cases. The first case is related negative concentrations of chemical reaction rates when we have some low negative concentrations in numerical simulations (i.e. $v_{i}^{f}<0$ and/or $v_{i}^{b}<0$). In this case the negative reaction rates are assumed to be a very small positive number. The second case is about the irreversible chemical reactions where the reversible (backward) reaction rate is zero ($v_{i}^{b}=0$). In this case, the reaction rate $v_{i}^{b}$ is assumed to be a very small positive number (chemically insignificant e.g. $10^{-30}$). The proposed steps are given below: 
\begin{enumerate}[noitemsep,nolistsep]
\item Calculate the numerical solution of concentration species for the original model of biochemical reaction networks (this can be calculated in numerical simulations using Matlab).
\item Calculate the maximum value of the suggested function (\ref{Ent23}) for each reaction i.e. 
$\gamma_{i}=$ Max$\Big\lbrace \mathcal{R}^{c}_{i}(t), \forall t\Big\rbrace$ , $i=1,2,...,n.$ 
\item Identify the non--important reactions (i.e. the least contributing reactions to the total entropy production); this can be identified by choosing the \textit{k}th reaction such that $\gamma_{k}^{min}=$ Min$\Big\lbrace \gamma_{1},\gamma_{2},...,\gamma_{n} \Big\rbrace$,  \quad $1\leq k \leq n.$   
\item Eliminate the non--important reaction (step 3) from the kinetic equations of the system when the reaction does not importantly change the model dynamics during the computational simulations. 

\item Lumping a set of isolated species with their neighbors, for instance if we have a model network with three species as bellow:
\begin{equation}  
\begin{array}{llll}
\quad A_{i}{ \overset{r_{1}}\longleftrightarrow} A_{k} \\
\quad \; \updownarrow r_{2} \\
 \quad \;  A_{j}
\end{array}
\end{equation}
and the reaction $r_{2}$ is non-important then $A_{j}$ will be isolated. We assume that the isolated species $A_{j}$ can be lumped either $A_{i}$ or $A_{k}$ such as
\newline $A^{*}_{i}=A_{i}+A_{j}$ or $A^{*}_{k}=A_{k}+A_{j}$  .

\item Calculate the difference between the full and reduced model using the function of deviation(\ref{SD}) at each reduction stages.   
\item If the value of deviation is within allowable limits then repeat the above steps (steps 2--6) for the new reduced model.  
\end{enumerate}
\noindent The above steps of the suggested algorithm of model reduction can be presented as a flowchart of model reduction as follows:
  \begin{center}     
    \resizebox{1 \linewidth}{!}{%
      \begin{tikzpicture}[node distance = 2.5cm, auto]
        \node[block,text width =7cm]  (A1){Calculate the numerical solutions of the original model};
        \node[block, below of=A1,text width =7cm]   (A2){ Compute $\gamma_{i}=$Max$\Big\lbrace \mathcal{R}_{i}^{c}(t),  \forall t \Big\rbrace$, $i$=1,2,...,n};
        \node[block, below of=A2](A3){Identify non--important \textit{k}th reaction s.t. \\ $\gamma_{k}^{min}=$ Min$\Big\lbrace\gamma_{1},\gamma_{2},...,\gamma_{n}\Big\rbrace$, \quad $1\leq k \leq n$};
        \node[block, below of=A3,text width =7cm]   (A4){Eliminate the non--important k\textit{th} reaction from the model};

        \node[block, below of=A4,text width =7cm ]   (L){Lumping isolated species and their neighbours with minimum error};        
        
        \node[block, below of=L,text width =4cm]   (A7){Calculate the value of deviation ($\mathcal{F}^{\mathcal{D}}$)};
       \node[decision, right  of=A7, node distance = 6cm ] (A8){$\mathcal{F}^{\mathcal{D}}\leqslant \rho\%$, $\rho \in(0,N_{0}]$, $N_{0}\in \mathbb{R}^{+}$};
        \node[block, below of=A8, node distance =4.5cm,text width =6cm ]   (A9){Reduced model with minimum parameters and variables};
      
        \path[line,line width=0.05cm] (A1) --          (A2); 
        \path[line,line width=0.05cm] (A2) --          (A3);
        \path[line,line width=0.05cm] (A3) --          (A4);
        \path[line,line width=0.05cm] (A4) --          (L);
        \path[line,line width=0.05cm] (L)  --          (A7);
        \path[line,line width=0.05cm] (A7) --          (A8);
        \path[line,line width=0.05cm] (A8) -- node{No}(A9);
        \path[line,line width=0.05cm] (A8) |- node{Yes}(A2);

      \end{tikzpicture}%
    }%
  \end{center}   
  \noindent \textbf{The flowchart of the relative contribution and lumping species algorithm of model reduction.}\\

\section{Applications}
The developed technique has a good step forward in model reductions. Particularly, this proposed algorithm has been used for model reductions in reversible chemical reactions for cell signaling pathways. We have applied this model reduction tool in Elongation Factors EF--Tu and EF-Ts signaling pathways and Dihydrofolate Reductase (DHFR) pathways in order to reduce the number of state variables and parameters.

\subsection{Elongation Factors EF--Tu and EF-Ts signalling Pathways}
\label{A53} 

 In this section, we apply the suggested algorithm to a model of biochemical reactions. The model was studied earlier in (Manchester, 2004, Schummer et al., 2007, Weiser et al., 2011). The model is for elongation factors EF--Tu and EF--Ts in cell signaling. There are 9 species and 14 reactions of the system. The species of the model are introduced in Table \eqref{table of variables of Elognation}.

\FloatBarrier 
\begin{table}
\caption{The set of state variables for Elongation Factors EF--Tu and EF-Ts signalling pathways.} 
\begin{tabular}{|c|c|p{11cm}|}
 \hline 
 No. & Species & Descriptions \\ 
 \hline 
 1 & GDP & guanosine diphosphate \\ 
 \hline 
 2 & GTP & guanosine triphosphate \\ 
 \hline 
 3 & Tu(EF$-$Tu) & elongation factor thermo unstable \\ 
 \hline 
 4 & Ts(EF-Ts) & elongation factor thermo stable \\ 
 \hline
 5 & Tu:GDP & elongation factor thermo unstable--guanosine diphosphate complex \\ 
 \hline 
 6 & Tu:GDP:Ts & elongation factor thermo unstable--guanosine diphosphate elongation factor thermo stable complex \\ 
 \hline 
 7 & Tu:Ts & elongation factor thermo unstable--elongation factor thermo stable complex \\ 
 \hline 
 8 & Tu:GTP & elongation factor thermo unstable--guanosine triphosphate complex \\ 
 \hline 
 9 & Tu:GTP:Ts & elongation factor thermo unstable--guanosine triphosphate elongation factor thermo stable complex \\ 
 \hline 
 \end{tabular}  
 \label{table of variables of Elognation}
 \end{table}
\FloatBarrier 
 
All reactions in the model are considered to be reversible: 
\begin{equation}  
\begin{array}{llll}
GDP+Tu \underset{k_{1}^{b}}{ \overset{k_{1}^{f}}{\rightleftharpoons}} Tu:GDP, \quad
Tu+Ts \underset{k_{2}^{b}}{ \overset{k_{2}^{f}}{\rightleftharpoons}} Tu:Ts , \\
Tu:GDP+Ts \underset{k_{3}^{b}}{ \overset{k_{3}^{f}}{\rightleftharpoons}} Tu:GDP:Ts , \quad 
Tu:Ts+GDP \underset{k_{4}^{b}}{ \overset{k_{4}^{f}}{\rightleftharpoons}} Tu:GDP:Ts, \\
Tu+GTP \underset{k_{5}^{b}}{ \overset{k_{5}^{f}}{\rightleftharpoons}} Tu:GTP ,\quad
Tu:GTP+Ts \underset{k_{6}^{b}}{ \overset{k_{6}^{f}}{\rightleftharpoons}} Tu:GTP:Ts, \\
Tu:Ts+GTP \underset{k_{7}^{b}}{ \overset{k_{7}^{f}}{\rightleftharpoons}} Tu:GTP:Ts,
\end{array}\label{Ent24}
\end{equation}  
where $k_{i}^{f}$ and $k_{i}^{b}$ for $i=1,2,...,7$ are the forward and backward reaction constants, respectively.

\noindent The chemical reactions (\ref{Ent24}) can be expressed as a system of ordinary differential equations:
\begin{equation}  
\begin{array}{llll}
\dfrac{d[GDP]}{dt}=-v_{1}-v_{4} , \quad  
\dfrac{d[Tu]}{dt}=-v_{1}-v_{2}-v_{5} , \\
\dfrac{d[Tu:GDP]}{dt}=v_{1}-v_{3}, \quad
\dfrac{d[Ts]}{dt}=-v_{2}-v_{3}-v_{6}, \\
\dfrac{d[Tu:Ts]}{dt}=v_{2}-v_{4}-v_{7}, \quad 
\dfrac{d[Tu:GDP:Ts]}{dt}=v_{3}+v_{4}, \\
\dfrac{d[GTP]}{dt}=-v_{5}-v_{7}, \quad 
\dfrac{d[Tu:GTP]}{dt}=v_{5}-v_{6},\\  
\dfrac{d[Tu:GTP:Ts]}{dt}=v_{6}+v_{7}, 
\end{array}\label{Ent25}
\end{equation} 
where $v_{1}=k_{1}^{f}[GDP](t)[Tu](t)-k_{1}^{b}[Tu:GDP](t),$ 
$v_{2}=k_{2}^{f}[Tu](t)[Ts](t)\\-k_{2}^{b}[Tu:Ts](t),$ 
$v_{3}=k_{3}^{f}[Tu:GDP](t)[Ts](t)-k_{3}^{b}[Tu:GDP:Ts](t),$ \\
$v_{4}=k_{4}^{f}[Tu:Ts](t)[GDP](t)-k_{4}^{b}[Tu:GDP:Ts](t),$ 
$v_{5}=k_{5}^{f}[Tu](t)[GTP](t)\\ -k_{5}^{b}[Tu:GTP](t),$ 
$v_{6}=k_{6}^{f}[Tu:GTP](t)[Ts](t)-k_{6}^{b}[Tu:GTP:Ts](t),$ \\
$v_{7}=k_{7}^{f}[Tu:Ts](t)[GTP](t)-k_{7}^{b}[Tu:GTP:Ts](t).$ 

And there are a set of data for state variables and parameters; see Tables \eqref{table of stationary values of Elognation} and
\eqref{table of parameter values of Elognation}.

\FloatBarrier 
\begin{table}
\caption{Stationary values of state variables for Elongation Factors EF--Tu and EF-Ts signalling pathways.} 
\begin{center}
\begin{tabular}{|c|c|c|}
 \hline 
 No. & Species & Stationary values \\ 
 \hline 
 1 & GDP & 0.2 \\ 
 \hline 
 2 & GTP & 0.6 \\ 
 \hline 
 3 & Tu(EF-Tu) & 0.3 \\ 
 \hline 
 4 & Ts(EF-Ts) & 0.8 \\ 
 \hline
 5 & Tu:GDP & 0\\ 
 \hline 
 6 & Tu:GDP:Ts & 0\\ 
 \hline 
 7 & Tu:Ts & 0\\ 
 \hline 
 8 & Tu:GTP &0\\ 
 \hline 
 9 & Tu:GTP:Ts & 0\\ 
 \hline 
 \end{tabular}  
 \end{center}
 \label{table of stationary values of Elognation}
 \end{table}
\FloatBarrier 

\FloatBarrier 
\begin{table}
\caption{Summary of parameter values for Elongation Factors EF--Tu and EF-Ts signalling pathways.}
\begin{center}
\begin{tabular}{|c|c|c|}
\hline 
No. & Parameters & Values \\ 
\hline 
1 & $k^f_{1}$ & 2 \\ 
\hline 
2 & $k^b_{1}$ & 0.002 \\ 
\hline 
3 & $k^f_{2}$ & 10 \\ 
\hline 
4 & $k^b_{2}$ & 0.03 \\ 
\hline 
5 & $k^f_{3}$ & 60 \\ 
\hline 
6 & $k^b_{3}$ & 350 \\ 
\hline 
7 & $k^f_{4}$ & 14 \\ 
\hline 
8 & $k^b_{4}$ & 125 \\ 
\hline 
9 & $k^f_{5}$ & 0.5 \\ 
\hline 
10 & $k^b_{5}$ & 0.03 \\ 
\hline 
11 & $k^f_{6}$ & 30 \\ 
\hline 
12 & $k^b_{6}$ & 60 \\ 
\hline 
13 & $k^f_{7}$ & 6 \\ 
\hline 
14 & $k^b_{7}$ & 85 \\ 
\hline 
\end{tabular} 
\end{center}
\label{table of parameter values of Elognation}
\end{table}
\FloatBarrier 
\subsection{Results and Discussions}   
\label{A54}
\begin{figure}[H] 
     \begin{center}
        \subfigure{%
            \label{fig:third}  
            \includegraphics[width=0.3\textwidth]{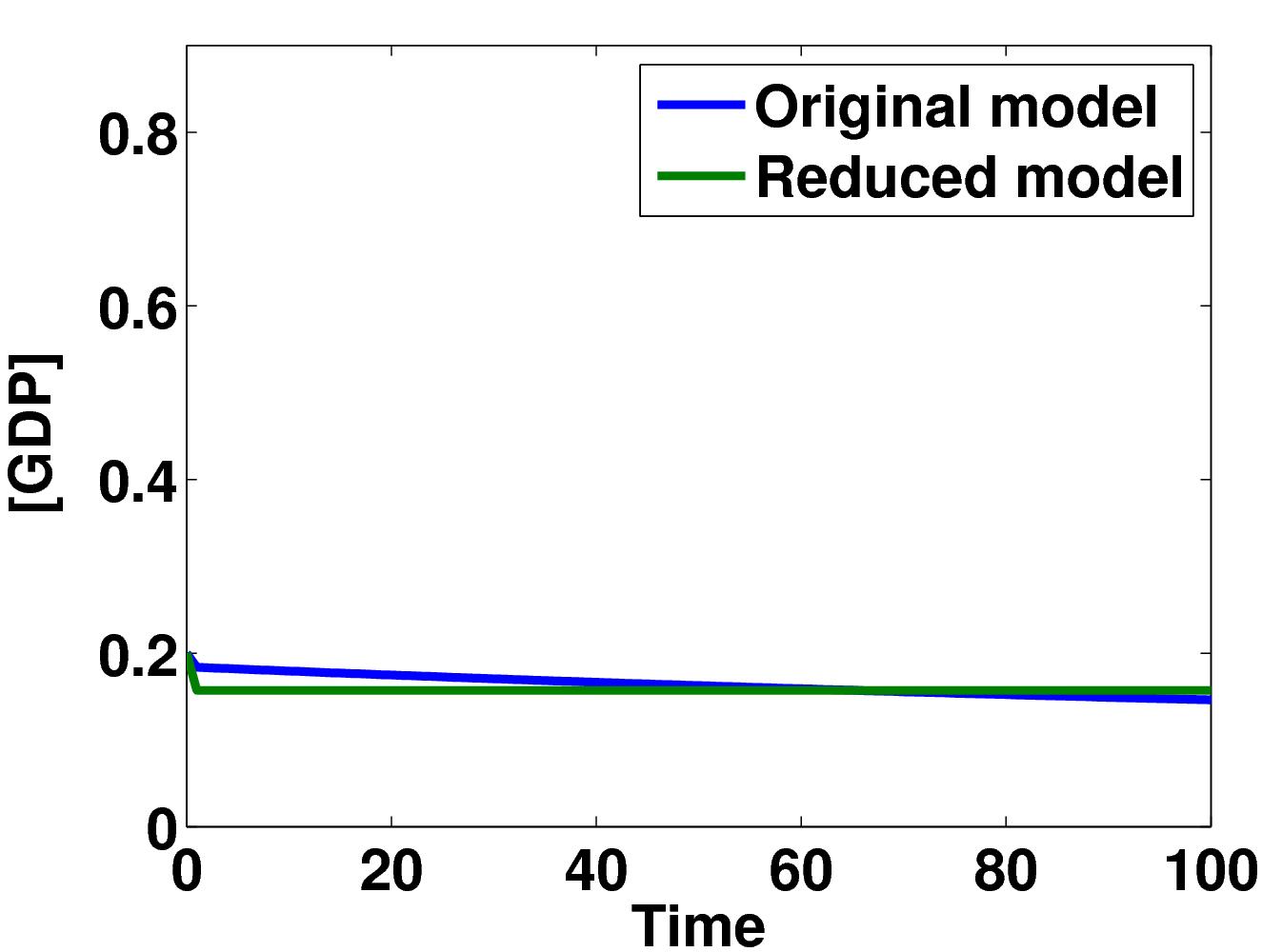}
        }%
                 \subfigure{%
            \label{fig:third}
            \includegraphics[width=0.3\textwidth]{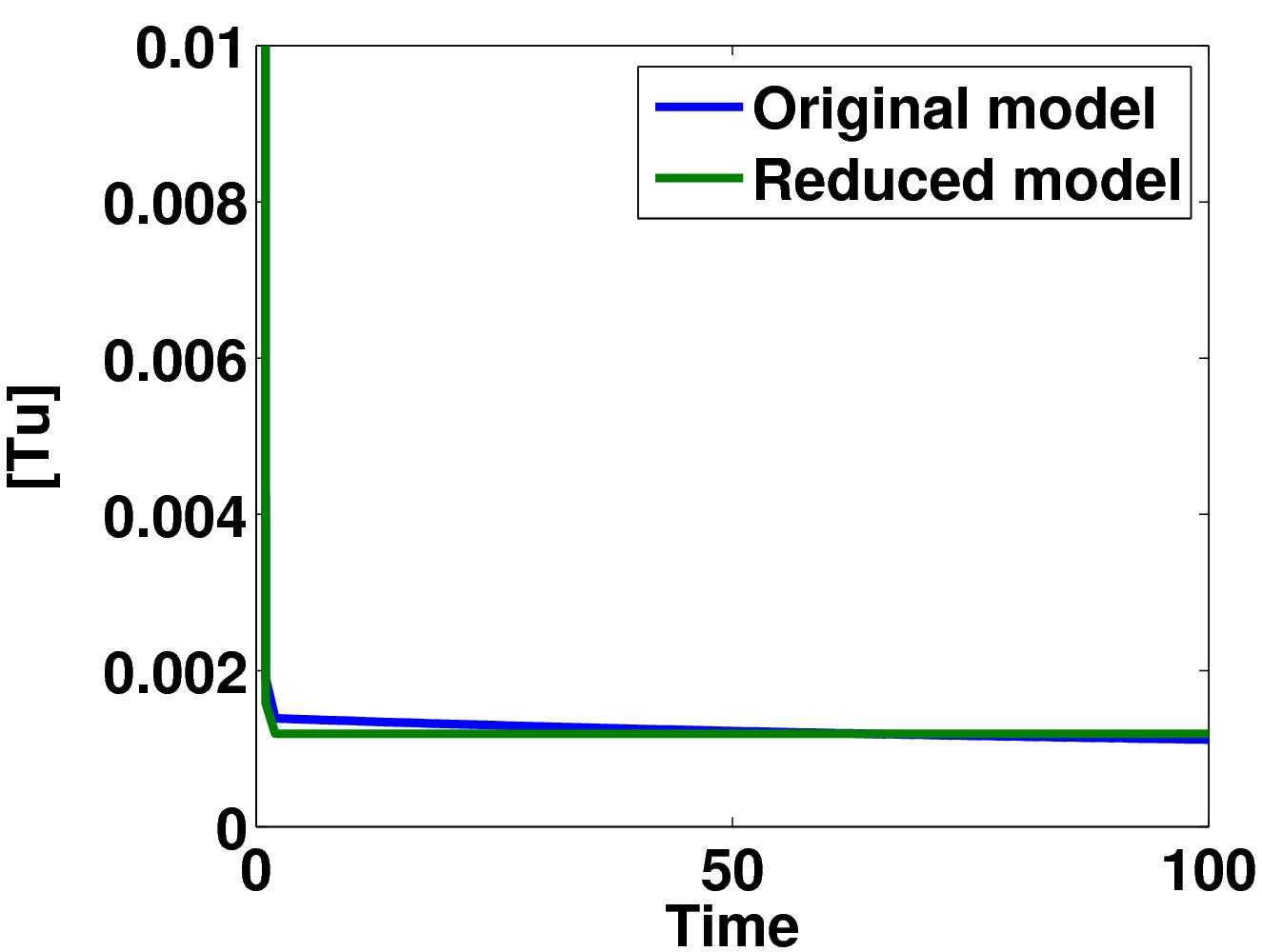}  
        }%
            \subfigure{%
            \label{fig:third}
            \includegraphics[width=0.3\textwidth]{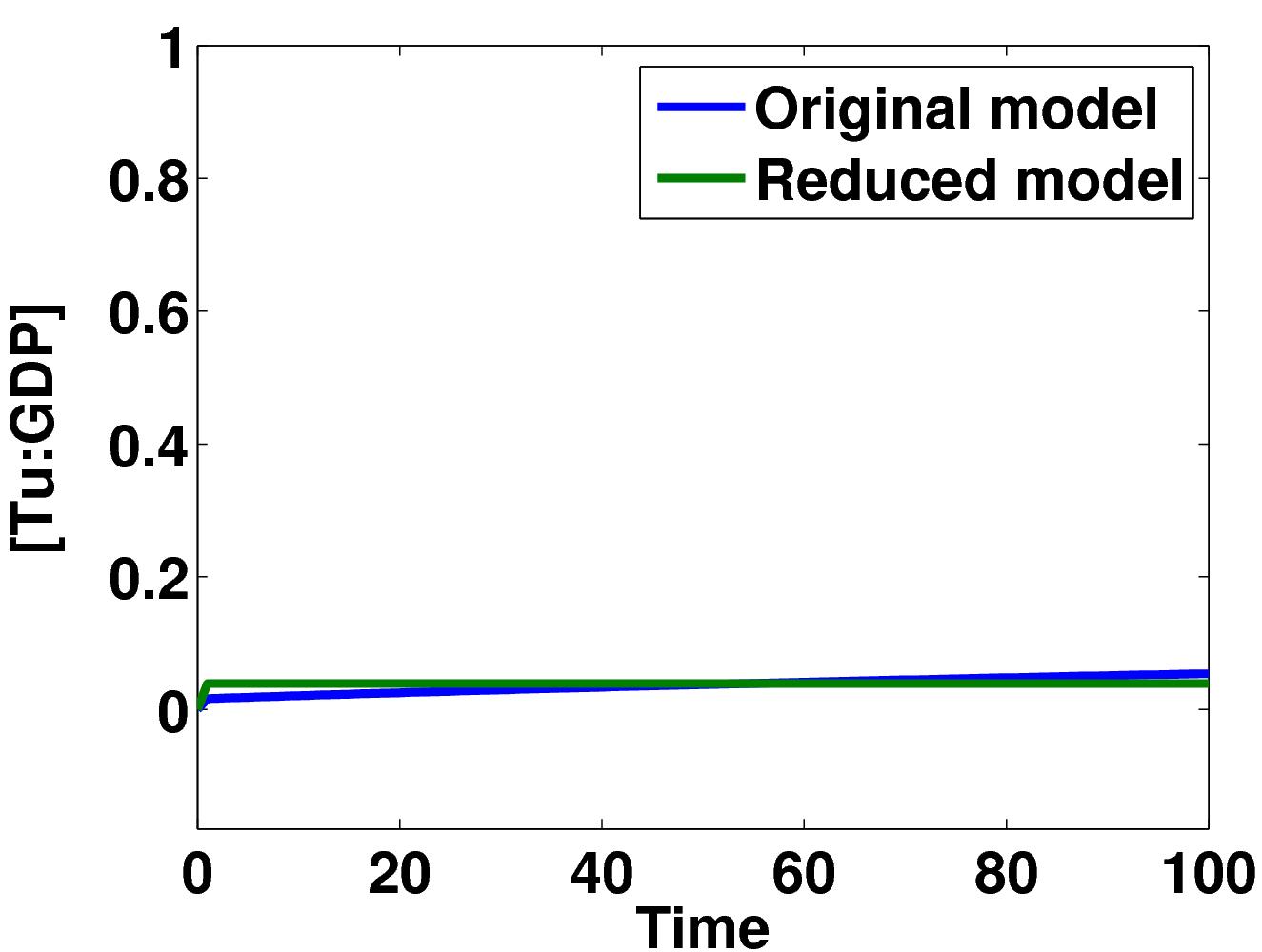}  
        }\\%
            \subfigure{%
            \label{fig:third}
            \includegraphics[width=0.3\textwidth]{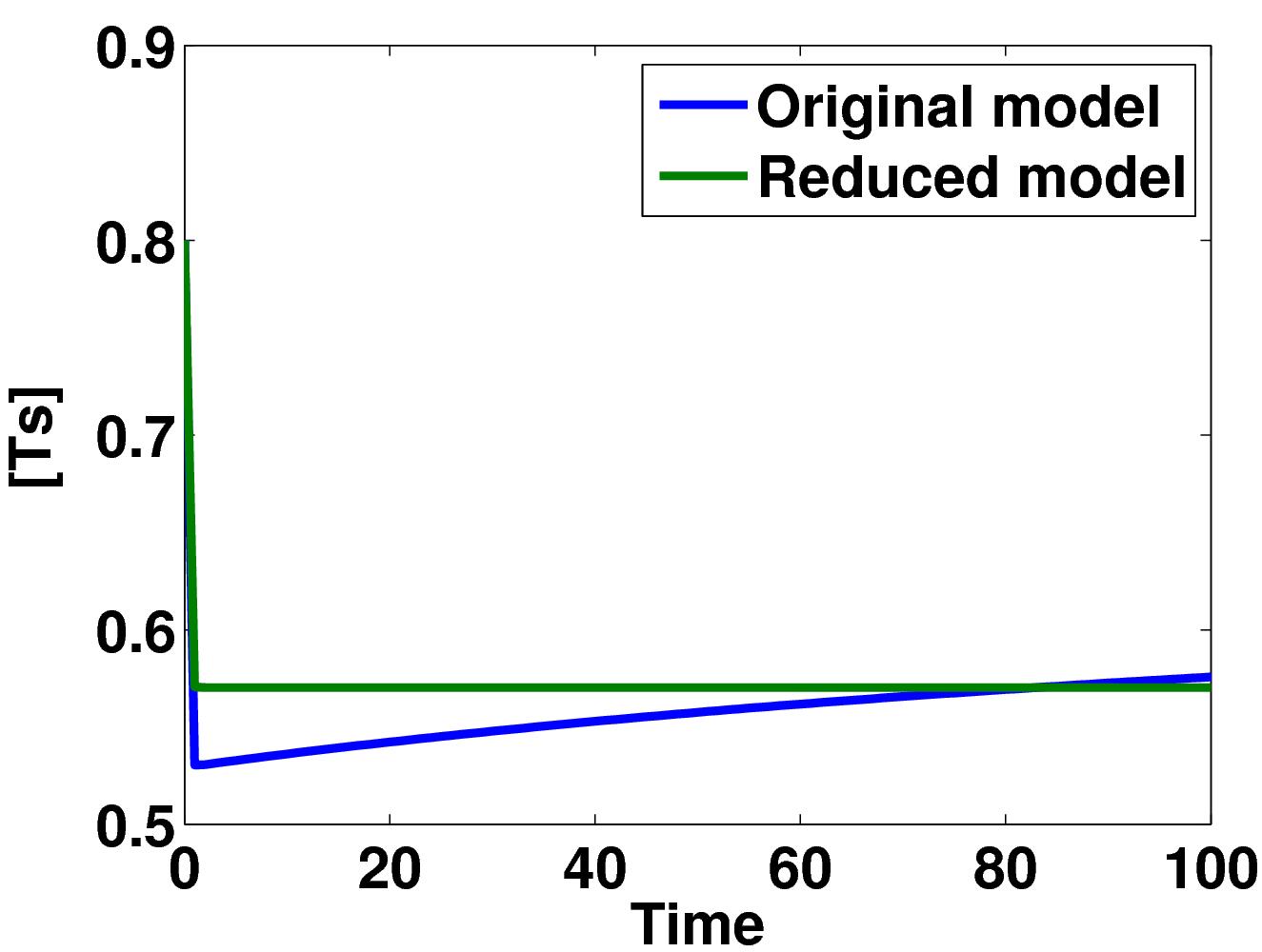}  
        }%
            \subfigure{%
            \label{fig:third}
            \includegraphics[width=0.3\textwidth]{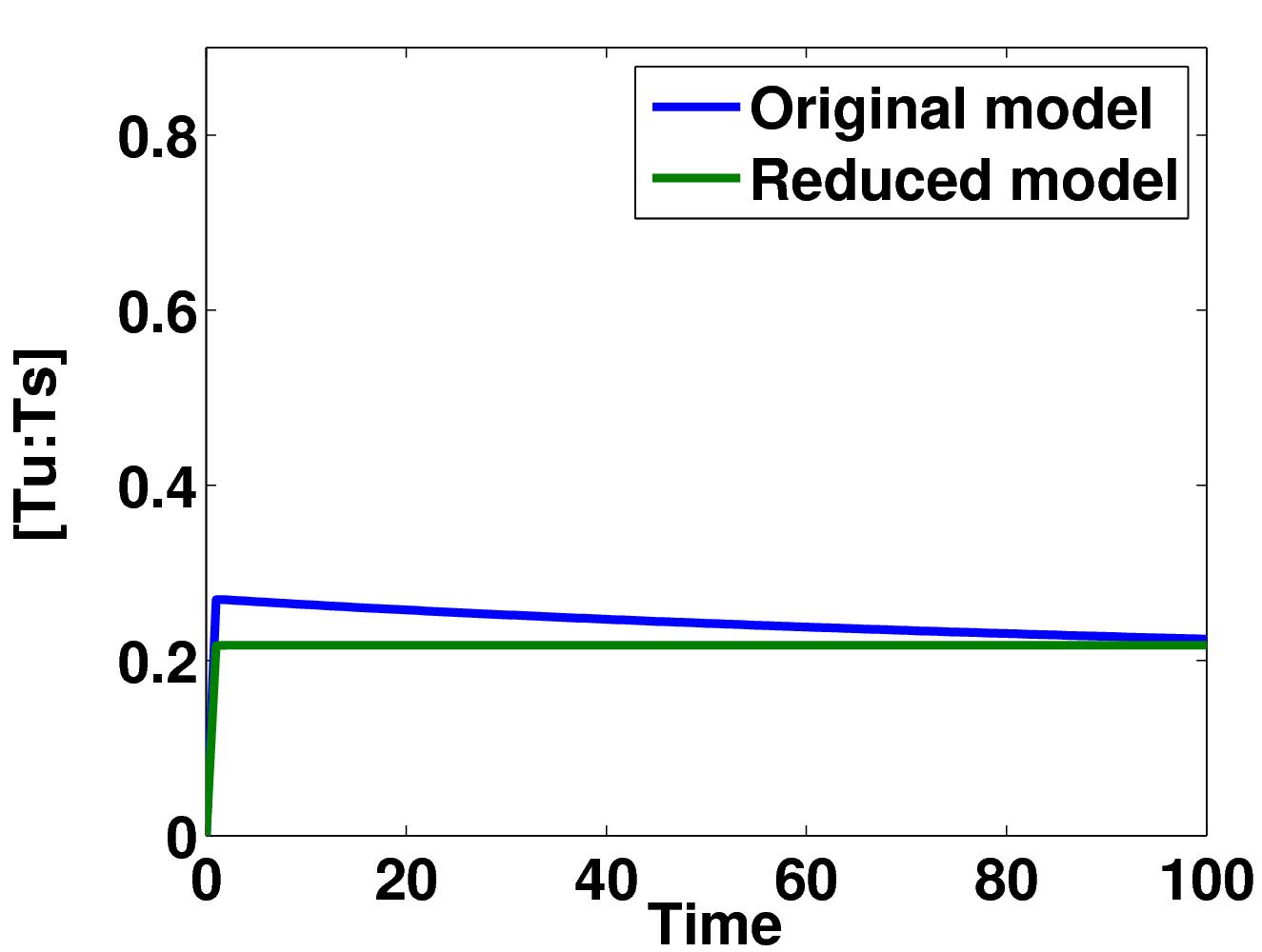}  
        }%
            \subfigure{%
            \label{fig:third}
            \includegraphics[width=0.3\textwidth]{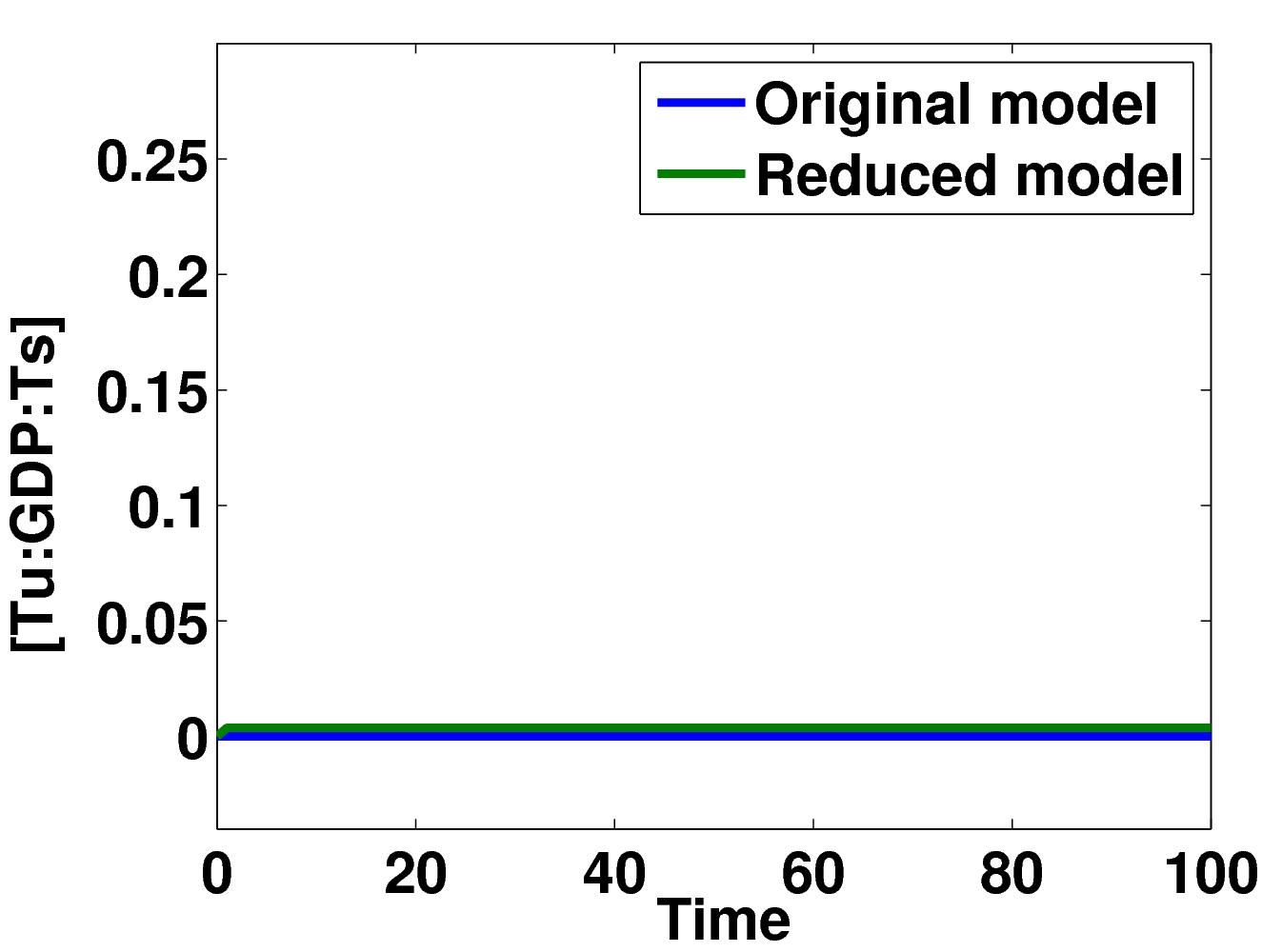}  
        }\\%
            \subfigure{%
            \label{fig:third}
            \includegraphics[width=0.3\textwidth]{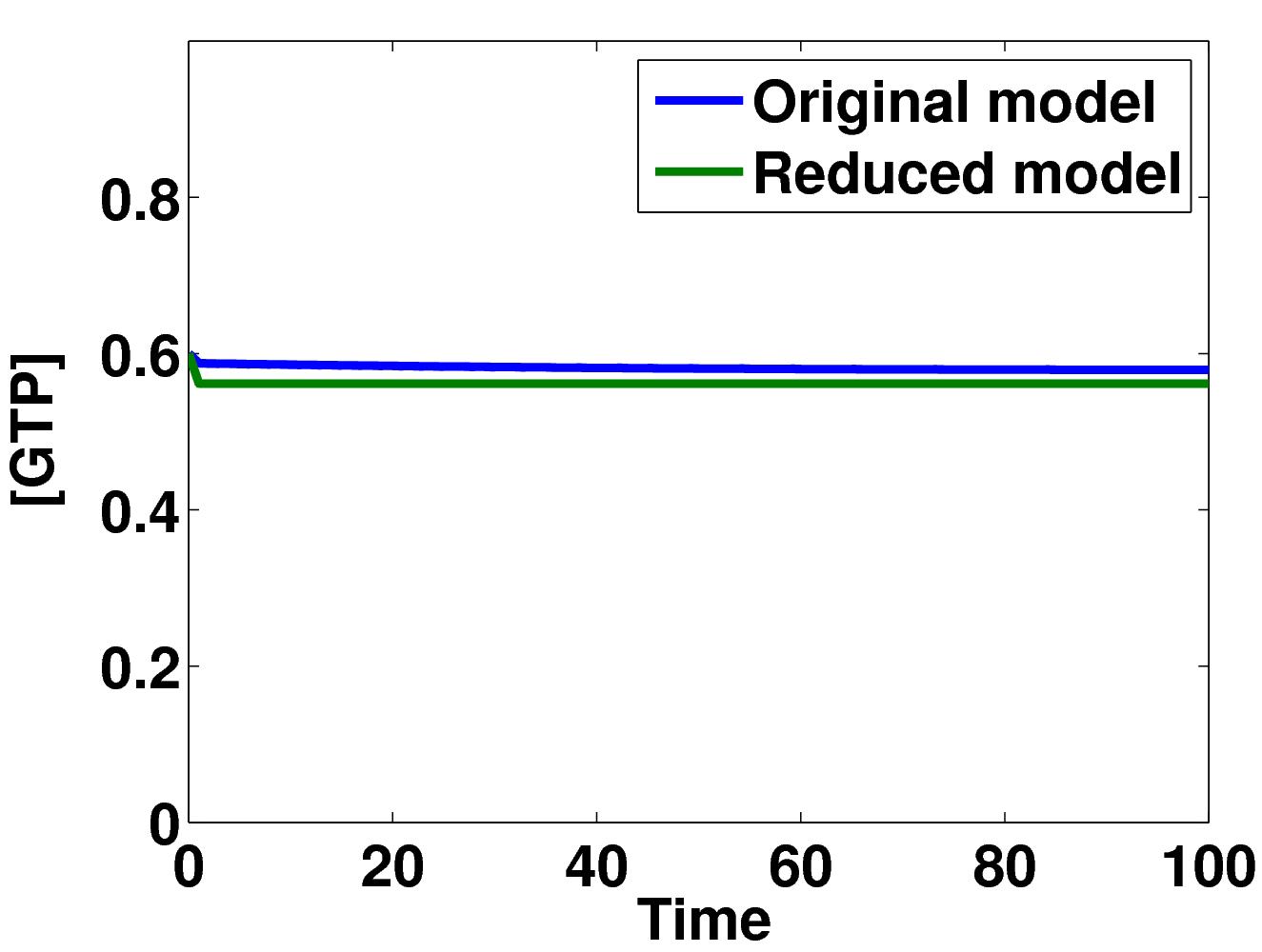}  
        }%
            \subfigure{%
            \label{fig:third}
            \includegraphics[width=0.3\textwidth]{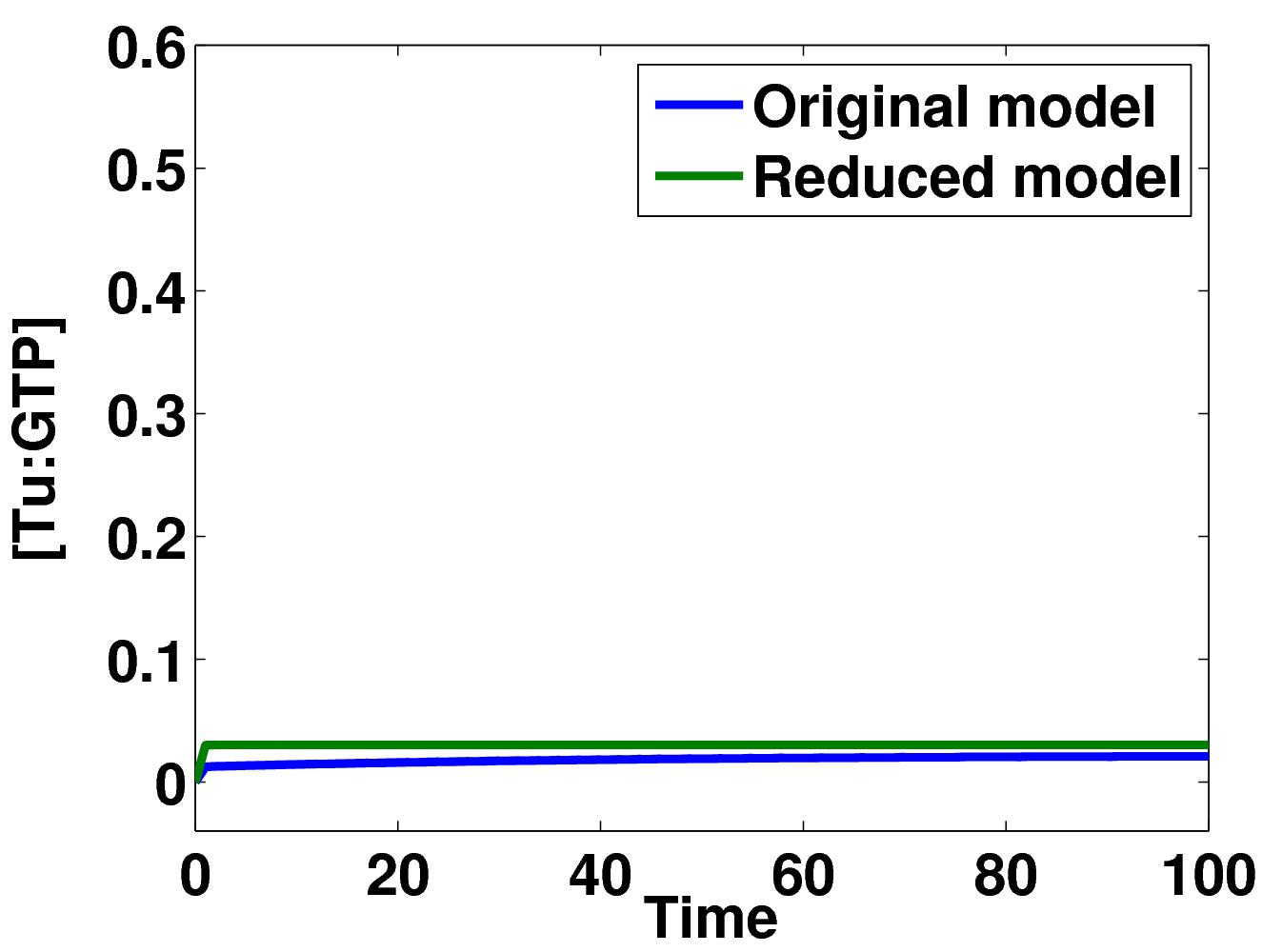}  
        }%
            \subfigure{%
            \label{fig:third}
            \includegraphics[width=0.3\textwidth]{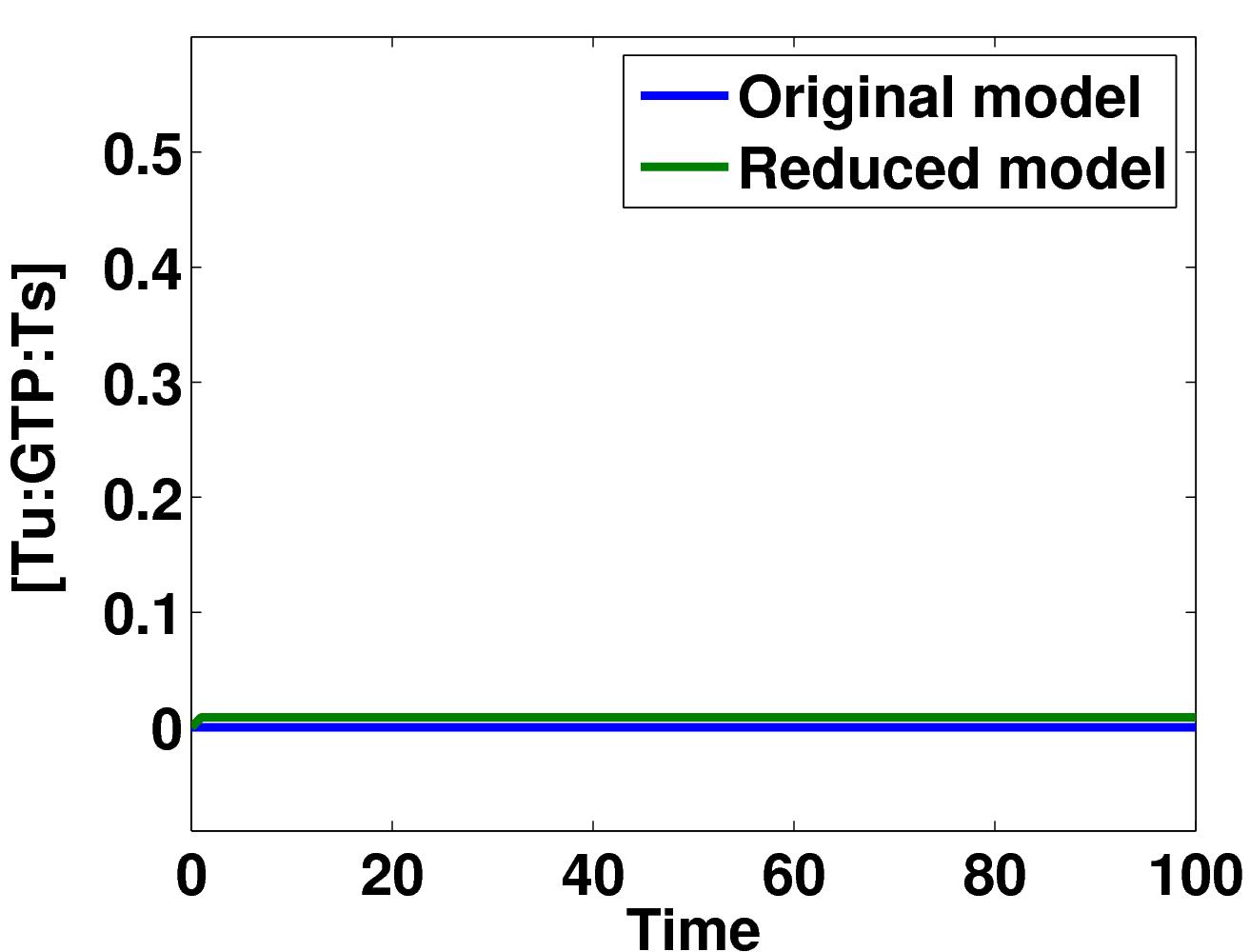}  
        }%
    \end{center}  
\caption {Numerical simulations of the full model (\ref{Ent25}) and reduced model (\ref{Ent26}) of elongation factors EF--Tu and EF--Ts signalling pathways, with the time interval $[0,100]$ in computational simulations.}  
   \label{fig:elognation}
\end{figure}

\begin{figure}[H] 
     \begin{center}
        \subfigure{%
            \label{fig:third}  
            \includegraphics[width=0.9\textwidth]{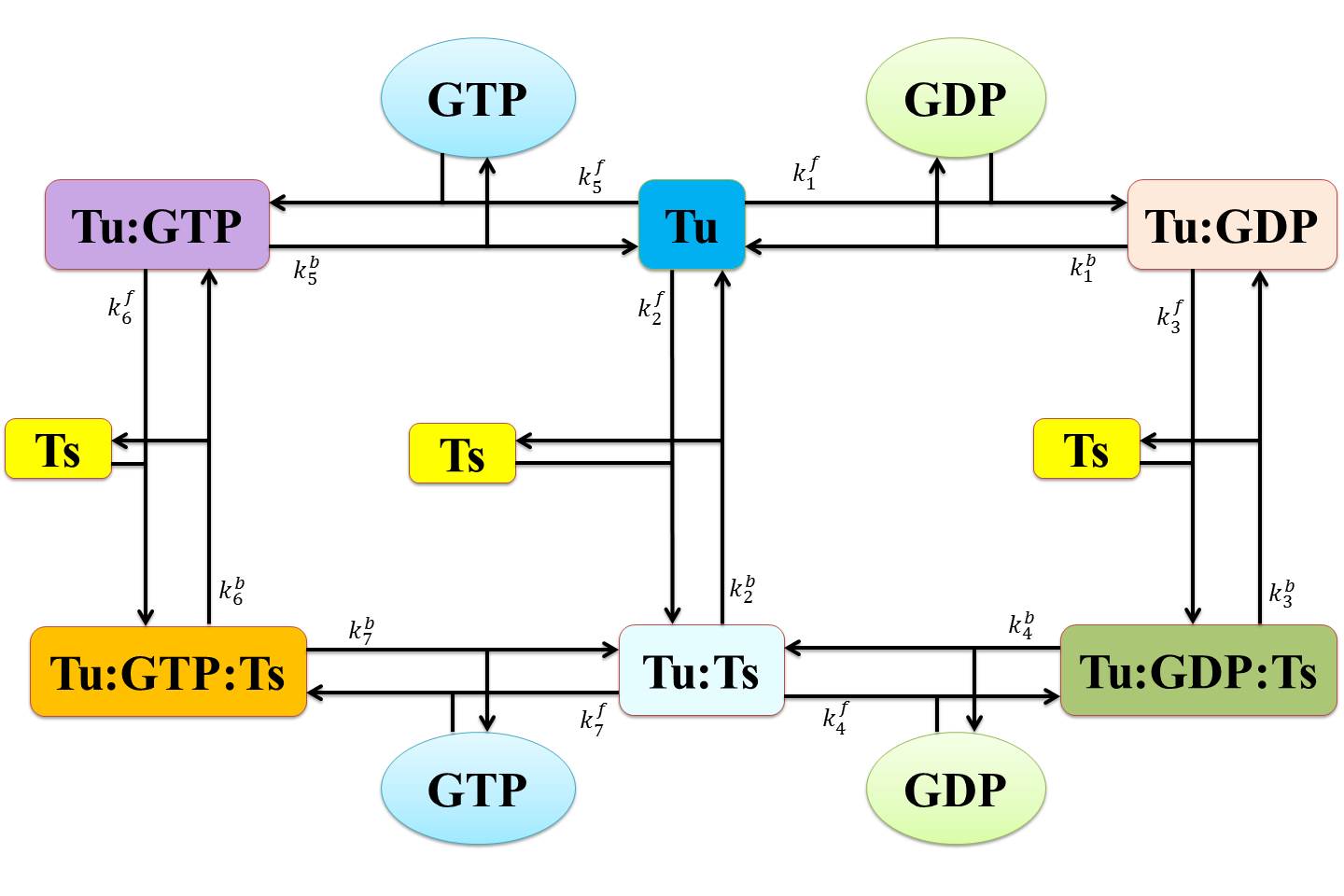}
        }
    \end{center}  
\caption {The original model of elongation factors EF-Tu and EF-Ts signalling pathways.}    
   \label{TuTs1}   
\end{figure}

\begin{figure}[H] 
     \begin{center}
        \subfigure{%
            \label{fig:third}  
            \includegraphics[width=0.9\textwidth]{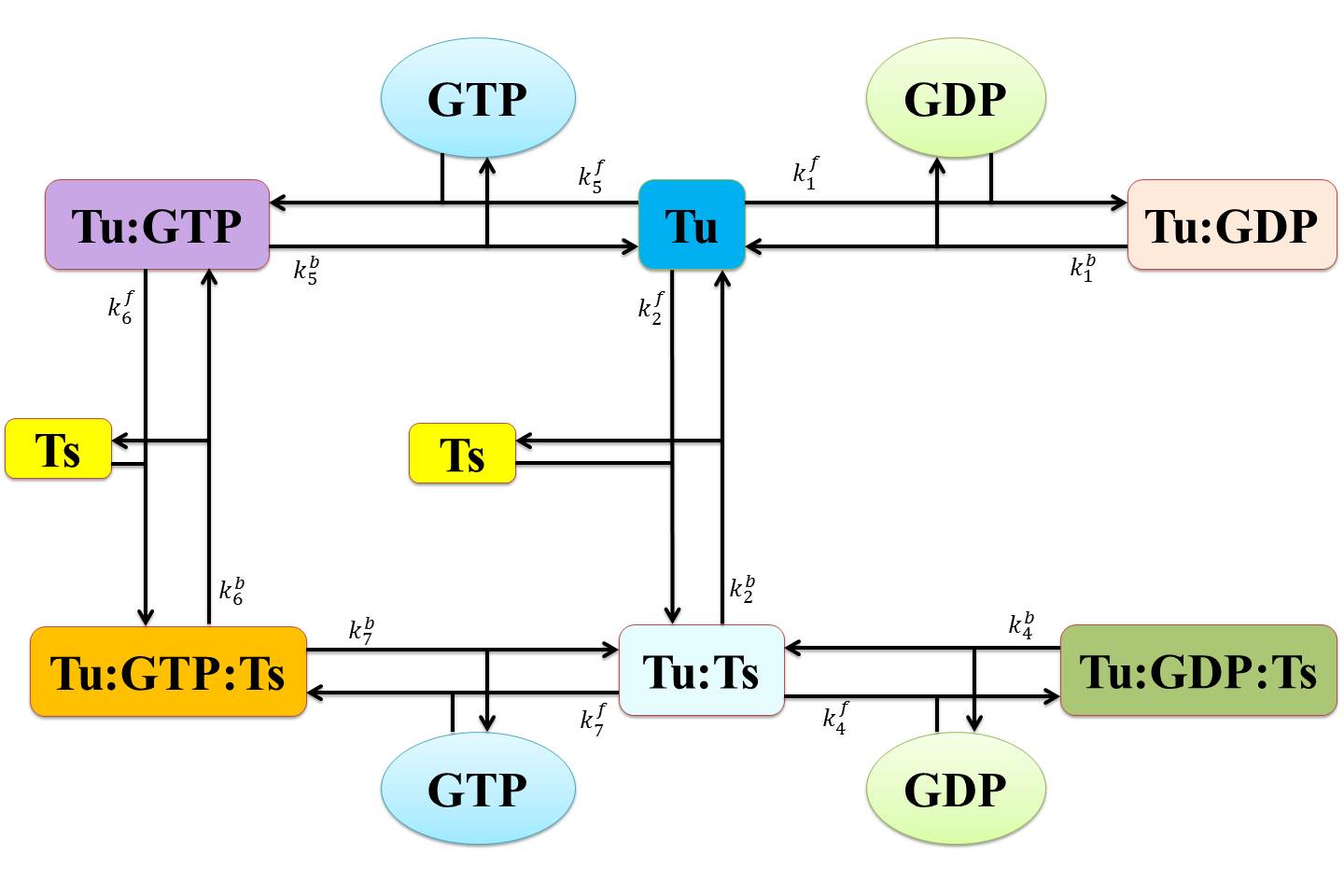}
        }
    \end{center}  
\caption {The structure of the model of elongation factors EF-Tu and EF-Ts after eliminating reaction 3.}  
   \label{TuTs2} 
\end{figure}

\begin{figure}[H] 
     \begin{center}
        \subfigure{%
            \label{fig:third}  
            \includegraphics[width=0.9\textwidth]{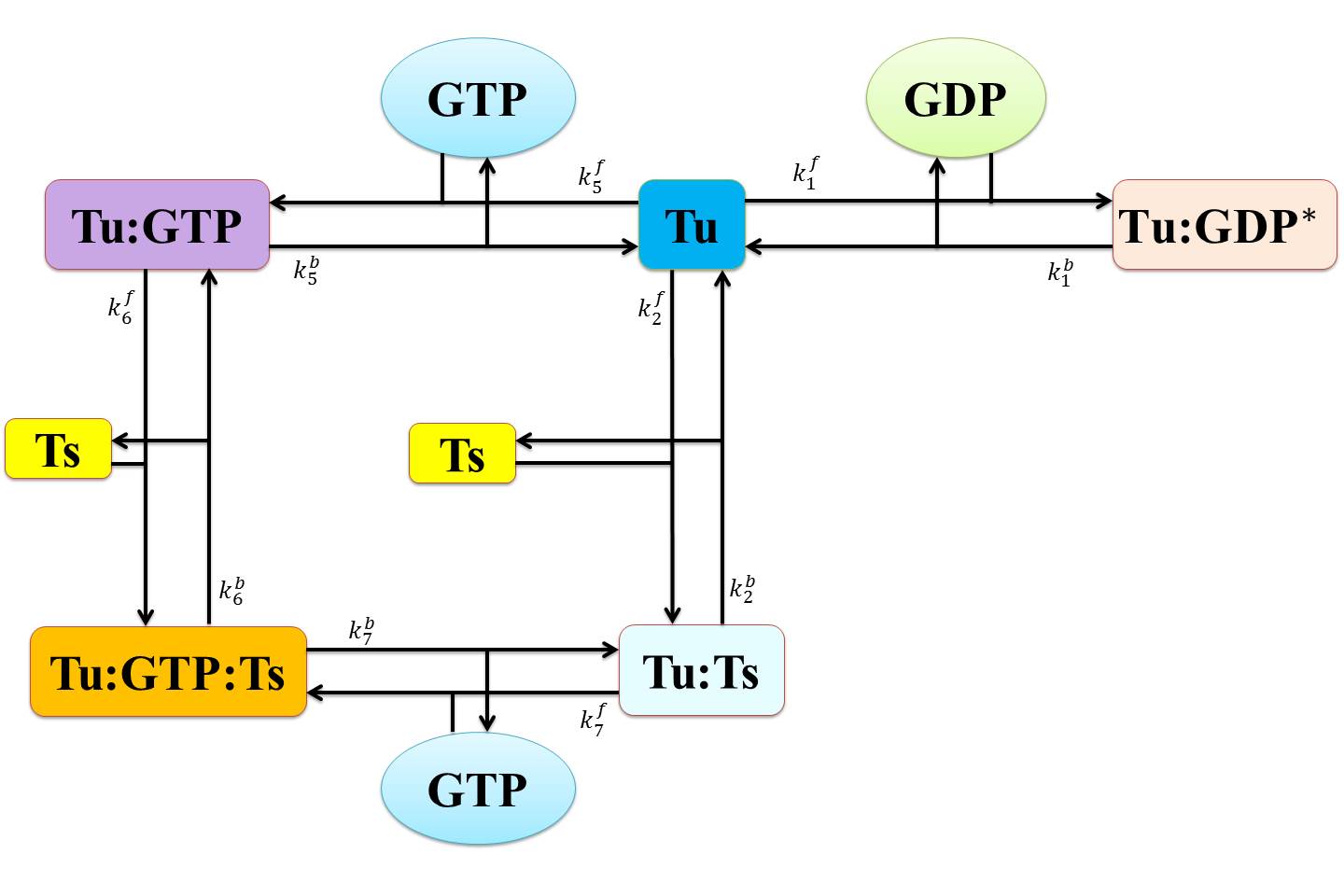}
        }
    \end{center}  
\caption {The structure of the model of elongation factors EF-Tu and EF-Ts after eliminating reactions 3 and 4.}  
   \label{TuTs3} 
\end{figure}

\begin{figure}[H] 
     \begin{center}
        \subfigure{%
            \label{fig:third}  
            \includegraphics[width=0.9\textwidth]{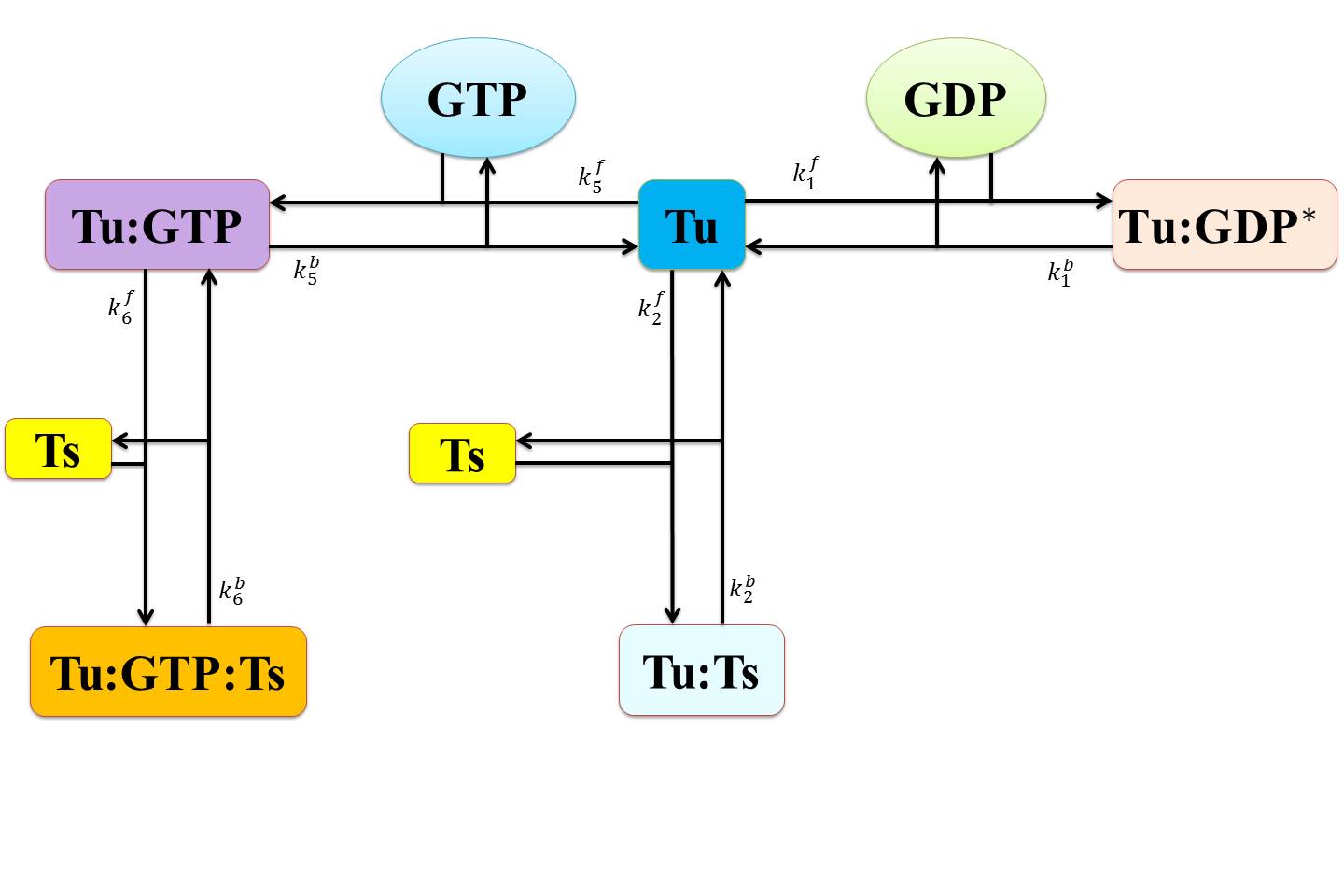}
        }
    \end{center}  
\caption {The structure of the model of elongation factors EF-Tu and EF-Ts after eliminating reactions 3, 4 and 7. }  
   \label{TuTs4} 
\end{figure}

\begin{figure}[H] 
     \begin{center}
        \subfigure{%
            \label{fig:third}  
            \includegraphics[width=0.9\textwidth]{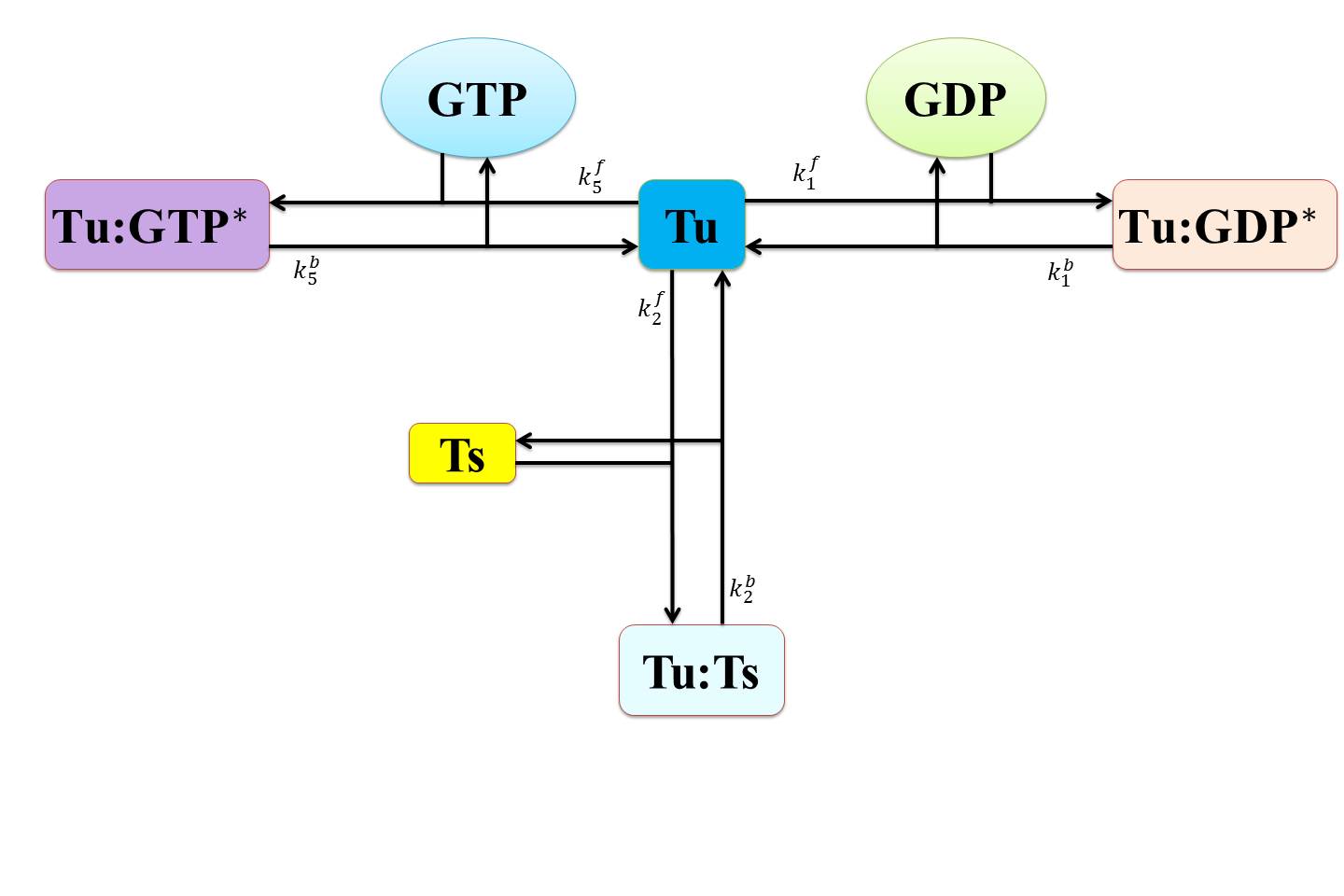}
        }
    \end{center}  
\caption {The structure of the model of elongation factors EF-Tu and EF-Ts after eliminating reactions 3, 4, 6 and 7. }  
   \label{TuTs5} 
\end{figure}

We applied the suggested approach to reduce the kinetic model of elongation factors EF--Tu and EF--Ts signalling pathways. As we studied before the number of reactions and species of the model are 14 and 9 and then reduced to 6 and 7, respectively. The steps of the model reduction are identify and shown in  Figures (\ref{TuTs1}--\ref{TuTs5}). The difference between the original and reduced model is calculated at each reduction step; see Table \eqref{tableENT1}. To check that the approximate solution is within allowable limits or not, which is an important task in model reduction we have to calculating the value of deviation. In Figure \eqref{fig:elognation}, blue lines and green lines represent the approximate solutions of the original and reduced model; the approximate solutions are computed using Matlab. The stages of model reduction here are based on eliminating some non--important reactions and lumped of isolated species. According to the value of relative contribution of entropy production for each reaction, firstly Reaction 3 is contributing least to the total entropy production. This reaction can be removed from the model. Then, we identify that Reaction 4 is contributing least among the remaining reactions. Therefore, Reaction 4 can be eliminated from the model; since [Tu:GDP:Ts] disappear then we lumped [Tu:GDP:Ts] with [Tu:GDP]. Similarly, we detect another two non--important reactions (Reactions 6 and 7) during the computational simulations, when $v_{6}=0$ then [Tu:GTP:Ts] disappear so we lumped [Tu:GTP:Ts] with [Tu:GTP]; this is detected by using the proposed algorithm. 

After applying the relative contribution algorithm to the model to eliminate non--important reactions and lumping isolated species, the reduced model takes the form: 
\begin{equation}  
\begin{array}{llll}
\dfrac{d[GDP]}{dt}=-k_{1}^{f}[GDP](t)[Tu](t)+k_{1}^{b}[Tu:GDP](t) , \\  
\dfrac{d[Tu:GDP]^{*}}{dt}=k_{1}^{f}[GDP](t)[Tu](t)-k_{1}^{b}[Tu:GDP](t), \\  
\dfrac{d[Ts]}{dt}=-k_{2}^{f}[Tu](t)[Ts](t)+k_{2}^{b}[Tu:Ts](t), \\  
\dfrac{d[Tu:Ts]}{dt}=k_{2}^{f}[Tu](t)[Ts](t)-k_{2}^{b}[Tu:Ts](t), \\
\dfrac{d[GTP]}{dt}=-k_{5}^{f}[Tu](t)[GTP](t)+k_{5}^{b}[Tu:GTP](t), \\
\dfrac{d[Tu:GTP]^{*}}{dt}=k_{5}^{f}[Tu](t)[GTP](t)-k_{5}^{b}[Tu:GTP](t),\\ 
\dfrac{d[Tu]}{dt}=-k_{1}^{f}[GDP](t)[Tu](t)+k_{1}^{b}[Tu:GDP](t)-k_{2}^{f}[Tu](t)[Ts](t)\\ \quad \quad \quad \quad+k_{2}^{b}[Tu:Ts](t)-k_{5}^{f}[Tu](t)[GTP](t)+k_{5}^{b}[Tu:GTP](t) ,
\end{array}\label{Ent26}
\end{equation} 

where $[Tu:GDP]^{*}=[Tu:GDP]+[Tu:GDP:Ts]$\;and\\$[Tu:GTP]^{*}=[Tu:GTP]+[Tu:GTP:Ts]$.
\newline
Then the system (\ref{Ent26}) has four independent stoichiometric conservation laws \\
$[GDP](t)+[Tu:GDP]^{*}(t)=[GDP]_{0},$ \\
$[Ts](t)+[Tu:Ts](t)=[Ts]_{0},$ \\
$[GTP](t)+[Tu:GTP]^{*}(t)=[GTP]_{0},$ \\
$[Tu](t)+[Tu:GDP]^{*}(t)+[Tu:Ts](t)+[Tu:GTP]^{*}(t)=[Tu]_{0}$.

In Figure \eqref{fig:elognation}, there are a good agreement between the original and reduced model in computational simulation. This means that our proposed technique plays a good role in model reduction.\\
Table \eqref{tableENT1} shows the value of deviation at each stage of reduction that helps one to test that the approximation of the model is sufficiently accurate for biochemical phenomena. The results here give a new perspective to the concept of model reduction based on entropy production and lumping of species, provide a new level of understanding of model comparison.\\
Our results show that the proposed algorithm is more developed compare to the previous algorithm (Khoshnaw, 2015), because the value of deviation in (Khoshnaw, 2015) between 0.53$\%$ and 9.14$\%$, so it is clear that the value of deviation of elongation in our technique is much smaller than the previous study as you can see in Table \eqref{tableENT1}.


\FloatBarrier 
\begin{table}   
\center
    \caption{The table shows the difference between the reduced and full model of elongation factors EF--Tu and EF--Ts signalling pathways at each reduction stage using the function of deviation and lumping of isolated species.} 
  
\begin{tabular}{|p{3cm}|p{3cm}|p{5.3cm}|p{3cm}|}
    \hline 
    Model Reduction Stages & Non-important reactions & Lumping of isolated species & Values of Deviation $\mathcal{F}^{\mathcal{D}}$ \\ 
    \hline 
    Stage One & Reaction three & No lump & $0.164\; \%$ \\ 
    \hline 
    Stage Two & Reaction four & $[Tu:GDP]^{*}=[Tu:GDP]+[Tu:GDP:Ts]$ & $2.792\; \%$ \\ 
    \hline 
    Stage Three & Reaction seven & No lump & $3.749\; \%$ \\ 
    \hline 
    Stage Four & Reaction six & $[Tu:GTP]^{*}=[Tu:GTP]+[Tu:GTP:Ts]$ & $3.676\; \%$ \\ 
    \hline 
    \end{tabular}     
\label{tableENT1}   
\end{table} 
\FloatBarrier 

\newpage \subsection{Dihydrofolate Reductase (DHFR) Pathways}

\begin{figure}[H] 
     \begin{center}
        \subfigure{%
            \label{fig:third}  
            \includegraphics[width=1\textwidth]{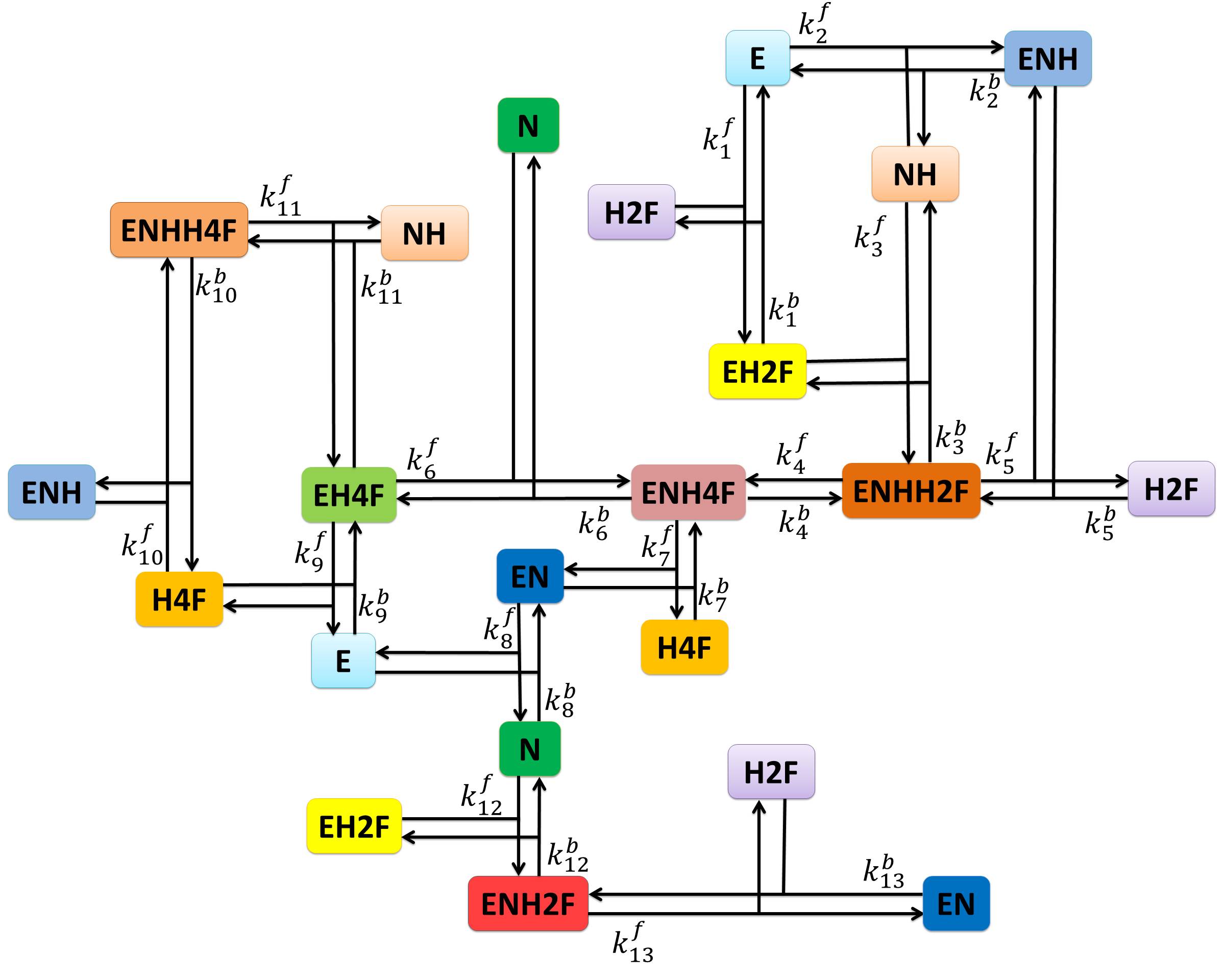}
        }
    \end{center}  
\caption {The Dihydrofolate reductase DHFR pathways.}  
   \label{DHRF model diagram} 
\end{figure}

Dihydrofolate Reductase  is one of the most important enzymes for DNA synthesis because it produces cofactor which is necessary in the building of DNA and other processes.  The enzyme DHFR catalyzes the reduction of 7,8-dihydrofolate (DHF) to the product 5,6,7,8-tetrahydrofolate (THF) by hydride transfer from the NADPH cofactor (Sittikornpaiboon et al., 2017),  And also reduces dihydrofolic acid to tetrahydrofolic acid, using NADPH (NADPH is the critical reducing agent and limiting factor in fatty acid synthesis)  as electron donor, which can be converted to the kinds of tetrahydrofolate cofactors used in 1-carbon transfer chemistry. In humans, the DHFR enzyme is encoded by the DHFR gene it is found in the region of chromosome 5.
Dihydrofolate reductase (DHFR) is a notable drug target for the design of anti-malarial , anti-bacterial , and anti-cancer drugs (Sittikornpaiboon et al., 2017).
Antifolate drugs, methotrexate (MTX) and trimetrexate, can tightly bind to DHFR and inhibit DNA synthesis and cell proliferation. On that account, antifolate drugs have been used as potent antitumor drugs.
Streptococcus pneumoniae is one of the clinically important Gram-positive bacterial pathogens (1, 2). The emergence of multidrug-resistant (MDR). pneumoniae strains has become a global concern. Resistance to trimethoprim/sulfamethoxazole (T/S) arises from mutations in the target enzyme dihydrofolate reductase (DHFR), whose activity is necessary for the maintenance of the cellular level of tetrahydrofolate that is essential for the biosynthesis of purines, some amino acids, and thymidine. Therefore, DHFR has long been a target for the discovery of novel antibacterial agents as well as anticancer drugs (Lee et al., 2009).
DHFR was active at each of the temperatures tested in the range ($12-55 \; C^{o}$), with the greatest activity detected at $37\; C^{o}$. The optimum pH for this enzyme was about pH 10.0, which is similar to the optimum pH for recombinant human DHFR. DHFRs from various sources can differ markedly in their affinities for DHFR inhibitors. DHFR inhibitors are in wide use as antibacterial and antiprotozoal agents (Wang et al., 2016).
There are many reaction change and product a new concentration in DHFR mechanism it can be clearly seen in the Figure \eqref{DHRF model diagram}, which has thirteen reversible reactions and consists of thirteen state variables (concentrations). The rate constants (rate of reaction) has great role on this chemical reaction.

We separate the complete reactions to three chemical reaction steps such as equation \eqref{DHFR model} for the easer making an ODE system  (Benkovic and Hammes-Schiffer, 2003, Craciun et al., 2006).

All reactions in the model are considered to be reversible: 
\begin{equation}  
\begin{array}{llll}
E+H2F \underset{k_{1}^{b}}{ \overset{k_{1}^{f}}{\rightleftharpoons}} EH2F, \quad
E+NH \underset{k_{2}^{b}}{ \overset{k_{2}^{f}}{\rightleftharpoons}} ENH , \\
EH2F+NH \underset{k_{3}^{b}}{ \overset{k_{3}^{f}}{\rightleftharpoons}} ENHH2F \underset{k_{5}^{b}}{ \overset{k_{5}^{f}}{\rightleftharpoons}} ENH+H2F  ,\\
\quad\quad\quad\quad\quad\quad\quad\quad k_{4}^{b}\upharpoonleft\downharpoonleft k_{4}^{f} \\
\quad\quad\quad\quad\quad\quad\quad\quad ENH4F\\

EH4F+N \underset{k_{6}^{b}}{ \overset{k_{6}^{f}}{\rightleftharpoons}} ENH4F \underset{k_{7}^{b}}{ \overset{k_{7}^{f}}{\rightleftharpoons}} EN+H4F ,\\

EN \underset{k_{8}^{b}}{ \overset{k_{8}^{f}}{\rightleftharpoons}} E+N ,\quad
EH4F \underset{k_{9}^{b}}{ \overset{k_{9}^{f}}{\rightleftharpoons}} E+H4F , \\

ENH+H4F \underset{k_{10}^{b}}{ \overset{k_{10}^{f}}{\rightleftharpoons}} ENHH4F \underset{k_{11}^{b}}{ \overset{k_{11}^{f}}{\rightleftharpoons}} EH4F+NH ,\\

EH2F+N \underset{k_{12}^{b}}{ \overset{k_{12}^{f}}{\rightleftharpoons}} ENH2F \underset{k_{13}^{b}}{ \overset{k_{13}^{f}}{\rightleftharpoons}} EN+H2F,
\end{array}\label{DHFR model}
\end{equation}   
where $k_{i}^{f}$ and $k_{i}^{b}$ for $i=1,2,...,13$ are the forward and backward reaction constants, respectively.

\noindent The chemical reactions (\ref{DHFR model}) can be expressed as a system of ordinary differential equations:  
\begin{equation}  
\begin{array}{llll}
\dfrac{d[E]}{dt}=-v_{1}-v_{2}+v_{8}+v_{9} , \quad  
\dfrac{d[H2F]}{dt}=-v_{1}+v_{5}+v_{13} , \\
\dfrac{d[EH2F]}{dt}=v_{1}-v_{3}-v_{12}, \quad
\dfrac{d[NH]}{dt}=-v_{2}-v_{3}+v_{11}, \\
\dfrac{d[ENH]}{dt}=v_{2}+v_{5}-v_{10}, \quad 
\dfrac{d[ENHH2F]}{dt}=v_{3}-v_{4}-v_{5}, \\
\dfrac{d[ENH4F]}{dt}=v_{4}+v_{6}-v_{7}, \quad 
\dfrac{d[EH4F]}{dt}=-v_{6}-v_{9}+v_{11},\\  
\dfrac{d[N]}{dt}=-v_{6}+v_{8}-v_{12}, \quad
\dfrac{d[EN]}{dt}=v_{7}-v_{8}+v_{13},\\
\dfrac{d[H4F]}{dt}=v_{7}+v_{9}-v_{10},\quad 
\dfrac{d[ENHH4F]}{dt}=v_{10}-v_{11},\\
\dfrac{d[ENH2F]}{dt}=v_{12}-v_{13},
\end{array}\label{DHFR ODE system}
\end{equation} 
where $v_{1}=k_{1}^{f}[E](t)[H2F](t)-k_{1}^{b}[EH2F](t),$ 
$v_{2}=k_{2}^{f}[E](t)[NH](t)\\-k_{2}^{b}[ENH](t),$ 
$v_{3}=k_{3}^{f}[EH2F](t)[NH](t)-k_{3}^{b}[ENHH2F](t),$ \\
$v_{4}=k_{4}^{f}[ENHH2F](t)-k_{4}^{b}[ENH4F](t),$ 
$v_{5}=k_{5}^{f}[ENHH2F](t)\\ -k_{5}^{b}[ENH](t)[H2F](t),$ 
$v_{6}=k_{6}^{f}[EH4F](t)[N](t)-k_{6}^{b}[ENH4F](t),$ \\
$v_{7}=k_{7}^{f}[ENH4F](t)-k_{7}^{b}[EN](t)[H4F](t).$ 
$v_{8}=k_{8}^{f}[EN](t)\\-k_{8}^{b}[E](t)[N](t),$ 
$v_{9}=k_{9}^{f}[EH4F](t)-k_{9}^{b}[E](t)[H4F](t),$ \\
$v_{10}=k_{10}^{f}[ENH](t)[H4F](t)-k_{10}^{b}[ENHH4F](t),$ 
$v_{11}=k_{11}^{f}[ENHH4F](t)\\ -k_{11}^{b}[EH4F](t)[NH](t),$ 
$v_{12}=k_{12}^{f}[EH2F](t)[N](t)-k_{12}^{b}[ENH2F](t),$ \\
$v_{13}=k_{13}^{f}[ENH2F](t)-k_{13}^{b}[EN](t)[H2F](t).$ 

And there are a set of data for state variables and parameters; see Tables \eqref{table of stationary values of DHFR} and
\eqref{table of parameter values of DHFR}.

\FloatBarrier 
\begin{table}
\caption{Summary of parameter values for DHFR pathways.}
\begin{center}
\begin{tabular}{|c|c|c|}
\hline 
No. & Parameters & Values \\ 
\hline 
1 & $k^f_{1}$ & 264 \\ 
\hline 
2 & $k^b_{1}$ & 14 \\ 
\hline 
3 & $k^f_{2}$ & 38 \\ 
\hline 
4 & $k^b_{2}$ & 1.7 \\ 
\hline 
5 & $k^f_{3}$ & 24 \\ 
\hline 
6 & $k^b_{3}$ & 19 \\ 
\hline 
7 & $k^f_{4}$ & 1360 \\ 
\hline 
8 & $k^b_{4}$ & 37 \\ 
\hline 
9 & $k^f_{5}$ & 94 \\ 
\hline 
10 & $k^b_{5}$ & 98 \\ 
\hline 
11 & $k^f_{6}$ & 0.7 \\ 
\hline 
12 & $k^b_{6}$ & 84 \\ 
\hline 
13 & $k^f_{7}$ & 46 \\ 
\hline 
14 & $k^b_{7}$ & 24 \\ 
\hline 
15 & $k^f_{8}$ & 32 \\ 
\hline 
16 & $k^b_{8}$ & 17 \\ 
\hline 
17 & $k^f_{9}$ & 5.1 \\ 
\hline 
18 & $k^b_{9}$ & 117 \\ 
\hline 
19 & $k^f_{10}$ & 14 \\ 
\hline 
20 & $k^b_{10}$ & 225 \\ 
\hline 
21 & $k^f_{11}$ & 100 \\ 
\hline 
22 & $k^b_{11}$ & 4.4 \\ 
\hline 
23 & $k^f_{12}$ & 20 \\ 
\hline 
24 & $k^b_{12}$ & 4.6 \\ 
\hline 
25 & $k^f_{13}$ & 110 \\ 
\hline 
26 & $k^b_{13}$ & 1.3 \\ 
\hline 
\end{tabular} 
\end{center}
\label{table of parameter values of DHFR}
\end{table}
\FloatBarrier 

\FloatBarrier 
\begin{table}
\caption{Stationary values of state variables for DHFR pathways.} 
\begin{center}
\begin{tabular}{|c|c|c|}
 \hline 
 No. & Species & Stationary values \\ 
 \hline 
 1 & E & 0.02 \\ 
 \hline 
 2 & H2F & 0.03 \\ 
 \hline 
 3 & EH2F & 0 \\ 
 \hline 
 4 & NH & 0.08 \\ 
 \hline
 5 & ENH & 0\\ 
 \hline 
 6 & ENHH2F & 0\\ 
 \hline 
 7 & ENH4F & 0\\ 
 \hline 
 8 & EH4F & 0.05 \\ 
 \hline 
 9 & N & 0.06 \\ 
 \hline 
 10 & EN & 0\\ 
 \hline 
 11 & H4F & 0 \\ 
 \hline 
 12 & ENHH4F & 0\\ 
 \hline 
 13 & ENH2F & 0\\ 
 \hline 
 \end{tabular}  
 \end{center}
 \label{table of stationary values of DHFR}
 \end{table}
 \FloatBarrier 
 \subsection{Results and Discussions}   
\label{DHFR result and disscussion}
\begin{figure}[H] 
     \begin{center}
        \subfigure{%
            \label{fig:third}  
            \includegraphics[width=0.3\textwidth]{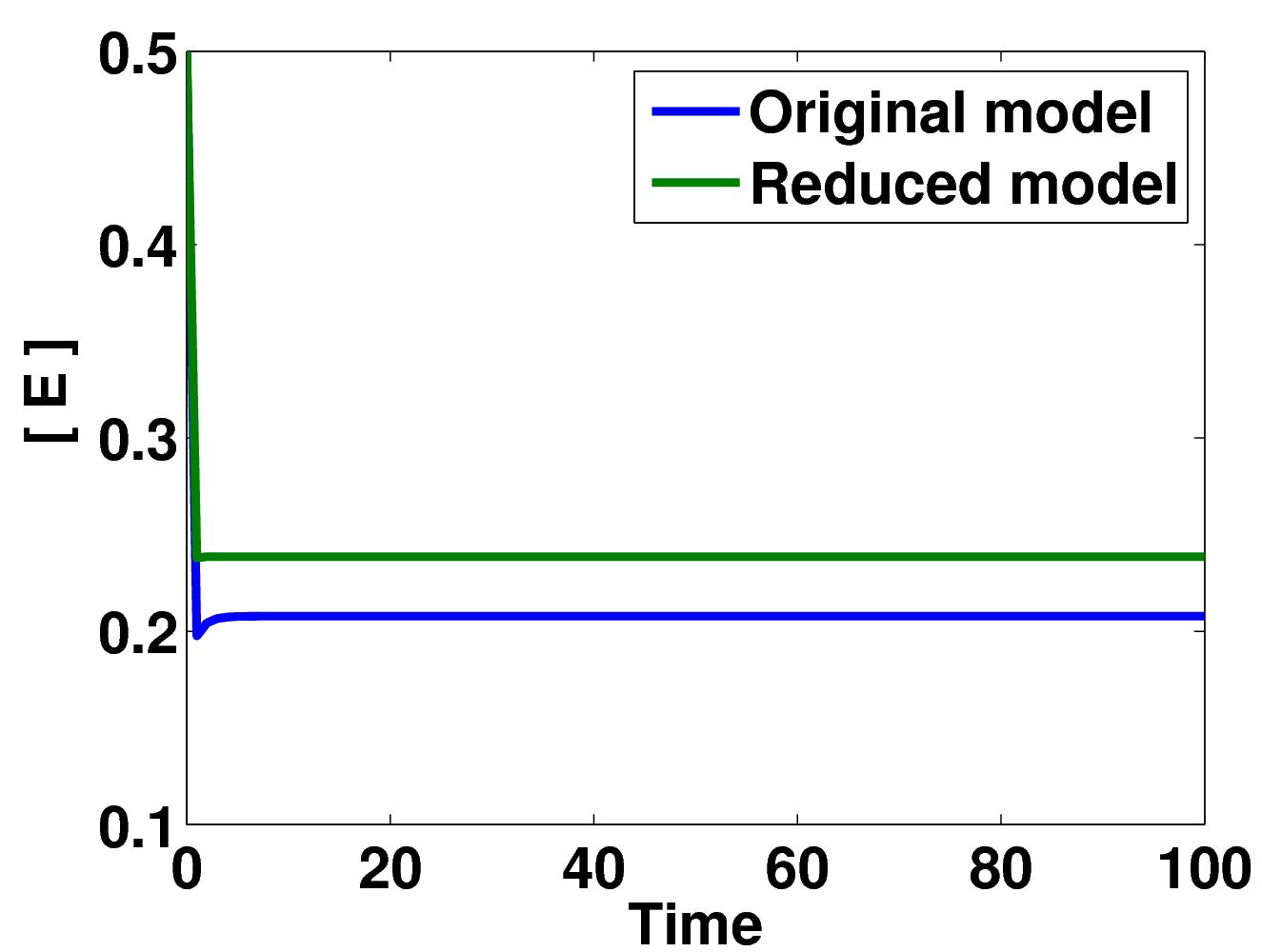}
        }%
                 \subfigure{%
            \label{fig:third}
            \includegraphics[width=0.3\textwidth]{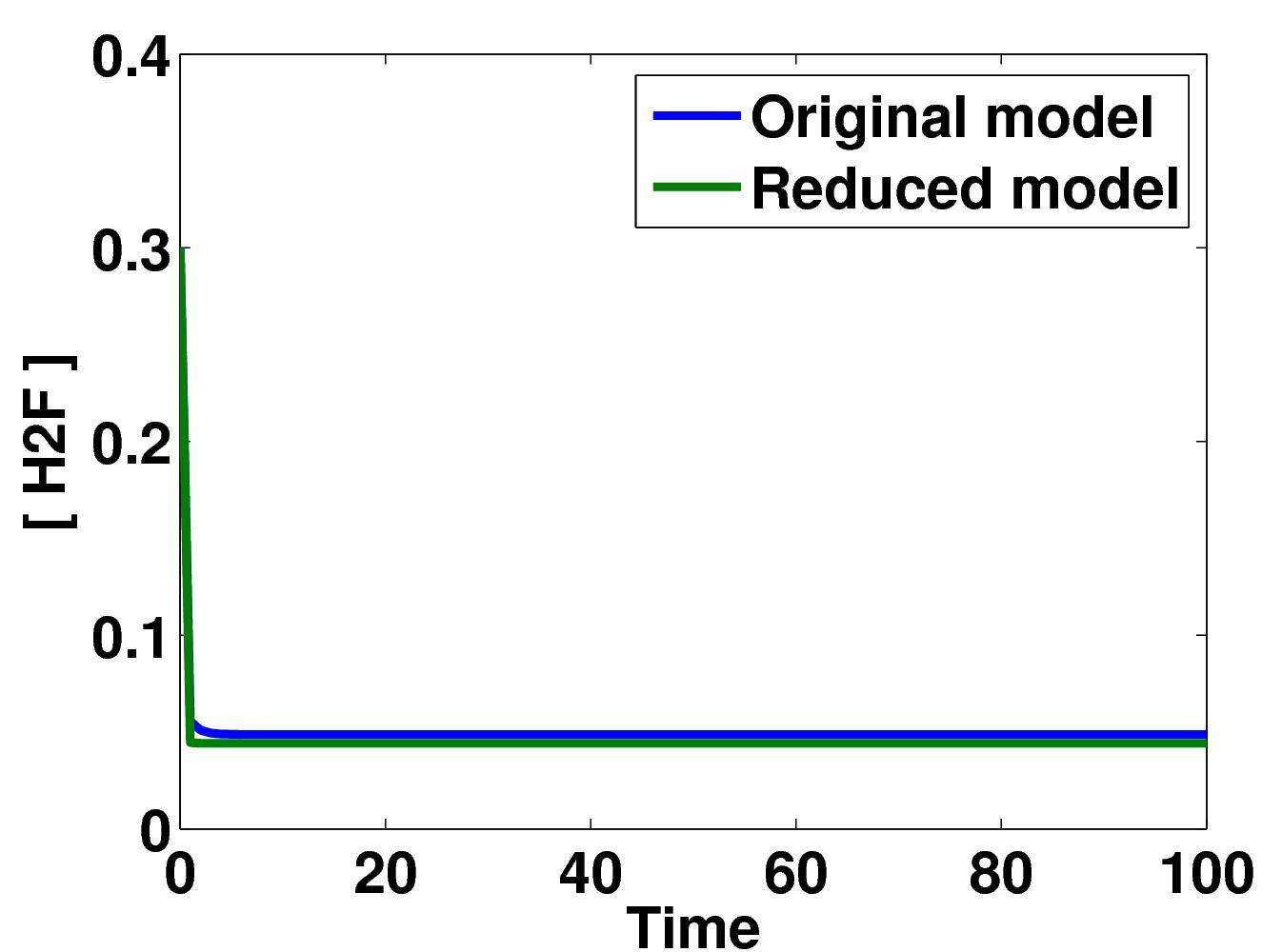}  
        }%
            \subfigure{%
            \label{fig:third}
            \includegraphics[width=0.3\textwidth]{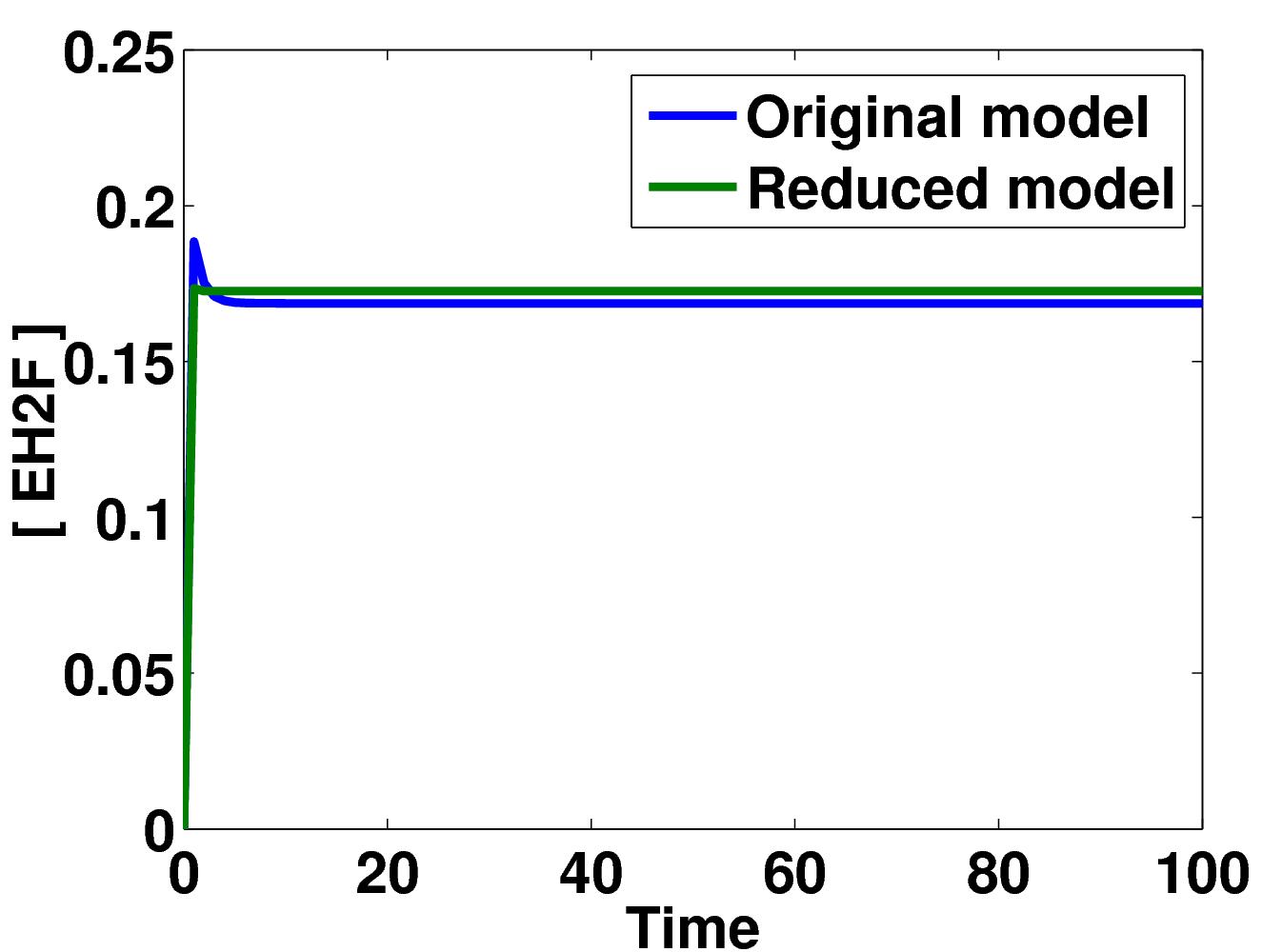}  
        }\\%
            \subfigure{%
            \label{fig:third}
            \includegraphics[width=0.3\textwidth]{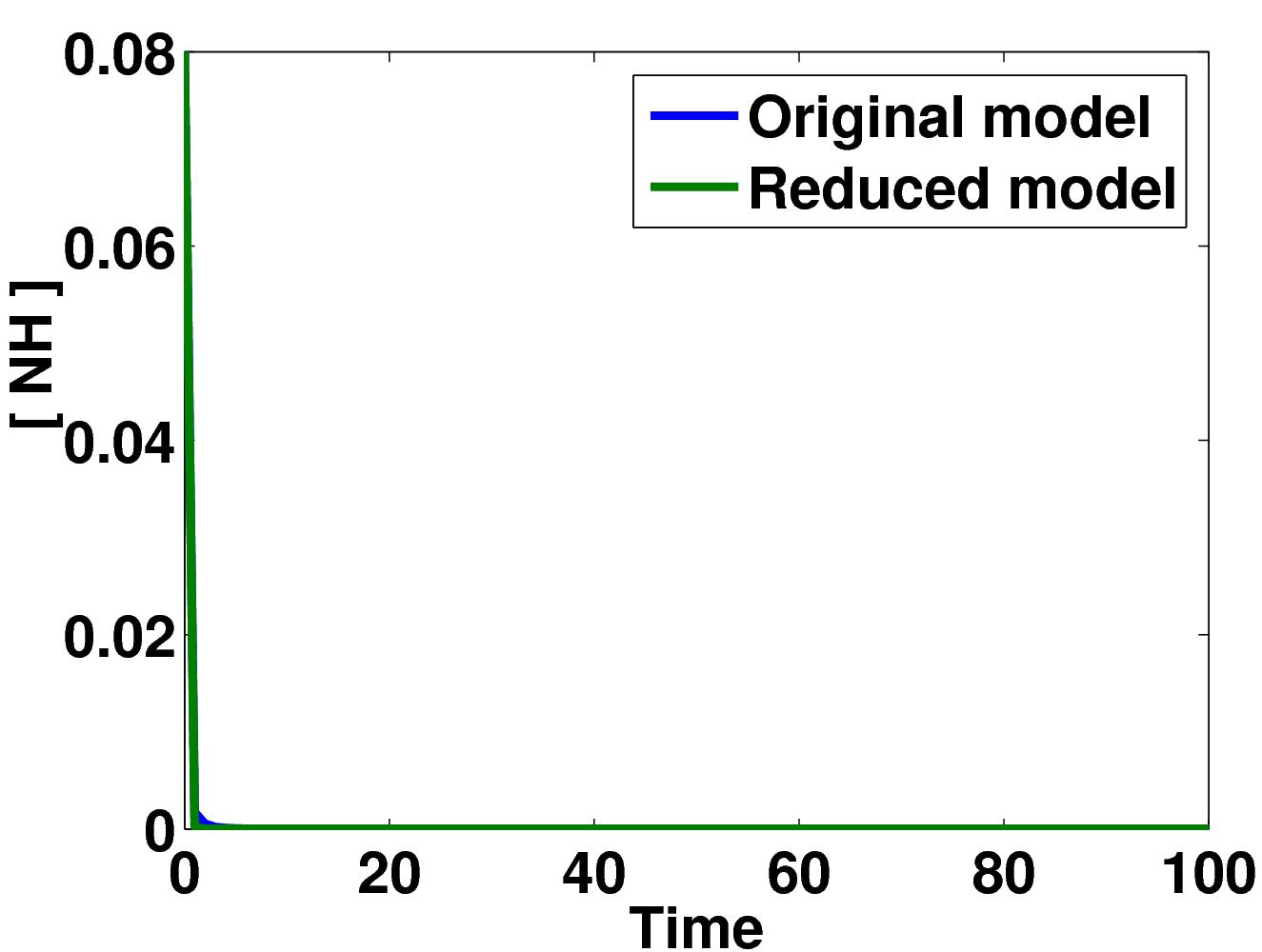}  
        }%
            \subfigure{%
            \label{fig:third}
            \includegraphics[width=0.3\textwidth]{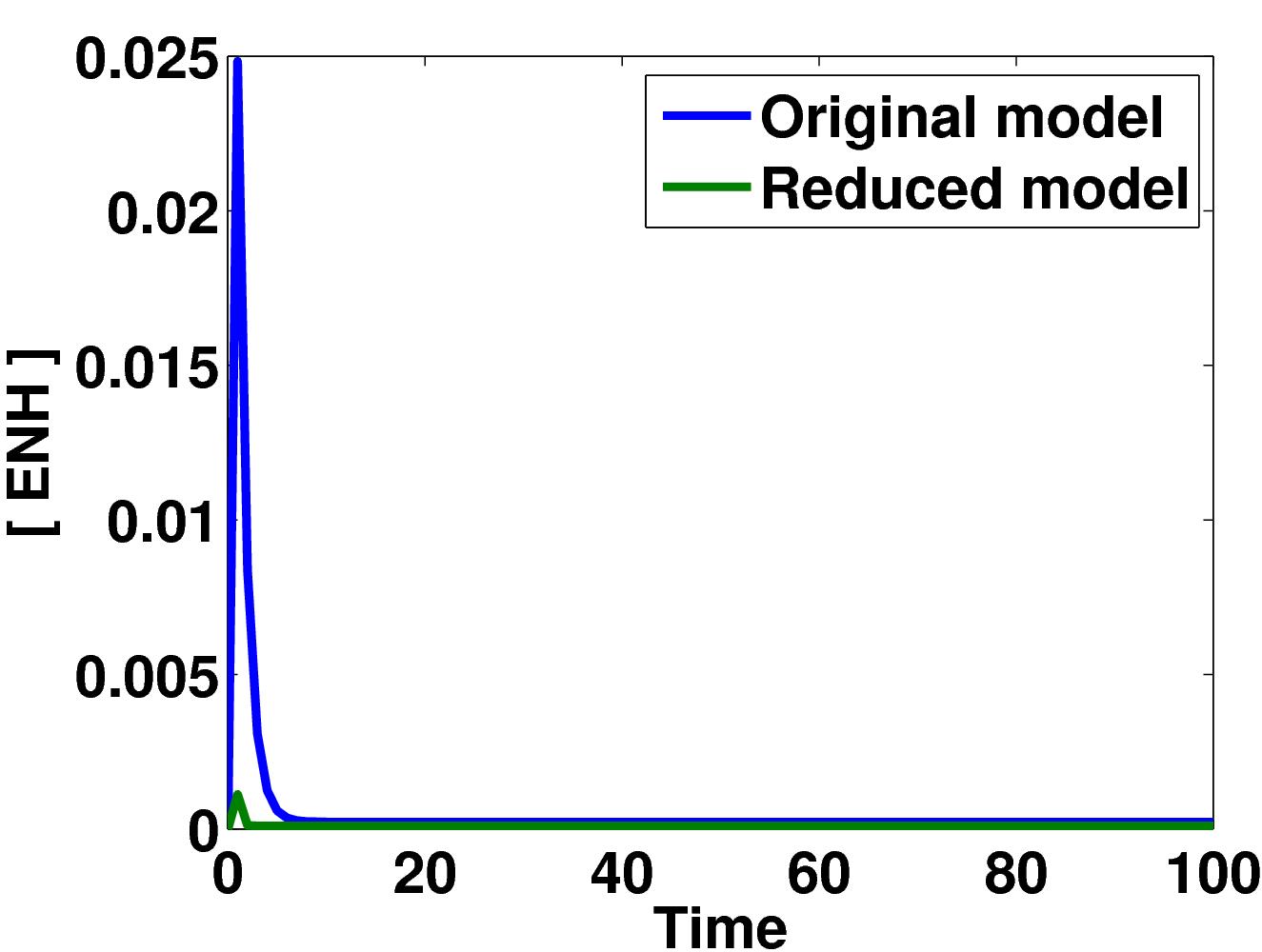}  
        }%
            \subfigure{%
            \label{fig:third}
            \includegraphics[width=0.3\textwidth]{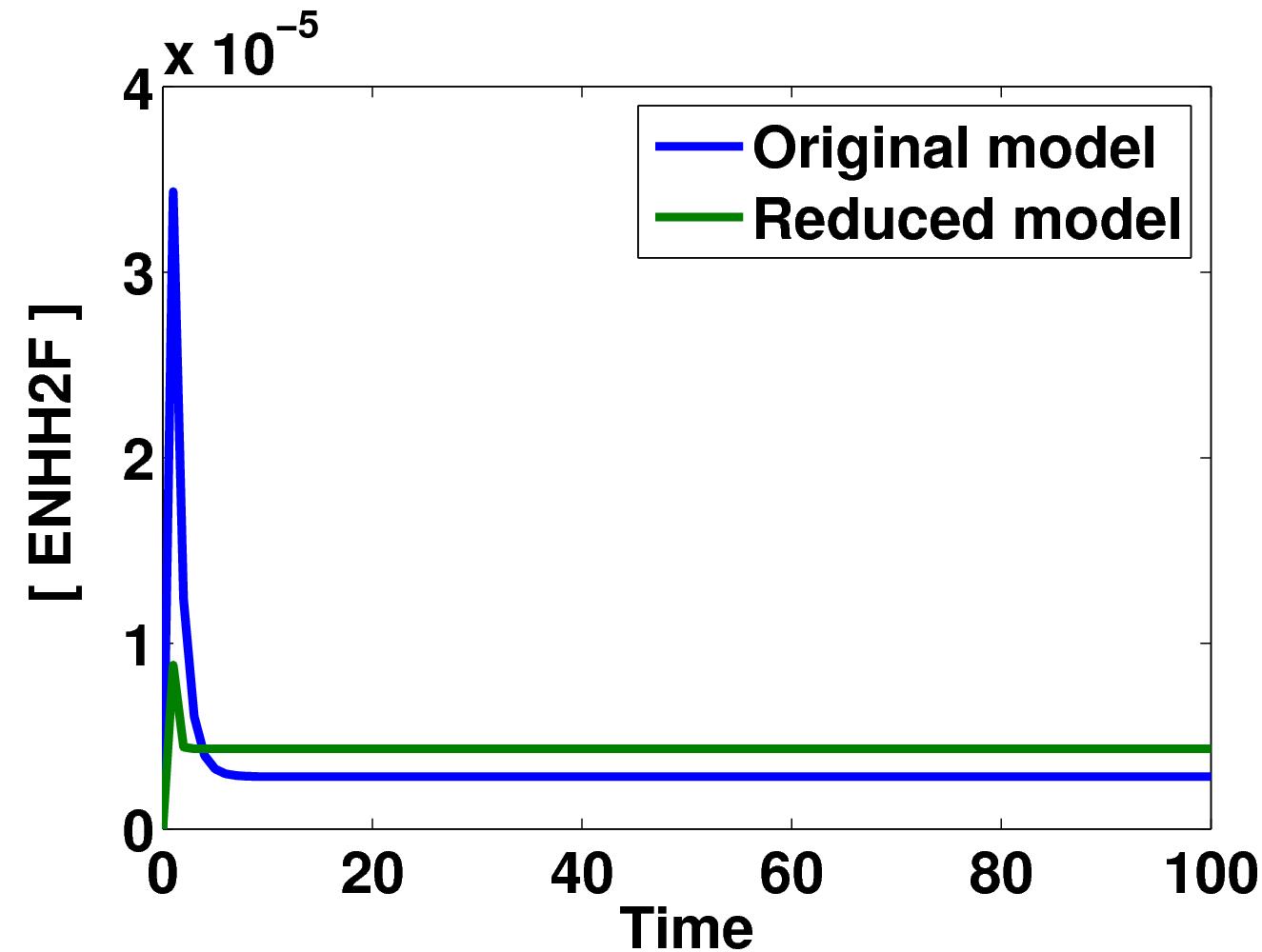}  
        }\\%
            \subfigure{%
            \label{fig:third}
            \includegraphics[width=0.3\textwidth]{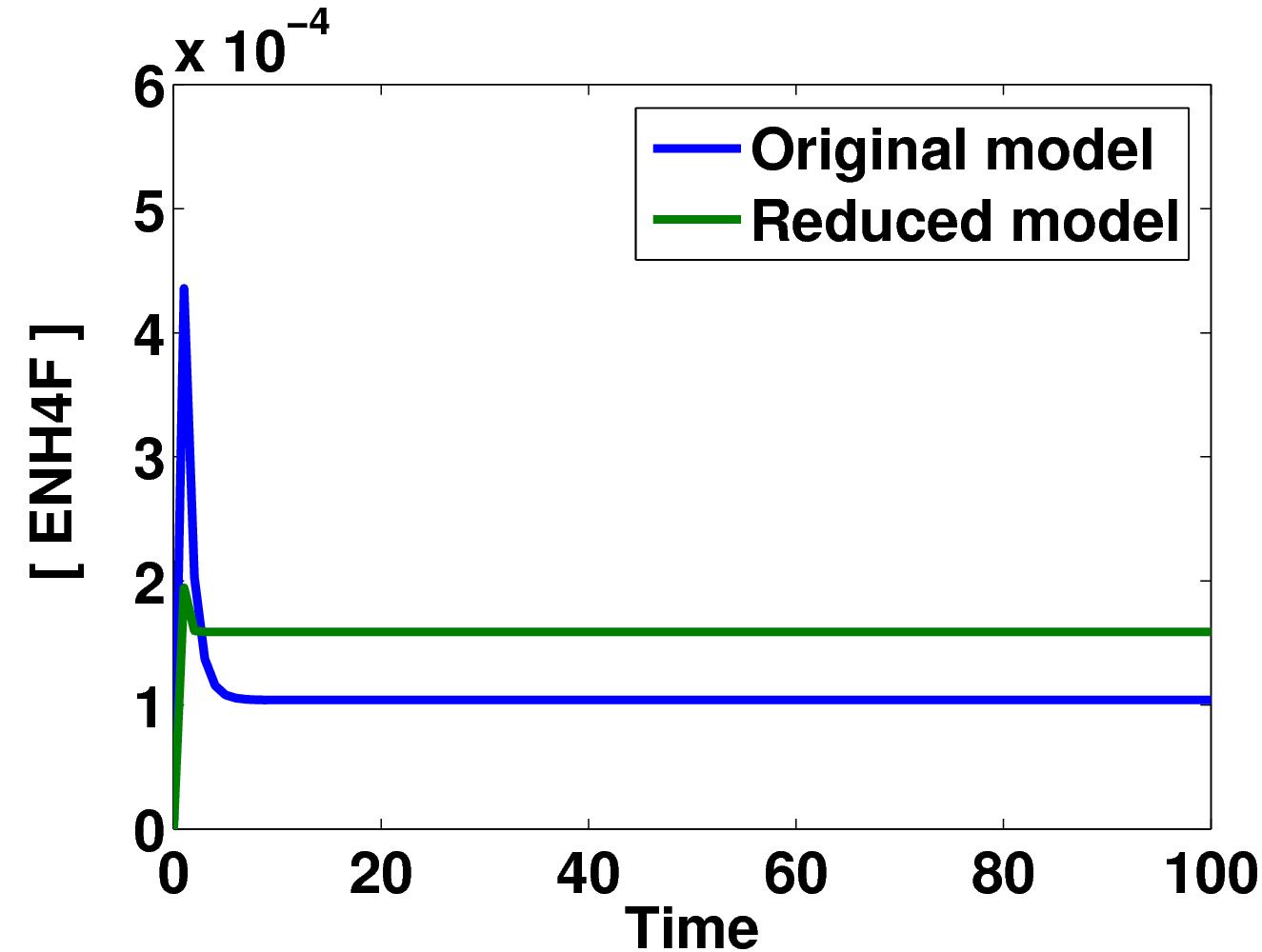}  
        }%
            \subfigure{%
            \label{fig:third}
            \includegraphics[width=0.3\textwidth]{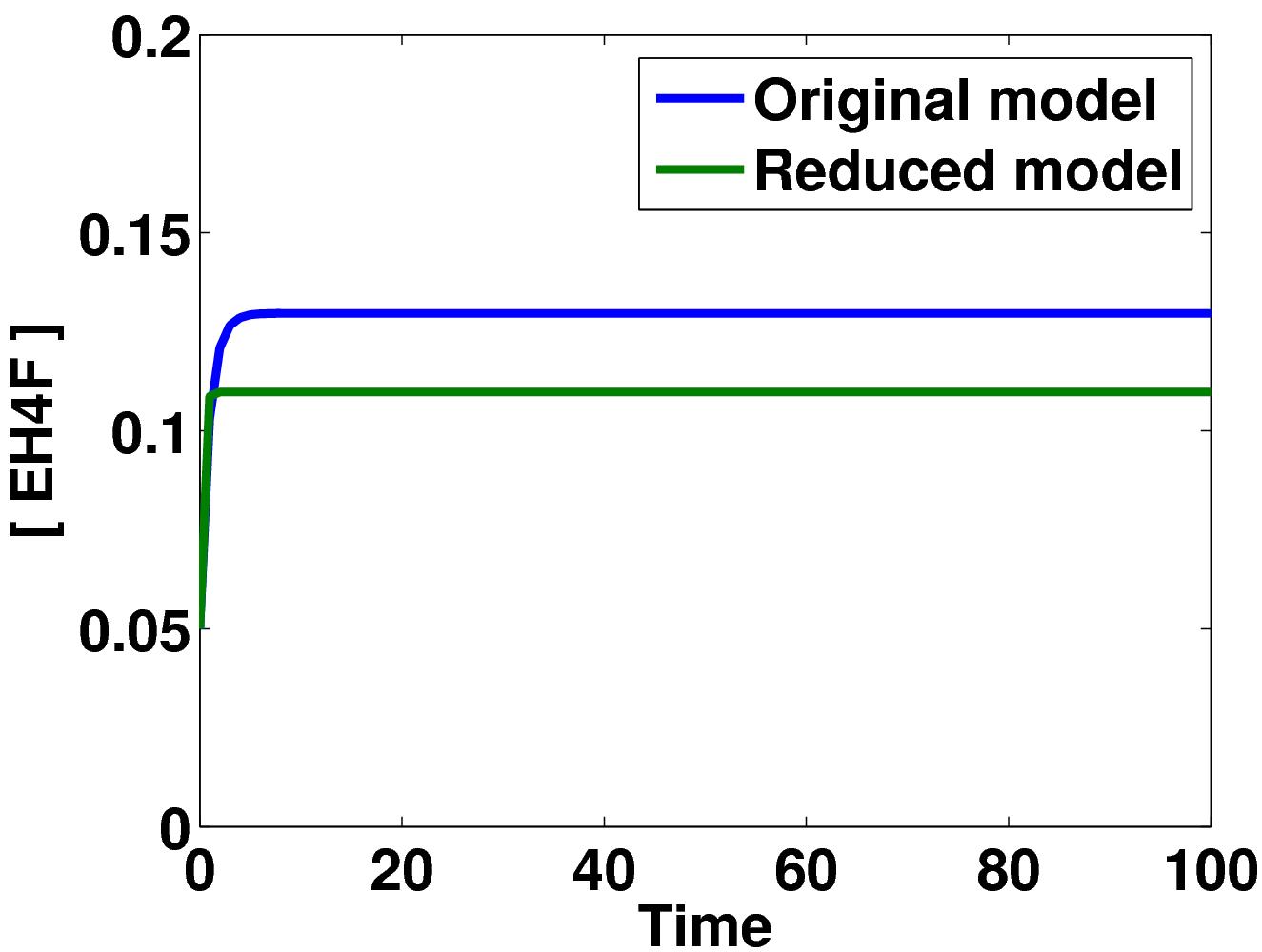}  
        }%
            \subfigure{%
            \label{fig:third}
            \includegraphics[width=0.3\textwidth]{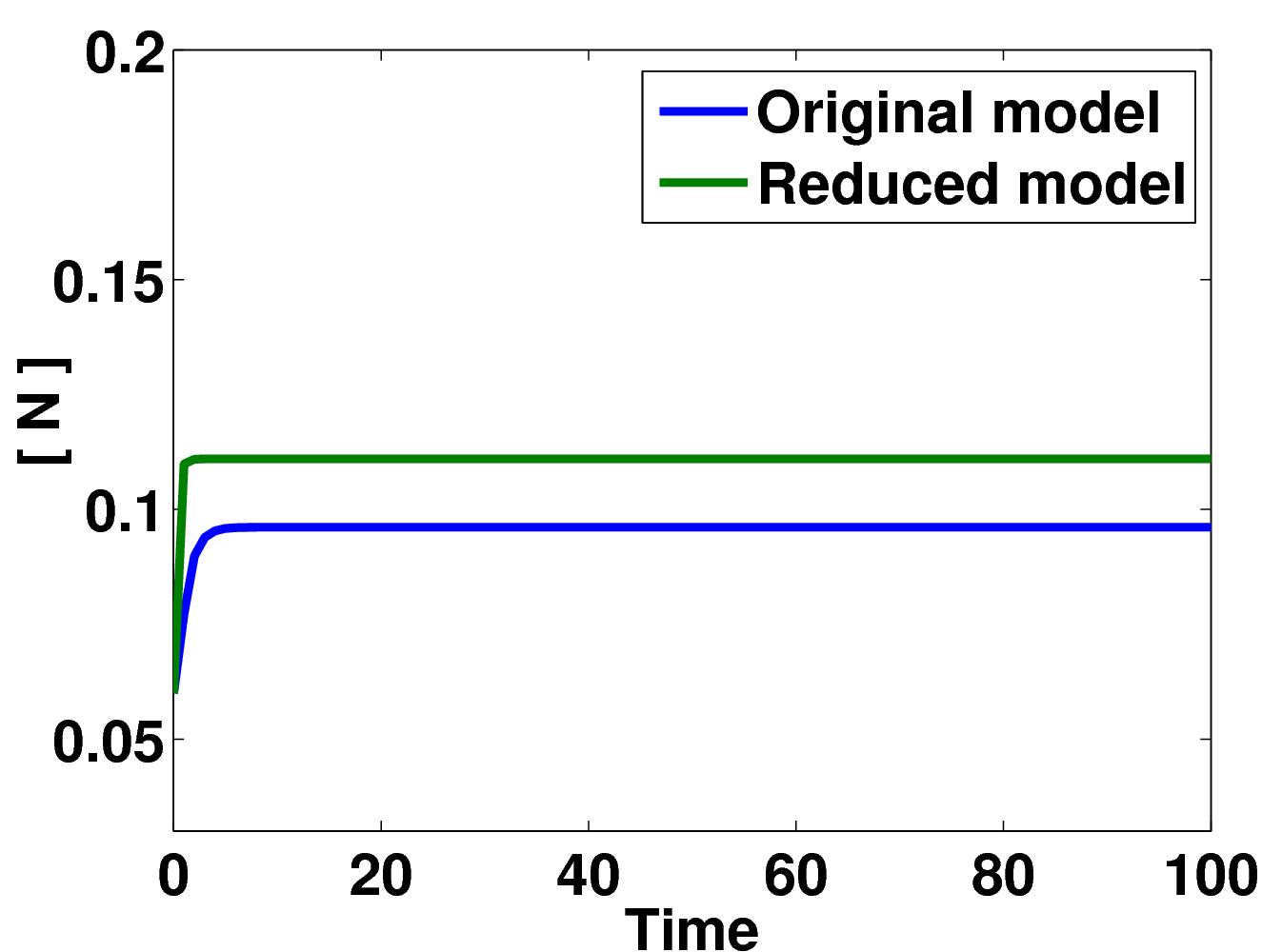}  
        }\\%
                    \subfigure{%
            \label{fig:third}
            \includegraphics[width=0.3\textwidth]{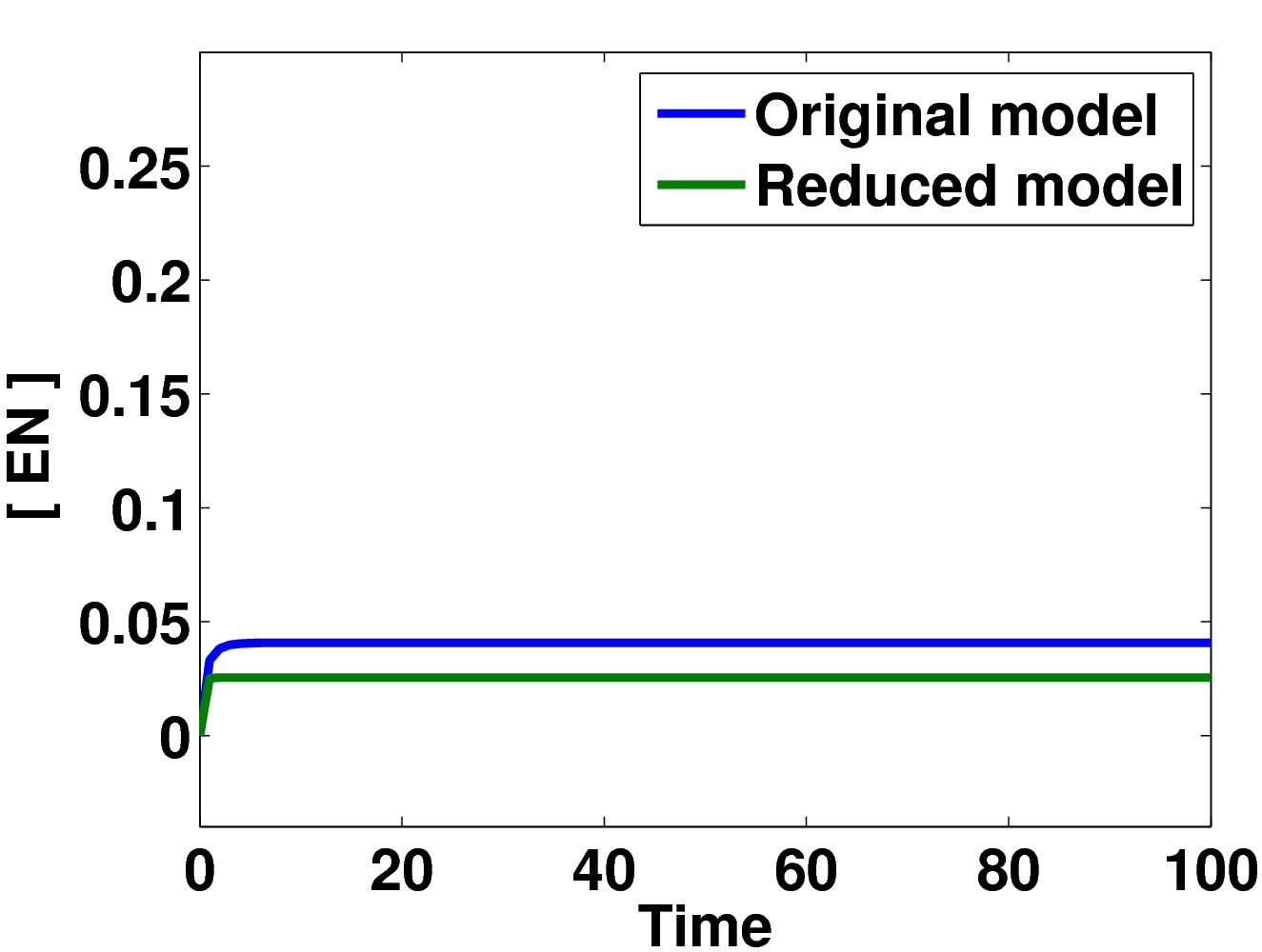}  
        }%
            \subfigure{%
            \label{fig:third}
            \includegraphics[width=0.3\textwidth]{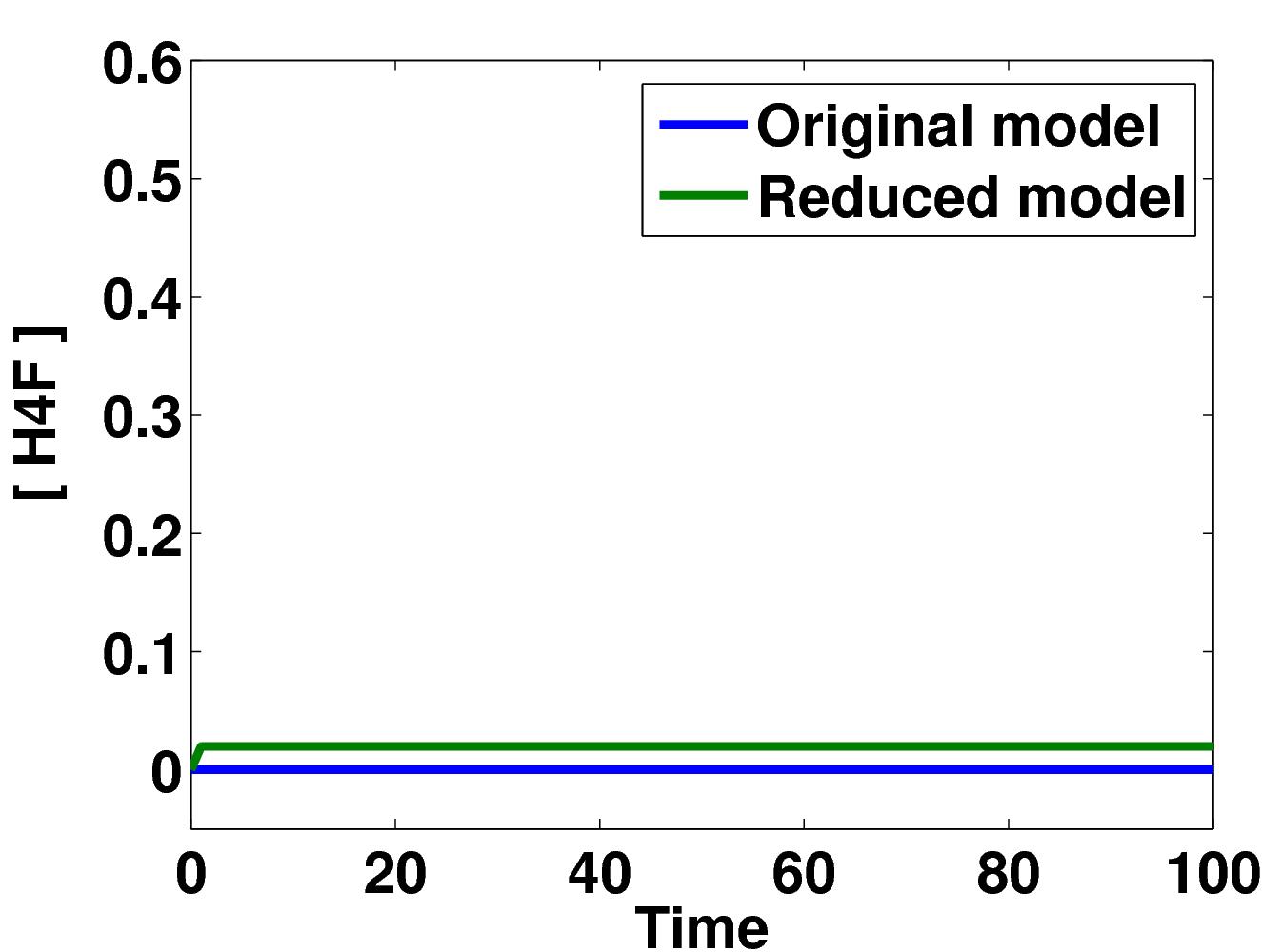}  
        }%
            \subfigure{%
            \label{fig:third}
            \includegraphics[width=0.3\textwidth]{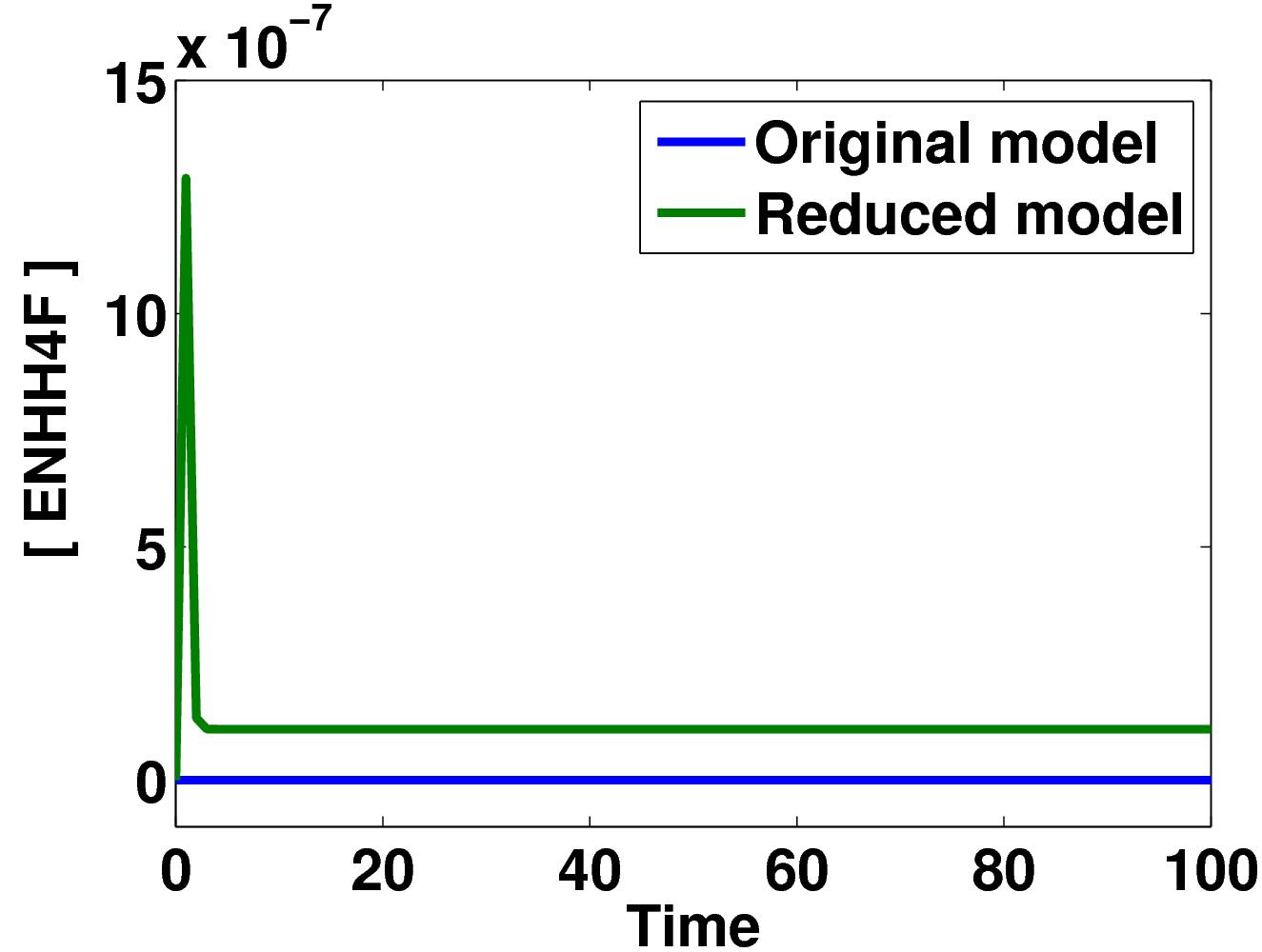}  
        }\\%
            \subfigure{%
            \label{fig:third}
            \includegraphics[width=0.3\textwidth]{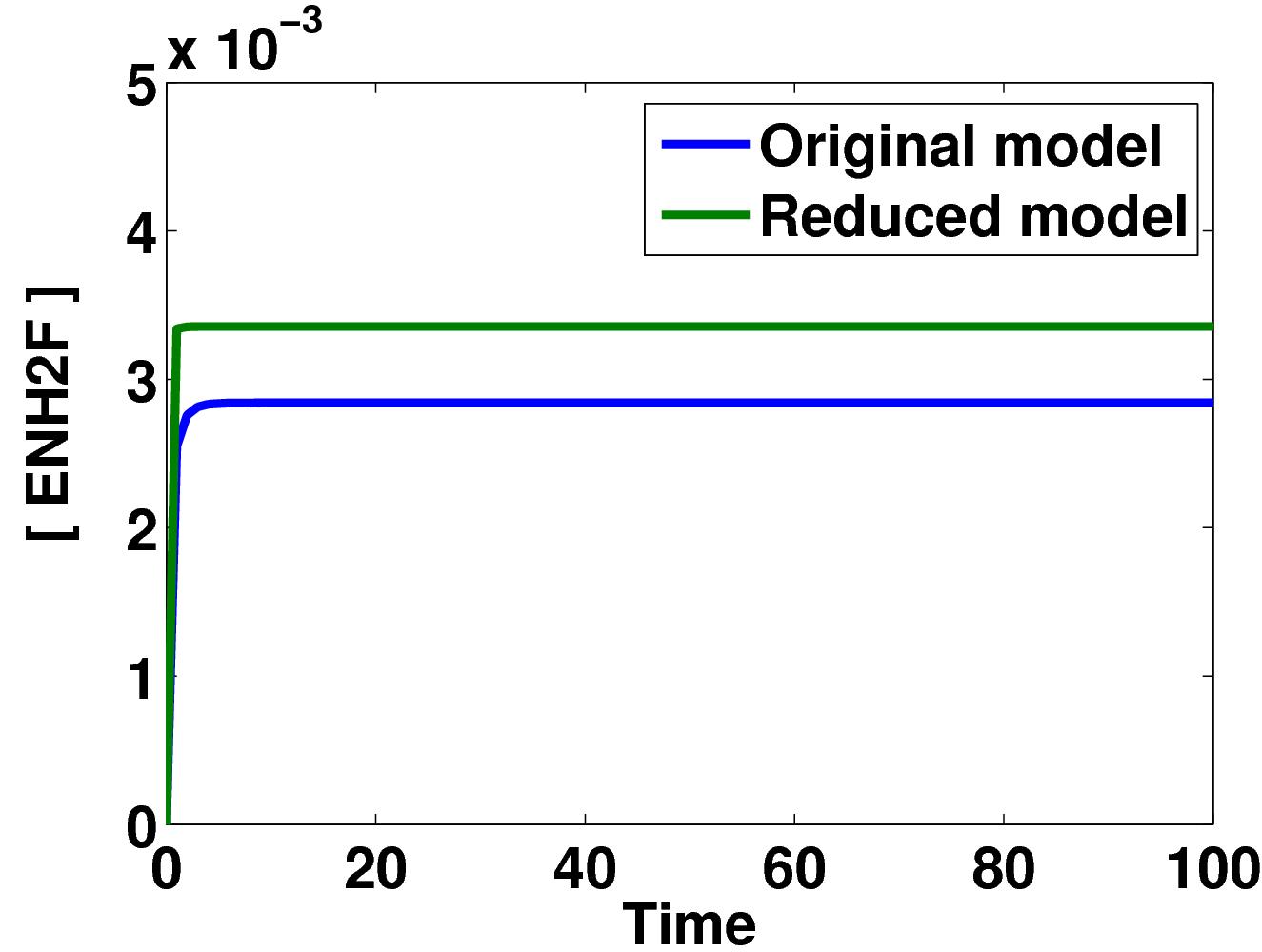}  
        }%
    \end{center}  
\caption {Numerical simulations of the full model \eqref{DHFR ODE system} and reduced model (\ref{Ent26}) of the dihydrofolate reductase (DHFR) pathways, with the time interval $[0,100]$ in computational simulations.}  
   \label{fig solution of DHFR}
\end{figure}

We applied the suggested approach to reduce the kinetic model of Dihydrofolate reductase(DHFR), as we studied before the Dihydrofolate reductase(DHFR) model contain 13 species with 26 parameters, since this model has 13 reversible reactions.\\
Then after applied our technique the model reduced to 11 and  16 species and parameters respectively. The difference between the full and reduced model is calculated at each stages see Table \eqref{table DHFR}. Calculating the value of deviation  is an important task in model reduction to check that the approximate solution is within allowable limits or not. We are calculating the full and reduced model approximate solutions see Figure \eqref{fig solution of DHFR}, the blue lines and green lines represent the approximate solutions of the original and reduced model which is computed by Matlab programming. The stage of model reduction here are based on eliminating some non important reactions and lumped of isolated species. According to the value of relative contribution of entropy production for each reactions, the first time reaction 10 is contributing least to the total entropy production; see Figure \eqref{DHRFdiagram1}. This reaction can be eliminated from the model. Then, in the second time reaction 11 is contributing least among the remaining reactions. Therefore, reaction 11 can be excluded from the model; and since the species [ENHH4F] disappear then we lumped [ENHH4F] with one of its neighbors so we  selected [ENH] for lumping, then $[ENH]^*=[ENH]+[ENHH4F]$. Similarly, we detect another three non important reactions (Reactions 9, 7, and 3) during the computational simulations, but when $v_{7}=0$ then [H4F] disappear so we lumped with [EN]; this is detected by using the proposed algorithm. And we stopped after eliminate reaction 3 because after that reaction 4 is contributing least among the remaining reactions, but we computed the value of deviation that is $13.71\%$, and it is large number for error so we stopped here.\\
After applying the relative contribution algorithm to the model to eliminate non--important reactions and lumping isolated species, the reduced model takes the form: 

\begin{equation}  
\begin{array}{llll}
\dfrac{d[E]}{dt}=-v_{1}-v_{2}+v_{8} , \quad  
\dfrac{d[H2F]}{dt}=-v_{1}+v_{5}+v_{13} , \\
\dfrac{d[EH2F]}{dt}=v_{1}-v_{12}, \quad
\dfrac{d[NH]}{dt}=-v_{2}, \\
\dfrac{d[ENH]}{dt}=v_{2}+v_{5}, \quad 
\dfrac{d[ENHH2F]}{dt}=-v_{4}-v_{5}, \\
\dfrac{d[ENH4F]}{dt}=v_{4}+v_{6}, \quad 
\dfrac{d[EH4F]}{dt}=-v_{6},\\  
\dfrac{d[N]}{dt}=-v_{6}+v_{8}-v_{12}, \quad
\dfrac{d[EN]}{dt}=-v_{8}+v_{13},\\
\dfrac{d[H4F]}{dt}=0,\quad 
\dfrac{d[ENHH4F]}{dt}=0,\\
\dfrac{d[ENH2F]}{dt}=v_{12}-v_{13},
\end{array}\label{DHFR ODE reduced system}
\end{equation} 
where $v_{1}=k_{1}^{f}[E](t)[H2F](t)-k_{1}^{b}[EH2F](t),$ 
$v_{2}=k_{2}^{f}[E](t)[NH](t)\\-k_{2}^{b}[ENH](t),$ \\
$v_{4}=k_{4}^{f}[ENHH2F](t)-k_{4}^{b}[ENH4F](t),$ 
$v_{5}=k_{5}^{f}[ENHH2F](t)\\ -k_{5}^{b}[ENH](t)[H2F](t),$ 
$v_{6}=k_{6}^{f}[EH4F](t)[N](t)-k_{6}^{b}[ENH4F](t),$ \\
$v_{8}=k_{8}^{f}[EN](t)\\-k_{8}^{b}[E](t)[N](t),$ 
$v_{12}=k_{12}^{f}[EH2F](t)[N](t)-k_{12}^{b}[ENH2F](t),$ \\
$v_{13}=k_{13}^{f}[ENH2F](t)-k_{13}^{b}[EN](t)[H2F](t).$ 

In Figure \eqref{fig solution of DHFR} there are a good agreement between the original and reduced model in computational simulation because the value of deviation (Total error) very small which is showed in Table \eqref{table DHFR}. This means that our proposed technique plays an important role in model reduction.\\
 
\FloatBarrier 
\begin{table}   
\center
    \caption{The table shows the difference between the reduced and full model of DHFR pathways at each reduction stage using the function of deviation and lumping of isolated species.} 
  
\begin{tabular}{|p{2.7cm}|p{3cm}|p{5.3cm}|p{3cm}|}
    \hline 
    Model reduction stages & Non-important reactions & Lumping of isolated species & Values of Deviation $\mathcal{F}^{\mathcal{D}}$ \\ 
    \hline 
    Stage One & Reaction ten & No lump & $1.23*10^{-11} \;\%$\\ 
    \hline 
    Stage Two & Reaction eleven &$[ENH]^{*}=[ENH]+[ENHH4F]$ & $1.01*10^{-3} \;\%$ \\ 
    \hline 
    Stage Three & Reaction nine & No lump &$9.98*10^{-4} \;\%$ \\ 
    \hline 
    Stage Four & Reaction seven & $[EN]^{*}=[EN]+[H4F]$ &$6.56*10^{-3} \;\%$ \\ 
    \hline
    Stage Five & Reaction three & No lump &$1.76*10^{-2} \;\%$ \\ 
    \hline 
    \end{tabular}     
\label{table DHFR}   
\end{table} 
\FloatBarrier 


\begin{figure}[H] 
     \begin{center}
        \subfigure{%
            \label{fig:third}  
            \includegraphics[width=0.8\textwidth]{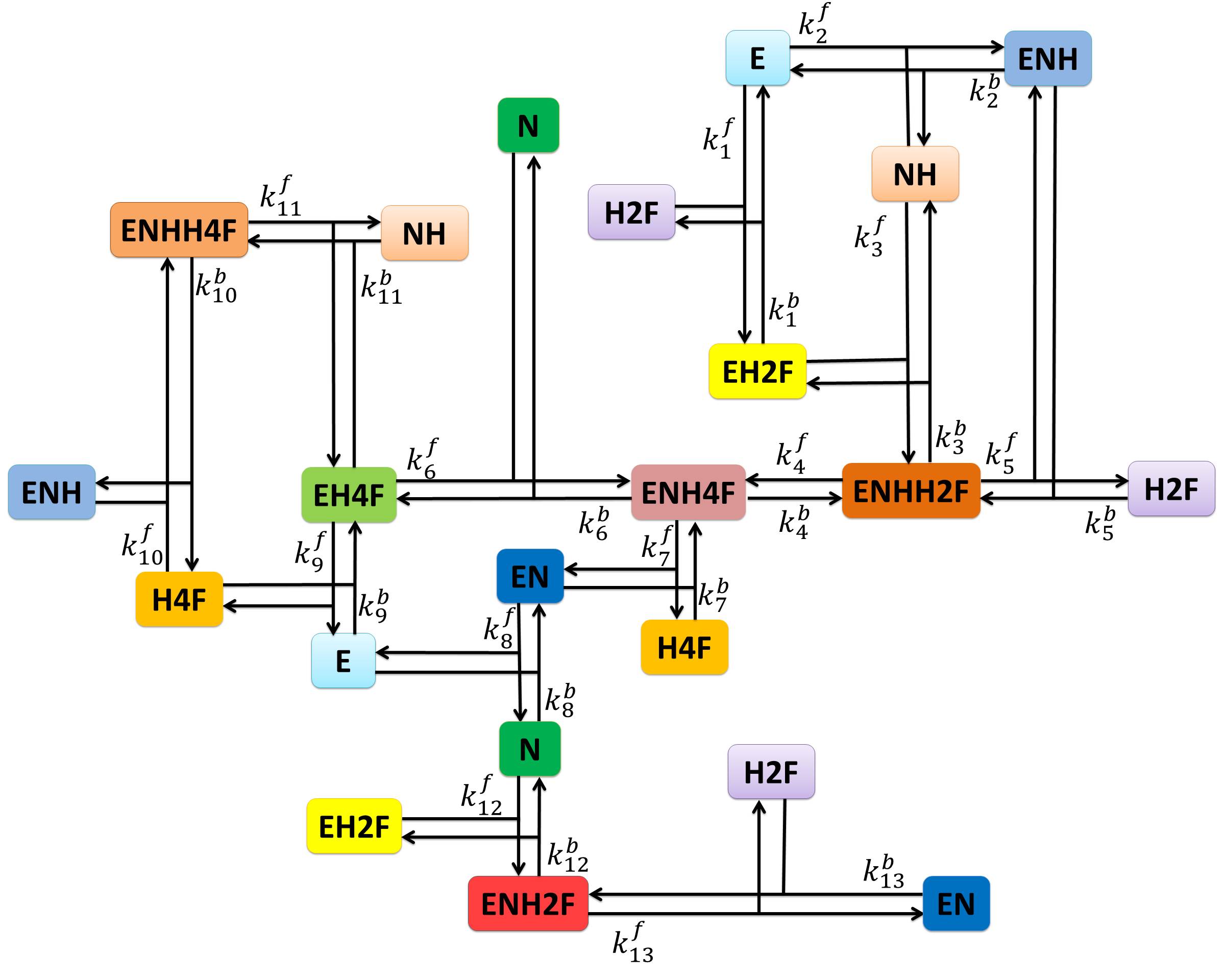}
        }
    \end{center}  
\caption {The original model of DHFR pathways.}  
   \label{DHRFdiagram} 
\end{figure}

\begin{figure}[H] 
     \begin{center}
        \subfigure{%
            \label{fig:third}  
            \includegraphics[width=0.8\textwidth]{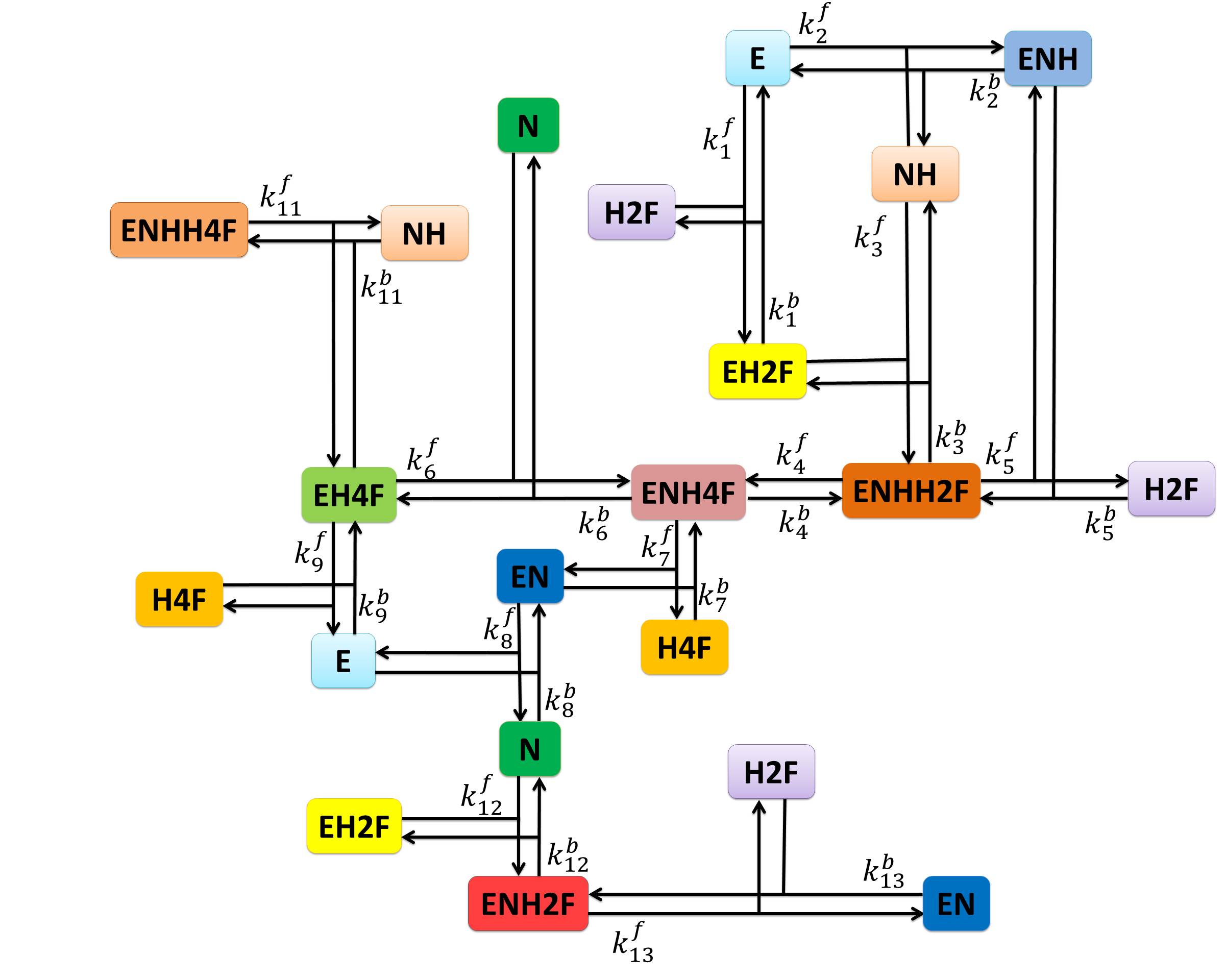}
        }
    \end{center}  
\caption {The structure of the model DHFR after eliminating reaction 10.}  
   \label{DHRFdiagram1} 
\end{figure}

\begin{figure}[H] 
     \begin{center}
        \subfigure{%
            \label{fig:third}  
            \includegraphics[width=0.8\textwidth]{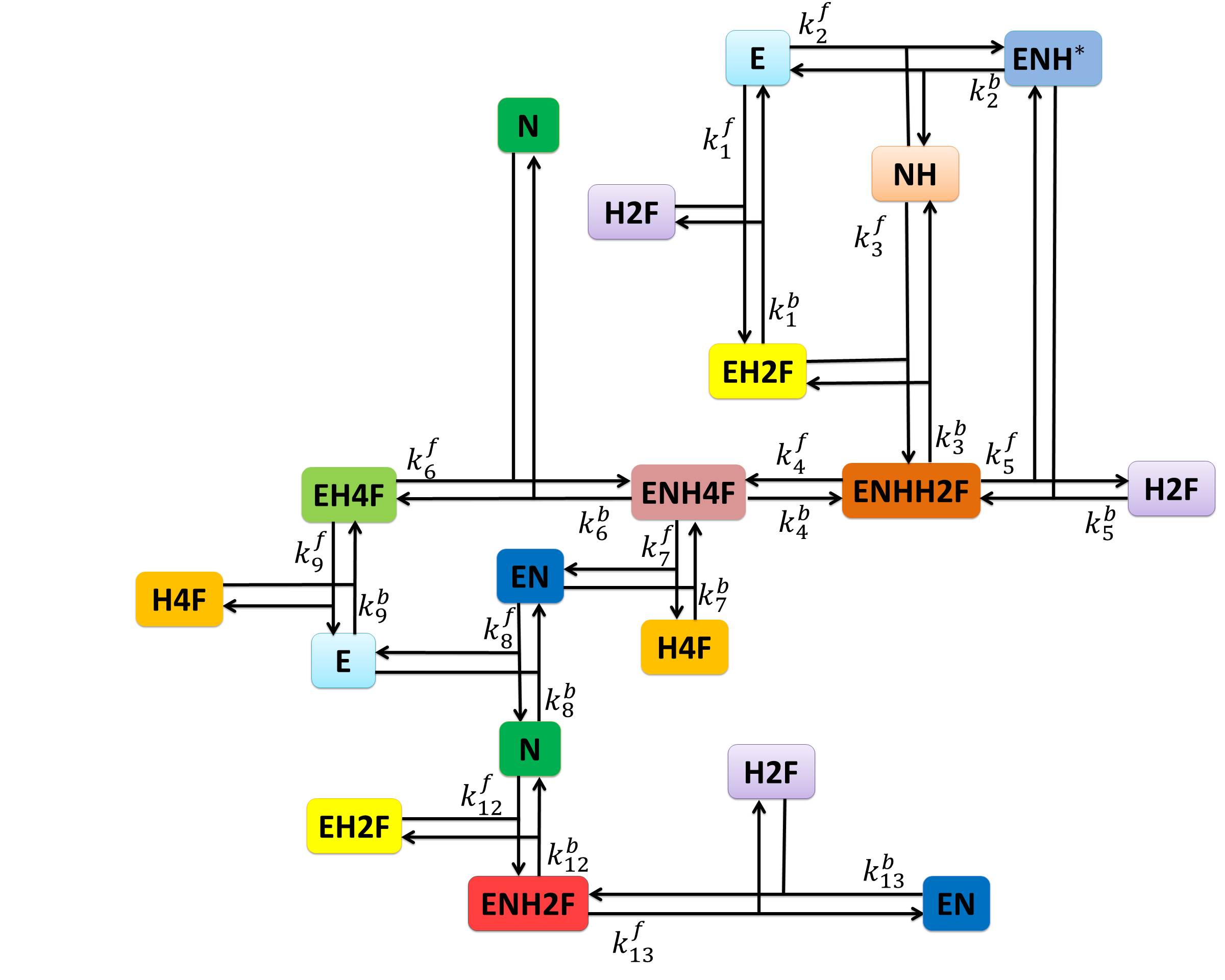}
        }
    \end{center}  
\caption {The structure of the model DHFR after eliminating reactions 10 and 11. }  
   \label{DHRFdiagram2} 
\end{figure}

\begin{figure}[H] 
     \begin{center}
        \subfigure{%
            \label{fig:third}  
            \includegraphics[width=0.8\textwidth]{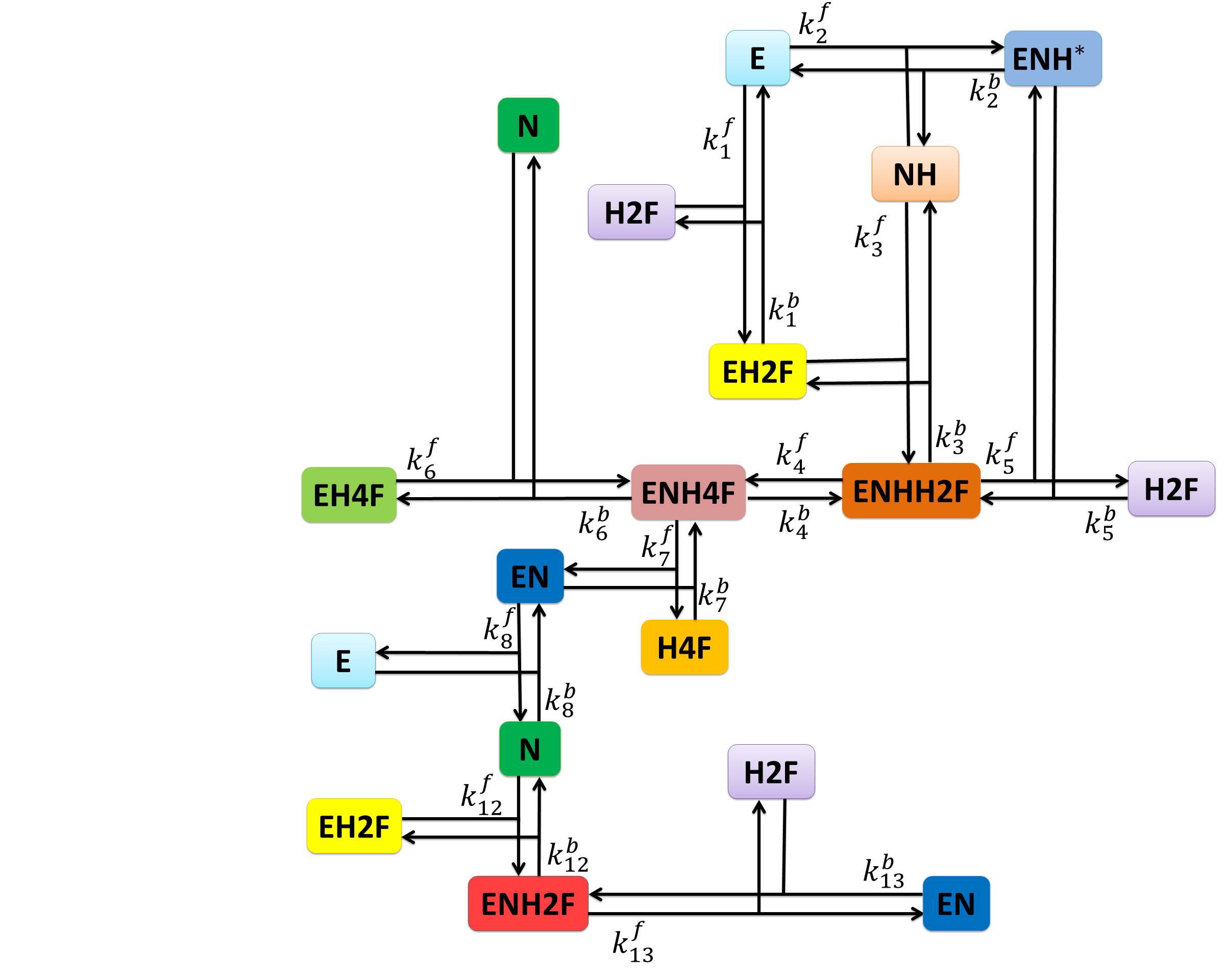}
        }
    \end{center}  
\caption {The structure of the model DHFR after eliminating reactions 9, 10, and 11. }  
   \label{DHRFdiagram3} 
\end{figure}

\begin{figure}[H] 
     \begin{center}
        \subfigure{%
            \label{fig:third}  
            \includegraphics[width=0.8\textwidth]{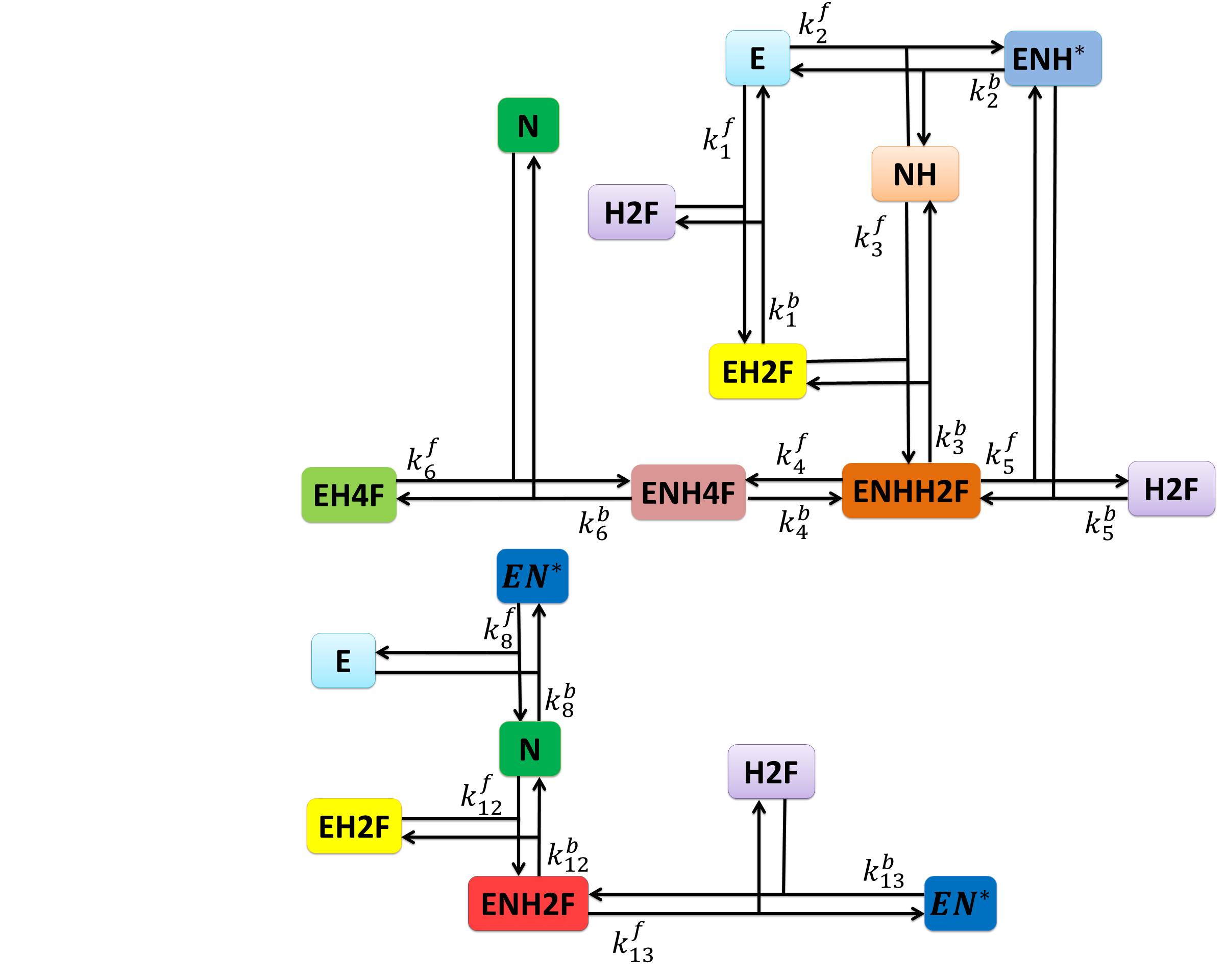}
        }
    \end{center}  
\caption {The structure of the model DHFR after eliminating reactions 7, 9, 10, and 11. }  
   \label{DHRFdiagram4} 
\end{figure}  

\begin{figure}[H] 
     \begin{center}
        \subfigure{%
            \label{fig:third}  
            \includegraphics[width=0.8\textwidth]{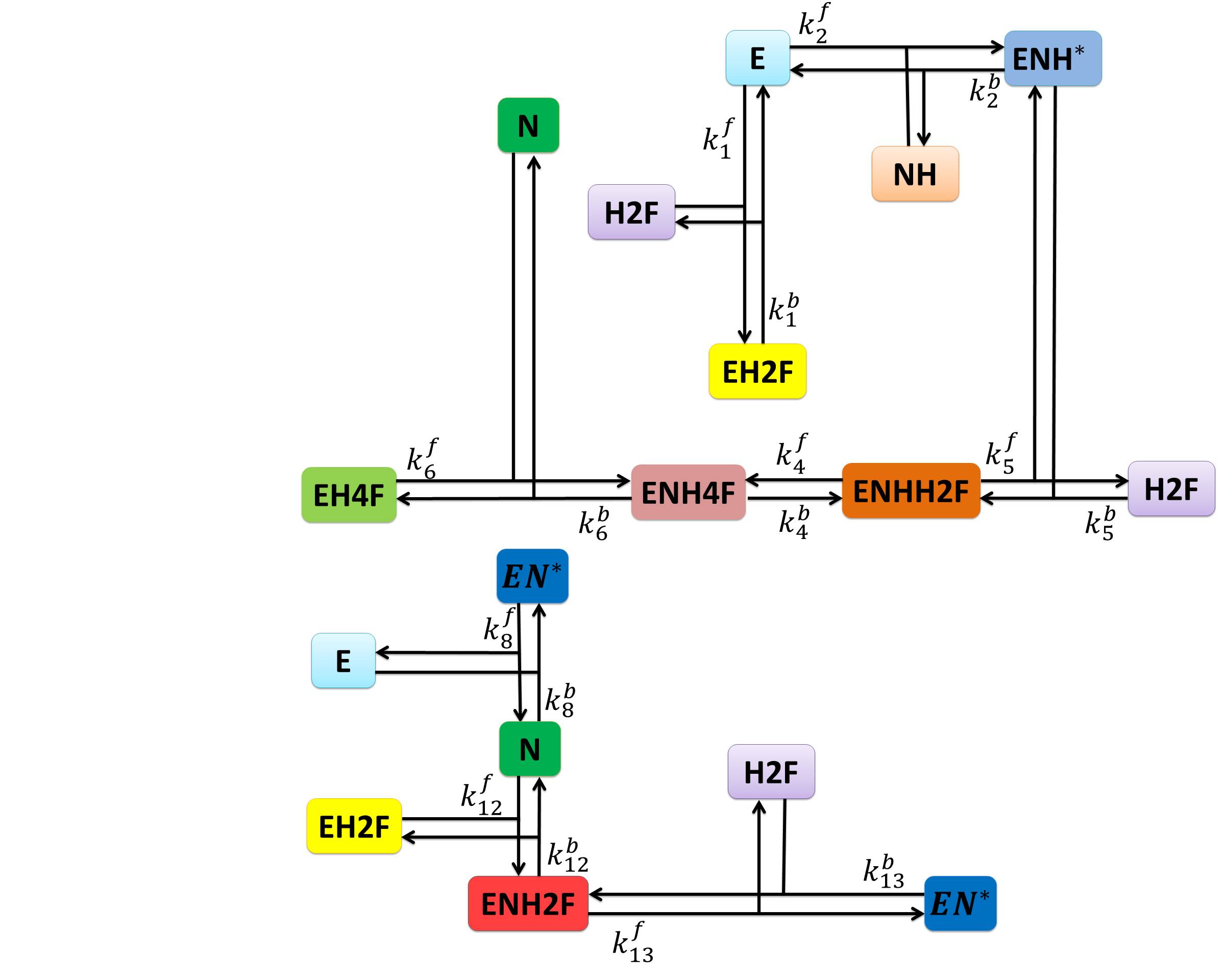}
        }
    \end{center}   
\caption {The structure of the model DHFR after eliminating reactions 3, 7, 9, 10, and 11. }  
   \label{DHRFdiagram5} 
\end{figure} 


\include{chapter[6]}
\chapter{Conclusions and Future Works}   
\pdfbookmark[0]{Conclusion and Future Works}{Conclusion and Future Works} 

\section{Conclusions}
Sometimes the process of changing the complex biochemical models to mathematical models (system of ordinary differential equations) give nonlinear complex models with high dimension of elements (variables and parameters). This becomes a hard task to understand the dynamic behaviors of variables and solving such models are also  difficult analytically. Therefore, we need techniques of model reduction to simplify complex models to smaller size and then we can simply study dynamic behaviors of such models. In addition, the reduced model solutions are close to the full model solutions.

In this thesis, we have worked on four techniques of model reduction. The first two methods are QSSA and QEA which are classical techniques and the most common techniques for biochemical reactions. The last two tools for model reductions are lumping of compartments and entropy production analysis. They are an important role in minimizing the number of elements for high dimension cell signalling pathways.   

In chapter two, we used QSSA on the model of non-linear Protein Translation Pathways. The model is nonlinear system of ordinary differential equations and it is difficult to find some analytical approximate solutions that is why the QSSA is required for simplifying  the model equations. We found some interesting results here. Firstly,  the model has three conservation lows and this is used for removing three equations. Secondly, we separated the model equations into slow and fast subsystems by defining a small parameter $\epsilon$, this is based on scaling of model variables. Then based on QSSA method when $\epsilon \rightarrow 0$, the fast subsystems approached to zero, we found fast variables depend on the slow variables. Furthermore, the slow manifolds are calculated and they are close to approximate solutions when the small parameter $\epsilon$ becomes smaller and smaller. Another important result here is that the approximate solutions of the full model and the reduced model are compared using SBedit Toolbox for Matlab for initial conditions. The approximate solutions of the reduced model and slow manifold are also computed in computational simulations, see Figures\eqref{fig:approximate solution with manifold of xy,xz}. We also calculated the analytical approximate solutions for all model variables using the slow-fast subsystems and conservation laws. Finally, we applied elasticity and control coefficient algorithms on the model, this shows that how variables, fluxes and species are depended on the parameters, and it also shows that how sensitive reactions are changed with reactants.\\

In chapter three, the QEA method has used as another tool for model reduction of reversible biochemical reaction models. This is mainly used here for simple and complex chemical enzymatic reactions. First of all, we applied the method on the simple model \eqref{chemical reaction 1}, which includes five variables, six parameters, three reversible reactions with two conservation lows. We supposed that the first reaction in this model becomes quasi equilibrium when the equilibrium is fast. By introducing new variables, the model equations are separated into slow and fast subsystems, then the system has been reduced to three variables and three parameters. After that the model equations were simplified and slow manifolds were calculated. The analytical approximate solutions calculated for the reduced model, this is based on the conservation laws of fast subsystems.\\    

Furthermore, we applied the QEA technique on a complex enzyme reaction network that includes $7$ variables, $10$ parameters with $3$ conservation lows. The model can not be solved analytically, therefore some approaches for model simplifications are required. In this situation, we supposed three cases of the model simplifications. 
In the first case, we supposed the first reaction becomes quasi equilibrium when the equilibrium is fast. Then, we simplified the model equations and calculated some analytical approximated solutions and slow manifolds. The second case for the model is that we supposed the first and third reaction become quasi equilibrium when the equilibriums are fast. Similarly, the reduced model was obtained that includes $5$ variables, $5$ parameters with two fast equations. Again, we computed the some approximate solutions for the model. In the last case, we supposed the first, third and fourth reactions become quasi equilibrium when the equilibriums are fast. Then, all previous procedures are applied in order to minimizing the number of elements and calculate approximate solutions. We concluded that for complex chemical reaction networks identifying fast reactions becomes an issue for model reduction and this is sometimes impossible because it may has many possibilities. Therefore, we proposed an algorithm for solving such problems. According to this algorithm, fast and slow reactions for complex chemical networks are easily separated. This helped us for identifying fast reactions first then applying the QEA technique.\\

In the next chapter, the powerful technique was introduced which is called lumping of compartments. The proposed approach has a great role in model reduction particularly for complex cell signalling pathways. In this study, we applied the method on three biochemical reaction models, the first two models are linear and nonlinear chemical chains. Their variables are reduced based on lumping some components. Results show that there are a good agreement between the original models and the simplified models, see Figures \eqref{fig:solution of linear simple example} and \eqref{fig:solution of nonlinear simple example}. The last model in this chapter is called ERK signalling pathways that consists of $11$ variables with $11$ parameters. Again, the lumping species here has a good step forward to minimizing the number of species. We have worked on $6$ different cases of model reduction on this model as shown in Table \eqref{table lumping of ERK}. It can be seen that the cases $5$ and $6$ are better than the other cases in terms of the number of elements and the value of deviation. Interesting, we found that case $5$ includes only $7$ variables and the value of deviation is only $9.87\;\% $. In addition, in case $6$ the remaining variables are $8$ and the total difference between the original and reduced model is only $4.54\;\% $. The approximate solutions of the original and reduced models for case $6$ illustrated in Figure \eqref{fig:solution of ERK}. It can be concluded that there are a good agreement of the dynamic behavior of variables for the original and reduced models.

At the end of this chapter, we developed the technique of lumping of species to lumping of parameters. We introduced the general formula for the proposed technique based on lumping of parameters. Firstly, we applied the method on a simple linear model, we found that the technique has a good role for reducing such models. Then, we worked in a complex model that is called NF-$\kappa$B signal transduction pathways. The model includes $29$ variables and $37$ parameters. We reduced the model in four different stages as shown in Table \eqref{table:lumping of NF-kB}. The model elements are reduced from $37$ to $8, 9, 13$ and $20$ for stages $1, 2, 3$ and $4$ respectively. The total difference between the original and reduced models are calculated using the function of deviation formula. This was given $2.55 * 10^{-11}\;\% $ and $2.98 * 10^{-8}\;\% $, and the approximate solutions of the full and reduced models  are shown in Figure \eqref{fig:solution of inflammatory}.\\

Finally, we worked in another powerful technique of model reduction. This is called entropy production analysis. We reviewed the method from the previous studies. Simply, the technique is based on neglecting non-important reactions that less contributed in total entropy production. Then, we developed the algorithm based on neglecting and lumping ideas. According to our improvement for the technique, at the stage that when the non important reactions are disappeared, we lumped such reactions with their neighbors. The developed method was applied on some cell signalling pathways in systems biology. Firstly, we applied on the elongation factors EF-Tu and EF-Ts signalling pathways. The model includes $9$ variables with $14$ parameters, the model reduced to $6$ and $7$ variables and parameters respectively. The model reduction stages are shown in Figures (\ref{TuTs1}--\ref{TuTs5}). The total errors between the original models and the simplified models are shown in Table \eqref{tableENT1}, and the numerical simulations are shown in Figure \eqref{fig:elognation}. The results showed that our developed algorithm is much better compared with previous algorithm (khoshnaw, 2015), because the value of deviation in (khoshnaw, 2015) between $0.53\;\%$ and $9.14\;\%$ while in our study is only between $0.164\;\%$ and $3.676\;\%$.
 We have also applied the suggested algorithm on a complex model of Dihydrofolate Reductase (DHFR) pathways. The model includes $13$ variables and $26$ parameters, after applied our developed technique the model reduced to $11$ and $16$ variables and parameters respectively. We reduced the model in $5$ stages and the total differences between the full and reduced model are between $1.23 * 10^{-11}\;\%$ and $1.76 * 10^{-2}\;\%$ as explained in Table \eqref{table DHFR}. Calculating the total error (difference) between the original and reduced models is an important task to check the approximate solutions is within allowable limits or not. Results are computed using Matlab programming as shown in Figure \eqref{fig solution of DHFR}. 

\newpage
\section{Future Works}  

The techniques of model reduction that have been studied here are great tools for minimizing the number of elements in systems biology. They give us a good step forward to understand the model dynamics and calculate some approximate solutions. Therefore, such techniques can be further studied and improved. The following  suggestions are recommended for future studies:
\begin{enumerate}
\item The QSSA method can be used on more complex models of chemical reactions in order to separate their equations into slow and fast subsystems.
\item The QEA method can be also used to reduce the number of elements of dihydrofolate reductase (DHFR) pathways model.
\item Our suggested technique (Lumping of Parameters) can be applied on further complex cell signalling pathways to reduce the number of parameters.
\item The entropy production analysis with lumping of isolated species  can also be applied on some other reversible biochemical reaction networks. 
\end{enumerate}
\newpage 
\pdfbookmark[0]{References}{References}
\include{References}      
mmm


\newpage  
\maketitle  
\pdfbookmark[0]{Appendix}{appendix} 
\begin{appendices} 
\chapter{Dynamic System Simulations Using Systems Biology Toolbox (SBToolbox) for Matlab}

\section{\large{SBToolbox file for Matlab containing the competitive enzymatic reactions}}
\label{A.3}
\footnotesize{********** MODEL NAME\\
The dimensionless form of competitive enzymatic reactions and its reduced model\\
********** MODEL NOTES \\
w1=w01=w001=w0001; w2=w02=w002=w0002; b1=$\beta_{1}$; b2=$\beta_{2}$; b3=$\beta_{3}$; b4=$\beta_{4}$; b5=$\beta_{5}$; e=e1=e2=e3=$\epsilon$;\\
u1r and u2r are concentrations of the reduced model \\ 
********** MODEL STATES\\
d/dt(u1)=-u1*(1-w1-w2)+b1*w1\\
d/dt(u2)=-b2*u2*(1-w1-w2)+b3*w2\\
d/dt(u1r)=(-(b3*b4+b4*b6)*u1r-(b3+b6)*(u1r)$^2$)/((b1*b3+b1*b6+b3*b4+b4*b6)+(b3+b6)*u1r\\
+(b1*b2*b5+b2*b4*b5)*u2r)\\
d/dt(u2r)=(-b2*(b1*b3+b1*b6+b3*b4+b4*b6)*u2r+b3*(b1*b2*b5+b2*b4*b5)*u2r)\\
/((b1*b3+b1*b6+b3*b4+b4*b6)+(b3+b6)*u1r+(b1*b2*b5+b2*b4*b5)*u2r)\\
d/dt(w1)=1/e * (u1*(1-w1-w2)-(b1+b4)*w1) \\
d/dt(w01)=1/e1 * (u1*(1-w01-w02)-(b1+b4)*w01) \\    
d/dt(w001)=1/e2 * (u1*(1-w001-w002)-(b1+b4)*w001) \\
d/dt(w0001)=1/e3 * (u1*(1-w0001-w0002)-(b1+b4)*w0001) \\
d/dt(w2)=1/e * (b2*b5*u2*(1-w1-w2)-(b3+b6)*w2)\\   
d/dt(w02)=1/e1 * (b2*b5*u2*(1-w01-w02)-(b3+b6)*w02)\\    
d/dt(w002)=1/e2 * (b2*b5*u2*(1-w001-w002)-(b3+b6)*w002)\\    
d/dt(w0002)=1/e3 * (b2*b5*u2*(1-w0001-w0002)-(b3+b6)*w0002)\\                                                                                                                                                                       
u1(0)=1\\
u2(0)=1\\
u1r(0)=1\\
u2r(0)=1\\
w1(0)=0\\
w01(0)=0\\
w001(0)=0\\
w0001(0)=0\\
w2(0)=0\\
w02(0)=0\\
w002(0)=0\\
w0002(0)=0\\
********** MODEL PARAMETERS\\
b1=1.3\\
b2=1.2\\
b3=0.9\\
b4=1.3\\
b5=1.1\\
b6=1.8\\
e=0.5\\
e1=0.2 \\ 
e2=0.08   \\
e3=0.006\\
********** MODEL VARIABLES\\
M1=(b3+b6)*u1/((b1*b3+b1*b6+b3*b4+b4*b6)+(b3+b6)*u1+(b1*b2*b5+b2*b4*b5)*u2)\\
M2=(b1*b2*b5+b2*b4*b5)*u2/((b1*b3+b1*b6+b3*b4+b4*b6)+(b3+b6)*u1+(b1*b2*b5+b2*b4*b5)*u2)\\
********** MODEL REACTIONS\\
********** MODEL FUNCTIONS\\ 
********** MODEL EVENTS\\
********** MODEL MATLAB FUNCTIONS}\\

\section{\large{SBToolbox file for Matlab containing the iterative equations of simple enzymatic reactions}}
\label{A.4}
\footnotesize{********** MODEL NAME\\
Iterative model of simple enzymatic reactions\\
********** MODEL NOTES\\
e=$\epsilon$; a=$\alpha$; b=$\beta$; \\ 
u1, u2, u3 and u4 are iterations of u; v1, v2, v3 and v4 are iterations of v\\
********** MODEL STATES \\
d/dt(u)=-u+a*v+u*v \\
d/dt(v)=1/e *(u-b*v +u*v) \\
d/dt(u1)=-u1+a*v1 \\
d/dt(v1)=1/e*(u1-b*v1)  \\
d/dt(u2)=-u2+a*v2+u1*v1 \\
d/dt(v2)=1/e*(u2-b*v2 +u1*v1) \\
d/dt(u3)=-u3+a*v3+u2*v2 \\
d/dt(v3)=1/e*(u3-b*v3 +u2*v2) \\
d/dt(u4)=-u4+a*v4+u3*v3\\
d/dt(v4)=1/e*(u4-b*v4 +u3*v3)\\
u(0)=1\\
v(0)=0\\
u1(0)=1\\
v1(0)=0\\
u2(0)=1\\
v2(0)=0\\
u3(0)=1\\
v3(0)=0\\
u4(0)=1\\
v4(0)=0\\
********** MODEL PARAMETERS\\
e=0.5\\
a=1.11\\
b=2.88\\   
********** MODEL VARIABLES\\
$D1=sqrt((u1-u2)^2+(v1-v2)^2)$\\
$D2=sqrt((u2-u3)^2+(v2-v3)^2)$\\
$D3=sqrt((u3-u4)^2+(v3-v4)^2)$\\
$D11=sqrt((u1-u)^2+(v1-v)^2)$\\
$D22=sqrt((u2-u)^2+(v2-v)^2)$\\   
$D33=sqrt((u3-u)^2+(v3-v)^2)$\\
$D44=sqrt((u4-u)^2+(v4-v)^2)$\\
********** MODEL REACTIONS\\
********** MODEL FUNCTIONS\\
********** MODEL EVENTS\\
********** MODEL MATLAB FUNCTIONS}\\ 

\end{appendices}

\begin{thebibliography}{1} 

\bibitem{Tikhonov1952}  
A. N. Tikhonov, \textit{Systems of differential equations containing a small parameter multiplying the derivative}, Mat. Sb., 31 \textbf{(1952)} 575--586. (In Russian)
\bibitem{Fenichel} N. Fenichel, \textit{Geometric singular perturbation theory for ordinary differential equations}, Journal of differential equations, 31 \textbf{(1979)} 53--98.

\bibitem{Tikhonov1952}  
A. N. Tikhonov, \textit{Systems of differential equations containing a small parameter multiplying the derivative}, Mat. Sb., 31 \textbf{(1952)} 575--586. (In Russian)
\bibitem{Fenichel} N. Fenichel, \textit{Geometric singular perturbation theory for ordinary differential equations}, Journal of differential equations, 31 \textbf{(1979)} 53--98.

\bibitem{Tikhonov1952}  
A. N. Tikhonov, \textit{Systems of differential equations containing a small parameter multiplying the derivative}, Mat. Sb., 31 \textbf{(1952)} 575--586. (In Russian)
\bibitem{Fenichel} N. Fenichel, \textit{Geometric singular perturbation theory for ordinary differential equations}, Journal of differential equations, 31 \textbf{(1979)} 53--98.

\bibitem{Tikhonov1952}  
A. N. Tikhonov, \textit{Systems of differential equations containing a small parameter multiplying the derivative}, Mat. Sb., 31 \textbf{(1952)} 575--586. (In Russian)
\bibitem{Fenichel} N. Fenichel, \textit{Geometric singular perturbation theory for ordinary differential equations}, Journal of differential equations, 31 \textbf{(1979)} 53--98.

   
\bibitem{Tikhonov1952}  
A. N. Tikhonov, \textit{Systems of differential equations containing a small parameter multiplying the derivative}, Mat. Sb., 31 \textbf{(1952)} 575--586. (In Russian)
\bibitem{Fenichel} N. Fenichel, \textit{Geometric singular perturbation theory for ordinary differential equations}, Journal of differential equations, 31 \textbf{(1979)} 53--98.

\bibitem{Tikhonov1952}  
A. N. Tikhonov, \textit{Systems of differential equations containing a small parameter multiplying the derivative}, Mat. Sb., 31 \textbf{(1952)} 575--586. (In Russian)
\bibitem{Fenichel} N. Fenichel, \textit{Geometric singular perturbation theory for ordinary differential equations}, Journal of differential equations, 31 \textbf{(1979)} 53--98.

\bibitem{Tikhonov1952}  
A. N. Tikhonov, \textit{Systems of differential equations containing a small parameter multiplying the derivative}, Mat. Sb., 31 \textbf{(1952)} 575--586. (In Russian)
\bibitem{Fenichel} N. Fenichel, \textit{Geometric singular perturbation theory for ordinary differential equations}, Journal of differential equations, 31 \textbf{(1979)} 53--98.

\bibitem{Tikhonov1952}  
A. N. Tikhonov, \textit{Systems of differential equations containing a small parameter multiplying the derivative}, Mat. Sb., 31 \textbf{(1952)} 575--586. (In Russian)
\bibitem{Fenichel} N. Fenichel, \textit{Geometric singular perturbation theory for ordinary differential equations}, Journal of differential equations, 31 \textbf{(1979)} 53--98.

\end{thebibliography}
\end{document}